\documentclass[aps,prb,amsmath,amssymb,
floatfix,nofootinbib,eqsecnum,
reprint%,draft
]{revtex4-2}

\usepackage{graphicx}% Include figure files
\usepackage{dcolumn}% Align table columns on decimal point
\usepackage{bm}% bold math
\usepackage{hyperref}% add hypertext capabilities
\usepackage{braket}
\usepackage{xcolor}
\usepackage[caption=false]{subfig}
\usepackage{comment}
\usepackage{empheq}

\hypersetup{breaklinks=true}

\usepackage{pst-all}	%call the pstricks package
\usepackage{bbm}
\usepackage{revsymb}
\usepackage{yfonts}
\usepackage{tensor} % for pre subscript
\usepackage{tikz}

\usetikzlibrary{external}
\makeatletter
\@ifclasswith{revtex4-2}{draft}{\PSTricksOff}{}
\makeatother

\renewcommand{\Tr}{\mathrm{Tr}}
\newcommand{\qTr}{\widetilde{\mathrm{Tr}}}
\newcommand{\trho}{\widetilde{\rho}}
\newcommand{\wt}{\widetilde}
\newcommand{\kket}[1]{\ket{#1}\rangle}
\newcommand{\bbra}[1]{\langle\bra{#1}}

\newcommand{\cH}{\mathcal{H}}
\newcommand{\cD}{\mathcal{D}}

\newcommand{\hex}{\psline(0,.5)(-.3,.2)(-.3,-.2)(0,-0.5)(.3,-.2)(.3,.2)(0,.5)\psline[ArrowInside=->](-.3,-.2)(-.3,.2)\rput(0,0){$\odot$}}
\newcommand{\hexbig}{\psline(0,.7)(-.5,.4)(-.5,-.4)(0,-0.7)(.5,-.4)(.5,.4)(0,.7)\psline[ArrowInside=->](-.5,-.4)(-.5,.4)\rput(0,0){$\odot$}}

\begin{document}

%\preprint{APS/123-QED}

\title{Entanglement in tripartitions of topological orders: a diagrammatic approach}

\author{Ramanjit Sohal}
\author{Shinsei Ryu}
\affiliation{ Department of Physics, Princeton University, Princeton, New Jersey, 08544, USA}

\begin{abstract}
Recent studies have demonstrated that measures of tripartite entanglement can probe data characterizing topologically ordered phases to which bipartite entanglement is insensitive. Motivated by these observations, we compute the reflected entropy and logarithmic negativity, a mixed state entanglement measure, in tripartitions of bosonic topological orders using the anyon diagrammatic formalism. We consider tripartitions in which three subregions meet at trijunctions and tetrajunctions. In the former case, we find a contribution to the negativity which distinguishes between Abelian and non-Abelian order while in the latter, we find a distinct universal contribution to the reflected entropy.
Finally, we demonstrate that the negativity and reflected entropy are sensitive to the $F$-symbols for configurations in which we insert an anyon trimer, for which the Markov gap, defined as the difference between the reflected entropy and mutual information, is also found to be non-vanishing.
\end{abstract}

\date{\today}

\maketitle

\tableofcontents

\section{Introduction \label{sec:introduction}}

Tools from quantum information have become a mainstay in probing the properties of quantum phases of matter. Indeed, understanding the patterns of entanglement of a many-body system allows for detecting phases which are otherwise imperceptible to local probes, prime examples of which are provided by topologically ordered systems. In this endeavor, a central role has been played by the entanglement entropy. Given the ground state of a two-dimensional gapped Hamiltonian, $\ket{\psi}$, and a bipartition of space into a region $A$ and its complement, one may form the reduced density matrix $\rho_A = \mathrm{Tr}[\rho]$, where $\rho = \ket{\psi}\bra{\psi}$,
and compute the entanglement entropy as $S_A = -\mathrm{Tr}(\rho_A \ln \rho_A)$, which obeys an area law
\begin{align}
	S_A = \alpha L - \gamma + \dots \, ,
\end{align}
where $\alpha$ is a non-universal constant, $L$ the length of the entanglement cut, and $\gamma = \ln \mathcal{D}$ the topological entanglement entropy (TEE) \cite{Kitaev2006,Levin2006}, a universal quantity which is non-zero in topologically ordered systems and where $\mathcal{D}$ is the total quantum dimension, a number characterizing the anyon content of the phase.

\begin{figure}
  \centering
\subfloat[]{%
    \includegraphics[width=0.45\linewidth]{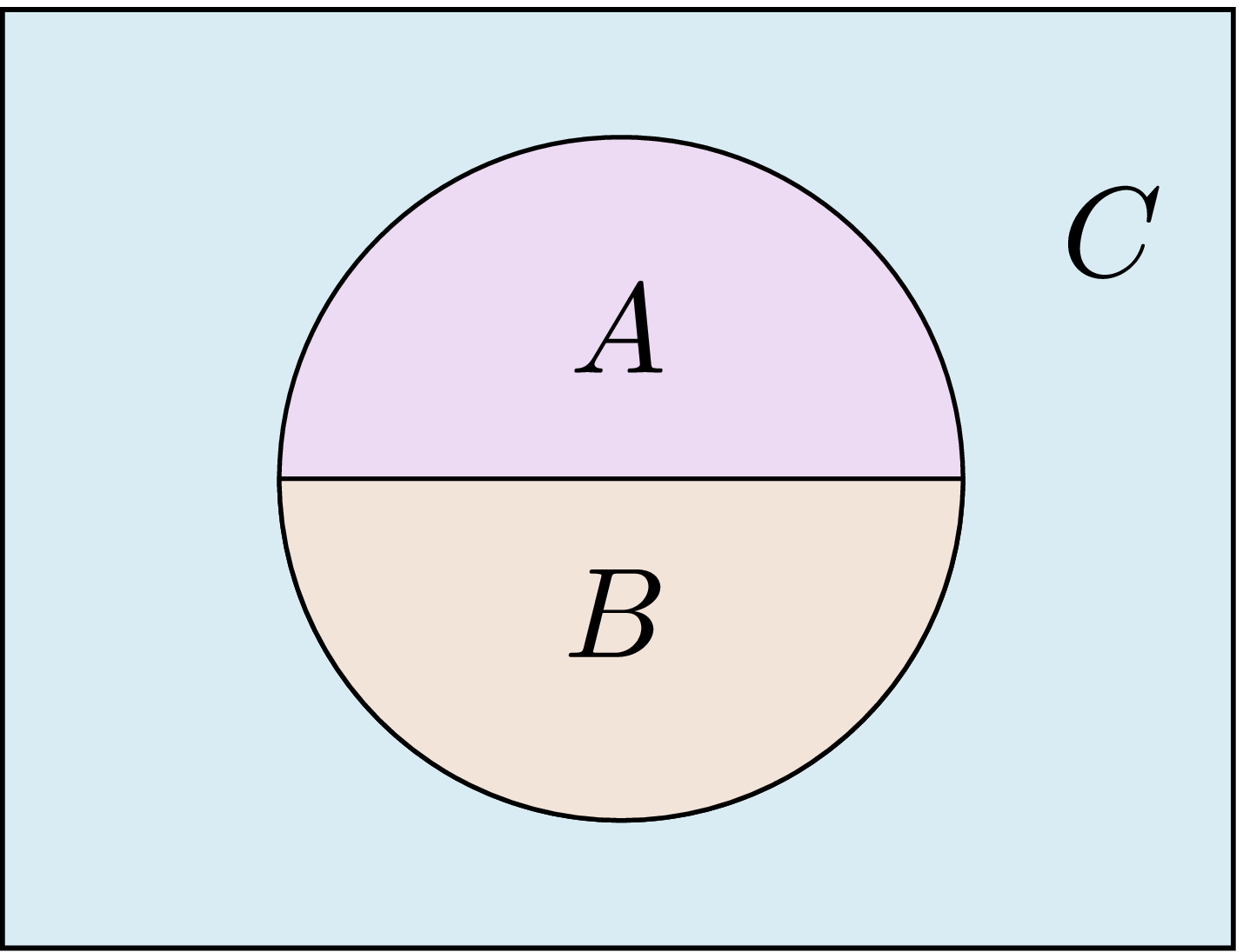}
}
\subfloat[]{%
    \includegraphics[width=0.45\linewidth]{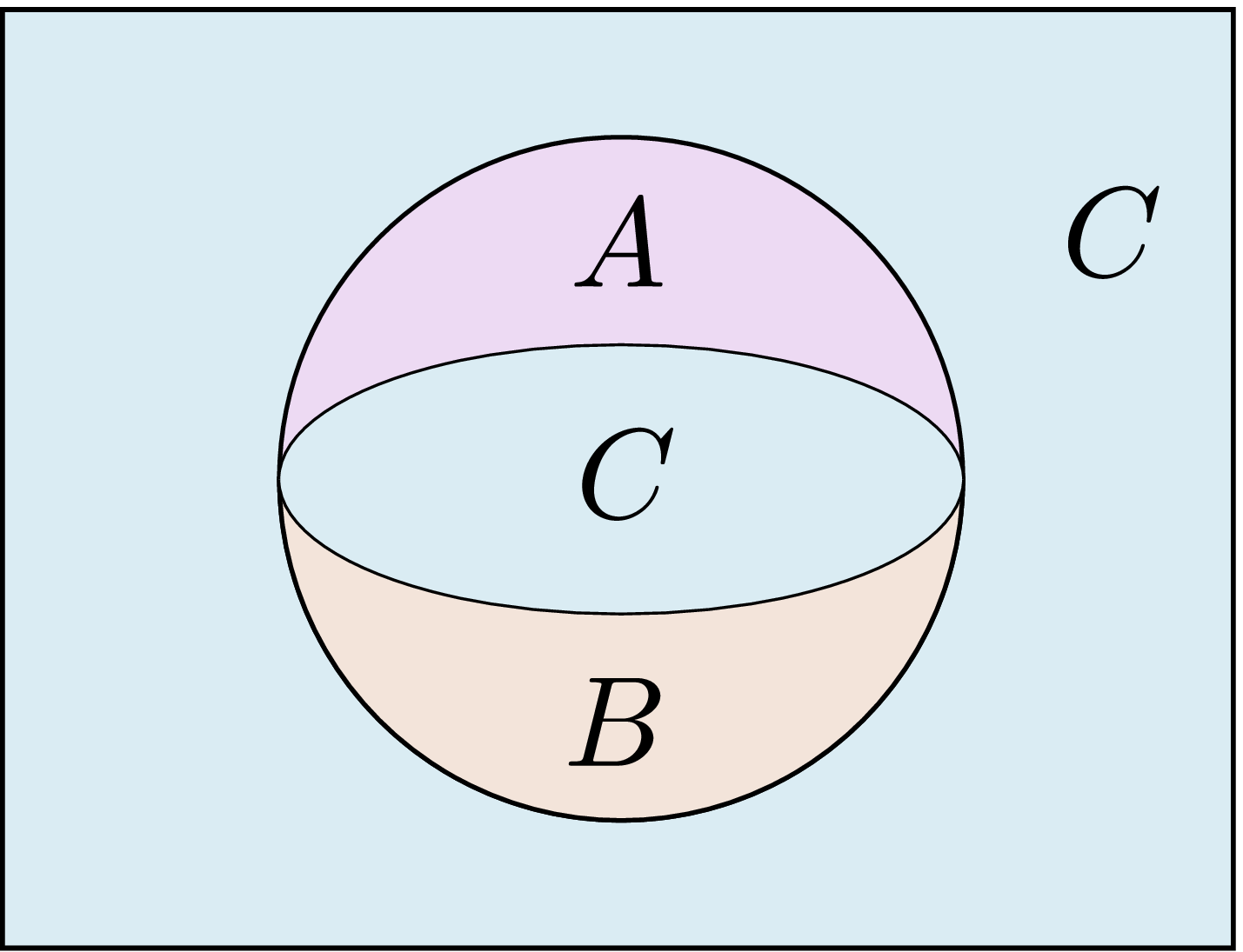}
}

\subfloat[]{%
    \includegraphics[width=0.45\linewidth]{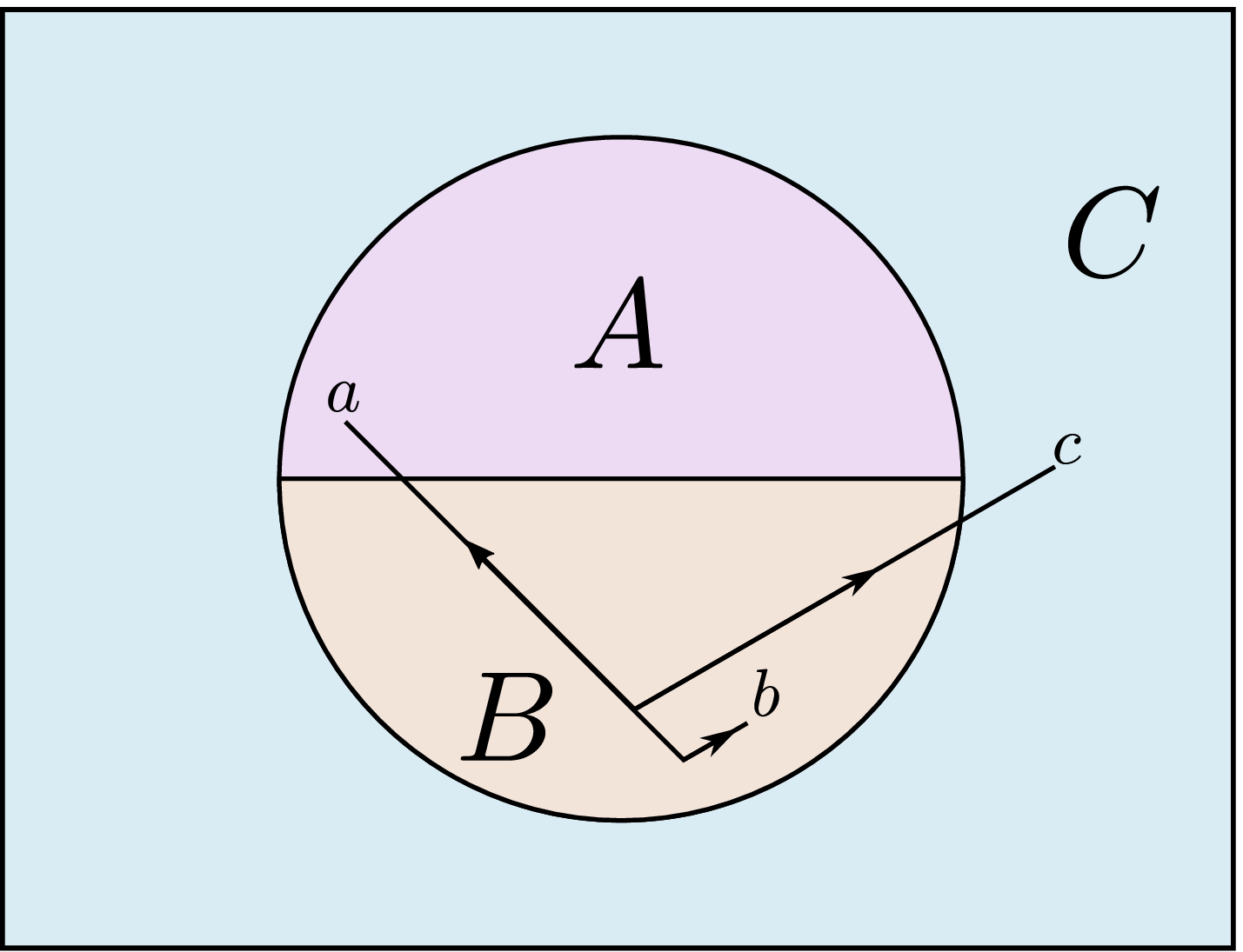}
}
    \caption{(a) Tripartition in which regions $A$, $B$, and $C$ meet at two trijunctions. (b) Tripartition in which regions $A$, $B$, and $C$ meet at two tetrajunctions. (c) Same tripartion as (a) in which we have inserted a trimer of anyons, $a$, $b$, and $c$.} \label{fig:partitions}
\end{figure}

While the entanglement entropy thus serves as a probe of topological order, it is limited in that it is only a good entanglement measure for \emph{pure} states and %, loosely speaking, is 
%only a good measure 
of bipartite entanglement. Moreover, the TEE is incapable of distinguishing Abelian from non-Abelian order, by virtue of the fact that multiple topological orders can have the same total quantum dimension -- for instance, both the toric code and Ising order have $\mathcal{D}=2$. 
It is thus natural to ask whether other entanglement measures, capable of detecting multipartite entanglement, can be computed and extract additional data defining a topological order.
Indeed, recent studies have shown that entanglement patterns of tripartitions of topological phases, like those in Fig. \ref{fig:partitions}, can do precisely this. 
Two key quantities studied in this context are the entanglement negativity \cite{Peres1996,Zyczkowski1998,Eisert1999,Vidal2002,Plenio2005} and reflected entropy \cite{Dutta2021}.

While we defer a detailed definition, the negativity provides a good measure of bipartite entanglement in a mixed state. 
In particular, the negativity has been studied extensively in topologically ordered phases of matter \cite{Lee2013,Castelnovo2013,Wen2016,Wen2016b,Lu2020,Lim2021,Lu2022}, previously finding use, for instance, in charactering topological order at finite temperature \cite{Lu2020} and distinguishing Abelian and non-Abelian order \cite{Wen2016}. The reflected entropy, originally introduced in the context of holography, has served as a probe of tripartite entanglement \cite{Akers2020,Hayden2021}. Indeed, the Markov gap, defined as the difference between the reflected entropy and mutual information has been shown to detect beyond GHZ-state like entanglement \cite{Siva2021a}. Moreover, for a tripartition of the form of Fig. \ref{fig:partitions}(a), the Markov gap has been conjectured to be quantized, in the thermodynamic limit and after an appropriate minimization procedure, in topological phases to $(c_+/3)\ln 2$, where $c_+$ is the ``minimal central charge" \cite{Siva2021b}. On the basis of this latter conjecture, the vanishing of the Markov gap thus serves as a signal for topological phases with gappable edges%, which is generally believed to be the class of topological states which can be described via tensor networks
.\footnote{See also Refs. \cite{Kim2022a,Kim2022b,Zou2022,Fan2022}, which extract the chiral central charge using a quantity known as the modular commutator.}% and Ref. \cite{Kato2016} for an earlier study of multipartite correlations in topological orders.}

The computation of these tripartite correlation measures is a challenging task,  though significant progress has already been made. The reflected entropy was computed in Chern-Simons theories using both boundary state and surgery methods in Ref. \cite{Berthiere2021} for tripartitions, but not those involving trijunctions, like that of Fig. \ref{fig:partitions}(a). The authors of Ref. \cite{Liu2021} investigated the negativity and Markov gap employing the boundary state approach in the tripartition of Fig. \ref{fig:partitions}(a), leveraging techniques from string field theory, focusing on chiral, non-fractionalized states (the extension to generic chiral topological orders is studied in Ref. \cite{Liu2023}). %, namely Chern insulators and chiral superconductors. 
Ref. \cite{Siva2021b} instead demonstrated the vanishing of the Markov gap in string-net states. 

The goal of the present work is to extend the computation of negativity and reflected entropy for a tripartition to generic (bosonic) topological orders, within the anyon diagrammatic approach of Ref. \cite{Bonderson2017}. 
As we shall review, the underlying strategy is to represent the ground state (or excited state, in the presence of anyons) in terms of an anyon diagram, for which the desired entanglement quantities may be computed. This formalism provides a useful heuristic picture of entanglement in a topological phase as arising from anyon-antianyon pairs condensed at the entanglement cut, with the topological contributions to entanglement quantities coming from the non-local constraint that the net anyon charge in a subregion is fixed. 
In particular, we apply the proposal of Ref. \cite{Shapourian2020} for an implementation of the partial transpose in anyon diagrams to define the negativity and a simple diagrammatic implementation of the canonical purification to compute the reflected entropy.

Leveraging this approach, we characterize the entanglement structure of generic topological orders by investigating the partitions depicted in Fig. \ref{fig:partitions}.
The main results we obtain are as follows. First, in the tripartition of Fig. \ref{fig:partitions}(a),  in which all three regions $A$, $B$, and $C$ meet at trijunctions, we find that the Markov gap indeed vanishes for non-chiral topological orders (as we will discuss, the diagrammatic approach cannot be used to extract the Markov gap of chiral orders). Interestingly, we find that the negativity for a tripartition possesses a universal contribution which is sensitive to whether the underlying phase is non-Abelian. While a similar behavior was observed for superpositions of degenerate ground states of Chern-Simons theories on a torus \cite{Wen2016}, we emphasize that we here find this to be true for the unique ground state on a sphere. %In fact, we illustrate that this additional universal contribution can be understood via analogy with the computations of Ref. \cite{Wen2016} on the torus. %Along this line, we will see that the diagrammatic computations make explicit the way the trijunctions are responsible for this additional contribution.
As a comparison, we also compute the reflected entropy and negativity in a tripartition in which $A$, $B$, and $C$ meet at \emph{tetrajunctions}, as shown in Fig. \ref{fig:partitions}(b). Here we find that the negativity and Markov gap both vanish. %Interestingly, the mutual information -- and hence reflected entropy -- take quantized forms which, again, can be understood in analogy with the computations of entanglement of Ref. \cite{Wen2016} on the torus. 

Finally, we compute the negativity and reflected entropy for a tripartition in which each region supports a single anyon, as shown in Fig. \ref{fig:partitions}(c). We find that both quantities depend in a complicated way on the $F$-symbols. While we are unable to simplify the expressions for these quantities, 
we demonstrate with some examples that the Markov gap is non-vanishing when all three anyons are non-Abelian, indicating a non-trivial tripartite entanglement structure. % of multi-anyon systems. 
Lastly, we remark that while the diagrammatic computation we employ very closely mirrors the computation of entanglement in the string-net formalism \cite{Levin2006} as noted in Ref. \cite{Bonderson2017}, our work requires an appropriate definition of partial transpose and canoncial purification in the diagrammatic formalism and our computation of the negativity and Markov gap in the presence of excitations goes beyond previous studies.

\begin{table}%[hbt]
\caption{\label{tab:results}%
Summary of main results for the entanglement, as measured by the topological contributions to the negativity $\mathcal{E}_{\mathrm{top}}$, reflected entropy $S_{R,\mathrm{top}}$, and the Markov gap $h$ in the three setups presented in Fig. \ref{fig:partitions}.
}
\begin{ruledtabular}
\begin{tabular}{cccc}
 &
(a) Trijunction &
(b) Tetrajunction &
(c) Trimer \\
\colrule
$\mathcal{E}_{\mathrm{top}}$ & $-\ln \cD + \frac{1}{2}\ln\sum_a \frac{d_a^3}{\cD^2}$ & $0$ & Eq. \eqref{eq:anyon-negativity} \\
$S_{R,\mathrm{top}}$ &  $-\ln \cD$ & $\sum_a \frac{d_a^2}{\cD^2}\ln \frac{d_a}{\cD} $ & Eq. \eqref{eq:anyon-renyi-markov} \\
$h$ & $=0$ & $=0$ & $>0$
\end{tabular}
\end{ruledtabular}
\end{table}
We summarize these results in Table \ref{tab:results}. The balance of this paper is structured as follows. In Section \ref{sec:anyon-entanglement}, we review the entanglement quantities of interest and their computation in the anyonic diagrammatic formalism. In Section \ref{sec:diagrammatic-approach} we review the computation of the entanglement entropy following Ref. \cite{Bonderson2017}, and confirm that this formalism yields the expected expressions for the negativity and reflected entropy for a bipartition. % in \ref{sec:bipartite}. 
In Section \ref{sec:trijunction} we move to the focus of this work and compute the negativity and Markov gap for the tripartition in Fig. \ref{fig:partitions}(a). In Sections \ref{sec:tetrajunction} and \ref{sec:tripartite-insertions}, we repeat this analysis for the partitions in Figs \ref{fig:partitions}(b) and (c). Finally, in Section \ref{sec:discussion}, we discuss our results and conclude. The Appendixes contain reviews of anyon diagrams and additional computational details. % employed in our computations.

\section{Entanglement in Anyon Models \label{sec:anyon-entanglement} }

As explained in the Introduction, we employ a diagrammatic formalism to compute the entanglement quantities of interest. We review the basics of anyon models and their diagrammatic representation in Appendix \ref{sec:app-anyon-models}. The key point is that a density matrix describing a set of anyons in fixed positions can be represented by a branching tree diagram. Operations such as the partial trace and partial transpose have analogues in the constrained Hilbert spaces describing these anyon states and allow for defining analogues of entanglement quantities, as we review in the following subsections. We note that when computing the entanglement of topological phases, rather than just of anyons within a topological phase, we will need to consider anyons on punctured manifolds of non-zero genus. We review the additional rules governing such diagrams in Appendix \ref{sec:app-anyon-models-nonzero-genus}. 

\subsection{Entanglement Entropy}

As noted in the Introduction, given a density matrix $\rho$, the entanglement entropy of region $A$ is defined by $$S_A = -\mathrm{Tr}(\rho_A \ln \rho_A),$$ where $\rho_A = \mathrm{Tr}[\rho]$ is the reduced density matrix for region $A$. We can also define the entanglement entropy as the $\alpha \to 1$ limit of the R\'enyi entropies, $$S_A^{(\alpha)} = - \frac{1}{1-\alpha} \ln \mathrm{Tr}[\rho_A^\alpha].$$
The corresponding quantities for anyon models are defined in a completely analogous manner, replacing the trace with the quantum trace defined in Appendix \ref{sec:app-anyon-models}. We then define  the anyonic R\'enyi and entanglement entropies as, respectively,
\begin{align}
	\widetilde{S}_A^{(\alpha)} = - \frac{1}{1-\alpha} \ln \qTr[\trho_A^\alpha]
\end{align}
and
\begin{align}
	\widetilde{S}_A = -\mathrm{\qTr}(\trho_A \ln \trho_A) = \lim_{\alpha \to 1} \widetilde{S}_A^{(\alpha)} \, .
\end{align}
As an example, let us consider the pure state describing a superposition of pairs of anyons and anti-anyons fusing to the vacuum:
\begin{equation}
\ket{\psi} =\sum_{a} \frac{\sqrt{d_a}}{\mathcal{D}}
\begin{pspicture}[shift=-0.3](0,-0.6)(1.4,0)%[shift=0](0,0)(0.,1.5)
	\scriptsize
	\psline[ArrowInside=->](0.8,-0.6)(0.2,0)\rput(0.2,0.2){$a$}
	\psline[ArrowInside=->](0.8,-0.6)(1.4,0)\rput(1.4,0.2){$\overline{a}$}
\end{pspicture}  \label{eq:dimer-example}
\end{equation}
The density matrix is given by
\begin{align}
	\trho = \sum_{a , a'} \frac{\sqrt{d_a d_{a'} }}{\mathcal{D}^2}
\begin{pspicture}[shift=-0.55](0,-1.6)(1.4,0)%[shift=0](0,0)(0.,1.5)
	\scriptsize
	\psline[ArrowInside=->](0.8,-0.6)(0.2,0)\rput(0.2,0.2){$a$}
	\psline[ArrowInside=->](0.8,-0.6)(1.4,0)\rput(1.4,0.2){$\overline{a}$}	%%
	\psline[ArrowInside=-<](0.8,-0.9)(0.2,-1.5)\rput(0.2,-1.7){$a'$}
	\psline[ArrowInside=-<](0.8,-0.9)(1.4,-1.5)\rput(1.4,-1.7){$\overline{a}'$}
\end{pspicture} 
\end{align}
Suppose we take the left and right leaves of the diagram to correspond to regions $A$ and $B$, respectively. The reduced density matrix is then
\begin{align}
	\trho_A = \sum_{a} \frac{d_a}{\mathcal{D}^2}
\begin{pspicture}[shift=-0.55](0,-0.6)(1.0,-0.6)%[shift=0](0,0)(0.,1.5)
	\scriptsize
	\psline[ArrowInside=->](0.5,-0.6)(0.5,0.6)\rput(0.75,0){$a$}
\end{pspicture} 
\end{align}
Taking the $\alpha^{th}$ power of this density matrix, we find the $\alpha^{th}$ R\'enyi entropy to be
\begin{align}
	\wt{S}_A^{(\alpha)} = \frac{1}{1-\alpha} \ln \sum_a \left( \frac{d_a^{1+\alpha}}{\mathcal{D}^{2\alpha}} \right) \, . \label{eq:dimer-example-renyi}
\end{align}
The replica limit $\alpha \to 1$ yields the entanglement entropy,
\begin{align}
	\wt{S}_A = - \sum_a \frac{d_a^2}{\mathcal{D}^2} \ln \frac{d_a}{\mathcal{D}^2} \, .
\end{align}
We recall that if the net topological charge of $\trho_A$ is trivial, as in this example, then the quantum trace coincides with the usual trace. In particular, in the cases we consider, the anyonic entanglement defined above will likewise coincide with the usual entanglement entropy, and so we will often drop the tildes. Finally, we note that we will also be interested in the mutual information, defined as a linear combination of entanglement entropies:
\begin{align}
	\wt{I}(A:B) = \wt{S}_A + \wt{S}_B - \wt{S}_{AB} \, .
\end{align}

%\raman{Should add example computation}

\subsection{Negativity}

Let us first review the definition of the (bosonic) partial transpose and logarithmic negativity in unconstrained Hilbert spaces. The logarithmic negativity is given by the logarithm of the trace norm of the density matrix partially transposed on subsystem $A$:
\begin{align}
	\mathcal{E}(A:B) = \ln || \rho_{AB}^{T_A} ||_1,
\end{align}
where we recall that the trace norm is defined as 
\begin{align}
	|| \rho_{AB}^{T_A} ||_1 = \Tr \sqrt{\rho_{AB}^{T_A} (\rho_{AB}^{T_A})^\dagger}.
\end{align}
For a density matrix
\begin{align}
	\rho = \sum_{ijkl} \rho_{ijkl} \ket{e_A^{(i)} e_B^{(j)} }\bra{e_A^{(k)} e_B^{(l)}} ,
\end{align}
the partial transpose is simply defined as
\begin{align}
		\rho^{T_A} = \sum_{ijkl} \rho_{ijkl} \ket{e_A^{(k)} e_B^{(j)} }\bra{e_A^{(i)} e_B^{(l)}} \, .
\end{align}
As the square root may not be straightforward to compute, % (even for the finite-dimensional matrices considered here), 
we can instead employ a replica trick to compute the negativity \cite{Calabrese2012,Calabrese2013}:
\begin{align}
	\mathcal{E}(A:B) = \lim_{n_e \to 1} \ln \Tr (\rho_{AB}^{T_A})^{n_e},
\end{align}
where the analytic computation is done from values $n_e \in 2\mathbb{Z}$.
The negativity, as its name would suggest, measures the number of negative eigenvalues of $\rho^{T_A}$. A necessary condition for separability %, irrespective of whether $\rho$ describes a pure or mixed state, 
is that $\rho^{T_A}$ has only positive eigenvalues and hence that the negativity vanishes. For that reason, the negativity serves as a good bipartite entanglement measure for mixed states, unlike the entanglement entropy. 

Extending the partial transpose to anyonic systems is a non-trivial task.
We may write a general density matrix as $\trho \in V_{A_1' \dots A_n' B_1' \dots B_n'}^{A_1 \dots A_n B_1 \dots B_n} = \sum_{\substack{ a_1 \dots a_n , b_1 \dots b_n \\ a_1' \dots a_n' b_1' \dots b_n'} } V^{a_1 \dots a_n , b_1 \dots b_n}_{a_1' \dots a_n' , b_1' \dots b_n'}$,
\begin{align}
\trho =
\begin{gathered}
	\includegraphics[height=6em]{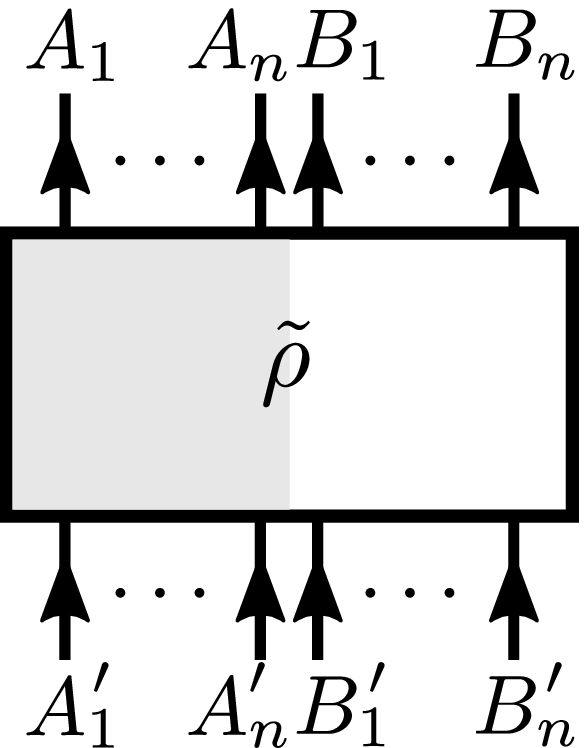} 
\end{gathered} 
\end{align}
where the capital anyon charges denote direct sums over all anyon charges and the box collects all internal lines and vertices. 
We employ the definition of the anyonic partial transpose  introduced in Ref. \cite{Shapourian2020}, 
in which the authors proposed to implement the partial transpose in the following way:
\begin{align}
\label{eq:p_transpose}
\begin{gathered}
	\includegraphics[height=6em]{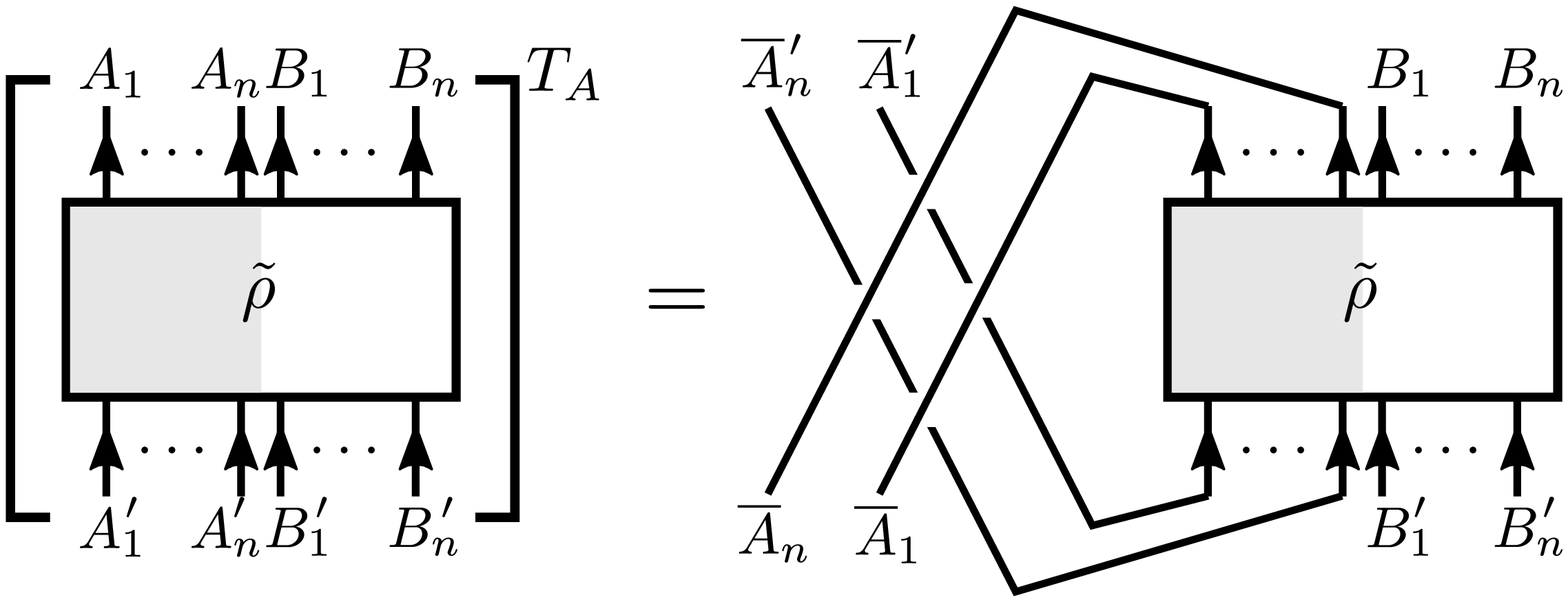} 
\end{gathered}
\end{align}
That is, we bend the output (input) legs of anyons in region $A$ into output (input) legs. Note that the ordering of the anyon charges in region $A$ are reversed. Additionally, there is an ambiguity in how the top and bottom lines are braided and this definition of the partial transpose is, in general, neither trace nor Hermiticity preserving. 
These caveats are, nevertheless, immaterial in the computation of the negativity, which depends only on the singular values of $\trho^{T_A}$ and hence the eigenvalues of the manifestly Hermitian operator $\sqrt{\trho^{T_A} (\trho^{T_A})^\dagger}$. 
In particular, when implementing the replica trick, we will instead have to compute
\begin{align}
	\mathcal{E}(A:B) = \lim_{n_e \to 1} \ln \Tr (\trho_{AB}^{T_A}(\trho_{AB}^{T_A})^\dagger)^{n_e/2} \, .
\end{align}
We refer the reader to Ref. \cite{Shapourian2020} for proofs that the negativity defined with this partial transpose yields a consistent entanglement measure.

As an example, let us compute the negativity of Eq. \eqref{eq:dimer-example}. The partially transposed density matrix is given by
\begin{align}
	\trho^{T_A} = \sum_{a , a'} \frac{\sqrt{d_a d_{a'} }}{\mathcal{D}^2} \varkappa_a \varkappa_{a'}^*
\begin{pspicture}[shift=-0.6](-1.2,0)(0.,1.7)
        \scriptsize
        % braid
        \psline(0,0)(-1,1.5)
        \psline[border=2pt](-1,0)(0,1.5)
        \psline[ArrowInside=->](0,0)(-0.33,0.5)\rput(-1,1){$\bar a$}
        \psline[ArrowInside=->](-1,0)(-0.66,0.5)\rput(0,1){$\bar a'$}
    \end{pspicture}
\end{align}
where $\varkappa_a$ is the Frobenius-Schur indicator for $a$, which satisfies $|\varkappa_a| = 1$ and arises from ``straightening" out the charge line via an $A$-move (see Appendix \ref{sec:app-anyon-models}). Note that this operator is indeed not Hermitian. We compute
\begin{align}
	\trho^{T_A}(\trho^{T_A})^\dagger = \sum_{a , a'} \frac{d_a d_{a'}}{\mathcal{D}^4}
\begin{pspicture}[shift=-1](0,-1)(1.0,1)
	\scriptsize
	\psline[ArrowInside=->](0.1,-0.7)(0.1,0.7)\rput(0.3,0){$a$}
	\psline[ArrowInside=->](0.9,-0.7)(0.9,0.7)\rput(1.1,0){$a'$}
\end{pspicture} 
\end{align}
This operator is diagonal, and so we can directly read off the singular values. We thus find the negativity to be
\begin{align}
	\mathcal{E} = \ln \sum_{a,a'} \frac{( d_a d_{a'})^{3/2}}{\mathcal{D}^2} = 2\ln \sum_a \frac{d_a^{3/2}}{\cD} \, .
\end{align}
Comparing with Eq. \eqref{eq:dimer-example-renyi}, we see that the negativity is precisely equal to the $\alpha=1/2$ R\'enyi entropy, as expected for a pure state.

\subsection{Reflected Entropy}

The final correlation measure we consider is the reflected entropy, defined as follows. 
Given a bipartite density matrix $\rho_{AB}$ defined on the Hilbert space $\cH_A \otimes \cH_B$, we can construct its canonical purification $\kket{\sqrt{\rho_{AB}}}$ on the doubled Hilbert space $(\cH_A \otimes \cH_{A^*}) \otimes (\cH_B \otimes \cH_{B^*})$. As a purification, this state satisfies $\Tr_{A^*B^*}[\kket{\sqrt{\rho_{AB}}} \bbra{\sqrt{\rho_{AB}}}] = \rho_{AB}$. 
%In the case that 
Explicitly, if $\rho_{AB} = \sum_n p_n \ket{n} \bra{n}$ is written in a diagonal form, we have simply $\kket{\sqrt{\rho_{AB}}} = \sum_n \sqrt{p_n}\ket{n}\ket{n}^*$, where we take the $CPT$ conjugate of the second ket.  The reflected entropy is then defined as the entanglement entropy of region $A \cup A^*$,
\begin{align}
	S_R(A:B) = S_{AA^*},
\end{align}
computed in the state $\kket{\sqrt{\rho_{AB}}}$. We note that the canonical purification is not unique, as one may always act on the purifying space by a unitary operator $U$; the state $ (\mathbf{1}_{AB}\otimes U_{A^*}U_{B^*})\kket{\sqrt{\rho_{AB}}}$ is an equally good purification, 
where $\mathbf{1}_{AB}$ is the identity on $AB$, and $U_{A^*,B^*}$ are unitaries acting on $A^*$ and $B^*$, respectively. Such unitary rotations, naturally, do not affect the reflected entropy.

In general, the canonical purification may not be straightforward to obtain. However, we can compute the reflected entropy via a double replica trick, with one replica accounting for the square root in the purification and the other the von Neumann limit. Setting $\alpha \in 2\mathbb{Z}^+$, we compute $\rho_{AB}^{\alpha/2}$ to find $\kket{\rho_{AB}^{\alpha/2}}$, the canonical purification of $\rho_{AB}^\alpha$. We then trace out $\cH_B \otimes \cH_{B^*}$ to find $\rho_{AA^*}^{(\alpha)}$, from which we compute the R\'enyi reflected entropy by analytically continuing $\alpha$ to unity:
\begin{align}
	S_R^{(\beta)}(A:B) = \frac{1}{1-\beta} \lim_{\alpha\to 1} \ln \frac{\Tr(\rho_{AA^*}^{(\alpha)})^\beta}{(\Tr\rho_{AB}^{\alpha})^\beta}.
\end{align}
The reflected entropy is then obtained by taking $\beta \to 1$. 

We are primarily interested in computing a quantity known as the \emph{Markov gap}, defined as
\begin{align}
	h(A:B) = S_R(A:B) - I(A:B).
\end{align}
The Markov gap provides a rare example of a computable quantity which distinguishes different patterns of tripartite entanglement. In particular, it was shown in Ref. \cite{Siva2021a} that $h(A:B)$ vanishes for GHZ-like states and is non-zero for W states. Motivated, by this, the reflected entropy has been studied in, for instance, the context of many-body quenches and boundary/interface conformal field theories \cite{KudlerFlam2020,Zou2022a,Kusuki2020,Kusuki2021,Kusuki2022}. Moreover, in Ref. \cite{Siva2021b}, it was conjectured that for a topologically ordered ground state $\ket{\psi}$, the Markov gap computed in the tripartition of Fig. \ref{fig:partitions}(a) is sensitive to the central charge of the corresponding edge CFT. Specifically, if we consider the family of states $U\ket{\psi}$, where $U$ is a unitary operator acting locally near the trijunctions where regions $A$, $B$, and $C$ meet in Fig. \ref{fig:partitions}(a), then we may consider the optimized quantity
\begin{align}
	h(A:B)^{IR} = \min_U h(A:B) \, ,
\end{align}
where $h(A:B)$ is computed in the state $U\ket{\psi}$. The optimization over $U$ has the effect of removing spurious short-range short-range entanglement near the trijunctions. It was argued in Ref. \cite{Siva2021b} that this optimized quantity satisfies 
\begin{align}
	h(A:B)^{IR} = \frac{c_+}{3} \ln 2 + O(e^{-l / \xi}) \, ,
\end{align}
%for the tripartition configuration shown in Fig. \ref{fig:partitions}(a), 
where $\xi$ is the correlation length, $l$ a length scale for the perimeters of the subregions, and $c_+$ is the ``minimal" central charge of the edge theory.  This is the total central charge after gapping out as many edge modes as possible. 
Indeed, certain topological orders, despite having a vanishing chiral central charge, have a topological obstruction to gapping out all edge modes \cite{Levin2013,Wang2015}. As we will be performing our computations deep in the topological phase, that is, in the zero-correlation length limit, we expect the optimization to be unnecessary and hence $h^{IR} = h$.

In order to compute the reflected entropy in the diagrammatic formalism, we propose a straightforward implementation of the canonical purification. As an illustration, let us first consider the following simple example of a diagonal mixed density matrix,
\begin{align}
	\trho &= \frac{1}{d_c} \ket{a,b; c ,\mu} \bra{a,b;c,\mu} = \frac{1}{\sqrt{d_a d_b d_c}} 
		\begin{pspicture}[shift=-0.9](-0.2,0)(1.6,1.8)
        \scriptsize
        \psline[ArrowInside=->](0,0)(0.6,0.6)\rput(0,0.3){$a$}
        \psline[ArrowInside=->](1.2,0)(0.6,0.6)\rput(1.2,0.3){$b$}
        \rput(0.77,0.65){$\mu$}
        \psline[ArrowInside=->](0.6,0.6)(0.6,1.2)\rput(0.37,0.9){$c$}
        \rput(0.77,1.15){$\mu$}
        \psline[ArrowInside=->](0.6,1.2)(0,1.8)\rput(0,1.5){$a$}
        \psline[ArrowInside=->,](0.6,1.2)(1.2,1.8)\rput(1.2,1.5){$b$}
    \end{pspicture}
\end{align}
We define its canonical purification to be
\begin{align}
	\begin{split}
	\kket{\sqrt{\trho}} &= \frac{1}{\sqrt{d_a d_b}} 	
	\begin{pspicture}[shift=-1.9](-2.2,-1.3)(1.3,1.9)
        \scriptsize
        \psline[ArrowInside=->](0,0)(0.6,0.6)%\rput(0,0.3){$a$}
        \psline[ArrowInside=->](1.2,0)(0.6,0.6)%\rput(1.2,0.3){$b$}
        \rput(0.77,0.65){$\mu$}
        \psline[ArrowInside=->](0.6,0.6)(0.6,1.2)\rput(0.37,0.9){$c$}
        \rput(0.77,1.15){$\mu$}
        \psline[ArrowInside=->](0.6,1.2)(0,1.8)\rput(0,1.5){$a$}
        \psline[ArrowInside=->,](0.6,1.2)(1.2,1.8)\rput(1.2,1.5){$b$}
        \psline(0.,0)(-.6,.6)(-.6,1.8)\rput(-.8,1.7){$\bar a$}
        \psline(1.2,0)(0.,-1.2)(-1.8,.6)(-1.8,1.8)\rput(-2.,1.7){$\bar b$}
    \end{pspicture}
    \end{split}
\end{align}
We have simply dragged the input legs into a new output Hilbert space. Note that we did not take the square root of the normalization factors which enforce isotopy invariance of the diagram in order to ensure the correct normalization with respect to the quantum trace of the purified state. Indeed, one may readily check that $\qTr_{A^*B^*}[ \kket{\sqrt{\trho}} \bbra{\sqrt{\trho}}] = \trho$. Importantly, it is evident that in this implementation of the canonical purification the original Hilbert space and the purifying Hilbert space  do \emph{not} have a tensor product factorization. Indeed, the purification is not constructed by simply flipping bras into kets. This is necessitated by the fact that the net anyon charge of the purified state must be trivial to ensure that it is truly a pure state with respect to the quantum trace.\footnote{A similar feature holds for purifications of purely fermionic systems -- the fermion operators in the original and purifying systems anticommute and hence the Hilbert space does not factorize. 
See Ref.\ \cite{Liu_2021} and references therein for discussion.}

For more complicated density matrices, we will need to employ the replica trick to construct their canonical purifications and compute the reflected entropy.
For a general density matrix,
\begin{align}
\trho =
\psscalebox{.85}{
 \pspicture[shift=-1.](-1,-1.1)(1,1.2)
  \footnotesize
%%%%% Box:
  \psframe[linewidth=0.9pt,linecolor=black,border=0](-0.8,-0.5)(0.8,0.5)
  \rput[bl]{0}(-0.1,-0.1){\normalsize $\trho$}
  \rput[bl]{0}(-0.22,0.7){$\mathbf{\ldots}$}
  \rput[bl]{0}(-0.22,-0.75){$\mathbf{\ldots}$}
%%%%% Line connections:
  \psset{linewidth=0.9pt,linecolor=black,arrowscale=1.5,arrowinset=0.15}
  \rput[bl]{0}(0.5,1.1){$A_n$}
  \psline(0.6,0.5)(0.6,1)
  \rput[bl]{0}(-0.7,1.1){$A_1$}
  \psline(-0.6,0.5)(-0.6,1)
  \rput[bl]{0}(0.5,-1.27){$A_n'$}
  \psline(0.6,-0.5)(0.6,-1)
  \rput[bl]{0}(-0.7,-1.27){$A_1'$}
  \psline(-0.6,-0.5)(-0.6,-1)
%%%%% Arrows:
  \psline{->}(0.6,0.5)(0.6,0.9)
  \psline{->}(-0.6,0.5)(-0.6,0.9)
  \psline{-<}(0.6,-0.5)(0.6,-0.9)
  \psline{-<}(-0.6,-0.5)(-0.6,-0.9)
\endpspicture 
} \in V^{A_1 \dots A_n}_{A_1' \dots A_n'} = \sum_{\substack{ a_1 \dots a_n \\ a_1' \dots a_n'} } V^{a_1 \dots a_n}_{a_1' \dots a_n'} ,
\end{align}
in order to carry out the first step of the replica trick, we raise the density matrix to the power $\alpha/2$ with $\alpha \in 2\mathbb{Z}$ and then apply the operator-state map to obtain a state in the splitting space $V_0^{\bar{A}_n' \dots \bar{A}_1' ; A_1 \dots A_N}$
\begin{align}
	\kket{\trho^{\alpha/2}} = \psscalebox{.85}{
 \pspicture[shift=-2.2](-2.9,-2.6)(1,1.2)
  \footnotesize
%%%%% Box:
  \psframe[linewidth=0.9pt,linecolor=black,border=0](-0.8,-0.5)(0.8,0.5)
  \rput[bl]{0}(-0.2,-0.1){\normalsize $\trho^{\alpha/2}$}
  \rput[bl]{0}(-0.22,0.7){$\mathbf{\ldots}$}
  \rput[bl]{0}(-0.22,-0.75){$\mathbf{\ldots}$}
%%%%% Line connections:
  \psset{linewidth=0.9pt,linecolor=black,arrowscale=1.5,arrowinset=0.15}
  \rput[bl]{0}(0.5,1.1){$A_n$}
  \psline(0.6,0.5)(0.6,1)
  \rput[bl]{0}(-0.7,1.1){$A_1$}
  \psline(-0.6,0.5)(-0.6,1)
  \rput[bl]{0}(-1.4,1.1){$\bar A_n'$}
  \psline(0.6,-0.5)(0.6,-1)
  \rput[bl]{0}(-2.6,1.1){$\bar A_1'$}
  \psline(-0.6,-0.5)(-0.6,-1)
%%%%% Arrows:
  \psline{->}(0.6,0.5)(0.6,0.9)
  \psline{->}(-0.6,0.5)(-0.6,0.9)
  \psline{-<}(0.6,-0.5)(0.6,-0.9)
  \psline{-<}(-0.6,-0.5)(-0.6,-0.9)
%%%%%% Operator-State Map
  \psline(-0.6,-1)(-1.05,-1.3)
  \psline(-1.05,-1.3)(-1.5,-1)
  \psline(-1.5,-1.)(-1.5,1)
  \psline(0.6,-1)(-1.65,-2.35)
  \psline(-1.65,-2.35)(-2.7,-1.3)
  \psline(-2.7,-1.3)(-2.7,1)
\endpspicture 
}
\end{align}
Note that, as in the definition of the partial transpose, the anyon charges originally in the input space have their ordering reversed and are conjugated. Using the fact that $\trho^\dagger = \trho$ for any density matrix, one may check that $\qTr_{A^*B^*}[ \kket{\trho^{\alpha/2}} \bbra{\trho^{\alpha/2}}] = \trho^\alpha$. As such, we claim that this mapping provides, in the limit $\alpha \to 1$, a sensible definition of the canonical purification in the anyon diagrammatic formalism.
From here, the computation of the reflected entropy proceeds as usual -- we trace out $B\cup B^*$ to obtain $\trho_{AA^*}^{(\alpha)}$. The computation of the anyonic versions of the R\'enyi and von Neumann reflected entropies is then simply done using the prescription for the entanglement entropy provided above.

To illustrate the diagrammatic computation of the reflected entropy, let us carry it out for the state given in Eq. \eqref{eq:dimer-example}. 
The first step of the replica trick is trivial, since $\kket{\psi}$ is a pure state and thus $\trho^\alpha = \trho$ for all $\alpha \in \mathbb{Z}$. The canonical purification of $\trho^{\alpha/2}$ is given by,
\begin{align}
	\kket{\trho^{\alpha/2}} =  \sum_{a , a'} \varkappa_{a} \frac{\sqrt{d_a d_{a'} }}{\mathcal{D}^2}
\begin{pspicture}[shift=-0.3](0,-0.6)(2.8,0)%[shift=0](0,0)(0.,1.5)
	\scriptsize
	\psline[ArrowInside=->](0.8,-0.6)(0.2,0)\rput(0.2,0.2){$a$}
	\psline[ArrowInside=->](0.8,-0.6)(1.4,0)\rput(1.4,0.2){$\overline{a}$}%%
	\psline[ArrowInside=->](2.4,-0.6)(1.8,0)\rput(1.8,0.2){$a'$}
	\psline[ArrowInside=->](2.4,-0.6)(3.0,0)\rput(3.0,0.2){$\overline{a}'$}
\end{pspicture}
\end{align}
Here, we were able to remove the bends in the $a$ anyon line at the expense of a factor of its Frobenius-Schur indicator, $\varkappa_{a}$.
We remind ourselves that, in the above diagram, the anyons $a$ and $a'$ ($\overline{a}$ and $\overline{a}'$) belong to regions $A$ and $A^*$ ($B$ and $B^*$), respectively. Hence,
\begin{align}
	\begin{split}
	\trho_{AA^*} &= \qTr_{BB^*}[ \kket{\trho^{\alpha/2}} \bbra{\trho^{\alpha/2}}] = \sum_{a , a'} \frac{d_a d_{a'} }{\mathcal{D}^4} \begin{pspicture}[shift=-0.55](0,-0.7)(1.0,-0.7)%[shift=0](0,0)(-0.1,1.6)
	\scriptsize
	\psline[ArrowInside=->](0.1,-0.7)(0.1,0.7)\rput(0.3,0){$a$}
	\psline[ArrowInside=->](0.9,-0.7)(0.9,0.7)\rput(1.1,0){$a'$}
\end{pspicture} 
	\end{split}
\end{align}
Since the $\alpha \to 1$ limit is trivial, the $\beta^{th}$ R\'enyi reflected entropy is readily found to be
\begin{align}
	S_R^{(\beta)} = \frac{1}{1-\beta} \ln \sum_{a,a'} \frac{(d_a d_{a'})^{1+\beta} }{\mathcal{D}^{4\beta}},
\end{align}
which is simply twice the R\'enyi entanglement entropy, Eq. \eqref{eq:dimer-example-renyi}. We thus can trivially take the von Neumann limit $\beta \to 1$ to obtain
\begin{align}
	S_R = -2  \sum_a \frac{d_a^2}{\mathcal{D}^2} \ln \frac{d_a}{\mathcal{D}^2},
\end{align}
which is twice the von Neumann entropy, as is expected for a pure state.

\section{Diagrammatic Approach to Entanglement \label{sec:diagrammatic-approach} }

Having introduced the entanglement quantities of interest and their implementation in the anyon diagrammatic formalism, we proceed to their computation for generic (bosonic) topological orders, employing the methodology of Ref. \cite{Bonderson2017}. The main technical points to be addressed are how to represent the ground state of a topological phase in terms of an anyon diagram and how to incorporate the entanglement cuts in this representation. 
Both issues require first understanding how to define anyon diagrams on manifolds of non-zero genus, and so we turn first to understanding this before discussing the diagrammatic computation of entanglement.

\subsection{Anyon Diagrams on Non-zero Genus Manifolds \label{sec:app-anyon-models-nonzero-genus} }

Anyon diagrams defined on manifolds of non-zero genus were first discussed in Refs. \cite{Pfeifer2012,Pfeifer2014}.
Here we review this formalism following the conventions of Ref. \cite{Bonderson2017}. 
When dealing with topological phases on punctured manifolds with non-contractible cycles, we can have anyon lines which end at the punctures or wrap around these cycles. As a minimal example, let us consider a punctured torus. There are two bases in which to express states of anyons on such a manifold: the inside and outside bases,
\begin{align}
	\begin{gathered}
	\includegraphics[height=6em]{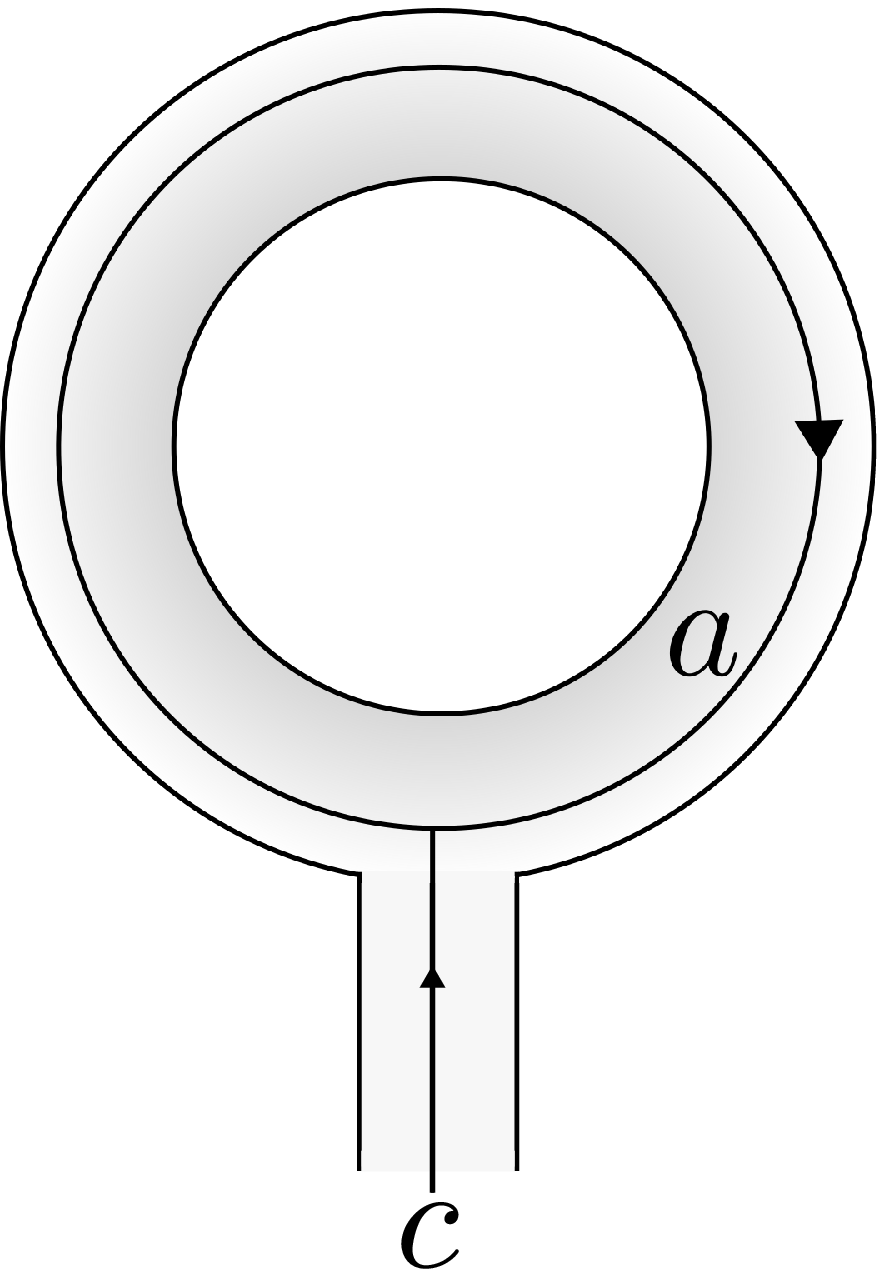}
	\end{gathered}  \quad \text{ and }  \quad
	\begin{gathered}
	\includegraphics[height=6em]{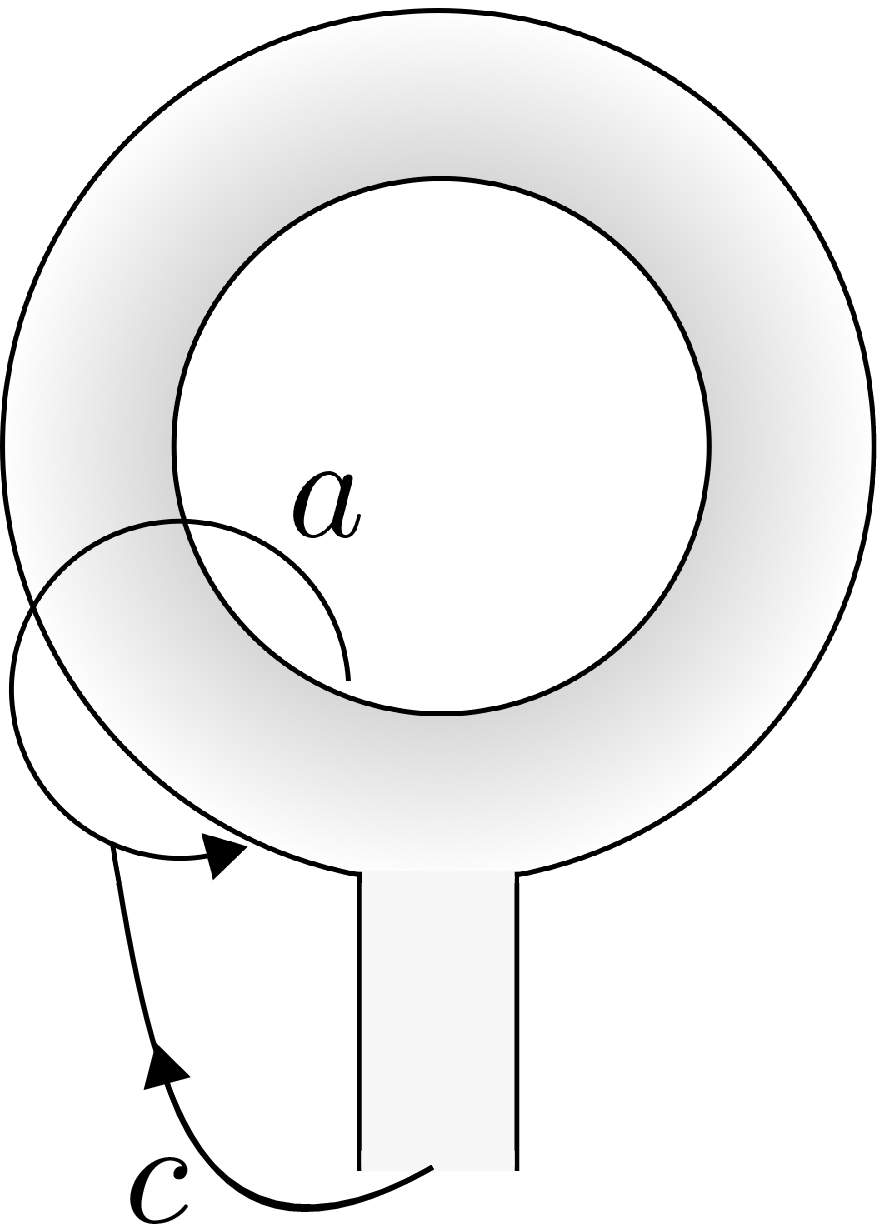}
	\end{gathered}
\end{align}
respectively, in which the anyon lines exist, as their names suggest, inside or outside the manifold and are permitted to wind around only one of the two cycles of the torus. We can change between these bases using a modular-$\mathcal{S}$ transformation:\footnote{Here, $\mathcal{S}_{ab}^{(c)}$ is the punctured $\mathcal{S}$ matrix. When $c$ is the vacuum, it reduces to the usual modular $\mathcal{S}$ matrix, which is the object which will appear in all of our calculations.}
\begin{align}
	\begin{gathered}
	\includegraphics[height=6em]{figures/inside-basis.eps}
	\end{gathered} =
	\sum_b \mathcal{S}_{ab}^{(c)} \quad
	\begin{gathered}
	\includegraphics[height=6em]{figures/outside-basis.eps}
	\end{gathered}
\end{align}
In diagrammatic notation, we denote a non-contractible cycle with a circled cross. Thus, the Hilbert space for, say, a punctured torus is spanned by
\begin{equation}
    \ket{(a);c,\mu}=d_c^{1/4}
    \begin{pspicture}[shift=-0.7](-0.2,-0.5)(1.2,1.2)
        \scriptsize
        \rput(0.5,0.5){$\otimes$}
        \psline[border=1.5pt](0.5,0)(1,0.5)(0.5,1)(0,0.5)(0.5,0)
        \psline[ArrowInside=->](0.5,0)(0,0.5)\rput(.05,0.25){$a$}
        \psline[ArrowInside=->](0.5,-0.5)(0.5,0)\rput(0.7,-0.25){$c$}\rput(0.7,0){$\mu$}
    \end{pspicture},
\end{equation}
Here, $a$ wraps around the non-contractible cycle while $c$ exits at the puncture. We enclose anyon charges wrapping around non-contractible cycles with parentheses. Note that we must specify whether this is in the inside or outside basis; in all of our subsequent computations, we will work primarily in the inside basis.

In order to be able to evaluate inner products and traces of diagrams involving anyons winding around non-contractible cycles, we make use of the fact that non-zero genus surfaces can be constructed via surgery from punctured spheres. As an example, consider the state on the torus with an anyon $a$ winding around a non-contractible cycle in the inside basis,
\begin{align}
    \ket{(a)}=
    \begin{pspicture}[shift=-.6](-0.2,-0.2)(1.2,1.2)
        \scriptsize
        \rput(0.5,0.5){$\otimes$}
        \psline[border=1.5pt](0.5,0)(1,0.5)(0.5,1)(0,0.5)(0.5,0)
        \psline[ArrowInside=->](0.5,0)(0,0.5)\rput(.05,0.25){$a$}
    \end{pspicture}. 
\end{align}
We can construct a torus by taking a sphere with two punctures (i.e. a cylinder) and gluing the two punctures together. Here, we reverse this process and cut open the torus to obtain a sphere with two punctures. We thus also cut the anyon loop, yielding a tree diagram of the anyon pair of $a$ and $\bar{a}$ with trivial net charge; explicitly, we define the ``cut" diagram as
\begin{align}
	\ket{(a)_{\mathrm{cut}}} = \frac{1}{\sqrt{d_a}}
    \begin{pspicture}[shift=-0.5](-0.7,-0.2)(.7,.7)
        \scriptsize
        \psline[ArrowInside=->](0,0)(.5,.5)\rput(.5,.7){$\bar{a}$}
        \psline[ArrowInside=->](0.,0.)(-.5,.5)\rput(-.5,.7){$a$}
    \end{pspicture}. 
\end{align}
Note that we have inserted a factor of $1/\sqrt{d_a}$ to ensure the cut diagram is properly normalized. For more general diagrams,  we must insert a factor of $d_{a}^{-1/4}$ for each new open $a$ line to obtain the properly normalized cut state on the punctured sphere. We then define the inner product
\begin{align}
	\braket{(a)|(b)} \equiv \braket{(a)_{\mathrm{cut}}|(b)_{\mathrm{cut}}} = \frac{1}{\sqrt{d_a d_b}}     
	\begin{pspicture}[shift=-.65](-0.7,-0.2)(.7,1.2)
        \scriptsize
        \psline[ArrowInside=-<](0,1)(.5,.5)\rput(.5,.7){$\bar{b}$}
        \psline[ArrowInside=-<](0.,1.)(-.5,.5)\rput(-.5,.7){$b$}
        \psline[ArrowInside=->](0,0)(.5,.5)\rput(.5,.3){$\bar{a}$}
        \psline[ArrowInside=->](0.,0.)(-.5,.5)\rput(-.5,.3){$a$}
    \end{pspicture} = \delta_{ab}.
\end{align}
We obtain the expected orthonormality of these basis states on the torus.

The same principle applies for computing inner products and traces in more complicated diagrams. We simply cut open each anyon loop encircling a contractible cycle, inserting a normalization factor of $d_{a}^{-1/4}$ for each $a$-line that is cut (equivalently, a factor of $d_a^{-1/2}$ for each new pair of $a$ leafs in the cut diagram). By mapping the anyon diagram on a manifold of non-zero genus to a punctured sphere, we can then apply the standard anyon diagrammatic rules reviewed in Appendix \ref{sec:app-anyon-models}.

\subsection{Constructing the Ground State Anyon Diagram}

\begin{figure}
  \centering
\subfloat[]{%
    \includegraphics[width=0.15\textwidth]{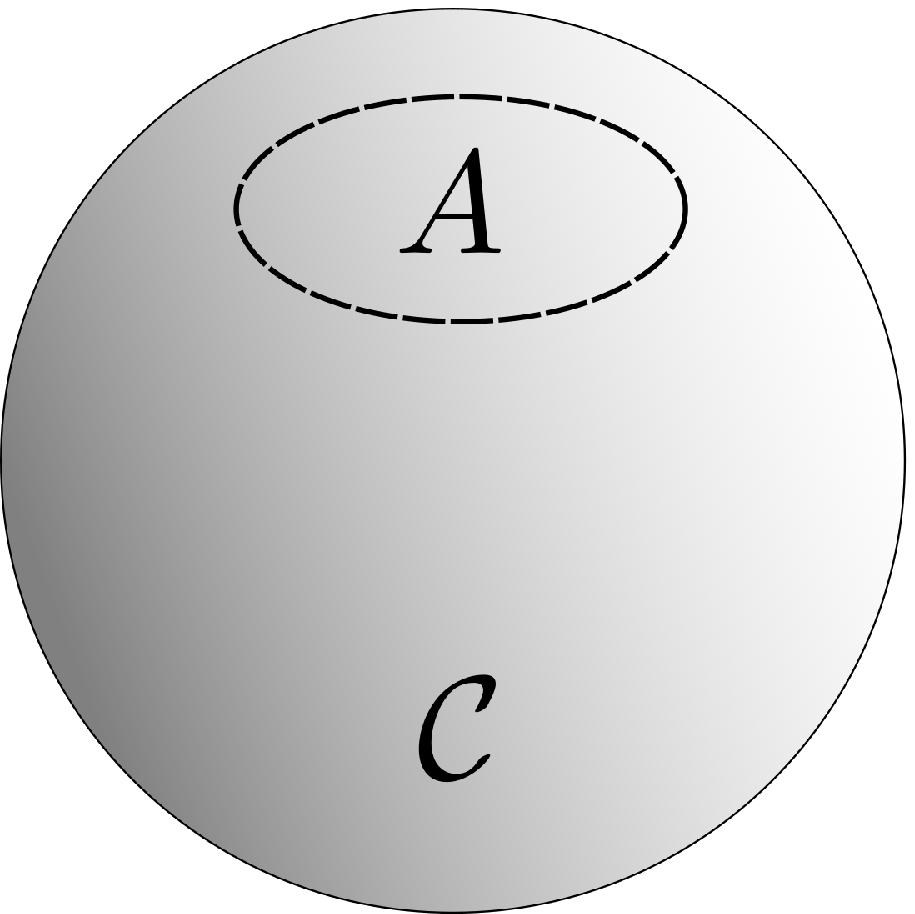}
}\hspace{6mm}
\subfloat[]{%
    \includegraphics[width=0.15\textwidth]{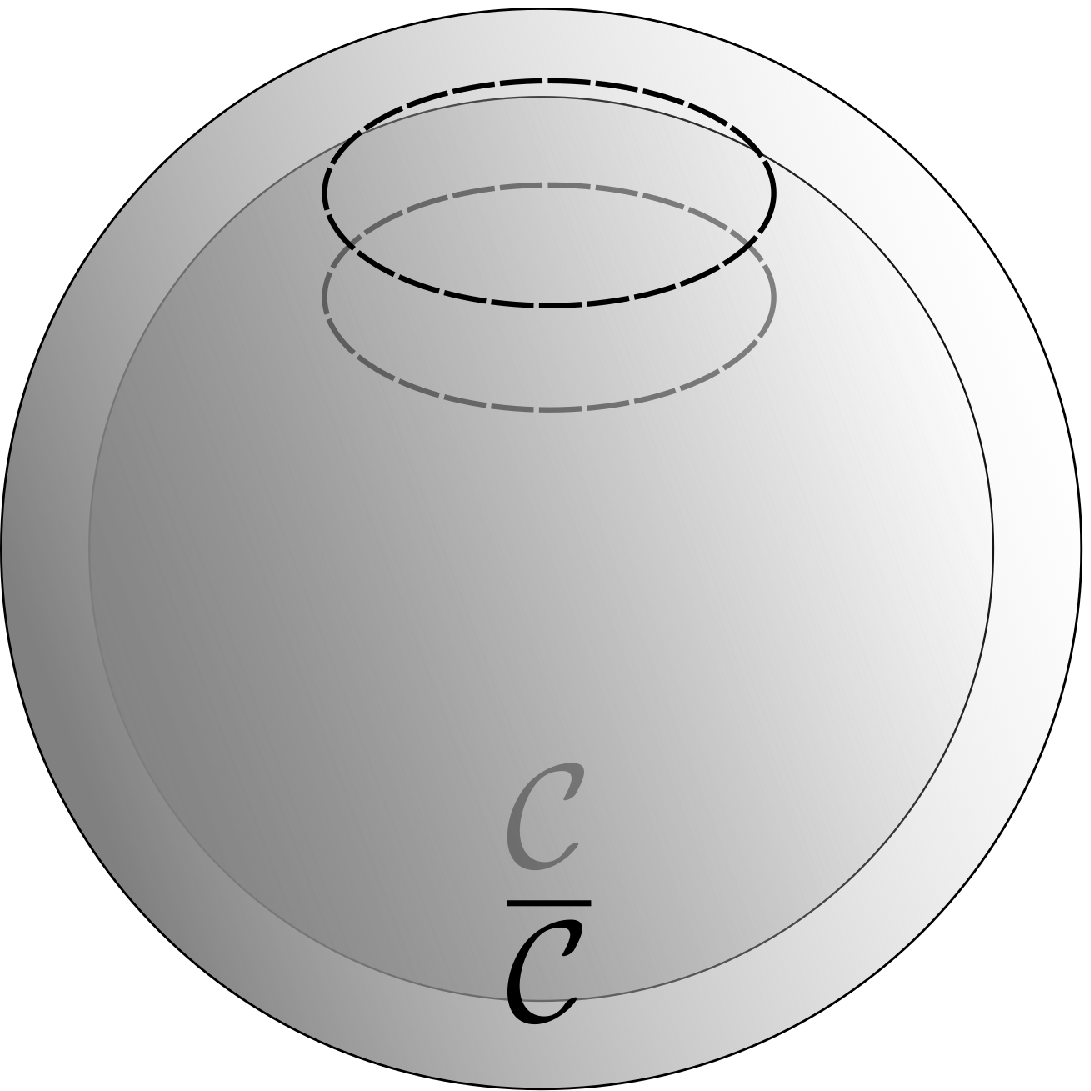}
}

\subfloat[]{%
    \includegraphics[width=0.36\textwidth]{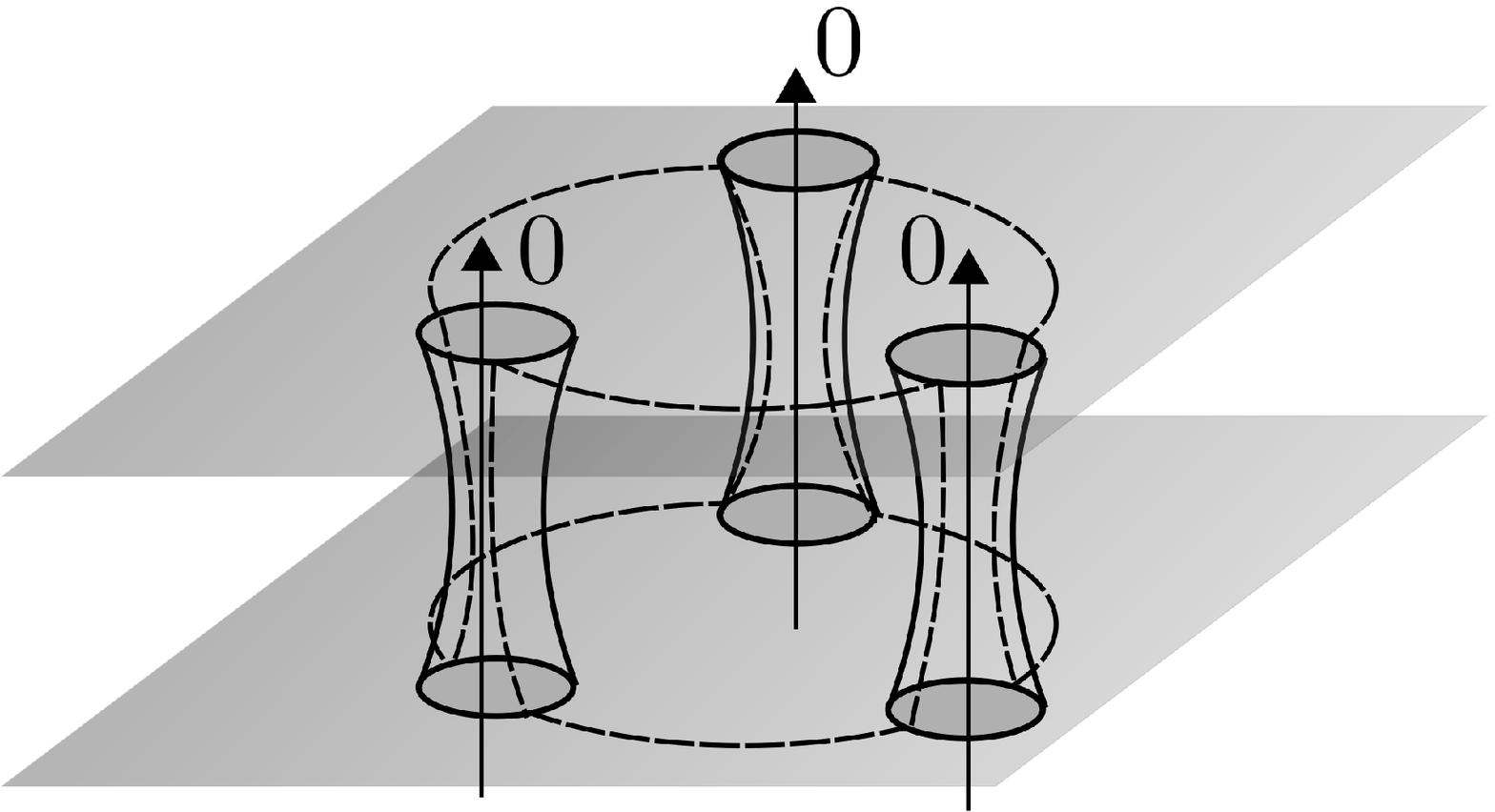}
}

\subfloat[]{%
    \includegraphics[width=0.28\textwidth]{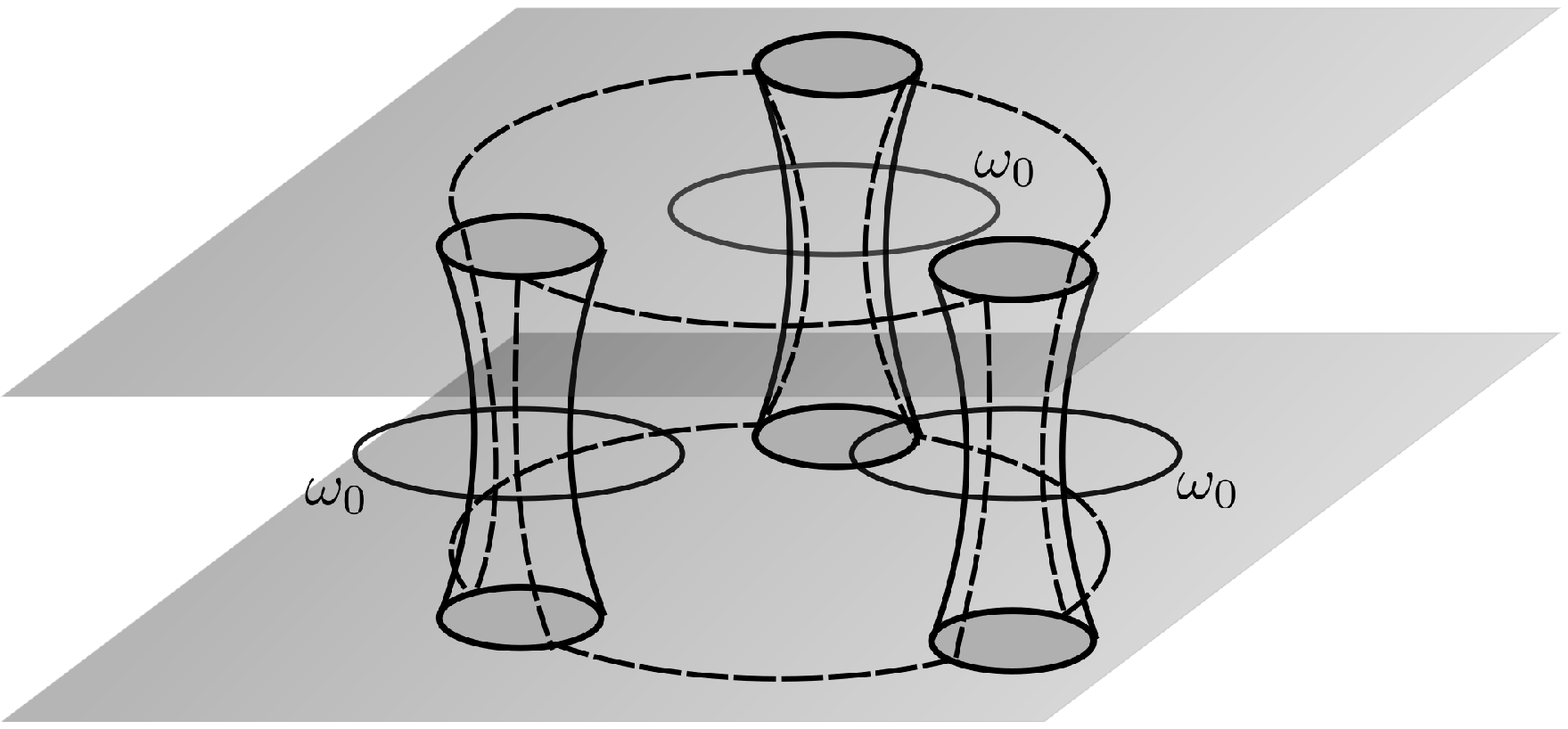}
}
\subfloat[]{%
    \includegraphics[width=0.13\textwidth]{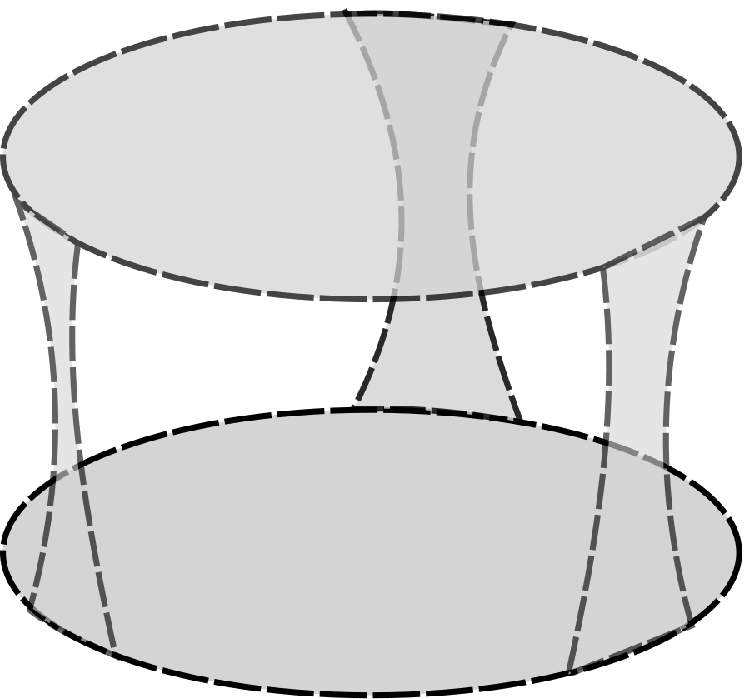}
}
    \caption{(a) Bipartition of the sphere, occupied by the topological phase $\mathcal{C}$ into a single connected region $A$ and its complement. (b) Introduction of a second sphere with the same bipartition and occupied by the time-reversed conjugate $\overline{\mathcal{C}}$. (c) Zoom in of region $A$, after introducing $n=3$ wormholes connecting the two manifolds. The arrows labelled $0$ indicate a trivial anyon flux passing through each wormhole. (d) The same configuration, after applying a modular $\mathcal{S}$ transformation to pass to the inside basis, such that the state is given by $\omega_0$ loops encircling each wormhole. (e) Region $A$ after cutting along the entanglement cut.} \label{fig:bipartition}
\end{figure}

%We are now prepared to spell out in more detail the methodology of Ref. \cite{Bonderson2017} for computing the entanglement of a topological phase.
We now proceed to compute these entanglement quantities for a bulk topological phase. The main technical points to be addressed are how to represent the ground state of a topological phase in terms of an anyon diagram and how to incorporate the entanglement cuts in this representation. To this end, we employ the method of Ref. \cite{Bonderson2017}, which is a generalization of the argument of Kitaev and Preskill \cite{Kitaev2006}. While the latter computed a specific linear combination of entanglement entropies in order extract only the TEE, the approach of the former allows for extracting the area law piece of the entanglement entropy. 
Let us consider a topological phase $\mathcal{C}$ on a closed manifold. For concreteness, we consider a bipartition of the sphere into regions $A$ and $\overline{A}$, as in Fig. \ref{fig:bipartition}(a), though the extension to tripartitions in subsequent sections follows straightforwardly. 
\begin{enumerate}
	\item First, we introduce a second sphere, which is occupied by the time reversed conjugate $\overline{\mathcal{C}}$ of $\mathcal{C}$, as shown in Fig. \ref{fig:bipartition}(b).
	\item Next, we adiabatically deform the two spheres to introduce $n$ wormholes connecting them along the entanglement cut, as shown in Fig. \ref{fig:bipartition}(c), forming a manifold of genus $n-1$. To form a wormhole, we effectively cut open a small hole on each sphere, yielding chiral edge states. These edges can be gapped out through a local tunneling interaction, gluing the spheres together. 
	As a result of the adiabatic nature of this process, each wormhole has \emph{trivial} anyon flux threading it; these trivial fluxes describe the state in the \emph{outside} basis.
	\item Locally, each wormhole throat has the topology of a punctured torus, with a trivial anyon charge threading the wormhole and at the puncture. We can then apply a change of basis, namely a modular $\mathcal{S}$-transformation, to each of these vacuum lines to obtain $\omega_0$ loops circling each wormhole, as shown in Fig. \ref{fig:bipartition}(d). This yields an anyon diagram in the \emph{inside} basis describing the ground state.
\end{enumerate}
Explicitly, an $\omega_0$ loop is defined as (see Appendix \ref{sec:app-anyon-models}),
\begin{equation}
\omega_0
\begin{pspicture}[shift=-.3](-.6,-.4)(.5,.3)
  \scriptsize
  \psellipse[linecolor=black,border=0](0,0)(.5,.3)
  \psline[ArrowInside=-<](-.1,-.279)(.075,-.2825)
  \end{pspicture}
  \equiv \sum_x \frac{d_x}{\mathcal{D}^2}
\begin{pspicture}[shift=-.3](-.6,-.4)(.6,.4)
\scriptsize
  \psellipse[linecolor=black,border=0](0,0)(.5,.3)  \rput(0,-.5){$x$}
  \psline[ArrowInside=-<](-.1,-.279)(.075,-.2825)
  \end{pspicture}.
\end{equation}
Note that this evaluates to unity, if the loop does not enclose a non-contractible cycle.
In the third step above, 
the modular transformation changing to the inside basis yields
\begin{align}
	\ket{(0);0}_{\mathrm{out}} = \sum_x \mathcal{S}_{0x} \ket{(x);0}_{\mathrm{in}} = \sum_x \frac{d_x}{\cD} \ket{(x);0}_{\mathrm{in}},
\end{align}
which is simply an $\omega_0$ loop encircling the wormhole throat, up to a factor of $\cD$. 

The introduction of the second sphere and the connecting wormholes is ultimately a theoretical trick used to express the ground state in terms of an anyon diagram, as we will spell out shortly. Since the process of creating the wormholes can be assumed to be adiabatic, the entanglement of state in connected sphere system should match that of the state in the original, unconnected sphere system. For entanglement cuts involving sharp corners, such as for the tripartitions we consider, this procedure of introducing wormholes may also be understood as a particular choice of regularization of the entanglement cut. The main distinction from the Kitaev-Preskill computation is the introduction of a large number, $n$, of wormholes along the entanglement cut. The number $n$ amounts to a measure of the length of the entanglement cut and hence allows for extracting the area law. Indeed, in the limit $n \to \infty$, we will find that all entanglement quantities have a term proportional to $n$, which we identify as the area law, and an $n$-independent term, which we identify as the topological contribution.

Through an abuse of notation, we let $A$ denote the union of $A$ on the original sphere and its copy on the second sphere.
In order to compute the desired entanglement quantities, we will in addition need to either trace out or apply a partial transpose to subregion $A$.
To do so, we physically cut the final genus $n-1$ manifold along the entanglement cut, such that $A$ and its complement form spheres with $n$ punctures [see Fig. \ref{fig:bipartition}(e)]. Following Ref. \cite{Bonderson2017} (see also Refs. \cite{Pfeifer2012,Pfeifer2014}), we can apply the same cutting procedure to the anyon diagram via the following steps:
\begin{enumerate}
	\item We first apply a sequence of $F$-moves such that there is a single anyon charge line threading each of the punctures of region $A$.
	\item We then cut open the state along each charge line threading the punctures. This splits the diagram into two pieces, one describing region $A$ and the other region $B$. 
	Following the prescription above, each cut charge line requires multiplying the state by a factor $d_a^{-1/2}$, where $d_a$ is the quantum dimension of the cut anyon line, to maintain the proper normalization. 
	These new open ``boundary anyon" lines label the topological charge of each puncture. 
\end{enumerate}
The upshot of this procedure is an anyon diagram representing the ground state, which is given by a sum of anyon operators which are tensor products of operators acting on $A$ and its complement, $\bar A$. Practically speaking, this allows us to apply the partial trace and partial transpose operations to each subregion, despite the lack of a tensor product factorization between $A$ and $\overline{A}$.
Although there are no anyon lines connecting $A$ and $\overline{A}$ in the cut diagram, the non-trivial entanglement between the two regions is encoded in the fact that the boundary anyon at each puncture of $A$ must equal the conjugate of the boundary anyon at the corresponding puncture of $\overline{A}$. 

An important caveat to this procedure is that we have really constructed  an anyon diagram representing the ground state of the doubled topological phase $\mathcal{C} \times \overline{\mathcal{C}}$. As such, the entanglement entropy, for instance, we compute in the ground state will be double that of the original phase of interest $\mathcal{C}$.

In the following subsection we work through this abstract procedure explicitly for the simple case of a bipartition of a topological order. 
This serves as a warm-up for the balance of the present work, in which we employ this diagrammatic procedure to compute the entanglement quantities of interest for a tripartition.

\subsection{Review of Bipartite Entanglement \label{sec:bipartite} }

In this section we compute entanglement quantities for a simple bipartition. 
We first briefly review the construction of the diagrammatic representation of the ground state. We then review the computation of the entanglement entropy and confirm that the proposed diagrammatic formulations of the negativity and reflected entropy yield the expected results.

As depicted in Fig. \ref{fig:bipartition}, following the general prescription detailed above, we begin by introducing a time-reversed copy of the system and insert $n$ wormholes connecting the entanglement cuts on the boundaries, yielding a manifold of genus $n-1$. In the outside basis, there is a trivial anyon charge threading each wormhole. We can then locally apply modular $\mathcal{S}$-transformations to change to the inside basis, in which the state is described by anyon lines circling the wormholes. Explicitly, the resulting state is a product of $\omega_0$ loops circling the wormholes:
\begin{align}
\ket{\psi} = \mathcal{D}^{n-1}
%%%
\begin{gathered}
\includegraphics[height=10em]{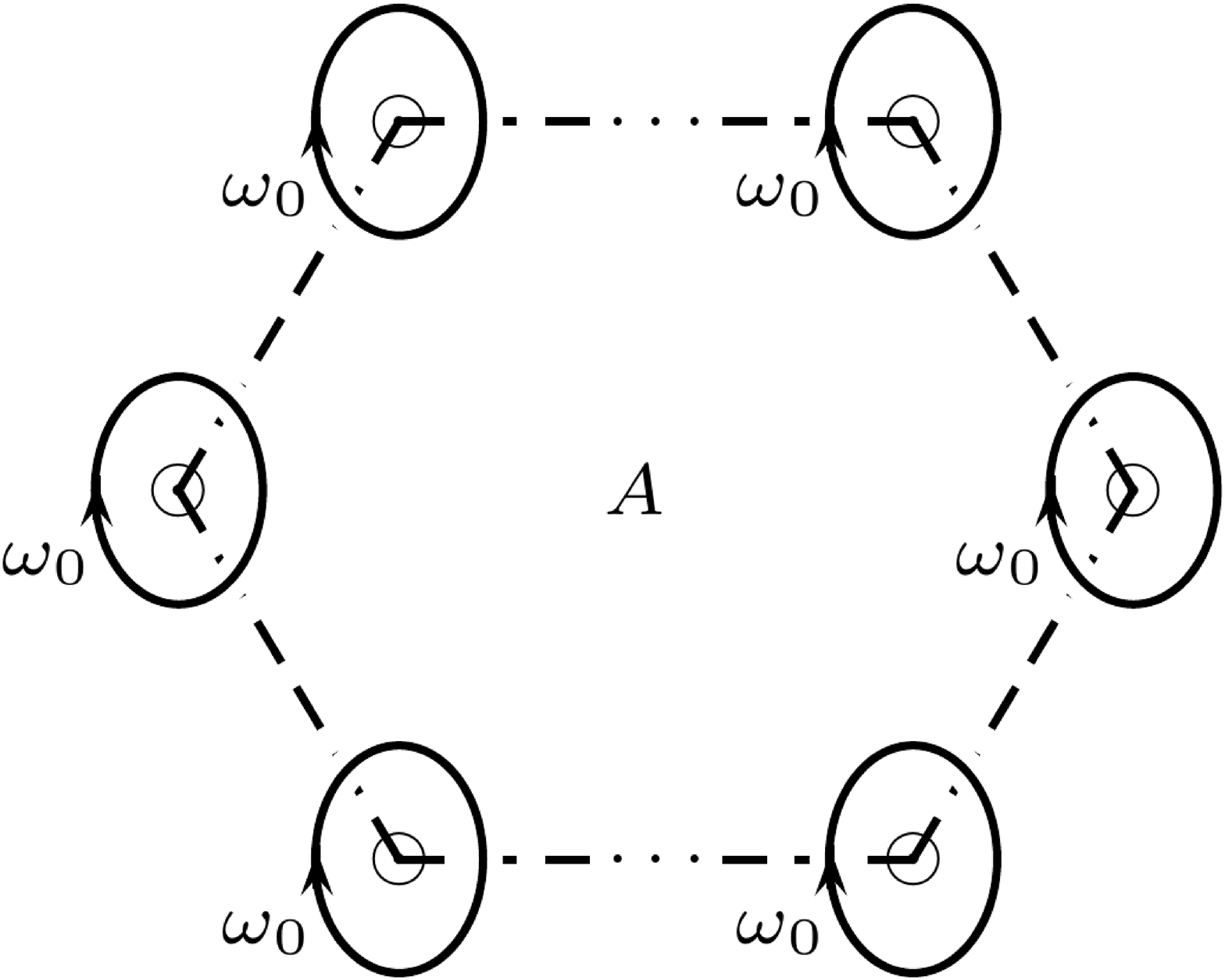} \end{gathered}  \label{eq:bipartite-omega0-loops}
\end{align}
The small circles encircled by $\omega_0$ loops correspond to the wormhole throats.
The $\mathcal{D}^{n-1}$ prefactor is included to ensure the state is properly normalized (the exponent is $n-1$ rather than $n$, as inserting $n$ wormholes between two spheres results in a genus $n-1$ manifold). 
The dashed line indicates the entanglement cut; the region enclosed by the cut corresponds to subregion $A$. This diagram amounts to a bird's-eye view of Fig. \ref{fig:bipartition}(c), which has $n=3$. 

Our goal now is to manipulate the diagram such that a single anyon line threads each component of the boundary of $A$, namely each portion of the dashed line connecting two wormhole throats. We begin by first deforming the entanglement cut, such that the $\omega_0$ loops are all placed on a single line: 
\begin{align}
\ket{\psi} &= \mathcal{D}^{n-1} \begin{gathered}
\includegraphics[height=5em]{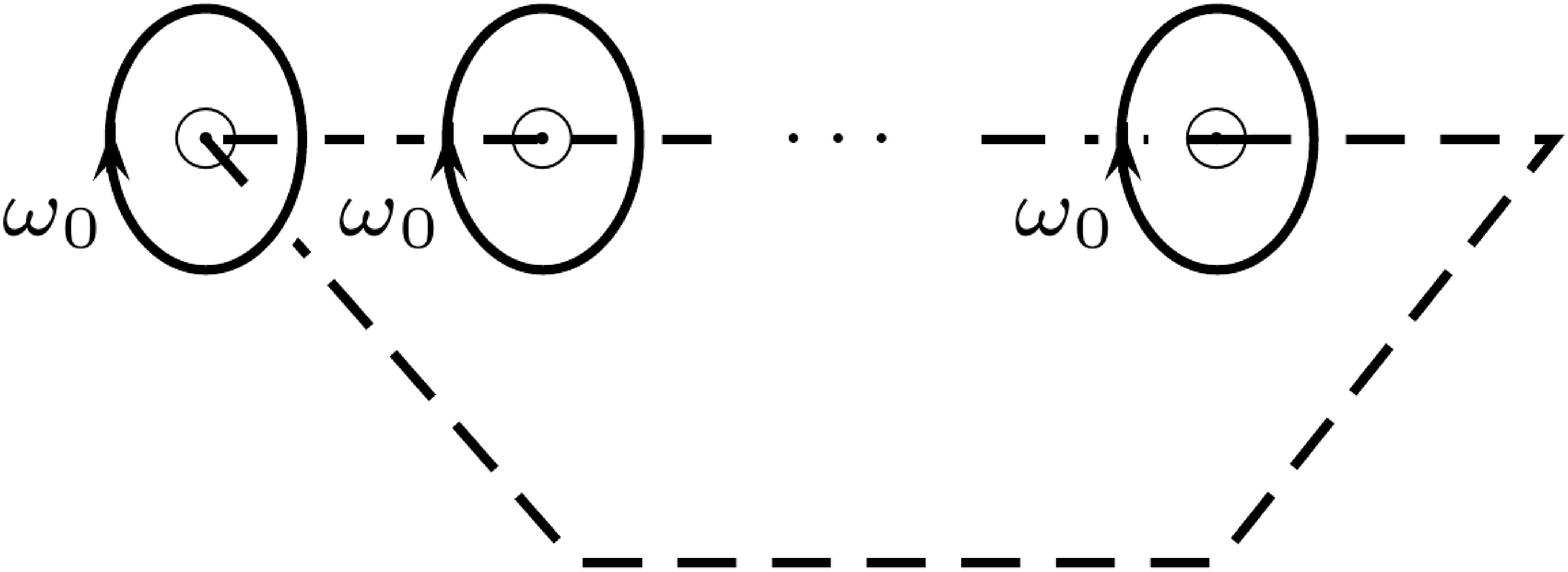} \end{gathered} 
\end{align}
Next, we make use of the \emph{handle-slide} property of the $\omega_0$ loops [see Eq. \eqref{eq:handle-slide}], which allows us to move any given anyon line through a contractible cycle which is encircled by an $\omega_0$ loop. In particular, we can expand the left-most $\omega_0$ loop, passing it through the others, until it encloses nothing (recall that we are working on the sphere). An $\omega_0$ loop enclosing nothing evaluates to unity, leaving us with,
\begin{align}
\ket{\psi} &= \mathcal{D}^{n-1} \begin{gathered}
\includegraphics[height=5em]{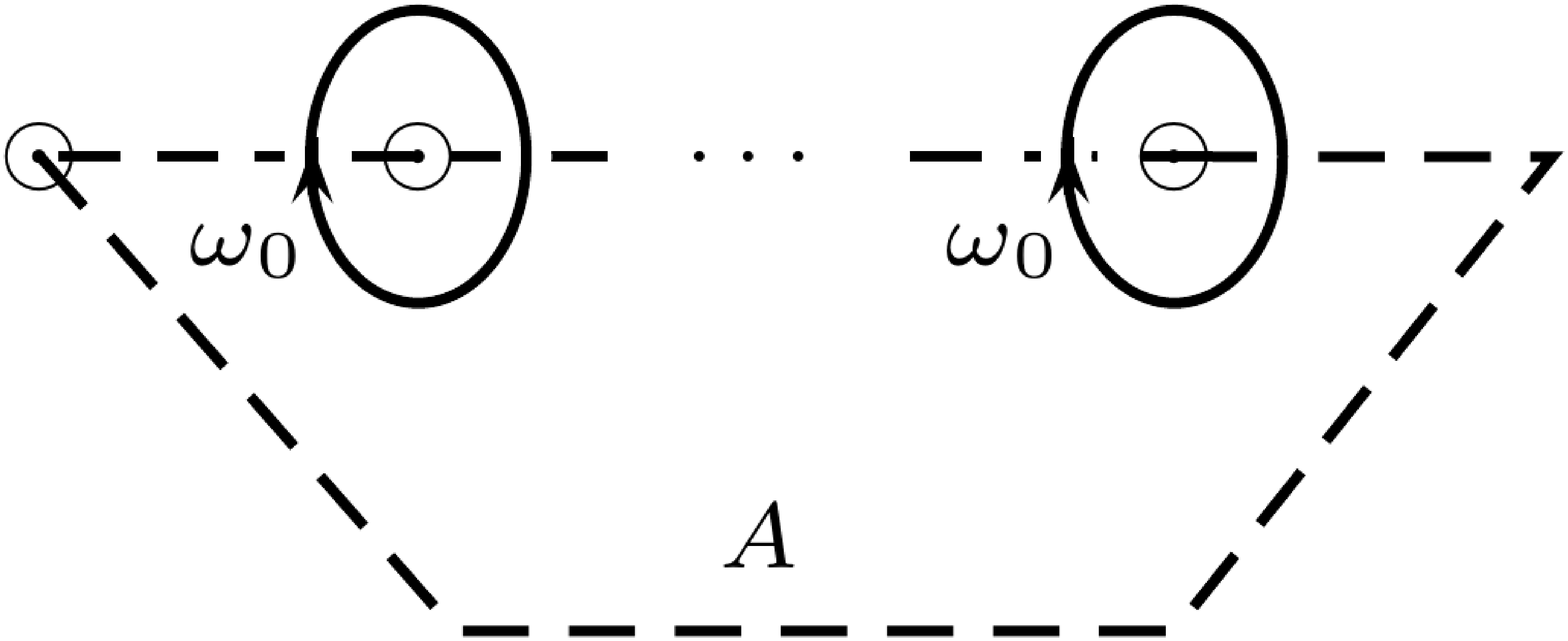} \end{gathered} 
\end{align}
Using the definition of the $\omega_0$ loops, we have
\begin{align}
\ket{\psi} &= \frac{1}{\mathcal{D}} \sum_{\vec{e}} \frac{d_{\vec{e}}}{\mathcal{D}^n} \begin{gathered}
\includegraphics[height=5em]{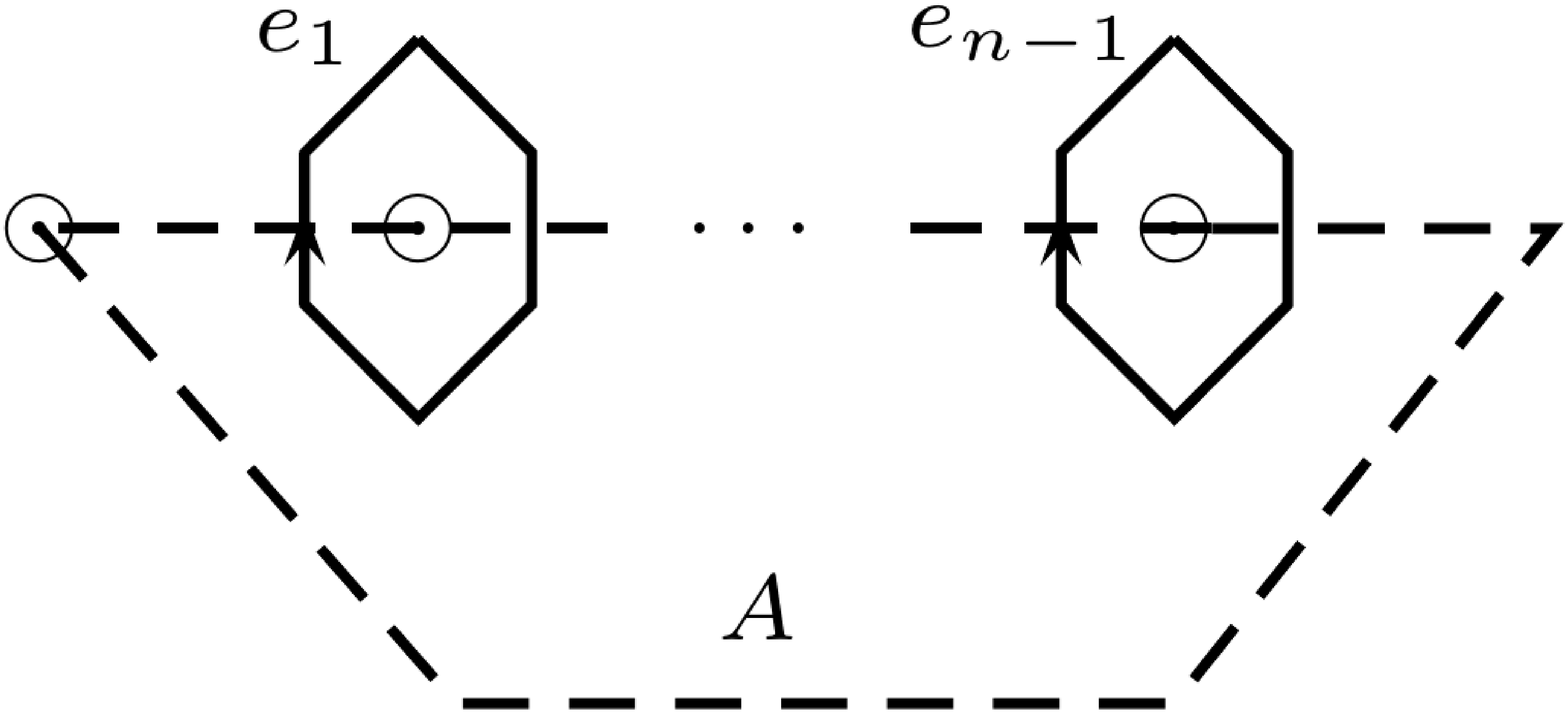} \end{gathered} 
\end{align}
where $\vec{e} = (e_1 , \dots , e_{n-1})$ and we have used the shorthand $d_{\vec{e}} = d_{e_1} \cdots d_{e_{n-1}}$. 

\begin{figure}
  \centering
    \includegraphics[width=0.35\textwidth]{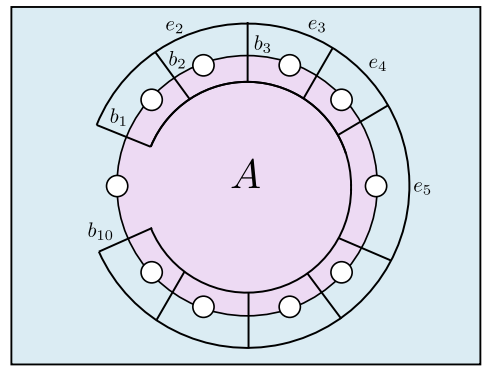}
    \caption{A view from the top down of the anyon diagram describing the ground state after applying $F$-moves and simplifying, as depicted in Eq. \eqref{eq:bipartite-diagram}.} \label{fig:bipartition-diagram}
\end{figure}

Next, by applying resolutions of identity [see Eq. \eqref{eq:resolution-identity}] to neighboring $e_j$ lines, we obtain
\begin{align}
	\ket{\psi}     &=\frac{1}{\mathcal{D}}\sum_{\vec{e}, \vec{b}} \frac{\sqrt{d_{\vec{b}}}}{\mathcal{D}^n}
\begin{gathered}
\includegraphics[height=8em]{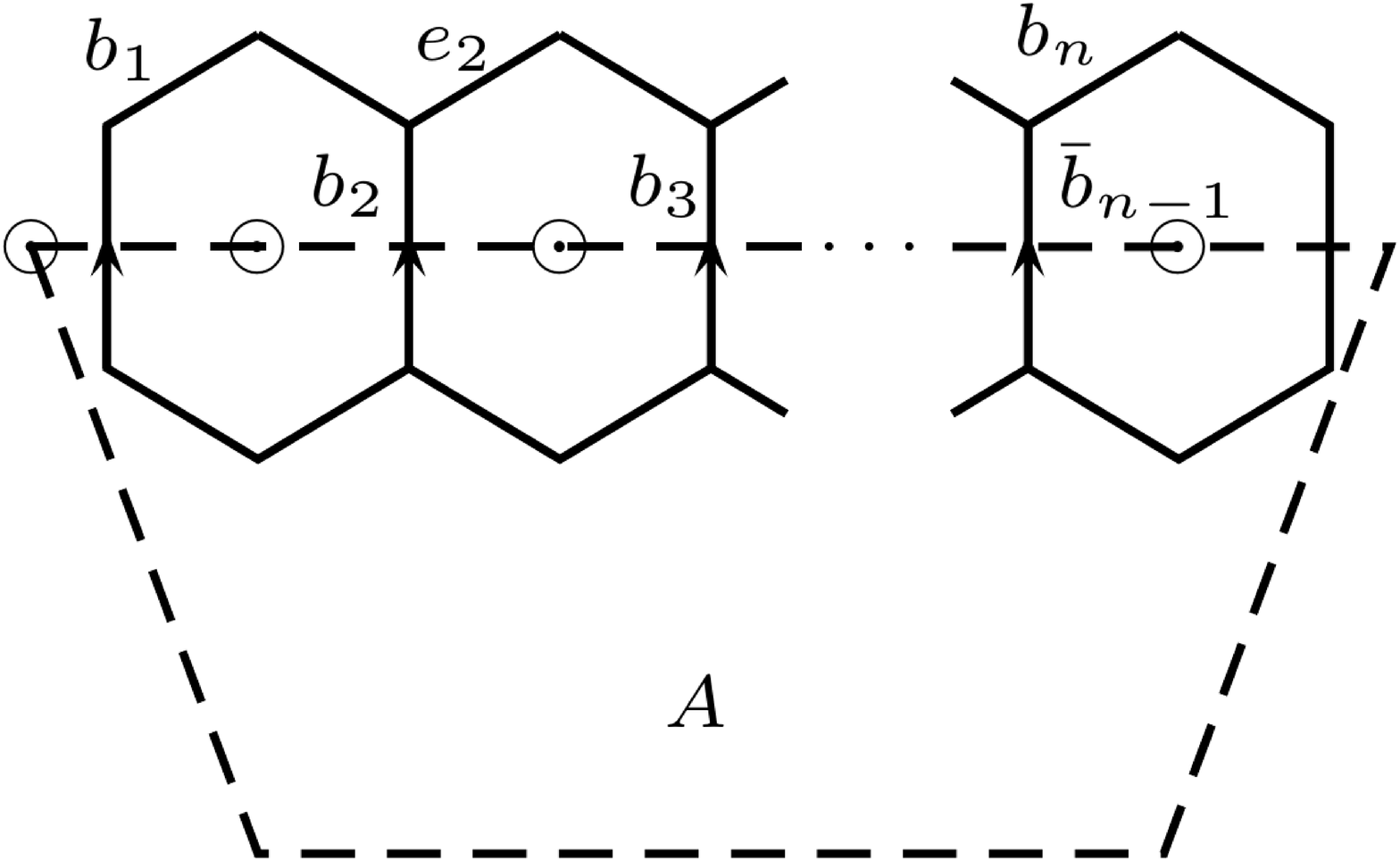} \end{gathered} 
\end{align}
Note that we have also relabeled $e_1 \mapsto b_1$ and $\bar{e}_{n-1} \mapsto b_n$ such that $\vec{b} = (b_1 , \dots , b_n)$ and $\vec{e} = (e_2 , \dots , e_{n-2})$. The upshot of these manipulations is that
the state is now in a form such that a single anyon line, one of the $b_j$, threads each component of the boundary between $A$ and $B$, as desired. For now, we have suppressed the fusion vertex labels to limit clutter. Finally, we can ``straighten" out the diagram into a tree-like form:
\begin{align}
\ket{\psi} &= \frac{1}{\mathcal{D}}\sum_{\vec{e}, \vec{b}} \frac{\sqrt{d_{\vec{b}}}}{\mathcal{D}^n}
  \begin{gathered}
\includegraphics[height=15em]{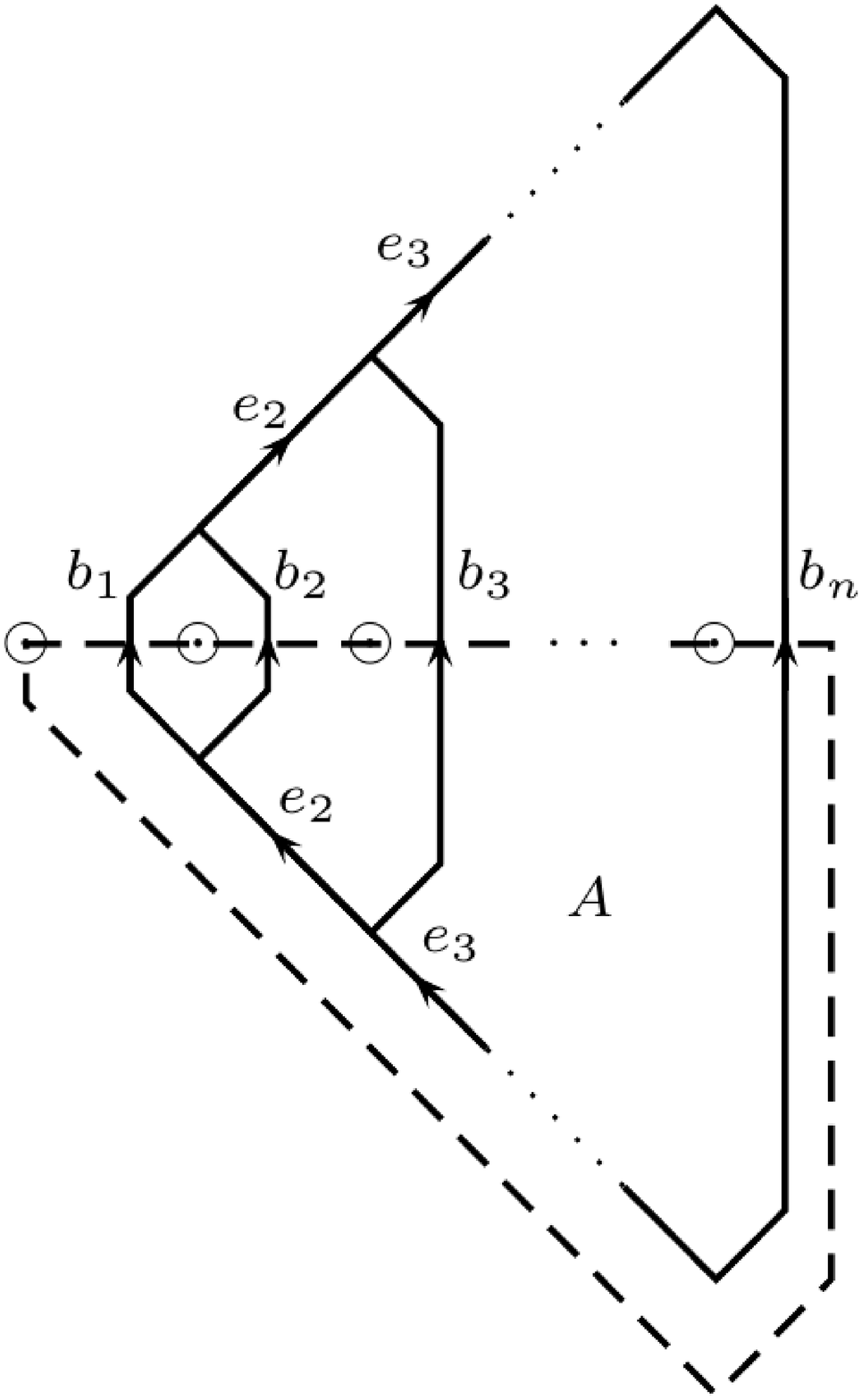} \end{gathered}  \label{eq:bipartite-diagram}
\end{align}
Straightening out the anyon lines requires applying a series of ``bending", or $A$-moves. The $A$-moves applied to vertices on the bottom of the hexagons are the Hermitian conjugates of those applied to the vertices on the top of the hexagons. As the $A$-symbols are unitary matrices, they cancel out in the final expression, yielding the tree diagram as presented.
For the sake of clarity, we present the resulting anyon diagram on top of the original bipartition in Fig. \ref{fig:bipartition-diagram} for $n=10$.%\footnote{We note that we could have arrived at this final expression without having used the handle-slide property in the first step, as was done in Ref. \cite{Bonderson2017}, which slightly modifies the intervening steps.}

\subsubsection{Entanglement Entropy}

Now, to compute the entanglement entropy, we must first construct the reduced density matrix for $A$. 
Following the prescription described above, we first cut the diagram along the entanglement cut:
\begin{align}
\begin{split}
   & |\psi_{\text{cut}}\rangle = \sum_{ \vec{b}, \vec{e}} \frac{1}{\mathcal{D}^{n-1}}
\begin{gathered}
\includegraphics[height=6em]{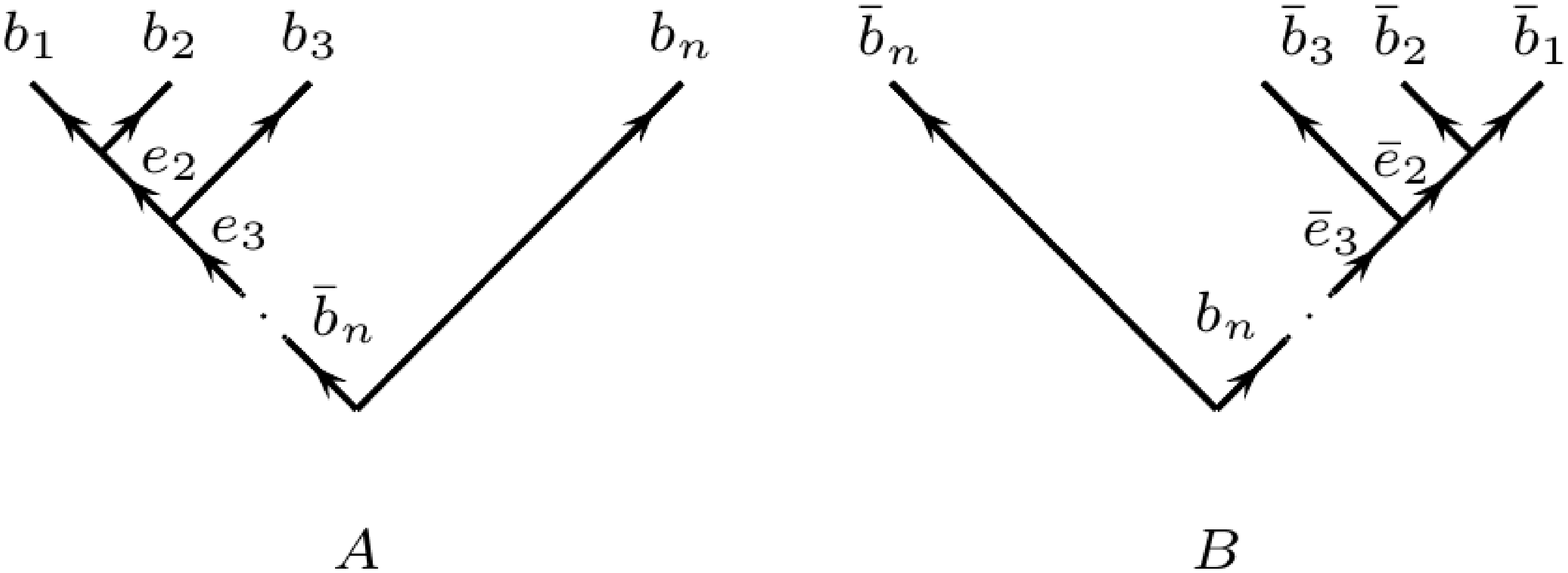} \end{gathered} 
\end{split}
\end{align}
Here and in what follows, a ``cut" label on a ket or density matrix will indicate that the corresponding diagram has been cut along the the relevant entanglement cut.
Recall that, for each new open $a$ line in the cut diagram, we must introduce a factor of $d_a^{-1/4}$ to maintain the correct normalization. Let us also pause here to highlight the physical meaning of this diagram. The $b_j$ anyons are the boundary charges; that is to say, the anyon charge of the $j^{th}$ puncture on the boundary between $A$ and $B$. The above diagram indicates that if a charge $b_j$ exits the $j^{th}$ puncture from region $A$ then, by conservation of topological charge, the conjugate charge $\overline{b}_j$ must exit the $j^{th}$ puncture from region $B$. Moreover, the diagram tells us that the net fusion channel of the $b_j$ anyons is trivial. This reflects the physical requirement that there is a vanishing net anyon charge in each of regions $A$ and $B$. 

From here, we may write out the full density matrix,
\begin{align}
\begin{split}
	&\trho_{\mathrm{cut}} = \sum_{ \substack{ \vec{b}, \vec{e} \\ \vec{b}', \vec{e}' } } \frac{1}{\mathcal{D}^{2n-2}}
\begin{gathered}
\includegraphics[height=10em]{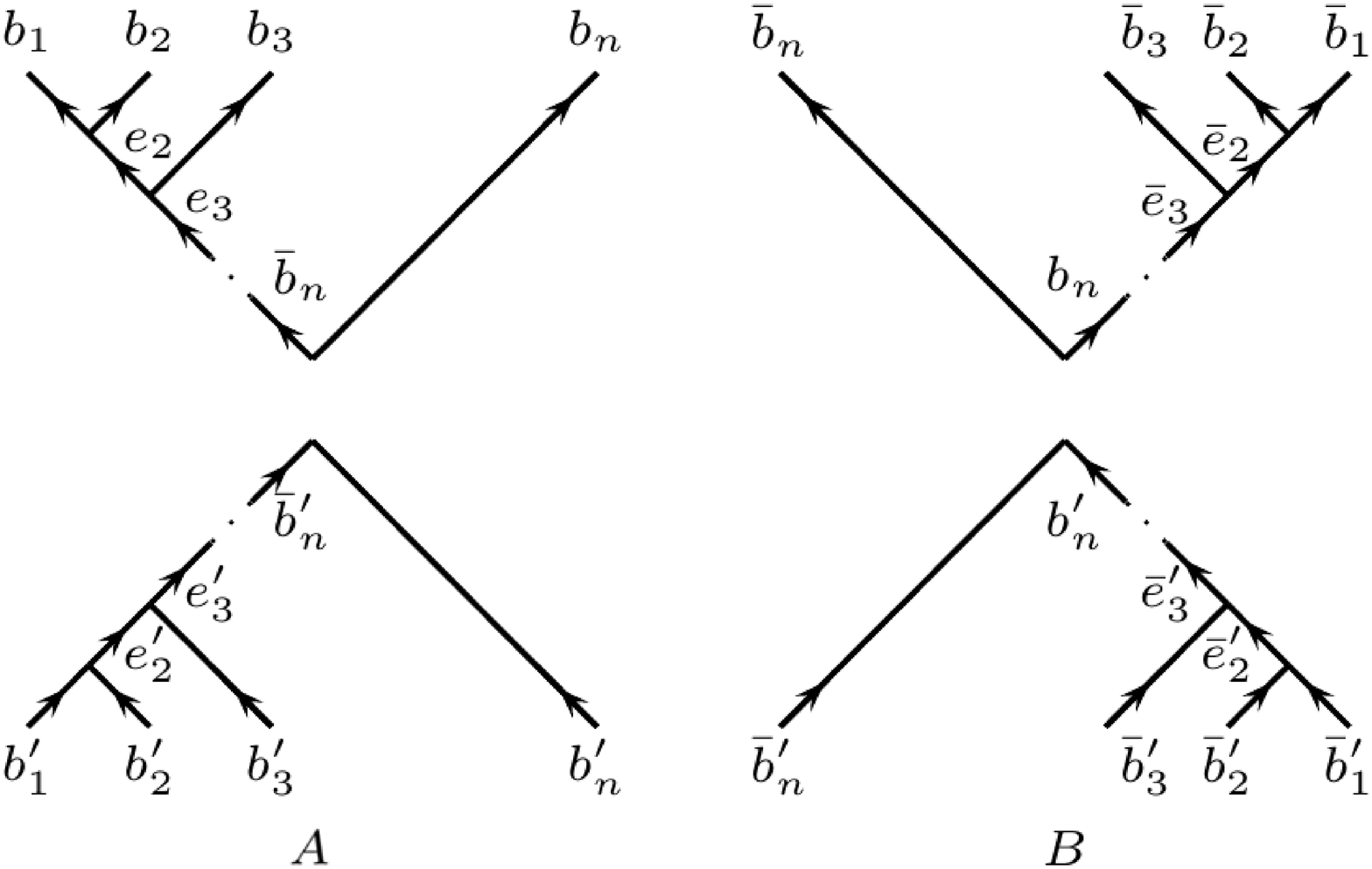} \end{gathered} 
\end{split}
\end{align}
and trace out $B$ using Eq. \eqref{eq:trace} to obtain the reduced density matrix for $A$:
\begin{equation}
\trho_A= \sum_{\vec{b}, \vec{e},\vec{\mu}} \frac{\sqrt{d_{\vec{b}}}}{\mathcal{D}^{2n-2}}
\begin{gathered}
\includegraphics[height=10em]{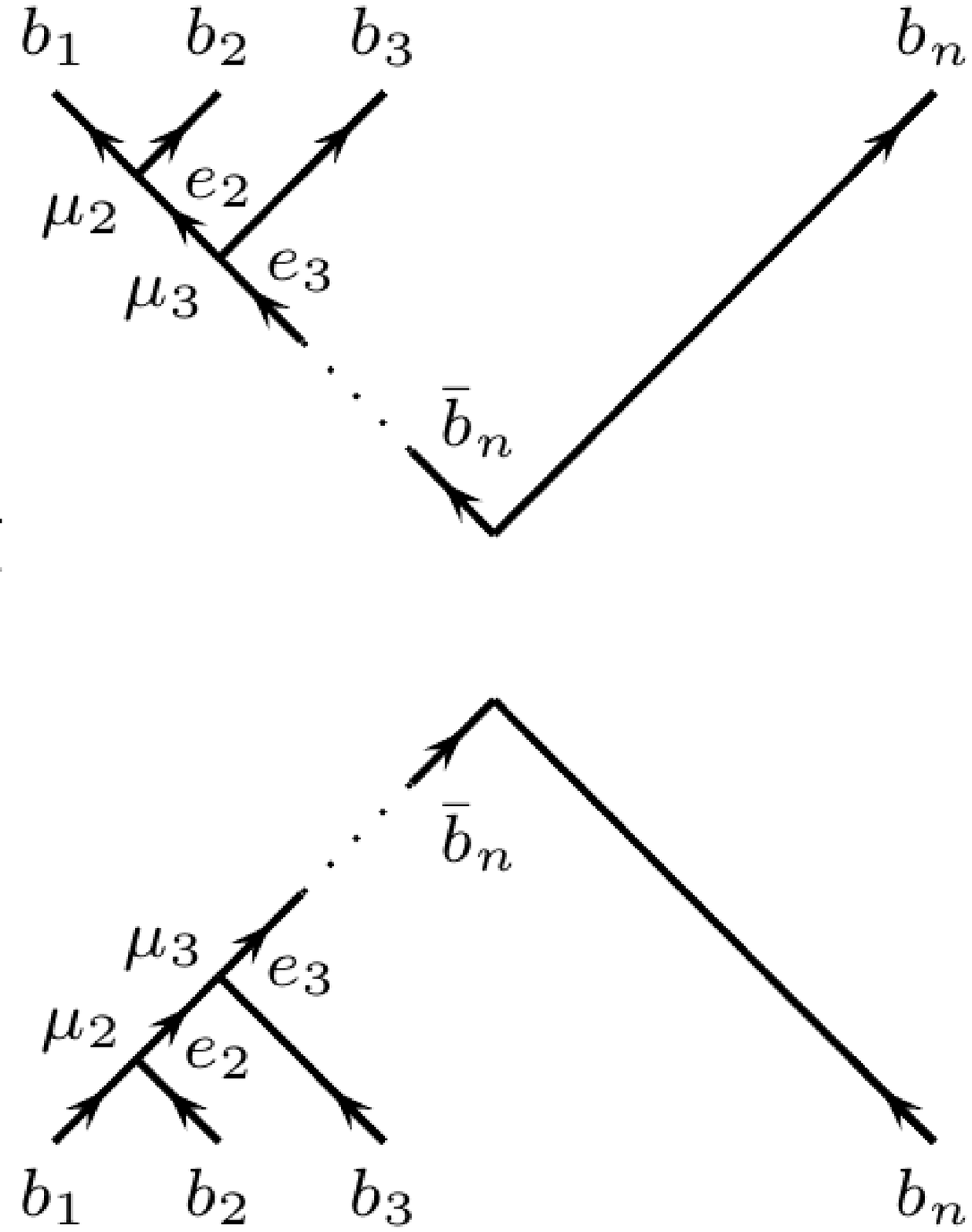}. \end{gathered} . \label{eq:bipartite-rhoA}
\end{equation}
Note that the partial trace introduces a factor $\sqrt{d_{\vec{b}}}$ and sets $b_j = b_j'$ and $e_j = e_j'$ for all $j$. As a result, the density matrix is now diagonal in this basis. Here we have restored the fusion vertex labels $\mu_j$ with $j \in \{ 2, \dots , n-1 \}$. For instance, we have that $e_2$ splits into $b_1$ and $b_2$ at $\mu_2$, $e_3$ splits into $e_2$ and $b_3$ at $\mu_3$, and so on. In order to evaluate the R\'enyi entropies, we evaluate the powers of the reduced density matrix. Multiplication of anyon diagrams is accomplished simply by stacking them. In particular, using Eq. \eqref{eq:overlap}, one readily finds,
\begin{equation}
\begin{split}
\left( \trho_A\right)^{\alpha}= \sum_{\vec{b},\vec{e},\vec{\mu}} &\frac{1}{\sqrt{d_{\vec{b}}}}\left( \frac{d_{\vec{b}}}{ \mathcal{D}^{2n-2}}\right)^{\alpha}
\psscalebox{.85}{\begin{pspicture}[shift=-2.6](0.5,-4.7)(3.3,.5)
        \scriptsize
        \psline[ArrowInside=->](.3,-.3)(0,0)\rput(0,.2){$b_1$}
        \psline[ArrowInside=->](.6,-.6)(.3,-.3)\rput(.6,-.35){$e_2$}
        \rput(.1,-.45){$\mu_2$}
        \psline[ArrowInside=->](.3,-.3)(.6,0)\rput(.6,.2){$b_2$}
        \psline[ArrowInside=->](.9,-.9)(.6,-.6)\rput(.9,-.65){$e_3$}
        \rput(.4,-.75){$\mu_3$}
        \psline(.6,-.6)(1.2,0)\psline[ArrowInside=->](.9,-.3)(1.2,0)\rput(1.2,.2){$b_3$}
        \psline[linestyle=dotted](.9,-.9)(1.3,-1.3)
        \psline[ArrowInside=->](1.6,-1.6)(1.3,-1.3)\rput(1.55,-1.2){$\bar{b}_{n}$}        \psline(1.6,-1.6)(3.2,0)\psline[ArrowInside=->](2.9,-.3)(3.2,0)\rput(3.2,.2){$b_n$}
        %%%
        \rput(0,.6){
        \psline[ArrowInside=->](0,-4.4)(.3,-4.1)\rput(0,-4.6){$b_1$}
        \psline[ArrowInside=->](.3,-4.1)(.6,-3.8)\rput(.6,-4.1){$e_2$}
        \rput(.1,-4){$\mu_2$}
        \psline[ArrowInside=->](.6,-4.4)(.3,-4.1)\rput(.6,-4.6){$b_2$}
        \psline[ArrowInside=->](.6,-3.8)(.9,-3.5)\rput(.9,-3.8){$e_3$}
        \rput(.4,-3.7){$\mu_3$}
        \psline(.6,-3.8)(1.2,-4.4)\psline[ArrowInside=->](1.2,-4.4)(.9,-4.1)\rput(1.2,-4.6){$b_3$}
        \psline[linestyle=dotted](.9,-3.5)(1.3,-3.1)
        \psline[ArrowInside=->](1.3,-3.1)(1.6,-2.8)\rput(1.55,-3.3){$\bar{b}_{n}$}
        \psline(3.2,-4.4)(1.6,-2.8)\psline[ArrowInside=->](3.2,-4.4)(2.9,-4.1)\rput(3.2,-4.6){$b_n$}
        }
    \end{pspicture}}.
\end{split}
\end{equation}
On evaluating the quantum trace, we find
\begin{equation} 
\begin{split}
\Tr \left( \trho_A\right)^\alpha & =\sum_{\vec{b},\vec{e},\vec{\mu}} \left( \frac{d_{\vec{b}}}{\mathcal{D}^{2n-2}}\right)^\alpha
= \sum_{\vec{b}} N_{b_1\dots b_n}^{0} \left( \frac{d_{\vec{b}}}{\mathcal{D}^{2n-2}}\right)^{\alpha}.
\end{split}
\end{equation}
In the second equality, we made use of the fact that the sum over $\vec{\mu}$ yields the fusion multiplicity of all the $\vec{b}$ anyons fusing to the identity (for instance, $\sum_{\mu_2} = N_{b_1 b_2}^{e_2}$) and we have used the shorthand notation
\begin{align}
	 N^{0}_{b_1\dots b_n} = N_{b_1 b_2}^{e_2}N_{e_2 b_3}^{e_3}\dots N^0_{\bar{b}_n,b_n}.
\end{align}
We then find the $\alpha$-R\'enyi entropy to be
\begin{equation}
S^{(\alpha)}_A = \frac{1}{1-\alpha} \ln\Big[ \sum_{\vec{b}} N^{0}_{b_1\dots b_n} \left( \frac{d_{\vec{b}}}{\mathcal{D}^{2n-2}}\right)^\alpha \Big], \label{eq:bipartition-renyi-unsimplified}
\end{equation}
Taking the replica limit $\alpha \to 1$, the entanglement entropy is then given by
\begin{align}
S_A &= -  \sum_{\vec{b}} N^{0}_{b_1\dots b_n}  \ln\Big[ \frac{d_{\vec{b}}}{\mathcal{D}^{2n-2}} \Big] 
\end{align}
Via repeated application of Eq. \eqref{eq:quantum-dimension-relation}, we have that
\begin{align}
	\sum_{\vec{b}, \vec{e}} N_{b_1 b_2}^{e_2}N_{e_2 b_3}^{e_3}\dots N^0_{\bar{b}_n,b_n} d_{b_1} \dots d_{b_n} = \mathcal{D}^{2(n-1)}.
\end{align}
Hence, the entanglement entropy reduces to
\begin{align}
	S_A = -n \sum_b \frac{d_b^2}{\mathcal{D}^2} \ln \left( \frac{d_b}{\mathcal{D}^2}\right) - 2\ln \mathcal{D} \label{eq:bipartition-vN}
\end{align}
The first term corresponds to the area law, as the length of the boundary is proportional to $n$, the number of punctures on the boundary between $A$ and $B$. 
The second term is the expected topological entanglement entropy. 
As explained above, this expression actually gives the entanglement entropy of $\mathcal{C} \times \overline{\mathcal{C}}$. To obtain the  entanglement entropy for just $\mathcal{C}$, we need only divide this expression by two.
 We note that, in the area law term, the argument of the sum is interpreted in Ref. \cite{Bonderson2017} as the entropy of a single ``boundary anyon". Indeed, the first term is a multiple of the entanglement entropy for superposition of anyon-antianyon pairs in Eq. \eqref{eq:dimer-example}. This leads to the appealing interpretation of the entanglement entropy as arising from the entanglement between anyon-antianyon pairs forming a condensate at the entanglement cut; the topological entanglement entropy arises from the global constraint that the net fusion channel of the anyons within a subregion must be trivial. %This is the entropy one obtains if one models the boundary as a condensate of $n$ anyon-antianyon pairs.

The expression Eq. \eqref{eq:bipartition-renyi-unsimplified} for the R\'enyi entropy does not make manifest the area law and topological contributions. Following Ref. \cite{Bonderson2017}, we can massage this expression to make these contributions more explicit. First, we introduce the matrix
\begin{align}
	[K_\alpha]_{ee'} \equiv \sum_b N^{e'}_{eb} d_b^\alpha. \label{eq:K-matrix}
\end{align}
It is clear that $K_\alpha$ is symmetric and has all real entries and is thus diagonalizable. Moreover, all its entries are strictly positive. Thus, by the Perron-Frobenius theorem, $K_\alpha$ has a unique eigenvector whose entries can be chosen to be all real and positive and whose corresponding eigenvalue is larger in magnitude than all other eigenvalues. We write
\begin{align}
	[K_\alpha]_{ee'} = \sum_\mu \kappa_{\alpha,\mu} [v_{\alpha,\mu}]_e [v_{\alpha,\mu}]_{e '}^*,
\end{align}
where we take $\mu=0$ to correspond to the above mentioned unique eigenvector. Now, we see that $[v_\alpha]_e = d_e /\mathcal{D}$ is a normalized eigenvector with all real entries and, as such, is the unique $\mu=0$ eigenvector with eigenvalue 
\begin{align}
	\kappa_{\alpha, 0} = \sum_e d_e^{1+\alpha}.
\end{align}
We then have,
\begin{align}
	\sum_{\vec{b}, \vec{e}} N_{b_1 b_2}^{e_2}N_{e_2 b_3}^{e_3}\dots N^0_{\bar{b}_n,b_n} d_{b_1}^\alpha \dots d_{b_n}^\alpha %&= \sum_{b_1, b_2, \vec{e}} [K_\alpha]_{b_1 e_2} [K_\alpha]_{e_2 b_3} \dots [K_\alpha]_{e_{n-2} \bar b_n} d_{b_1}^\alpha d_{b_n}^\alpha \\
	%&= \sum_{b_1, b_2, \vec{e}, a_1, a_2} N^0_{a_1 b_1} [K_\alpha]_{a_1 e_2} [K_\alpha]_{e_2 b_3} \dots [K_\alpha]_{e_{n-2} \bar a_2} N^{\bar b_n}_{\bar a_2 0}  d_{b_1}^\alpha d_{b_n}^\alpha \\
	&= [K^n_\alpha]_{00}.
\end{align}
%where we have used the fact $N_{0a}^{b} = \delta_{a b}$. 
Plugging this back into the expression for the R\'enyi entropy, we find
\begin{equation}
S^{(\alpha)}_A = \frac{1}{1-\alpha} \ln\Big[ [K^n_\alpha]_{00}\left( \frac{1}{\mathcal{D}^{2n-2}}\right)^\alpha \Big].
\end{equation}
Using the diagonal form of $K_\alpha$, we have that 
\begin{align}
	[K_\alpha^{n}]_{0 0} &= \frac{\kappa_{\alpha,0}^{n}}{\mathcal{D}^2} +  \sum_{\mu \neq 0} \kappa_{\alpha,\mu}^{n} [v_{\alpha,\mu}]_{0}[v_{\alpha,\mu}]_{0}^* \equiv  \frac{\kappa_{\alpha,0}^{n}}{\mathcal{D}^2}  e^{F (n,0,K_\alpha)}
\end{align}
where we have made use of the explicit form of the $\mu=0$ eigenvector and defined
\begin{align}
	F (n,c,K_\alpha) \equiv \ln\left(1 + \frac{\mathcal{D}^2}{d_{c}} \sum_{\mu \neq 0} \left(\frac{\kappa_{\alpha,\mu}}{\kappa_{\alpha,0}}\right)^{n} [v_{\alpha,\mu}]_{c}[v_{\alpha,\mu}]_{0}^* \right).
\end{align}
Since $\kappa_{\alpha,0}$ is the largest eigenvalue, we have that $F(n,c,K_\alpha) \to 0$ in the thermodynamic limit in which we take the number of wormholes $n \to \infty$. We thus find
\begin{align}
	S^{(\alpha)}_A = \frac{n}{1-\alpha} \ln \frac{\kappa_{\alpha,0}}{\mathcal{D}^{2\alpha}} - \ln \mathcal{D}^2 \label{eq:bipartition-renyi}
\end{align}
The first term is the area law term and the second the topological contribution. 

\subsubsection{Negativity}

Next, we employ the proposed anyonic partial transpose of Ref. \cite{Shapourian2020} to compute the negativity. 
We first note that the density matrix, after performing the cut along the entangling surface, has the form of a sum of tensor products between two anyonic operators,
\begin{align}
	\trho_{\mathrm{cut}} = \sum_{ \substack{ \vec{b} , \vec{e}, \vec{\mu} \\ \vec{b}' , \vec{e}' , \vec{\mu}' } } 	\frac{1}{\mathcal{D}^{2n-2}} \trho_A^{(\vec{b} , \vec{e} , \vec{\mu}, \vec{b}' , \vec{e}', \vec{\mu}' )} \otimes \trho_B^{(\vec{b} , \vec{e} , \vec{\mu} , \vec{b}' , \vec{e}', \vec{\mu}')},
\end{align}
where $\trho_A$ and $\trho_B$ act only on regions $A$ and $B$ and we have schematically represented their dependence on the anyons and fusion vertices in their superscripts. Hence, taking the partial transpose amounts to taking the full transpose of $\trho_A^{(\vec{b} , \vec{e} , \vec{\mu}, \vec{b}' , \vec{e}', \vec{\mu}' )}$. Explicitly,
\begin{align}
	\trho_{A}^{T} = 
\begin{gathered}
\includegraphics[height=13em]{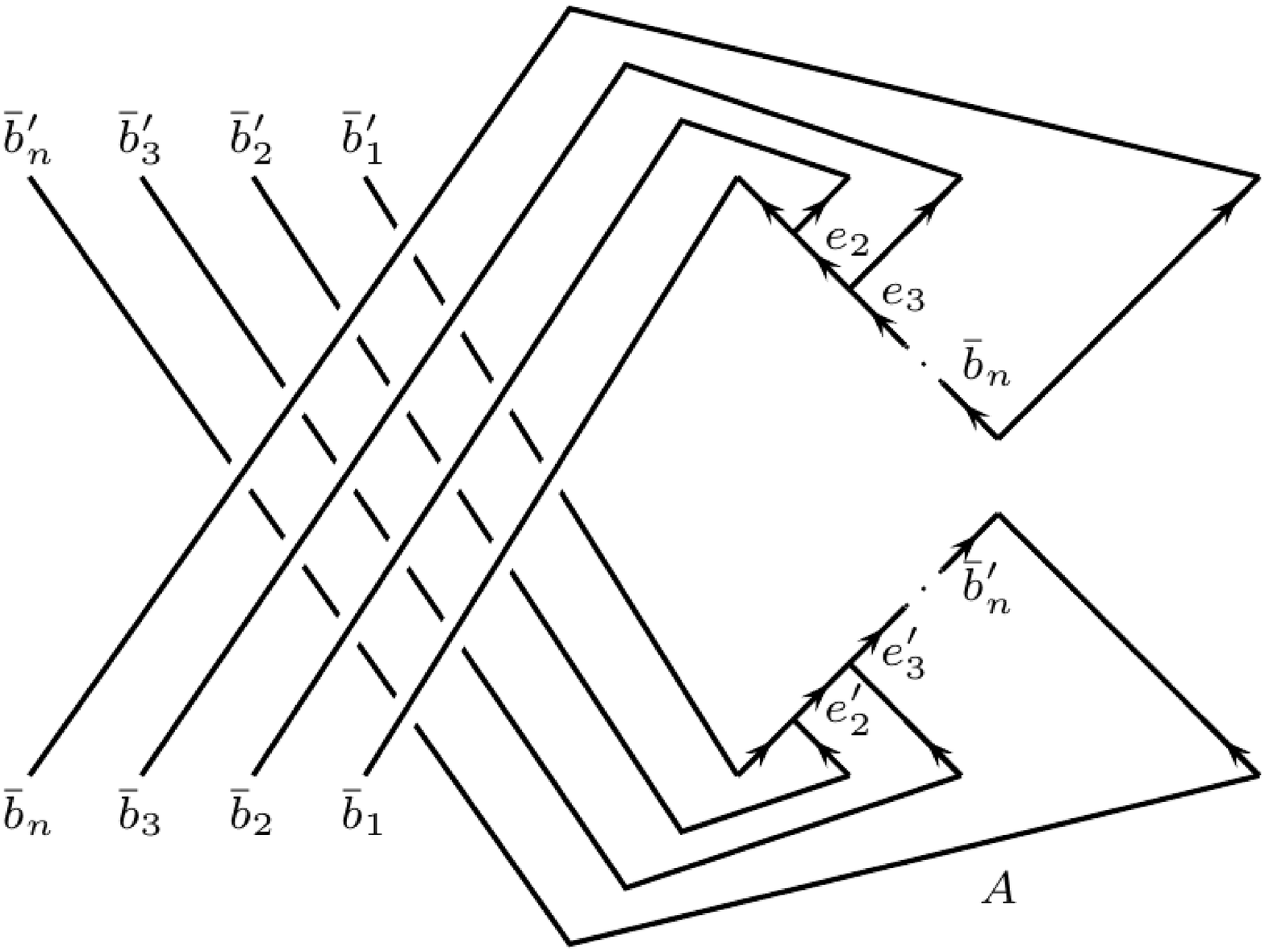} \end{gathered} 
\end{align}
Note that, in $\trho^{T_A}$, the bra of $A$ is labeled by $\bar b$ charges whereas the bra of $B$ is labeled by $\bar b'$ charges and vice versa for the kets. 
As noted above, $\trho^{T_A}_{\mathrm{cut}}$ need not be Hermitian, and so when applying the replica trick, rather than computing $(\trho^{T_A})^{n_e}$ for $n_e \in 2\mathbb{Z}$, we compute $(\trho^{T_A} (\trho^{T_A})^\dagger)^{n_e/2}$. We have,
\begin{align}
	\trho^{T_A}_{\mathrm{cut}} (\trho^{T_A}_{\mathrm{cut}})^\dagger = \sum_{ \substack{ \vec{b} , \vec{e}, \vec{\mu} \\ \vec{b}' , \vec{e}' , \vec{\mu}' } } 	\frac{1}{\mathcal{D}^{2n-2}} (\trho_A^{T} (\trho_A^{T})^\dagger) \otimes (\trho_B \trho_B^\dagger),
\end{align}
where we have suppressed the superscripts on $\trho_{A,B}^{(\vec{b} , \vec{e} , \vec{\mu}, \vec{b}' , \vec{e}', \vec{\mu}' )}$ (note that these do not represent the reduced density matrices for $A$ and $B$).  Explicitly,
\begin{align}
	(\trho_{A}^{T})^\dagger = 
	\begin{gathered}
\includegraphics[height=12em]{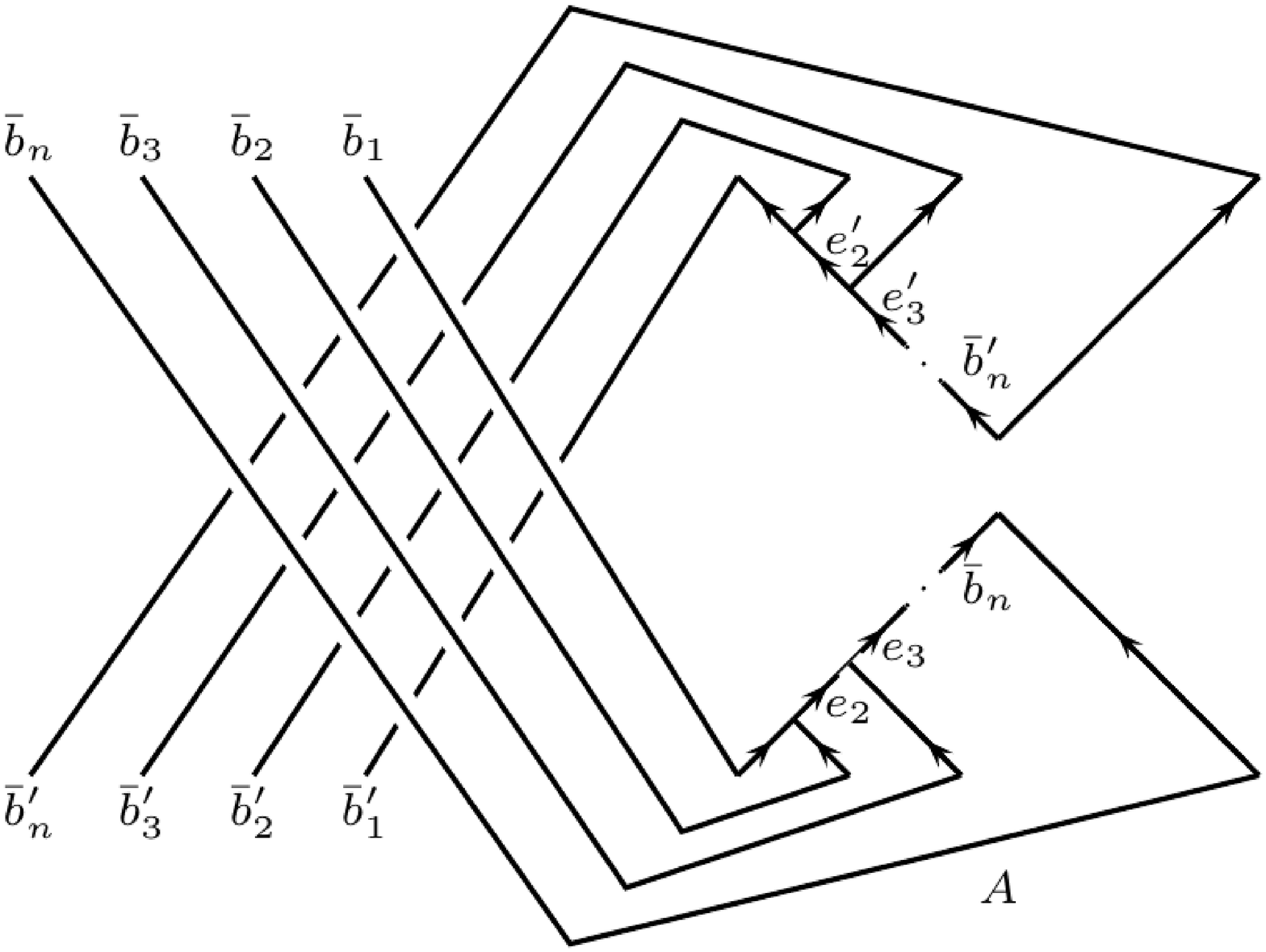} \end{gathered} 
\end{align}
Note that the $b_j$ anyon lines now lie on top of the $b_j'$ lines, as the Hermitian conjugate does not affect the ordering of lines appearing in braids. As such, when computing $\trho^{T_A} (\trho^{T_A})^\dagger$, we can apply Eq. \eqref{eq:trace} without having to apply any braiding operations in evaluating the $A$ part of the diagram while we again use Eq. \eqref{eq:overlap} in evaluating the $B$ part of the diagram. Additionally, the $A$-moves cancel.
Ultimately, we find
\begin{align}
%\begin{split}
	&\trho_{\mathrm{cut}}^{T_A}(\trho_{\mathrm{cut}}^{T_A})^\dagger = \sum_{ \substack{ \vec{b} , \vec{e} \\ \vec{b}' , \vec{e}' } } \frac{d_{\vec{b}} d_{\vec{b}'}}{\mathcal{D}^{4n-4}} \frac{1}{\sqrt{d_{\vec{b}} d_{\vec{b}'} }} \\
	\notag &\times 
		\begin{gathered}
\includegraphics[height=14em]{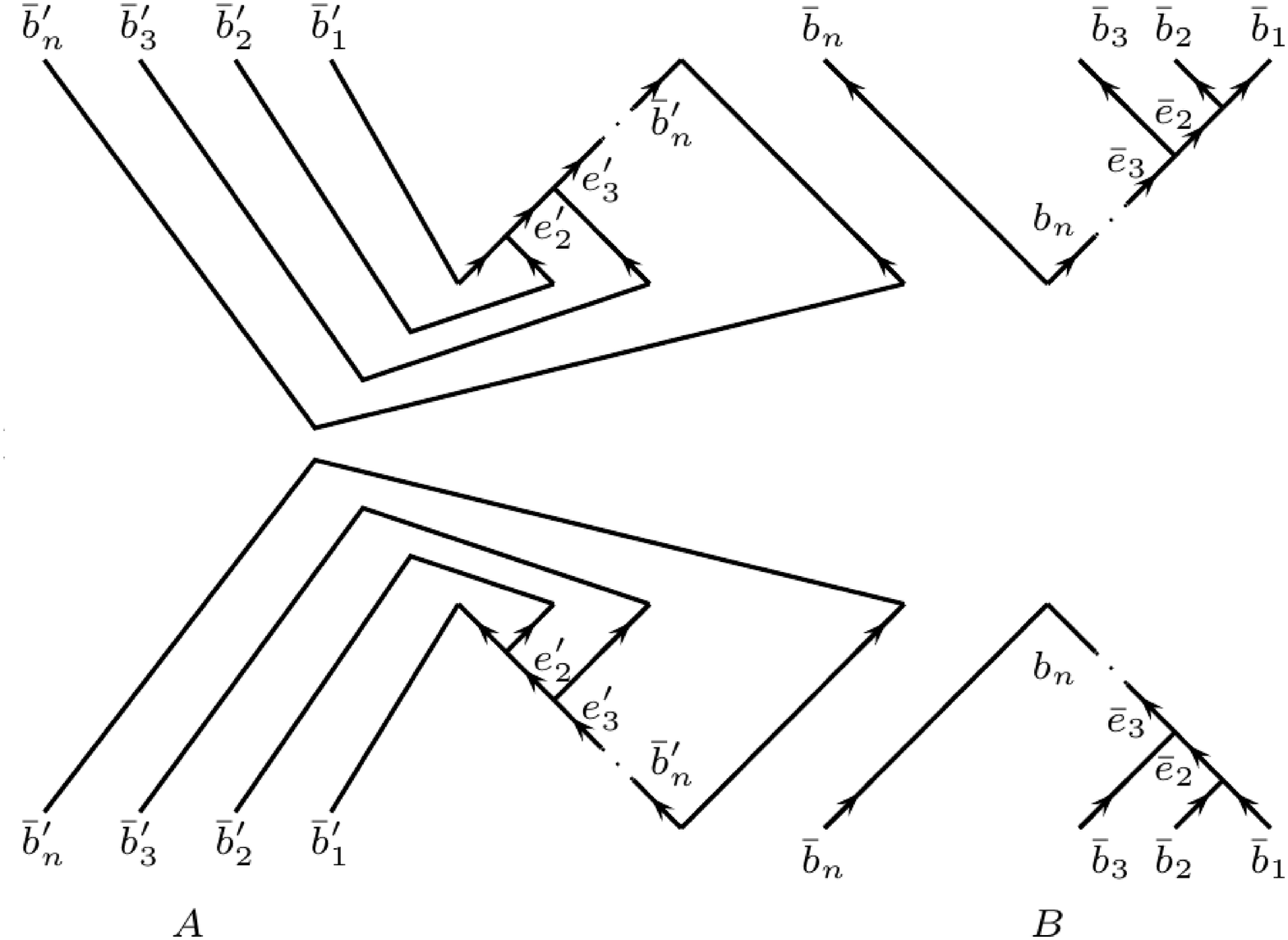} \end{gathered} 
%\end{split}
\end{align}
We see that this operator is now a tensor product of operators in the $A$ and $B$ subregions, each of which is diagonal in this basis. Indeed, since this operator is diagonal, by stacking diagrams, we may readily evaluate for $n_e \in 2 \mathbb{Z}$,
\begin{align}
	\Tr[(\trho_{\mathrm{cut}}^{T_A} (\trho_{\mathrm{cut}}^{T_A})^\dagger)^{n_e/2}] = \sum_{ \substack{ \vec{b}, \vec{e}, \vec{\mu} \\ \vec{b}', \vec{e}', \vec{\mu}' } } \left( \frac{d_{\vec{b}} d_{\vec{b}'}}{\mathcal{D}^{4n-4}} \right)^{n_e/2},
\end{align}
%We have that $\rho^{T_A} = (\rho^{T_A})^\dagger$. We compute
%\begin{align}
%	\rho^{T_A} (\rho^{T_A})^\dagger = 
%\end{align}
%Since this matrix is diagonal, we can read off the singular values and compute
where we have restored the fusion vertex labels. Analytically continuing $n_e \to 1$, we thus find for the logarithmic negativity,
\begin{align}
	\mathcal{E}(A:B) = 2 \ln \left( \sum_{ \vec{b} , \vec{e} , \vec{\mu} } \frac{\sqrt{d_{\vec{b}}}}{\mathcal{D}^{n-1}} \right).
\end{align}
Comparing with the results of the previous section, we see that the logarithmic negativity is precisely the $\alpha = 1/2$ R\'enyi entropy, as expected for a pure state. Explicitly, in the limit $n \to \infty$
\begin{align}
	\mathcal{E}(A:B) =  2n \ln \frac{\kappa_{1/2,0}}{\mathcal{D}} - \ln \mathcal{D}^2  \label{eq:bipartition-negativity}
\end{align}
We again find the area law term and a universal topological contribution expected for the phase $\mathcal{C} \times \overline{\mathcal{C}}$. Dividing this expression by two yields the corresponding negativity for just $\mathcal{C}$. This provides a nontrivial check that the proposal of Ref. \cite{Shapourian2020} indeed provides a consistent extension of the partial transpose to anyonic systems.

\subsubsection{Reflected Entropy}

Finally, let us compute the reflected entropy, using our proposal for the canonical purification. Proceeding with the double replica trick, we first note that $\trho^{\alpha/2} = \trho$ for $\alpha \in 2\mathbb{Z}$, since $\trho$ describes a pure state. We can thus directly construct the canonical purification as
\begin{align}
	&\kket{\trho^{\alpha/2}} = \sum_{ \substack{ \vec{b}, \vec{e}, \vec \mu \\ \vec{b}', \vec{e}', \vec{\mu}' } } \frac{1}{\mathcal{D}^{2n-2}} \begin{gathered} \includegraphics[height=8.5em]{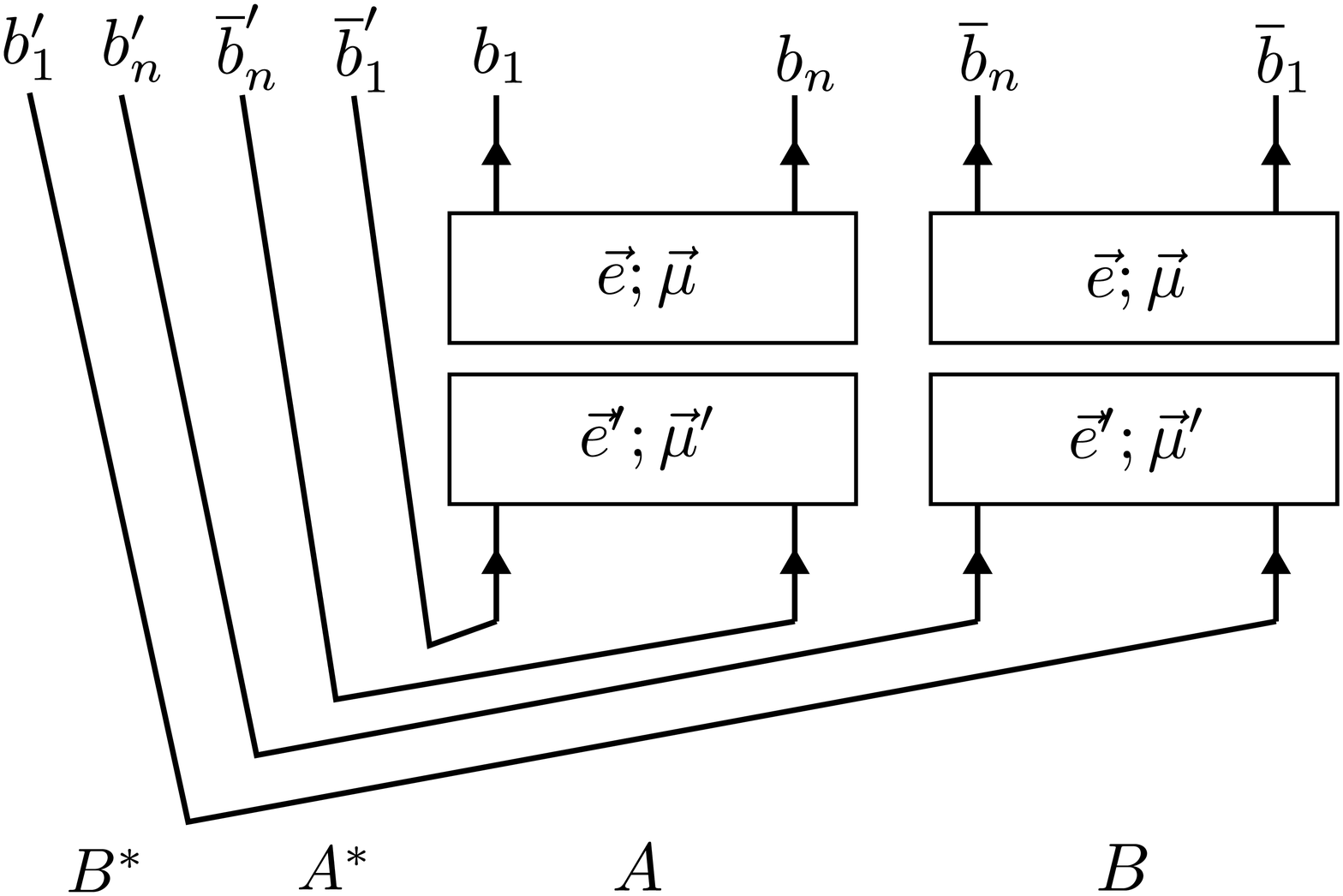} \end{gathered}
\end{align}
Here we have used a shorthand notation for the density matrix to represent the purification, in order to reduce clutter. The boxes with arrows exiting from the top and entering from the bottom represent the ``ket" and ``bra" parts, respectively, of the $A$ and $B$ regions of the diagram and are labeled by the internal anyon lines and fusion vertex labels.
%\end{widetext}
Forming the density matrix $\kket{\trho^{\alpha/2}}\bbra{\trho^{\alpha/2}}$ and tracing out $B$ and $B^*$, we find 
\begin{align}
	&\trho_{AA^*}^{(\alpha)} =  \sum_{ \substack{ \vec{b}, \vec{e} \\ \vec{b}', \vec{e}' } } \frac{\sqrt{d_{\vec{b}} d_{\vec{b}'}}}{\mathcal{D}^{4n-4}} \psscalebox{.85}{\begin{pspicture}[shift=-4.8](0.5,-9.4)(5.4,.3)
        \scriptsize
        %% ket
        %\rput(1.4,-2){A}
        \rput(2.4,0){
        \psline[ArrowInside=->](.3,-.3)(0,0)\rput(0,.2){$b_1$}
        \psline[ArrowInside=->](.6,-.6)(.3,-.3)\rput(.6,-.35){$e_2$}
        %\rput(.1,-.45){$\mu_2$}
        \psline[ArrowInside=->](.3,-.3)(.6,0)\rput(.6,.2){$b_2$}
        \psline[ArrowInside=->](.9,-.9)(.6,-.6)\rput(.9,-.65){$e_3$}
        %\rput(.4,-.75){$\mu_3$}
        \psline(.6,-.6)(1.2,0)\psline[ArrowInside=->](.9,-.3)(1.2,0)\rput(1.2,.2){$b_3$}
        \psline[linestyle=dotted](.9,-.9)(1.1,-1.1)
        \psline[ArrowInside=->](1.4,-1.4)(1.1,-1.1)\rput(1.35,-1.){$\bar{b}_n$}
        \psline(1.4,-1.4)(2.8,0)\psline[ArrowInside=->](2.5,-.3)(2.8,0)\rput(2.8,.2){$b_n$}}
        \rput(2.4,-1.8){\psline[ArrowInside=->](0,-1.4)(.3,-1.1)\psline(0,-1.4)(-1,1.8)\rput(-1,2){$\bar b_1'$}
        \psline[ArrowInside=->](.3,-1.1)(.6,-.8)\rput(.6,-1.05){$e_2'$}
        %\rput(.1,-.45){$\mu_2$}
        \psline[ArrowInside=->](.6,-1.4)(.3,-1.1)\psline(.6,-1.4)(-.3,-1.7)(-1.6,1.8)\rput(-1.6,2){$\bar b_2'$}
        \psline[ArrowInside=->](.6,-.8)(.9,-.5)\rput(.9,-.75){$e_3'$}
        %\rput(.4,-.75){$\mu_3$}
        \psline(.6,-.8)(1.2,-1.4)\psline[ArrowInside=->](1.2,-1.4)(.9,-1.1)\psline(1.2,-1.4)(-.6,-2)(-2.2,1.8)\rput(-2.2,2){$\bar b_3'$}
        \psline[linestyle=dotted](.9,-.5)(1.1,-0.3)
        \psline[ArrowInside=->](1.1,-0.3)(1.4,0.0)\rput(1.35,-0.4){$\bar{b}_n'$}
        \psline(1.4,0)(2.8,-1.4)\psline[ArrowInside=->](2.8,-1.4)(2.5,-1.1)\psline(2.8,-1.4)(-.9,-2.3)(-2.8,1.8)\rput(-2.8,2){$\bar b_n'$}}
        %% bra
        \rput(2.4,-5.5){\rput(-1.9,-3.8){$A^*$}
        \psline[ArrowInside=->](.3,-.3)(0,0)\psline(0,0)(-1.,-3.2)\rput(-1,-3.4){$\bar b_1'$}
        \psline[ArrowInside=->](.6,-.6)(.3,-.3)\rput(.6,-.35){$e_2'$}
        %\rput(.1,-.45){$\mu_2$}
        \psline[ArrowInside=->](.3,-.3)(.6,0)\psline(.6,0)(-.3,0.3)(-1.6,-3.2)\rput(-1.6,-3.4){$\bar b_2'$}
        \psline[ArrowInside=->](.9,-.9)(.6,-.6)\rput(.9,-.65){$e_3'$}
        %\rput(.4,-.75){$\mu_3$}
        \psline(.6,-.6)(1.2,0)\psline[ArrowInside=->](.9,-.3)(1.2,0)\psline(1.2,0)(-.6,.6)(-2.2,-3.2)\rput(-2.2,-3.4){$\bar b_3'$}
        \psline[linestyle=dotted](.9,-.9)(1.1,-1.1)
        \psline[ArrowInside=->](1.4,-1.4)(1.1,-1.1)\rput(1.35,-1.){$\bar{b}_n'$}
        \psline(1.4,-1.4)(2.8,0)\psline[ArrowInside=->](2.5,-.3)(2.8,0)\psline(2.8,0)(-.9,.9)(-2.8,-3.2)\rput(-2.8,-3.4){$\bar b_n'$}}
        \rput(2.4,-7.3){\rput(1.4,-2){$A$}\psline[ArrowInside=->](0,-1.4)(.3,-1.1)\rput(0,-1.6){$b_1$}
        \psline[ArrowInside=->](.3,-1.1)(.6,-.8)\rput(.6,-1.05){$e_2$}
        %\rput(.1,-.45){$\mu_2$}
        \psline[ArrowInside=->](.6,-1.4)(.3,-1.1)\rput(.6,-1.6){$b_2$}
        \psline[ArrowInside=->](.6,-.8)(.9,-.5)\rput(.9,-.75){$e_3$}
        %\rput(.4,-.75){$\mu_3$}
        \psline(.6,-.8)(1.2,-1.4)\psline[ArrowInside=->](1.2,-1.4)(.9,-1.1)\rput(1.2,-1.6){$b_3$}
        \psline[linestyle=dotted](.9,-.5)(1.1,-0.3)
        \psline[ArrowInside=->](1.1,-0.3)(1.4,0.0)\rput(1.35,-0.4){$\bar{b}_n$}
        \psline(1.4,0)(2.8,-1.4)\psline[ArrowInside=->](2.8,-1.4)(2.5,-1.1)\rput(2.8,-1.6){$b_n$}}
        %%%
    \end{pspicture}}
\end{align}
%We note that this is simply the direct product of two copies of the density matrix Eq. \eqref{eq:bipartite-rhoA}. 
Here, tracing out $BB^*$ had the effect of setting the anyon charges in the ket of $A$ ($\vec{b}$ and $\vec{e}$) to those in the bra and likewise for the $A^*$ part of the diagram. The partial trace also introduced a factor of $\sqrt{d_{\vec{b}} d_{\vec{b}'}}$. 
Proceeding with the second replica trick, we must compute the $\beta^{th}$ power of the above expression. Again, using Eqs. \eqref{eq:inner-product} and \eqref{eq:trace}, we find on restoring the vertex labels,
\begin{align}
	\Tr[(\rho_{AA^*}^{(\alpha)})^\beta] = \sum_{ \substack{ \vec{b}, \vec{e} , \vec{\mu} \\ \vec{b}', \vec{e}' , \vec{\mu}' } } \left( \frac{d_{\vec{b}} d_{\vec{b}'} }{\mathcal{D}^{4n-4}}\right)^\beta = \left( \sum_{\vec{b},\vec{e},\vec{\mu}} \left( \frac{d_{\vec{b}}}{\mathcal{D}^{2n-2}}\right)^\beta \right)^2
\end{align}
We thus immediately find the R\'enyi reflected entropies to be equal to twice the R\'enyi entanglement entropies and hence the von Neumann reflected entropy matches twice the entanglement entropy:
\begin{align}
	S_R = 2 S_{AA^*} \, .
\end{align}
This is as expected for a pure state and provides a check on our construction of the canonical purification.

\section{Trijunction \label{sec:trijunction} }

\begin{figure}
  \centering
\subfloat[]{%
    \includegraphics[width=0.24\textwidth]{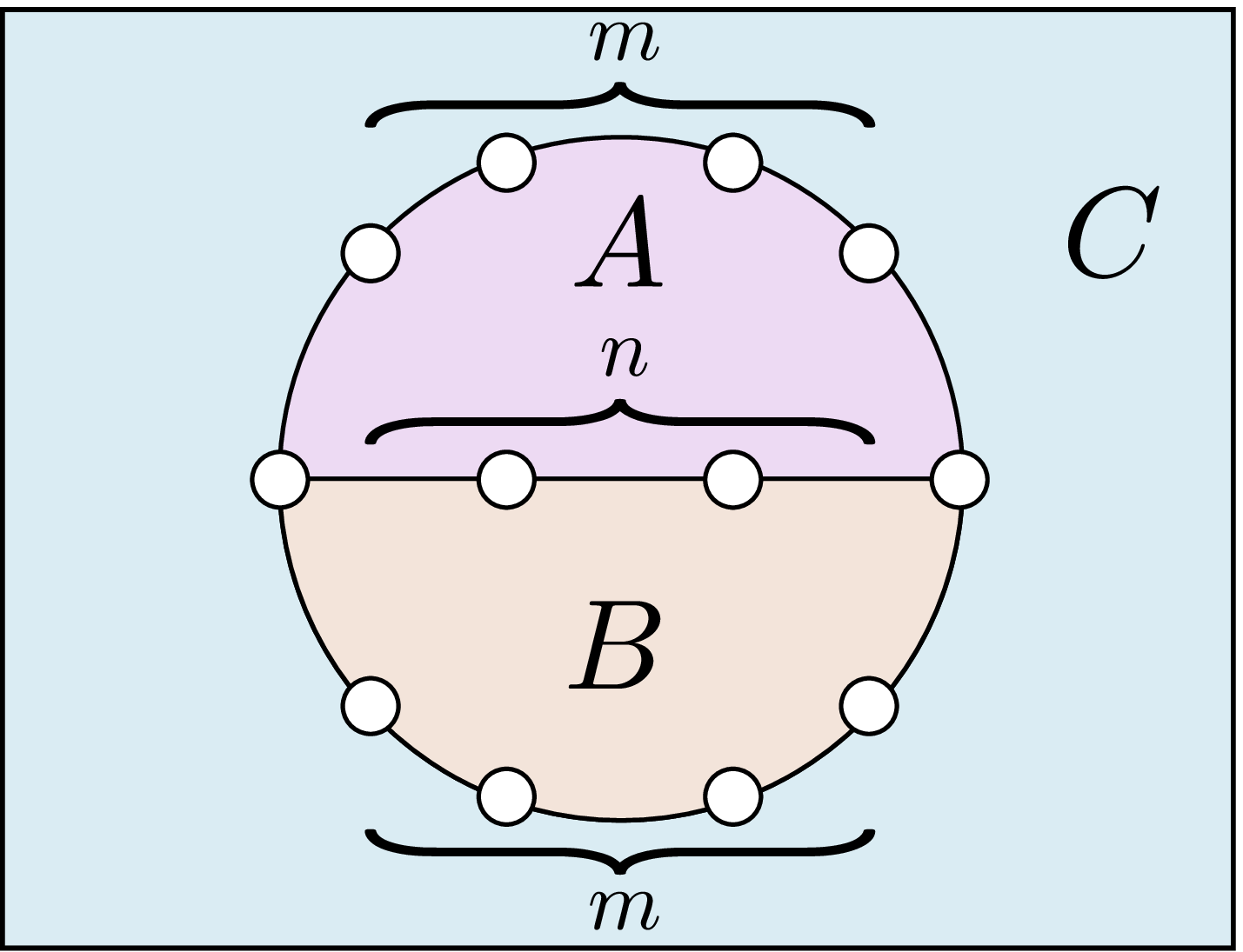}
}
\subfloat[]{%
    \includegraphics[width=0.24\textwidth]{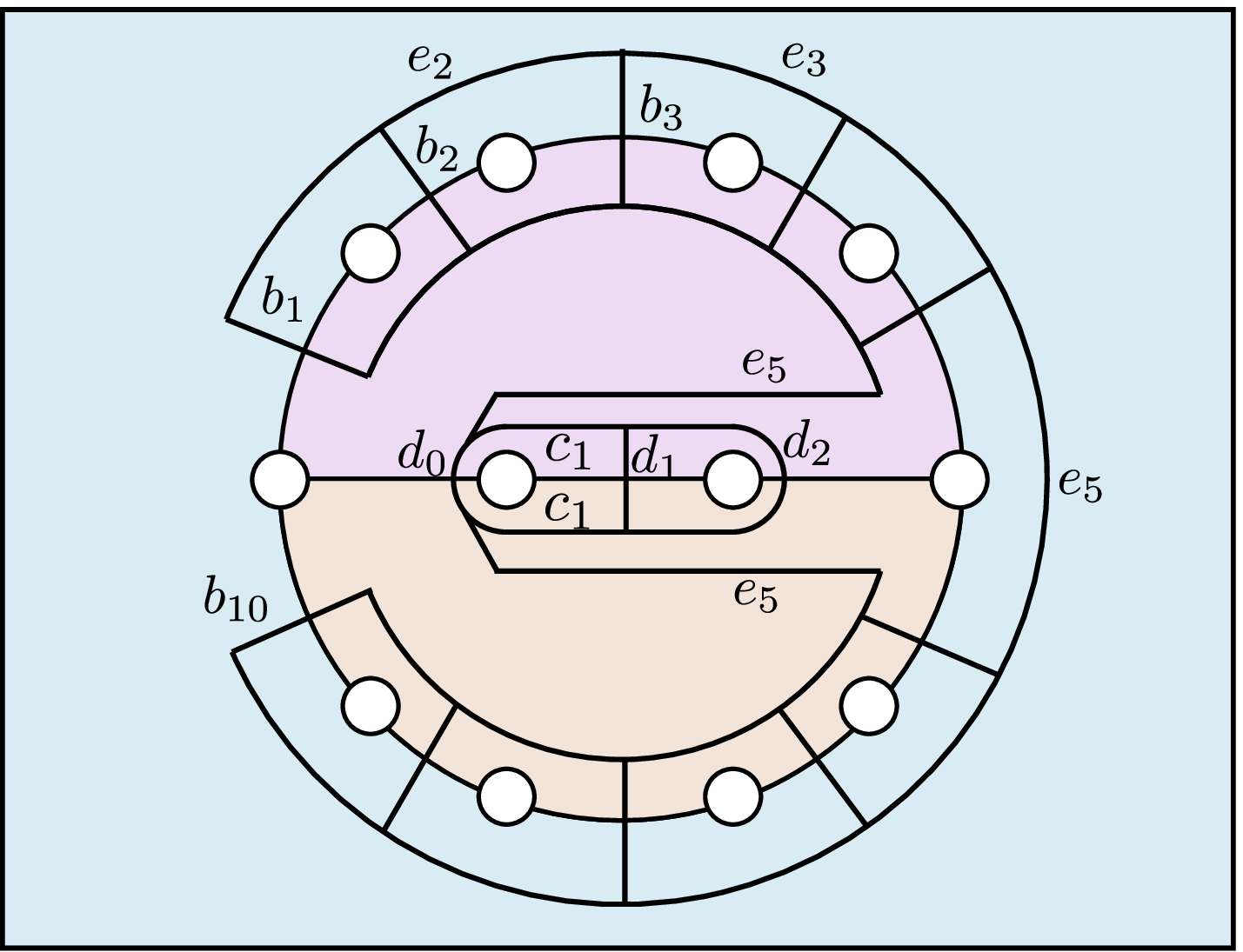}
}%0.36
    \caption{(a) Setup for the tripartition calculation, viewed from the top down, where the white circles indicate the wormholes connecting the sphere occupied by $\mathcal{C}$ with the other sphere occupied by the time-reversed conjugate $\overline{\mathcal{C}}$. There is one wormhole on each trijunction, $n$ wormholes on the boundary between $A$ and $B$ and $m$ wormholes on each of the boundaries of $A$ and $B$ with $C$ (the complement of $AB$). (b) Top-down view of the anyon diagram Eq. \ref{eq:diagram-tri-final-F} with $m=4$ and $n=2$.} \label{fig:general-tripartition}
\end{figure}

Having recovered the bipartite entanglement structure of a topological phase, we now move to the subject of the present work, namely tripartitions. Let us first consider a sphere partitioned into three regions $A$, $B$, and $C$, as shown in Fig. \ref{fig:general-tripartition}(a), where we display only a portion of the sphere, which locally looks flat. This particular configuration of the regions $A$, $B$, and $C$ is motivated by Ref. \cite{Siva2021b}, which conjectured that the Markov gap in such a configuration is given by $\frac{c_+}{3} \ln 2$,
where $c_+$ is the ``minimal" central charge of the corresponding edge theory, as noted above. 
As we have already emphasized, since the diagrammatic approach actually yields the ground state of a non-chiral, product theory $\mathcal{C}\times \overline{\mathcal{C}}$, we expect the Markov gap to vanish. Indeed, we will find this to be the case, providing a verification of the above conjecture, complementary to the string-net computation of Ref. \cite{Siva2021b}. 
Surprisingly, we will find a contribution to the negativity which distinguishes between Abelian and non-Abelian topological orders, and whose presence can be motivated by comparison to the earlier results of Ref. \cite{Wen2016}. 

As before, we begin by introducing a time reversed copy of the topological phase on a second manifold and nucleating a large number of wormholes $N = 2m+n+2$, with one centered at each trijunction, $2m$ additional wormholes along the boundary of $AB$ and $n$ wormholes along the boundary between $AB$, as depicted in Fig. \ref{fig:general-tripartition}(a) by the white circles. Note that it is important to insert wormholes at the trijunctions to ensure that, after cutting along the entanglement cut, each puncture connects exactly two of the subregions. The total number of wormholes on the boundary of $C$ is then $2m+2$ while the total number on the boundary of $A$ (and likewise $B$) is $m+n+2$. Hence, after cutting along the entanglement cut, regions $A$ and $B$ become topologically equivalent to spheres with $m+n+2$ punctures while $C$ is equivalent to a sphere with $2m+2$ punctures.

Once again, a trivial flux threads each wormhole, and so the application of modular $S$-transformations as change of bases at each wormhole yields an $\omega_0$ loop encircling each. This allows us to express the ground state in terms of $\omega_0$ loops in the inside basis as,
%\begin{widetext}
\begin{align}
&\ket{\psi} = \mathcal{D}^{N-1}
%%%
\begin{gathered}
\includegraphics[height=9em]{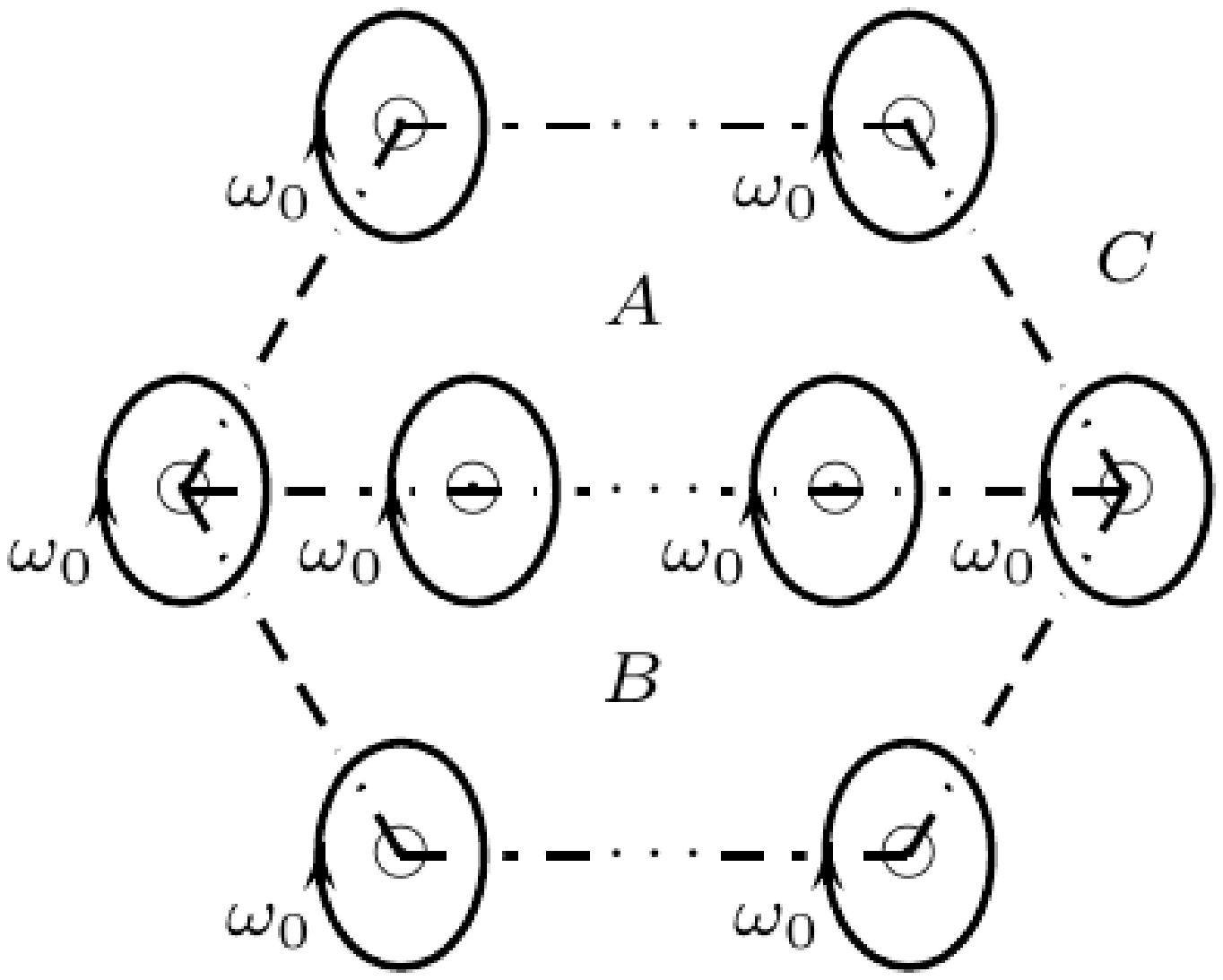} \end{gathered} 
\end{align}
The configuration of $\omega_0$ loops corresponds to the wormhole insertions of Fig. \ref{fig:general-tripartition}(a). For convenience, we deform the subregions and entanglement cuts to arrange the wormholes into two horizontal lines:
\begin{align}
 \ket{\psi} &= \mathcal{D}^{N-1}
 \begin{gathered}
\includegraphics[height=8em]{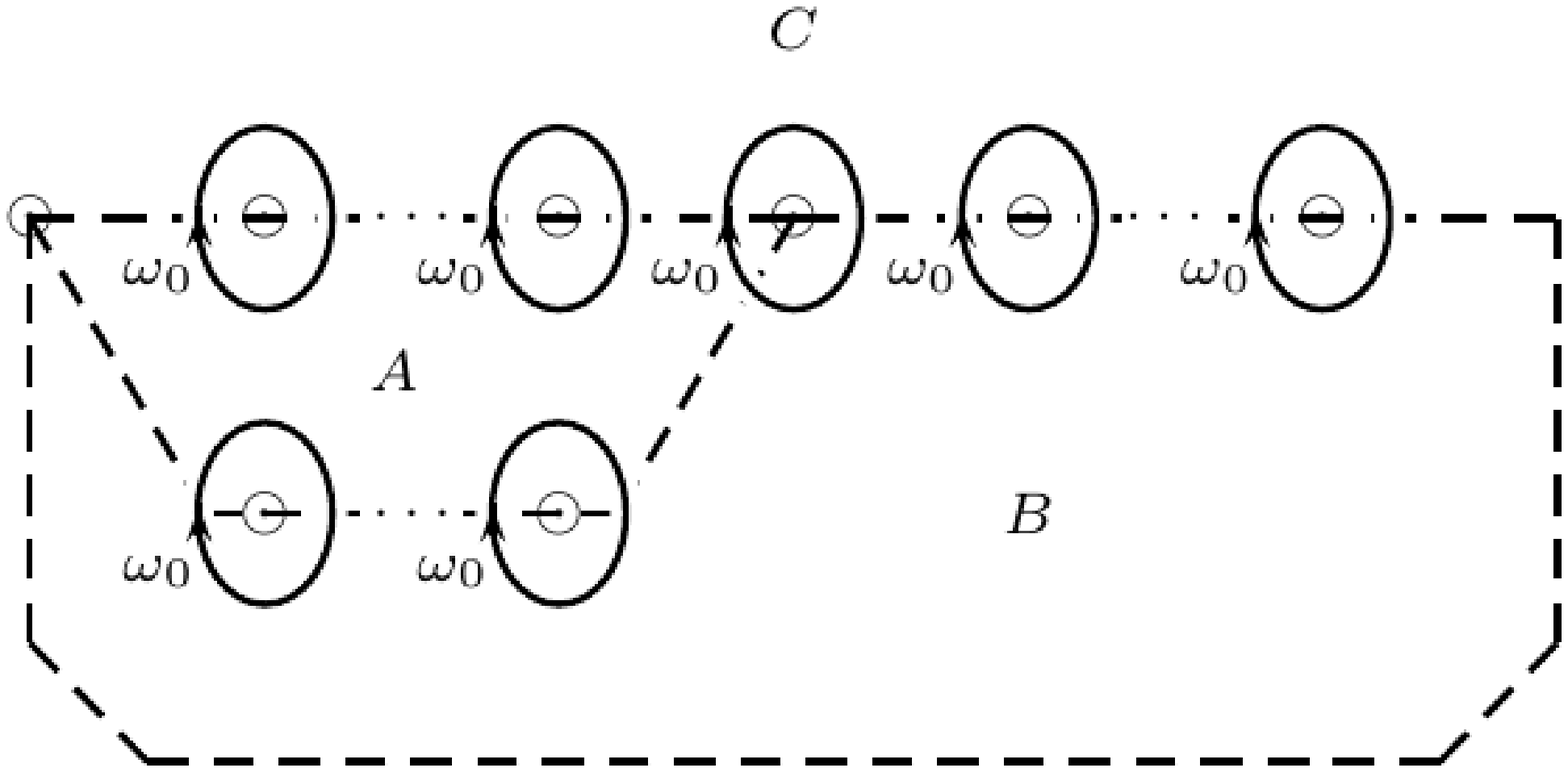} \end{gathered} 
\end{align}
%\begin{align}
%	\includegraphics[height=23em]{figures/trijunction/tri_ket_loops.eps}
%\end{align}
%\end{widetext}
%The configuration of $\omega_0$ loops in the first line corresponds to the wormhole insertions of Fig. \ref{fig:general-tripartition}(a). In the second equality, we deformed the entanglement cut and the 
In this expression, the positions of the wormholes are such that all the wormholes, of which there are a number $2m+2$, on the boundary of $AB$ with $C$ lie along a single horizontal line and the wormholes along the boundary between $A$ and $B$, of which there are $n$, lie along a second horizontal line. Additionally, as in the bipartite case, we used the handle-slide property to remove an $\omega_0$ loop located at one of the trijunctions -- this particular choice simplifies the diagrammatic manipulations.
Next, making use of the definition of the $\omega_0$ loops, we write
\begin{align}
&\ket{\psi} =  \sum_{\vec{e} , \vec{c}} \frac{d_{\vec{e}} d_{\vec{c}}}{\mathcal{D}^{N-1}} \begin{gathered} \includegraphics[height=8em]{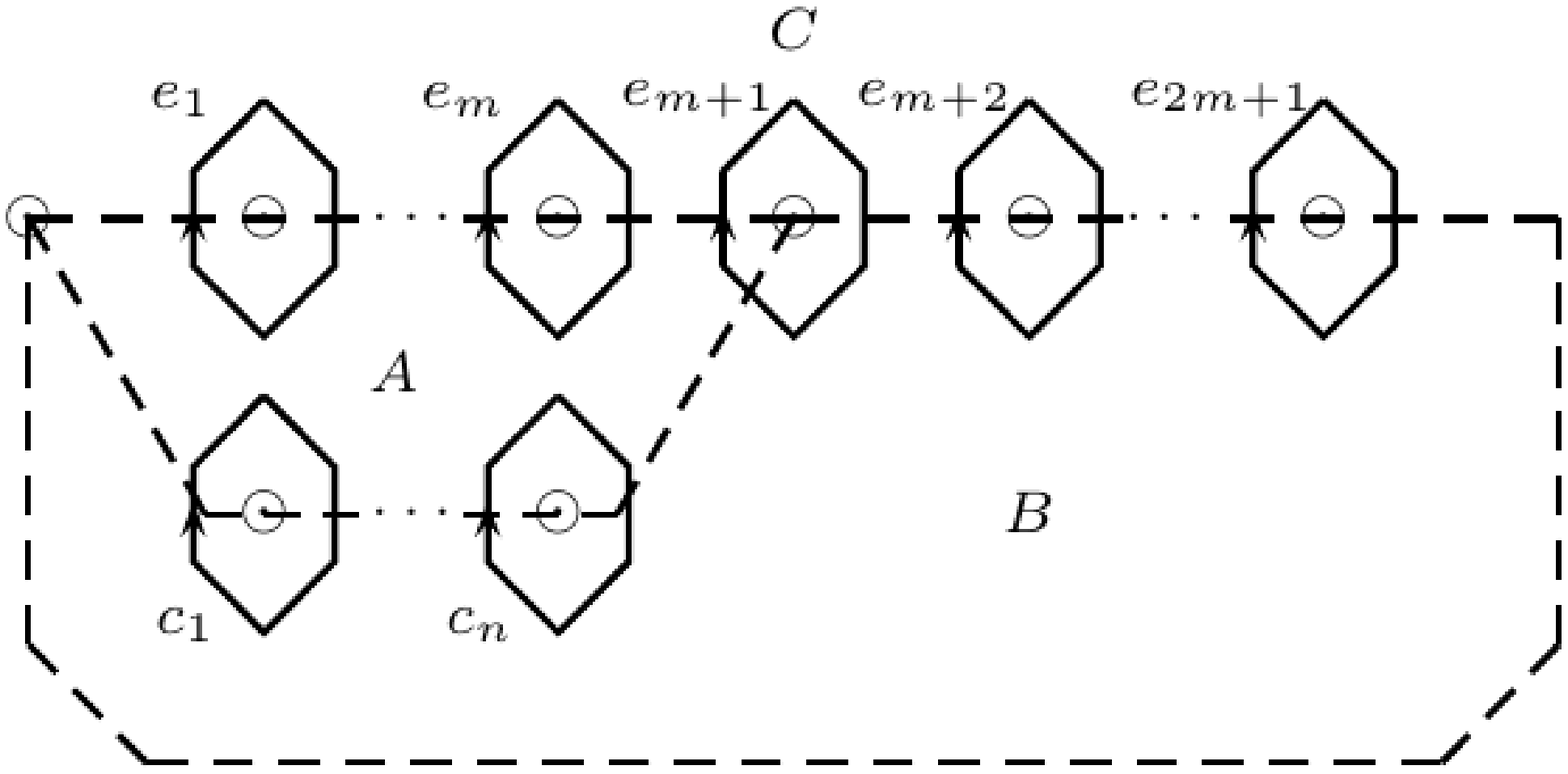} \end{gathered} 
\end{align}
Here the $\vec{e}$ ($\vec{c}$) anyons circle around wormholes lying on the boundary between $AB$ and $C$ ($A$ and $B$). There are therefore $2m+2$ $e_j$ anyon loops and $n$ $c_j$ anyon loops. Compared with the bipartite computation, the only new ingredients are the $\vec{c}$ loops around the wormholes along the $A$/$B$ boundary. 

We can apply the same sequence of manipulations as in the construction of the bipartite ground state of the preceding section to the line of $\vec{e}$ loops on the $AB$/$C$ boundary, yielding
\begin{align}
\ket{\psi} = \sum_{\vec{b},\vec{e},\vec{c}} \frac{\sqrt{d_{\vec{b}}}d_{\vec{c}}}{\mathcal{D}^{N-1}}
\begin{gathered}  \includegraphics[height=20em]{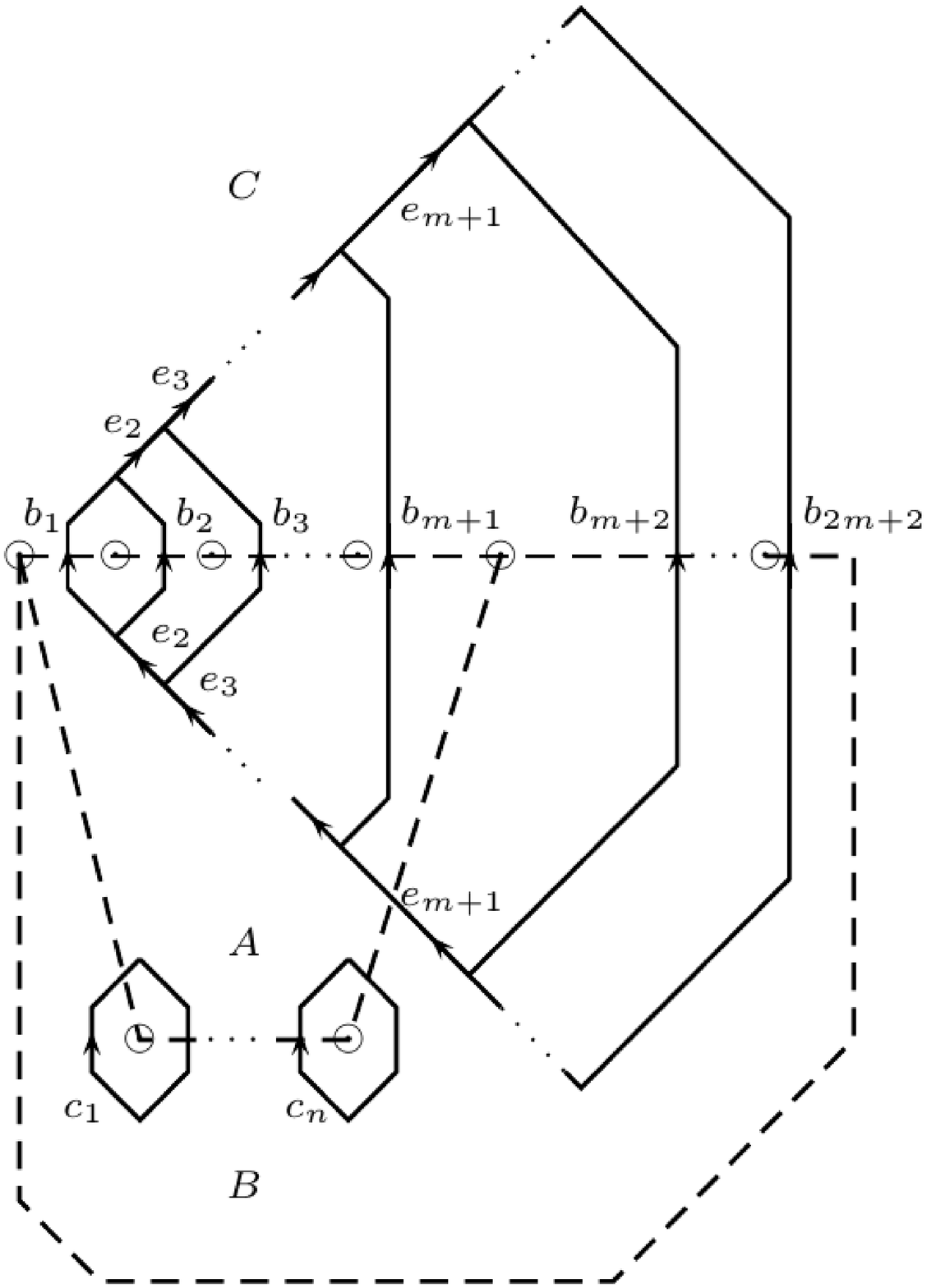}. \end{gathered} 
\end{align}
As before, we used resolutions of identity to join the $\vec{e}$ anyon loops, leading to the introduction of the $b_j$ anyon lines, with $j \in \{ 1, \dots , 2m+2 \}$, which thread the connected components of the boundary between $AB$ and $C$ (i.e. the segments between the wormholes).
Note that we have relabeled $e_1 , e_{2m+1} \mapsto b_1, b_{2m+2}$.

Next, we manipulate the $\vec{c}$ loops. 
Using the handle-slide property , we pass the $e_{m+1}$ line past all the $\vec{c}$ loops (which are still $\omega_0$ loops) to bring them within the region bounded by the $b_{m+1}$ and $b_{m+2}$ anyon lines for convenience. %:
%\begin{align}
%	\ket{\psi} = \sum_{a,\vec{b},\vec{e},\vec{d}} \frac{\sqrt{d_{\vec{b}}}d_{\vec{c}}}{\mathcal{D}^{N-1}} \begin{gathered}  \includegraphics[height=18em]{figures/trijunction/tri_ket_handle_slide.eps} \end{gathered}
%\end{align}
Now, focusing on the part of the diagram containing the $\vec{c}$ loops, we insert resolutions of identity to join the $\vec{c}$-loops together: %, we perform a series of manipulations to cast it in a tree-like form:
\begin{align}
&\sum_{\vec{c}} d_{\vec{c}}
\begin{pspicture}[shift=-.3](-0.3,-.5)(5.,1.)
 \scriptsize
 \psline[linestyle=dashed](0,0)(2.75,0)
 \rput(3.,0){$\dots$}
 \psline[linestyle=dashed](3.25,0)(5.9,0)
 \rput(.75,0){\hexbig}\rput(.3,.65){$c_1$}
 \rput(2.05,0){\hexbig}\rput(1.7,.65){$c_2$}
 \rput(3.9,0){\hexbig}\rput(3.45,.65){$c_{n-1}$}
 \rput(5.2,0){\hexbig}\rput(4.8,.65){$c_{n}$}
 \end{pspicture} \\
 \notag &=  \sum_{\vec{c} , \vec{d}}  
 \sqrt{d_{c_1} d_{c_{n}} d_{\vec{d}}}
 \begin{pspicture}[shift=-1.3](.2,-1.3)(3.2,1.3)
        \scriptsize
        \rput(1.1,0.){\hex}
        \rput(1.1,0.){$\odot$}
        \rput(1.7,0.){$\odot$}
        %\psline(0.8,0.2)(1.7,1.1)
        \psline(1.4,.8)(2,.2)(2,-.2)(1.4,-0.8)
        \psline[ArrowInside=->](2,-.2)(2,.2)\rput(2.2,.25){$d_3$}
        \rput(0.8,0.5){$c_1$}
        \psline[ArrowInside=->](1.1,.5)(1.4,.8)\rput(1.15,.8){$c_2$}
        \psline[ArrowInside=->](1.4,.8)(1.7,.9)\rput(1.45,1.1){$c_3$}
        \psline[linestyle=dotted](1.7,.9)(2,1.)
        \psline[ArrowInside=->](2.2,1.)(2.5,1.1)
        \psline(2.5,1.1)(2.8,1.)(2.8,-1.)(2.5,-1.1)
        \psline[ArrowInside=->](1.4,-.2)(1.4,0.2)\rput(1.6,0.25){$d_2$}
        \psline[ArrowInside=->](2.8,-.2)(2.8,0.2)
        \rput(3,0.25){$\bar{c}_{n}$}
        \psline[ArrowInside=->](1.4,-0.8)(1.1,-.5)\rput(1.45,-0.5){$c_2$}
        \psline[ArrowInside=->](1.7,-.9)(1.4,-0.8)\rput(1.75,-0.75){$c_3$}
        \psline[linestyle=dotted](1.7,-.9)(2,-1.)
        \psline[ArrowInside=->](2.5,-1.1)(2.2,-1.)
        \rput(2.2,0.){$\dots$}
        %\psline[linestyle=dashed](3.05,0.)(2.71,0.)
        %\psline[linestyle=dashed](1.9,0.)(.6,0.)
      \rput(2.61,0.){$\odot$}
      \psline[linestyle=dashed](2.61,0)(3.3,0)
		\psline[linestyle=dashed](0.4,0)(2.1,0)
    \end{pspicture}
\end{align}
In the first equality, we inserted resolutions of identity to combine adjacent $\vec{c}$ loops. This gives rise to the $d_j$ anyons, such that $c_j$ and $c_{j+1}$ fuse at vertex $\nu_{j}$ into $d_j$, which thread the connected components of the boundary between $A$ and $B$. These manipulations also result in the factor $\sqrt{d_{\vec{d}}} / (d_{c_2} \dots d_{c_{n-1}} \sqrt{d_{c_1}d_{c_n}})$ (only a single resolution of identity involving each of the $c_1$ and $c_n$ loops appears, while two such moves are applied to all the other $\vec{c}$ loops). In the second equality, we applied a series of $A$-moves to bend the diagram into a tree-like form. 

\begin{figure*}
  \centering
  \subfloat[]{%
  \includegraphics[width=0.9\textwidth]{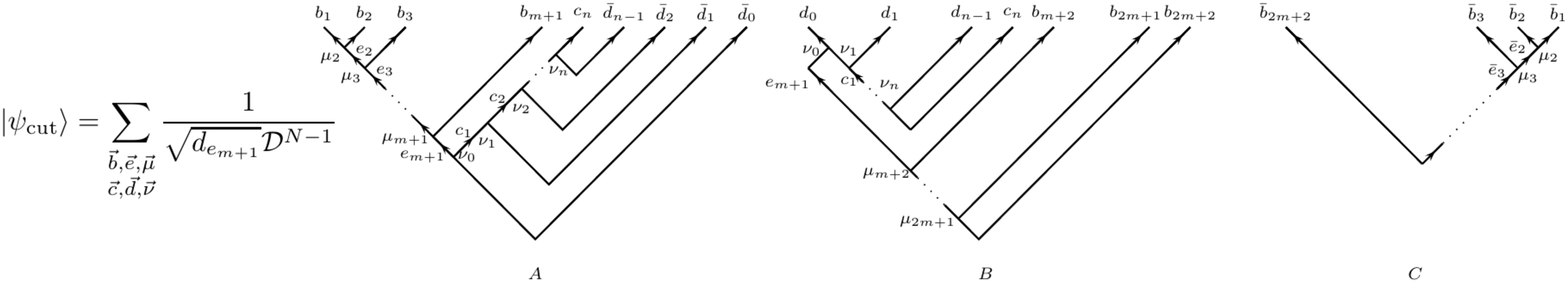}
  }
  
  \subfloat[]{%
  \includegraphics[width=0.85\textwidth]{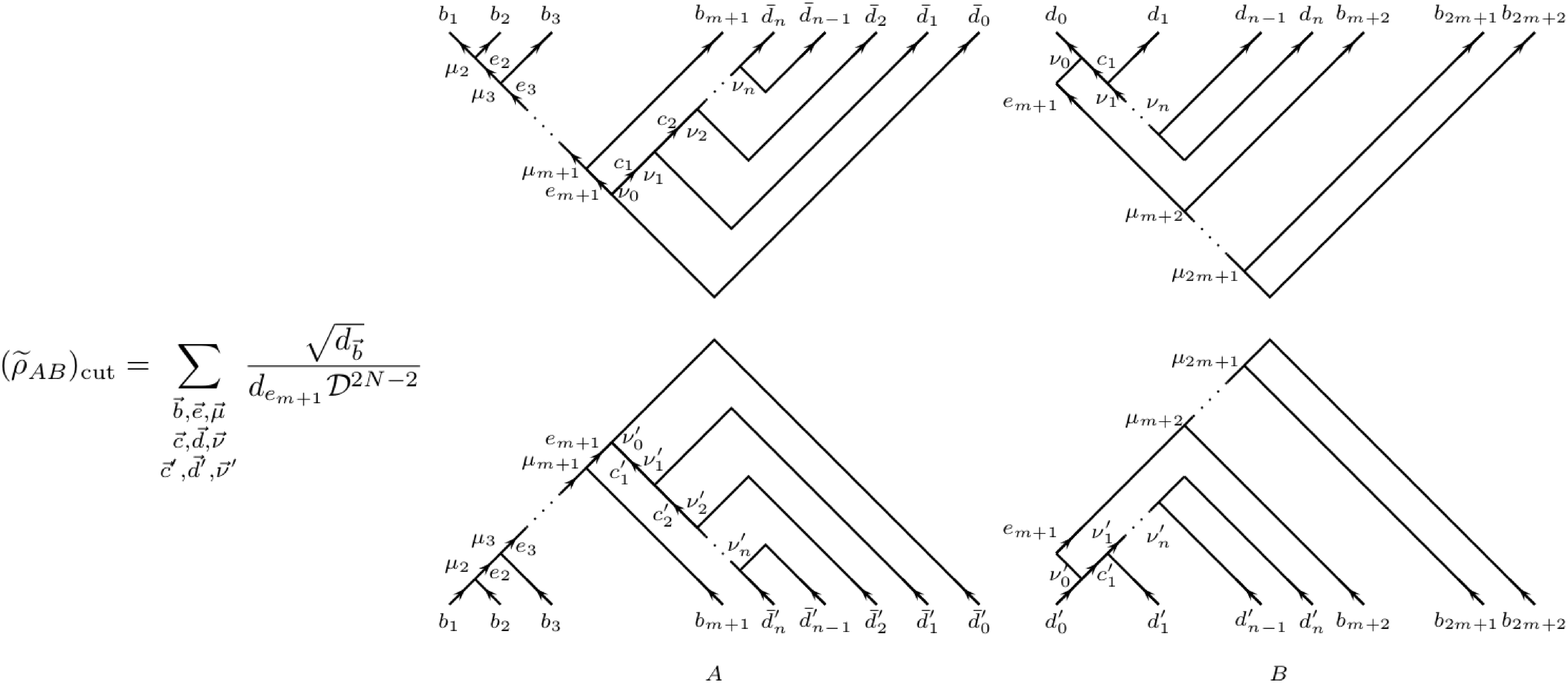}
  }
    \caption{(a) State for the tripartition of Fig. \ref{fig:general-tripartition} after cutting along the entanglement cuts.  (b) Cut reduced density matrix  for region $AB$. } \label{fig:trijunction-rhoAB}
\end{figure*}

Finally, we apply another resolution of identity to combine the $c_1$ line and $e_{m+1}$ line, which both thread a single boundary component on the boundary between $A$ and $B$, into a single $d_0$ anyon line. 
This gives rise to a factor of $\sqrt{d_0 / d_{c_1} d_{e_{m+1}}}$ and yields
\begin{align}
	\ket{\psi} = &\sum_{\vec{b},\vec{e},\vec{c},\vec{d}} \frac{\sqrt{d_{\vec{b}}d_{\vec{d}}} }{\sqrt{d_{e_{m+1}}} \mathcal{D}^{N-1}} \label{eq:diagram-tri-final-F} \\
	\notag &\times \begin{gathered}  \includegraphics[height=25em]{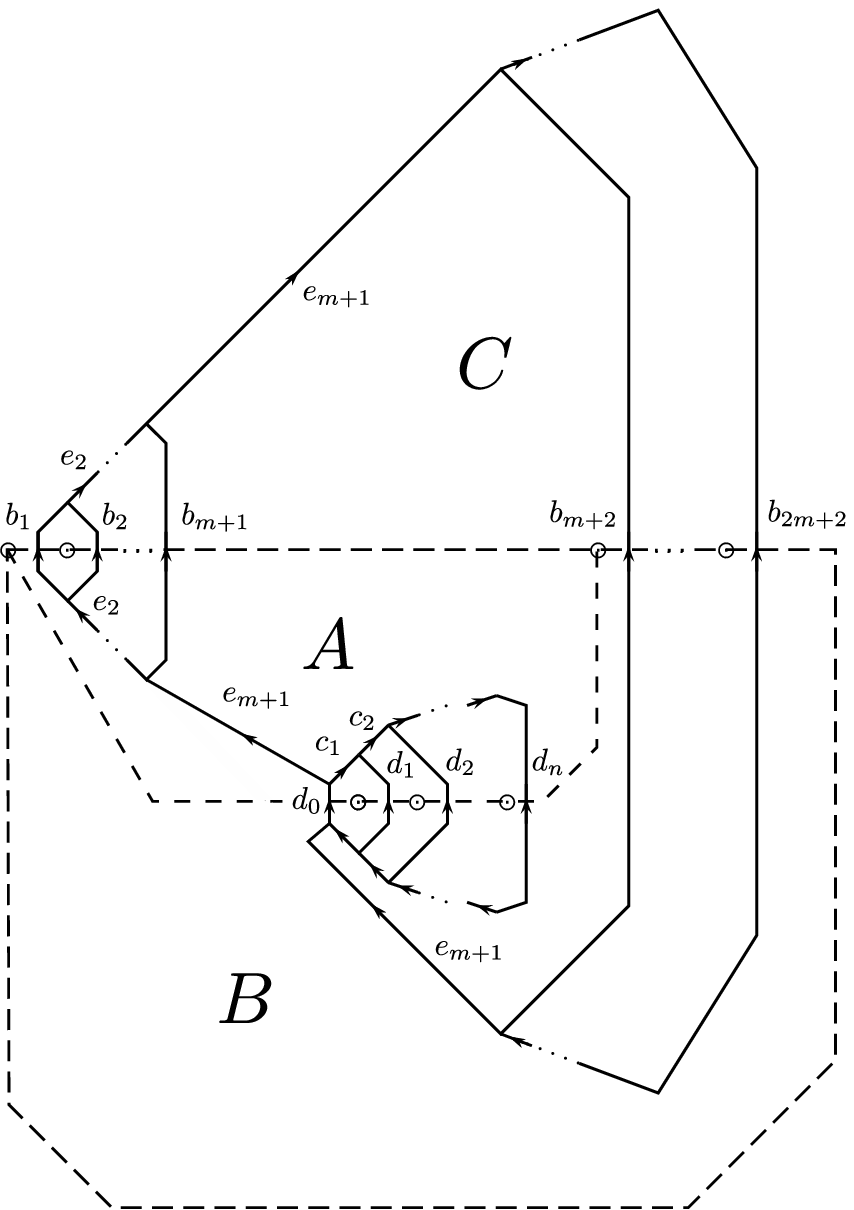}. \end{gathered} 
\end{align}
Note that we have relabeled $\bar c_n \mapsto d_n$ for convenience. With these manipulations there is now only a single anyon charge line threading each connected component of each boundary. In Fig. \ref{fig:general-tripartition}, we overlay the anyon diagram for $m=4$ and $n=2$ on top of the tripartition with the original topology of each subregion for clarity. %We are now prepared to compute $\trho_{AB}$.
In order to compute $\trho_{AB}$, we cut the diagram along all the entanglement cuts, leading to the state in Fig. \ref{fig:trijunction-rhoAB}(a), and then trace out the $C$ part of the diagram using Eq. \eqref{eq:trace}. The resulting $\trho_{AB}$ is given in Fig. \ref{fig:trijunction-rhoAB}(b). %We find,
Here we have restored the fusion vertex labels. 
Note that, as a result of the trace over $C$, both the bras and kets have $\vec{b}$ anyons (i.e. they are diagonal in the $\vec{b}$ indices) while the bras have $\vec{c}$ and $\vec{d}$ anyons and the kets have $\vec{c}'$ and $\vec{d}'$. 

Let us unpack this expression and remind ourselves of the notation. Once again, the $\vec{b}$ charges are the anyon lines threading the components of the boundary between $AB$ and $C$. The $\vec{e}$ charges connect the $\vec{b}$ lines and the $\vec{\mu}$ label the fusion vertices for these lines. For instance, $e_2$ splits into $b_1$ and $b_2$ at $\mu_2$, $e_3$ splits into $e_2$ and $b_3$ at $\mu_3$ and so on up to $\mu_{2m}$, at which point $\bar b_{2m+2}$ splits into $b_{2m+1}$ and $e_{2m}$. Recall that we have $b_j$ for $j \in \{1, \dots , 2m+2 \}$, $e_j$ for $j \in \{2, \dots , 2m \}$, and $\mu_j$ for $j \in \{2, \dots , 2m+1 \}$. We note further that the charges and vertices $\vec{b}_A = (b_{1} , \dots , b_{m+1})$, $\vec{e}_A = (e_{2} , \dots , e_{m})$, $\vec{\mu}_A = (\mu_{2} , \dots , \mu_{m+1})$ appear only in the $A$ subregion, while the charges and vertices $\vec{b}_B = (b_{m+2} , \dots , b_{2m+2})$, $\vec{e}_B = (e_{m+2} , \dots , e_{2m+2})$, $\vec{\mu}_B = (\mu_{m+2} , \dots , \mu_{2m+1})$ appear only in the $B$ subregion. The $e_{m+1}$ charge, however, appears in both $A$ and $B$. % subregions (and also in subregion $C$). 

Next, the $\vec{d}$ charges label the anyon lines threading the components of the boundary between $A$ and $B$. The $\vec{c}$ charges connect the $\vec{d}$ lines and the $\vec{\nu}$ label the fusion vertices for these lines. Explicitly, $c_1$ fuses with $d_1$ at $\nu_1$ into $c_2$, $c_2$ fuses with $d_2$ at $\nu_2$ into $c_3$, and so on up to $\nu_{n-1}$ where $c_{n-1}$ fuses with $d_{n-1}$ into $\bar{d}_n$. Altogether, we have $c_j$ for $j \in \{1, \dots , n-1 \}$, $d_j$ for $j \in \{0, \dots , n \}$, and $\nu_j$ for $j \in \{0, \dots , n-1 \}$. 
The same is true for the primed versions of these charges (i.e. $\vec{c}'$, $\vec{d}'$, and $\vec{\nu}'$). 
Focusing on the $d_0$ charge line, we observe from Eq. \eqref{eq:diagram-tri-final-F} that 
$e_{m+1}$ fuses with $\bar c_1$ into $d_0$, which then splits into $\bar c_1$ and $e_{m+1}$. Both of these vertices are indexed by the same label,  $\nu_0$. The same statements hold \textit{mutatis mutandi} for the primed version of the charge $d_0'$ and fusion vertex $\nu_0'$. % (note that there is no primed version of $e_{m+1}$). 
Here we pause to note that, upon inspection of these manipulations, the anyon $e_{m+1}$ resides precisely at one of the trijunctions, as shown in Fig. \ref{fig:general-tripartition}(b) for $m=5$. It is this anyon which will play a role in the novel contributions to the entanglement quantities of interest in this trijunction configuration.

In order to simplify manipulations involving this density matrix, we will use the following shorthand:
\begin{align}
	(\trho_{AB})_{\mathrm{cut}} = 
	%\\
	%&\notag 
	\sum_{\substack{ \vec{b},\vec{e}, \vec{\mu} \\ \vec{c},\vec{d},\vec{\nu} \\ \vec{c}',\vec{d}',\vec{\nu}'} } \frac{\sqrt{d_{\vec{b}}}}{d_{e_{m+1}} \mathcal{D}^{2N-2}} \begin{gathered} \includegraphics[height=10em]{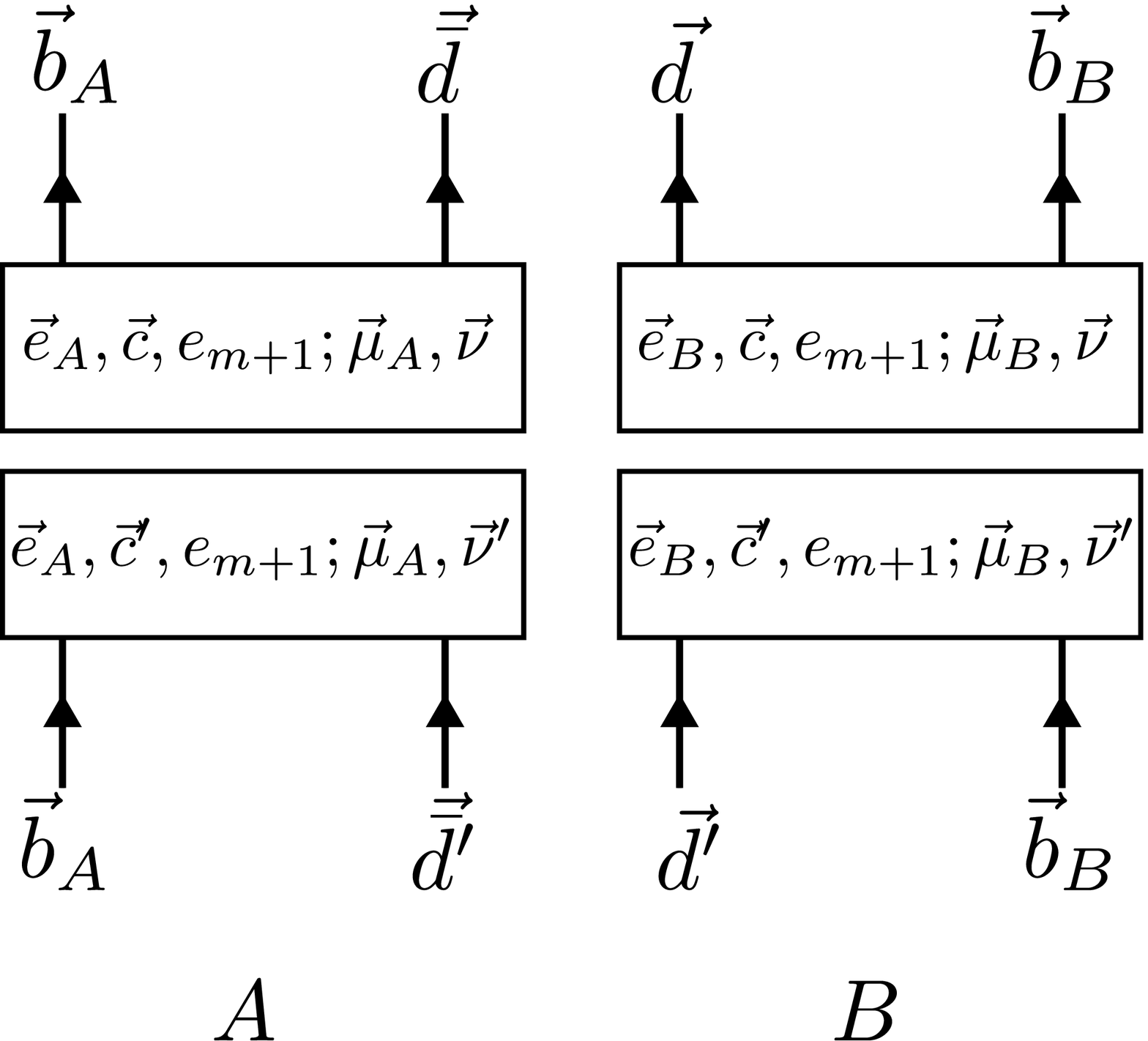} \end{gathered}
\end{align}
Here, the boxes with arrows exiting from the top and entering from the bottom represent the ``ket" and ``bra" parts, respectively, of the $A$ and $B$ regions of the diagram and are labeled by the internal anyon lines and fusion vertex labels. We denote the external anyon lines with the collective labels $\vec{b}_{A,B}$ (as defined above) and $\vec{d}$.
In this notation we have, for instance, the inner product:
\begin{align}
	\begin{gathered} \includegraphics[height=6em]{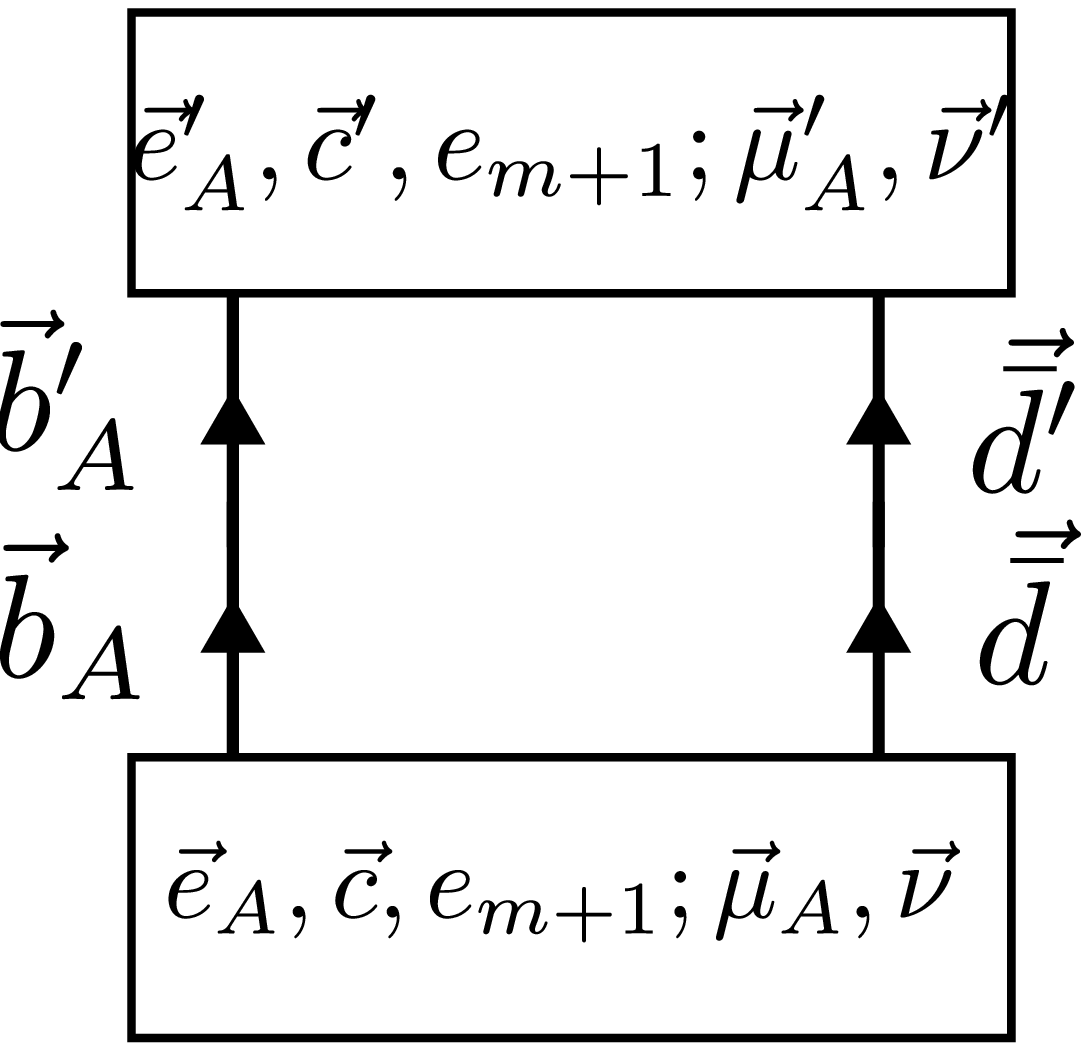} \end{gathered} = \sqrt{ d_{\vec{b}_A} d_{\vec{d}} } \delta_{\vec{e},\vec{e}'} \delta_{\vec{b}_A,\vec{b}_A'} \delta_{\vec{\mu}_A,\vec{\mu}_A'} \delta_{\vec{c},\vec{c}'} \delta_{\vec{d},\vec{d}'} \delta_{\vec{\nu},\vec{\nu}'}
\end{align}
 In the following computations, when we need to make use of the explicit fusion tree structure of the anyons, we may refer back to Fig. \ref{fig:trijunction-rhoAB}.
With the reduced density matrix for the ground state in hand, we can proceed to compute the entanglement quantities of interest.

\subsection{Mutual Information}

We begin with the mutual information.
Here,  we can simply use the results for the bipartite entanglement above and the fact that $S_A = S_B$ to find the known result,
\begin{align}
	I(A:B) = -2(n+1)\sum_b \frac{d_b^2}{\mathcal{D}^2} \ln (d_b/\mathcal{D}^2) - \ln \mathcal{D}^2  \label{eq:tripartition-MI} %\equiv S_{2n+2}
\end{align}
We see that the area law term of the mutual information is only sensitive to the length of the entanglement cut between $A$ and $B$, as expected, while the only subleading correction comes from the topological entanglement entropy. We again emphasize that this quantity is strictly the mutual information for the phase $\mathcal{C} \times \overline{\mathcal{C}}$ and is twice that for just $\mathcal{C}$. 

\begin{widetext}
\subsection{Reflected Entropy}

We now move on to compute the reflected entropy for the tripartition. Since $\trho_{AB}$ is not a diagonal matrix, we must employ the replica trick to compute $\kket{\sqrt{\rho_{AB}}}$. 
Making use of  Eq. \eqref{eq:overlap} and employing the relation
\begin{align}
	\notag \sum_{ \substack{\vec{c}' , \vec{d}' \vec{\nu}' } } d_{\vec{d}'} &= \sum_{\vec{c}' , \vec{d}' } N_{\bar{d}_0' c_1'}^{\bar{e}_{m+1}} N_{d_1' c_2'}^{c_1'} N_{d_2' c_3'}^{c_2'} \dots N_{d_{n-1}' d_n'}^{c_{n-1}'} d_{\vec{d}'} \\
	 &= d_{e_{m+1}} \mathcal{D}^{2n}  ,
\end{align}
where we can read off this structure of fusion coefficients from  Fig. \ref{fig:trijunction-rhoAB}, 
we find for $\trho_{AB}$ raised to a power $\alpha \in 2\mathbb{Z}$:
\begin{align}
	\notag (\trho_{AB})_{\mathrm{cut}}^{\alpha/2}  = \sum_{\substack{ \vec{b},\vec{e}, \vec{\mu} \\ \vec{c},\vec{d},\vec{\nu} \\ \vec{c}',\vec{d}',\vec{\nu}'} } & \frac{\sqrt{d_{\vec{b}}}}{d_{e_{m+1}} \mathcal{D}^{2N-2}} \left( \frac{\mathcal{D}^{2n} d_{\vec{b}}}{\mathcal{D}^{2N-2}} \right)^{\frac{\alpha}{2}-1} 
	%\\
	%&
	%\times
	\begin{gathered} \includegraphics[height=10em]{figures/tri-boxes.eps} \end{gathered} .
\end{align}
Following our prescription, we obtain the canonical purification by simply dragging the anyon lines in the input space into the output space:
\begin{align}
	\kket{(\trho_{AB})_{\mathrm{cut}}^{\alpha/2}} = \sum_{\substack{ \vec{b},\vec{e}, \vec{\mu} \\ \vec{c},\vec{d},\vec{\nu} \\ \vec{c}',\vec{d}',\vec{\nu}'} } & \frac{\sqrt{d_{\vec{b}}}}{\mathcal{D}^{2N-2} d_{e_{m+1}}} \left( \frac{\mathcal{D}^{2n} d_{\vec{b}} }{\mathcal{D}^{2N-2}} \right)^{\alpha/2-1} 
	%\\
	%\notag &\times
	\begin{gathered} \includegraphics[height=10em]{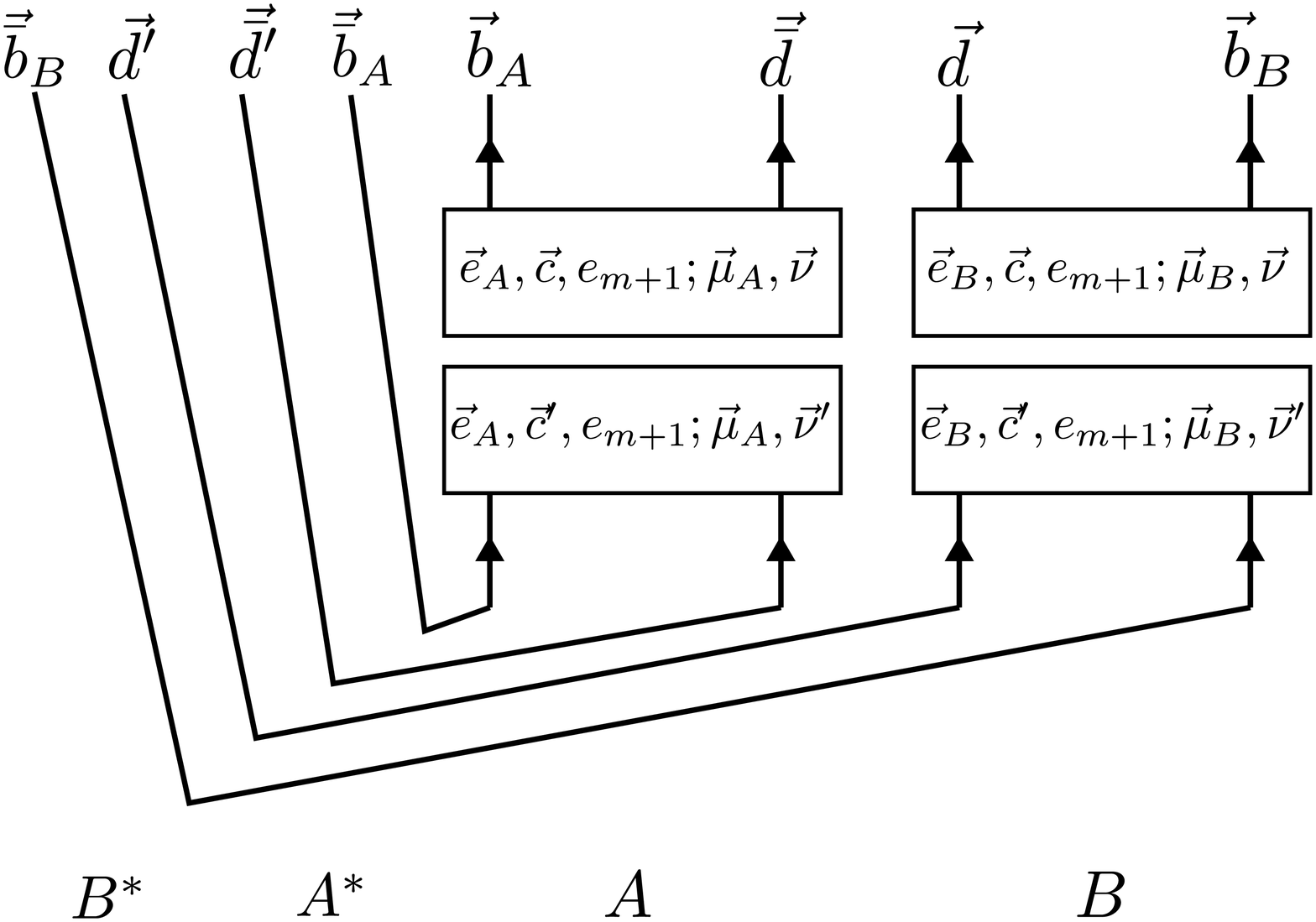} \end{gathered}
\end{align}
%Note that we have also relabeled $c_n \mapsto d_n$ for convenience. 
%\begin{widetext}
We then construct the density matrix $\kket{\trho^{\alpha/2}_{\mathrm{cut}}}\bbra{\trho^{\alpha/2}_{\mathrm{cut}}}$ and trace out $BB^*$ using Eq. \eqref{eq:trace}, to obtain
\begin{align}
	\trho_{AA^*}^{(\alpha)} =& \sum_{\substack{ \vec{b}_A,\vec{e}_A, \vec{\mu}_A \\ \vec{b}_A' , \vec{e}_A' , \vec{\mu}_A' \\ \vec{c},\vec{d}, \vec{\nu} \\ \vec{c}',\vec{d}',\vec{\nu}' \\ e_{m+1} } }  B_{\alpha} \frac{\sqrt{d_{\vec{b}_A} d_{\vec{b}_A'} d_{\vec{d}} d_{\vec{d}'} }}{\mathcal{D}^{4N-4} d_{e_{m+1}}^2} 
	%\\
	%\notag \times 
	%&
	\left( \frac{\mathcal{D}^{4n}d_{\vec{b}_A} d_{\vec{b}_A'} }{\mathcal{D}^{4N-4}} \right)^{\frac{\alpha}{2}-1} \begin{gathered} \includegraphics[height=14em]{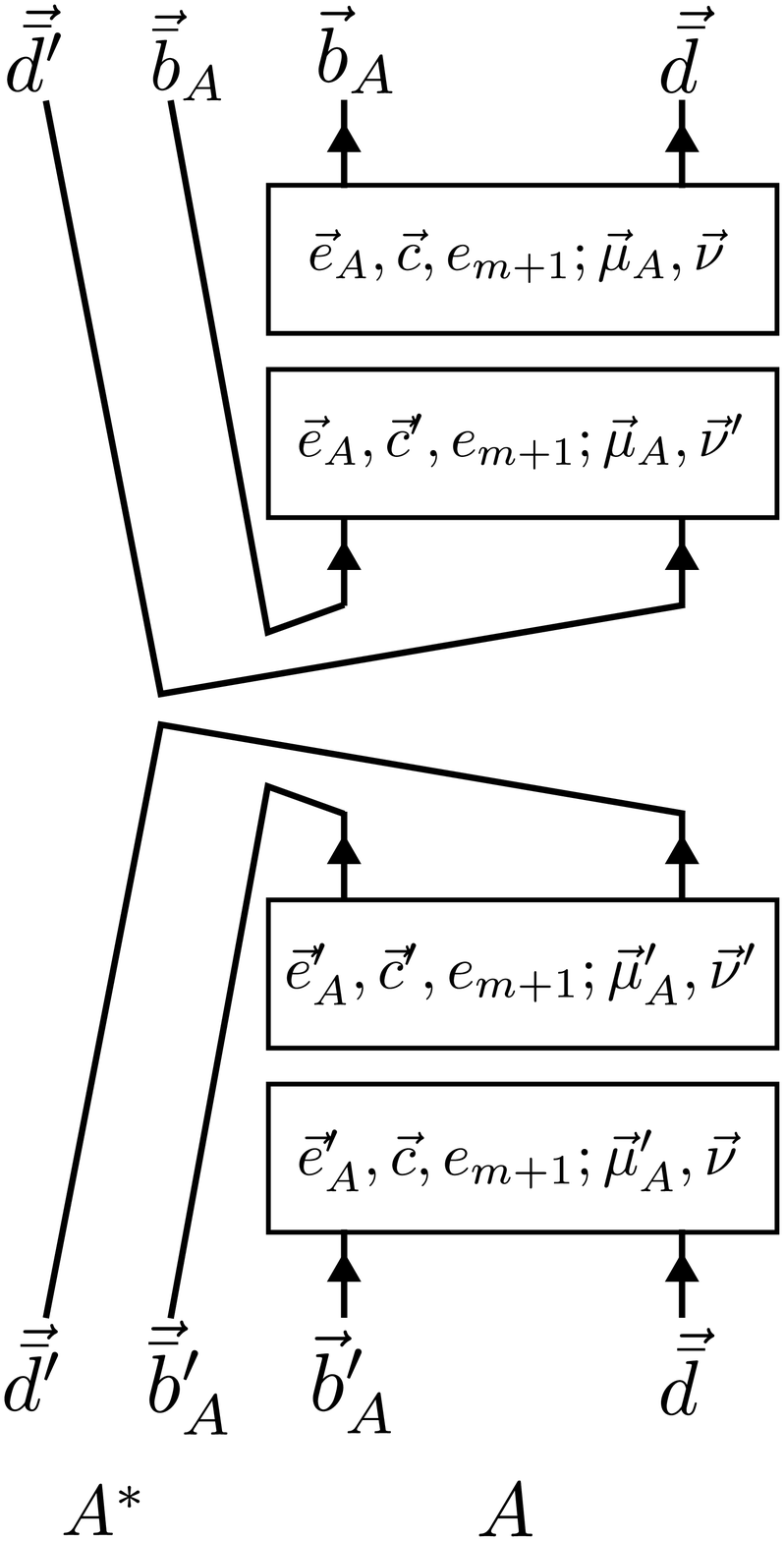} \end{gathered}
\end{align}
%\end{widetext}
Note the index structure of the anyon diagram -- it is now diagonal in the $\vec{d}$ and $\vec{d}'$ labels, while the ket part of the diagram involves unprimed $\vec{b}_A$ charges and the bra primed $\vec{b}_A'$ charges. Here we have also defined 
\begin{align}
	B_{\alpha} = \sum_{\vec{\mu}_B, \vec{b}_B, \vec{e}_B} d_{\vec{b}_B}^{\alpha},
\end{align}
and we recall that the ``$A$" and ``$B$" subscripts denote charges/vertices which belong to the $A$ and $B$ subregions, respectively.
We now apply the second replica trick.  We first compute %the square of the density matrix, from which we can deduce the form of an arbitrary power of the density matrix with 
for $\beta \in \mathbb{Z}$:
%\begin{widetext}
\begin{align}
	%\notag
	[\trho_{AA^*}^{(\alpha)}]^\beta  = &\sum_{\substack{ \vec{b}_A,\vec{e}_A, \vec{\mu}_A \\ \vec{b}_A' , \vec{e}_A' , \vec{\mu}_A' \\ \vec{c},\vec{d},\vec{\nu} \\ \vec{c}',\vec{d}',\vec{\nu}' \\  e_{m+1} } } \left[ \frac{B_{\alpha}}{d_{e_{m+1}}^2 \mathcal{D}^{4N-4}} d_{\vec{d}} d_{\vec{d}'} \left( \frac{\mathcal{D}^{4n}}{\mathcal{D}^{4N-4}} \right)^{\frac{\alpha}{2}-1} \right]^{\beta} 
	%\\
	%&\times   
	\frac{A_\alpha^{\beta-1} (d_{\vec{b}_A}d_{\vec{b}_A'})^{\frac{\alpha-1}{2}}}{\sqrt{d_{\vec{d}} d_{\vec{d}'}} }
	\begin{gathered} \includegraphics[height=14em]{figures/tri-boxes-purAAstr.eps} \end{gathered}
	%\sum_{\substack{ \vec{b}_A,\vec{e}_A, \vec{\mu}_A \\ \vec{b}_A' , \vec{e}_A' , \vec{\mu}_A' \\ \vec{c},\vec{d},\vec{\nu} \\ \vec{c}',\vec{d}',\vec{\nu}' \\  e_{m+1} } } \left[ \frac{A_{\alpha} B_{\alpha}}{d_{e_{m+1}}^2 \mathcal{D}^{4N-4}} d_{\vec{d}} d_{\vec{d}'} \left( \frac{\mathcal{D}^{4n}}{\mathcal{D}^{4N-4}} \right)^{\frac{\alpha}{2}-1} \right]^{\beta-1} \\
	%&\times   B_{\alpha} \frac{\sqrt{d_{\vec{b}_A} d_{\vec{b}_A'} d_{\vec{d}} d_{\vec{d}'} }}{\mathcal{D}^{4N-4} d_{e_{m+1}}^2} \left( \frac{\mathcal{D}^{4n}d_{\vec{b}_A} d_{\vec{b}_A'} }{\mathcal{D}^{4N-4}} \right)^{\frac{\alpha}{2}-1}
\end{align}
%\begin{align}
%	\includegraphics[height=15em]{figures/trijunction/tri_rho_canpur_mn_replica.eps}
%\end{align}
%\end{widetext}
%\end{widetext}
\end{widetext}
where we have defined
\begin{align}
	A_{\alpha} = \sum_{\vec{\mu}_A, \vec{b}_A, \vec{e}_A} d_{\vec{b}_A}^{\alpha}.
\end{align}
%While this is a rather involved expression, the main aspect of the anyon diagram to note is the index structure. Namely, the $A$ ($A^*$) part of the diagram is diagonal in $\vec{\bar{d}}$ ($\vec{\bar{d}}'$) charges, while both the bra of $A$ and $A^*$
We thus find, on taking the quantum trace,
%\begin{widetext}
\begin{align}
	\Tr[\rho_{AA^*}^{(\alpha)}]^\beta = & \sum_{\substack{\vec{c}, \vec{d} , \vec{\nu} \\ \vec{c}', \vec{d}',  \vec{\nu}' \\ e_{m+1} }} \left[ \frac{A_\alpha B_{\alpha}}{d_{e_{m+1}}^2 \mathcal{D}^{4N-4}}  \left( \frac{\mathcal{D}^{4n}}{\mathcal{D}^{4N-4}} \right)^{\frac{\alpha}{2}-1} d_{\vec{d}} d_{\vec{d}'} \right]^{\beta}  
\end{align}
To compute the reflected entropy, we take the limit $m \to 1$. In this limit, we have
\begin{align}
	\notag B_1 &= \sum_{ \substack{b_{m+2} \dots b_{2m+2} \\ e_{m+2} \dots e_{2m}} } N_{e_{m+1} b_{m+2}}^{e_{m+2}} \dots N_{e_{2m} b_{2m+1}}^{\bar b_{2m+2}} d_{b_{m+2}} \dots d_{b_{2m+2}} \\
	&= d_{e_{m+1}} \mathcal{D}^{2m},
\end{align}
and
\begin{align}
	\notag A_1 &= \sum_{ \substack{b_{1} \dots b_{m+1} \\ e_{2} \dots e_{m}} } N_{b_1 b_2}^{e_2}  \dots N_{e_{m} b_{m+1}}^{e_{m+1}} d_{b_{1}} \dots d_{b_{m+1}} \\
	&= d_{e_{m+1}} \mathcal{D}^{2m} .
\end{align}
Recalling that the $\vec{b}_{A,B}$ anyons are those connecting regions $A$ and $B$, respectively, with region $C$, we see that $A_1$ and $B_1$ effectively count the number of anyon configurations on the boundaries of $A$ and $B$ with $C$ allowed by fusion, and hence contribute to the area law entanglement between $AB$ and $C$.
Summing over the vertex labels $\vec{\nu}$ and $\vec{\nu}'$ yields fusion multiplicity factors $N_{\vec{d}}^{e_{m+1}}$ and $N_{\vec{d}'}^{e_{m+1}}$, respectively. Altogether, recalling that $N=2m+n+2$, we obtain
\begin{align}
	\lim_{\alpha \to 1} \Tr[\rho_{AA^*}^{(\alpha)}]^\beta = \sum_{ \substack{\vec{c}, \vec{d}, e_{m+1} \\ \vec{c}', \vec{d}'} } N_{\vec{d}}^{\bar{e}_{m+1}} N_{\vec{d}'}^{\bar{e}_{m+1}} \left(\frac{d_{\vec{d}} d_{\vec{d}'}}{\mathcal{D}^{4n+2}} \right)^{\beta},
\end{align}
where we have used the shorthand 
\begin{align}
	N_{\vec{d}}^{\bar{e}_{m+1}} = N_{\bar{d}_0 d_1 \dots d_n}^{\bar{e}_{m+1}} = N_{d_{n-1} d_n}^{c_{n-1}} N_{d_{n-2} d_{n-1}}^{c_{n-2}} \dots N_{c_{1} \bar{d}_0}^{\bar{e}_{m+1}}.
\end{align}
Note that the factor of $\mathcal{D}^{2m}$ arising from $A_1 B_1$ has canceled out. In particular, no contribution from the $\vec{b}$ anyons remains, and hence the R\'enyi reflected entropies do not have an area law contribution from the entanglement between $AB$ and $C$.

Finally, on taking the von Neumann limit $\beta \to 1$, we find for the reflected entropy,
\begin{align}
	S_R(A:B) &= - \sum_{ \substack{\vec{c}, \vec{d} \\ \vec{c}', \vec{d}' \\ e_{m+1} }} N_{\vec{d}}^{\bar{e}_{m+1}} N_{\vec{d}'}^{\bar{e}_{m+1}} \left(\frac{d_{\vec{d}} d_{\vec{d}'}}{\mathcal{D}^{4n+2}} \right) \ln \left(\frac{d_{\vec{d}} d_{\vec{d}'}}{\mathcal{D}^{4n+2}} \right) %\\
	%&= - \sum_{ \substack{\vec{c}, \vec{d},  e_{m+1} \\ \vec{c}', \vec{d}'}} N_{\vec{d}}^{e_{m+1}} N_{\vec{d}'}^{e_{m+1}} \left(\frac{d_{\vec{d}} d_{\vec{d}'}}{\mathcal{D}^{4n+2}} \right) \ln \left(\ln \frac{d_{\vec{d}'}}{\mathcal{D}^2} +  \ln \frac{d_{\vec{d}}}{\mathcal{D}^2} + \ln \mathcal{D}^2 \right).
\end{align}
Summing over the fusion multiplicities, we obtain
\begin{align}
	S_R(A:B) &= -(2n+2) \sum_a \frac{d_a^2}{\mathcal{D}^2} \ln \frac{d_a}{\mathcal{D}^2} - \ln \mathcal{D}^2 
	\nonumber \\
	&= I(A:B)
\end{align}
and hence the Markov gap vanishes. This constitutes a positive check on the conjecture of Ref. \cite{Siva2021b} since, as noted above, the minimal central charge in the doubled system $\mathcal{C} \times \overline{\mathcal{C}}$ vanishes. 

\subsection{Negativity}

We now proceed to a computation of the negativity for the tripartition for which, to our knowledge, no prediction exists in the literature.
Applying the partial transpose on region $A$, we obtain 
\begin{align}
	&(\trho_{AB})_{\mathrm{cut}}^{T_A} = \\
	&\notag \sum_{\substack{ \vec{b},\vec{e}, \vec{\mu} \\ \vec{c},\vec{d}, \vec{\nu} \\ \vec{c}',\vec{d}',\vec{\nu}'} } \frac{\sqrt{d_{\vec{b}}}}{d_{e_{m+1}} \mathcal{D}^{2N-2}} \begin{gathered} \includegraphics[height=10em]{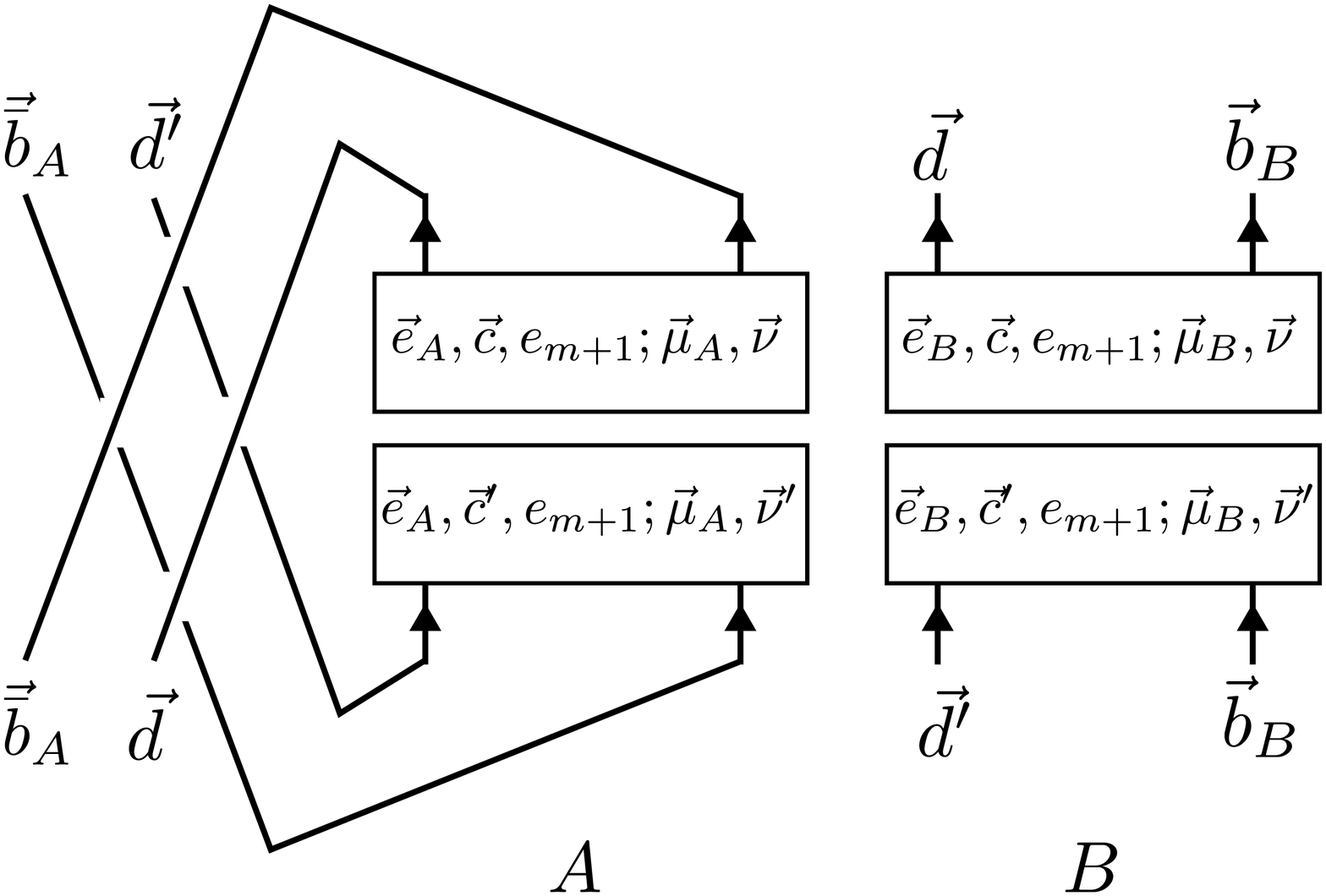} \end{gathered}
\end{align}
%Again, the only effect of the partial transpose is to conjugate and reverse the order of the anyon charges of the $A$ part of the diagram. 
Note that now the bra of $A$ has primed charges $\vec{d}'$ and $\vec{c}'$ while the ket has the unprimed charges; the opposite is true of the $B$ part of the diagram. Proceeding with the replica trick, we find for an even power $n_e \in 2\mathbb{Z}$,
%\begin{widetext}
\begin{align}
\notag [(\trho_{AB}^{T_A})_{\mathrm{cut}}  (\trho_{AB}^{T_A})_{\mathrm{cut}}^\dagger & ]^{\frac{n_e}{2}}  =  \sum_{\substack{ \vec{b},\vec{e}, \vec{\mu} \\ \vec{c},\vec{d}, \vec{\nu} \\ \vec{c}',\vec{d}',\vec{\nu}'} } \frac{1}{\sqrt{d_{\vec{d}} d_{\vec{d}'} d_{\vec{b}} } } \left( \frac{d_{\vec{b}}^2 d_{\vec{d}} d_{\vec{d}'}}{D^{4N-4} d_{e_{m+1}}^2} \right)^{\frac{n_e}{2}} \\
	&\times\begin{gathered} \includegraphics[height=9em]{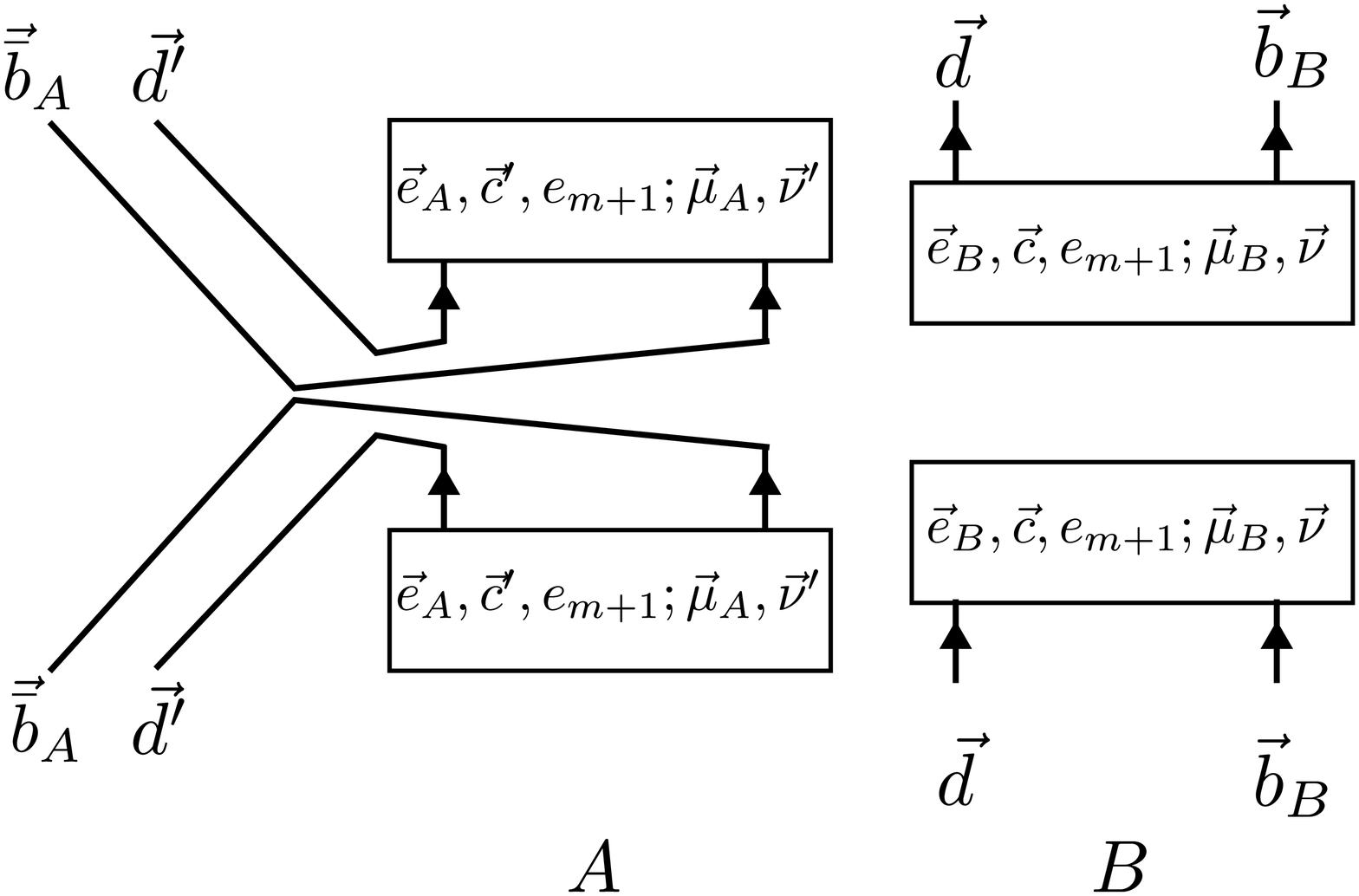} \end{gathered}
\end{align}
%\end{widetext}
Note that the matrix is now diagonal -- for instance, in the $A$ part of the diagram, both the bra and ket involve $\vec{d}'$ anyons, whereas in $\rho_{AB}^{T_A}$, the bra contained $\vec{d}'$ anyons and the ket $\vec{d}$ anyons. 
Taking the trace then yields,
\begin{align}
	\Tr [(\trho_{AB}^{T_A})_{\mathrm{cut}} & (\trho_{AB}^{T_A})^\dagger_{\mathrm{cut}}]^{\frac{n_e}{2}} = \sum_{\substack{ \vec{b},\vec{e}, \vec{\mu} \\ \vec{c},\vec{d}, \vec{\nu} \\ \vec{c}',\vec{d}',\vec{\nu}'} } \left( \frac{d_{\vec{b}}^2 d_{\vec{d}} d_{\vec{d}'}}{\mathcal{D}^{4N-4} d_{e_{m+1}}^2} \right)^{n_e / 2}
\end{align}
Analytically continuing $n_e \to 1$, we find
\begin{align}
	\mathcal{E}(A:B) = \ln\left[ \sum_{\vec{b}, \vec{e}, \vec{\mu} } \frac{d_{\vec{b}}}{\mathcal{D}^{2N-2} d_{e_{m+1}}} \left( \sum_{\vec{c},\vec{d},\vec{\nu}} \sqrt{d_{\vec{d}}} \right)^2 \right].
\end{align}
Now, we note that
\begin{align}
	\sum_{\substack{\vec{b}_A, \vec{e}_A, \vec{\mu}_A \\ \vec{b}_B, \vec{e}_B, \vec{\mu}_B} } d_{\vec{b}} = A_1 B_1 = \mathcal{D}^{4m} d_{e_{m+1}}^2,
\end{align}
using the notation introduced in the computation of the reflected entropy. 
So, recalling that $N = 2m +n + 2$, we obtain
\begin{align}
	\mathcal{E}(A:B) = \ln\left[ \sum_{e_{m+1}} \frac{d_{e_{m+1}}}{\mathcal{D}^{2(n+1)} } \left( \sum_{\vec{c},\vec{d},\vec{\nu}} \sqrt{d_{\vec{d}}} \right)^2 \right].
\end{align}
Observe that all dependence on the boundary of $AB$ with $C$ -- namely, the number of boundary components $2m+2$, the boundary charges $\vec{b}$ and $e_j$ for $j \neq m+1$ -- has dropped out. All that remains are the dependence on the length of the boundary between $A$ and $B$ (i.e. $n+1$ boundary components), the anyon charges contained entirely in $AB$ ($\vec{c}$, $\vec{d}$, $f$) and the charge line $e_{m+1}$ which appears at one of the trijunctions.
In order to evaluate the term in the parentheses, we follow the trick reviewed in Sec. \ref{sec:bipartite} to evaluate the R\'enyi entropies.
In terms of the $K$ matrix defined in Eq. \eqref{eq:K-matrix}, we have
\begin{align}
	\sum_{\vec{c},\vec{d},\vec{\nu}} d_{\vec{d}}^\alpha = \sum_{\vec{c},\vec{d}} N_{\vec{d}}^{e_{m+1}} d_{\vec{d}}^\alpha %&= \sum_{\vec{c},\vec{d}} N_{d_1 c_2}^{c_1} N_{d_2 c_3}^{c_2} \dots N_{d_{n-1} d_n}^{c_{n-1}} d_{\vec{d}}^\alpha \\
	%&= \sum_{\vec{c}', d_n} [K_\alpha]_{c_1c_2} [K_\alpha]_{c_2 c_3} \dots [K_\alpha]_{c_{n-1} d_n} d_{d_n}^\alpha \\
	%&= \sum_{\vec{c}', d_n,a} [K_\alpha]_{c_1c_2} [K_\alpha]_{c_2 c_3} \dots [K_\alpha]_{c_{n-1} a} N_{0a}^{d_n} d_{d_n}^\alpha \\
	%&= \sum_{\vec{c}', d_n,a} [K_\alpha]_{c_1c_2} [K_\alpha]_{c_2 c_3} \dots [K_\alpha]_{c_{n-1} a} [K_\alpha]_{a 0} \\
	&= [K_\alpha^{n+1}]_{e_{m+1} 0}.
\end{align}
Plugging this back into the expression for the negativity, we find
\begin{align}
	\mathcal{E}(A:B) = \ln\left[ \sum_{e_{m+1}} \frac{d_{e_{m+1}}}{\mathcal{D}^{2(n+1)} } \left( [K_{1/2}^{n+1}]_{e_{m+1} 0} \right)^2 \right].
\end{align}
Using the diagonal form of $K_\alpha$ and that $[v_{\alpha,0}]_e = d_e / \cD$, we have that 
\begin{align}
	[K_\alpha^{n+1}]_{e_{m+1} 0} 
	&=  \kappa_{\alpha,0}^{n+1} \frac{d_{e_{m+1}}}{\mathcal{D}^2}  e^{F(n+1,e_{m+1},\alpha)} 
\end{align}
Once again, $F(n+1,e_{m+1},\alpha) \to 0$ in the thermodynamic limit in which we take the number of wormholes $n \to \infty$. We thus find in this limit, 
\begin{equation}
	\mathcal{E}(A:B) = 2(n+1) \ln( \frac{\kappa_{\alpha,0}}{ \mathcal{D}}) - \ln \mathcal{D}^2 + \ln\left( \sum_{e_{m+1}} \frac{ d_{e_{m+1}}^3 }{ \mathcal{D}^2 } \right).
\end{equation}
As usual, we should identify this expression as the negativity for the topological order $\mathcal{C} \times \overline{\mathcal{C}}$. Dividing by two yields the negativity for just $\mathcal{C}$:
\begin{align}
	\mathcal{E}(A:B) = (n+1) \ln( \frac{\kappa_{\alpha,0}}{ \mathcal{D}}) - \ln \mathcal{D} + \frac{1}{2}\ln\left( \sum_{e_{m+1}} \frac{ d_{e_{m+1}}^3 }{ \mathcal{D}^2 } \right) \label{eq:tripartite-negativity}
\end{align}
This constitutes one of the main results of this work. The first term is an area law piece which is proportional to the length of the boundary between $A$ and $B$, as expected. 
The second term takes the usual form of the topological logarithmic negativity \cite{Wen2016}. The last term, however, appears to be a novel contribution.
It is worth noting that the $e_{m+1}$ anyon which appears in this expression appears precisely at the trijunction in the above diagrams, as indicated in Fig. \ref{fig:general-tripartition}(b), which suggests that this additional contribution should be associated with correlations arising from the trijunctions. We should be careful to note, however, that in our computation, we eliminated the $\omega_0$ loop at one of the trijunctions. This obscures the fact that we should interpret this additional contribution as being non-local and associated with \emph{both} trijunctions. 
Additionally, we see that this quantity \emph{vanishes} for purely Abelian topological order. We thus find that the negativity computed in this tripartite configuration allows for distinguishing Abelian and non-Abelian topological order. 

\section{Tetrajunction \label{sec:tetrajunction} }

\begin{figure}
  \centering
\subfloat[]{%
    \includegraphics[width=0.24\textwidth]{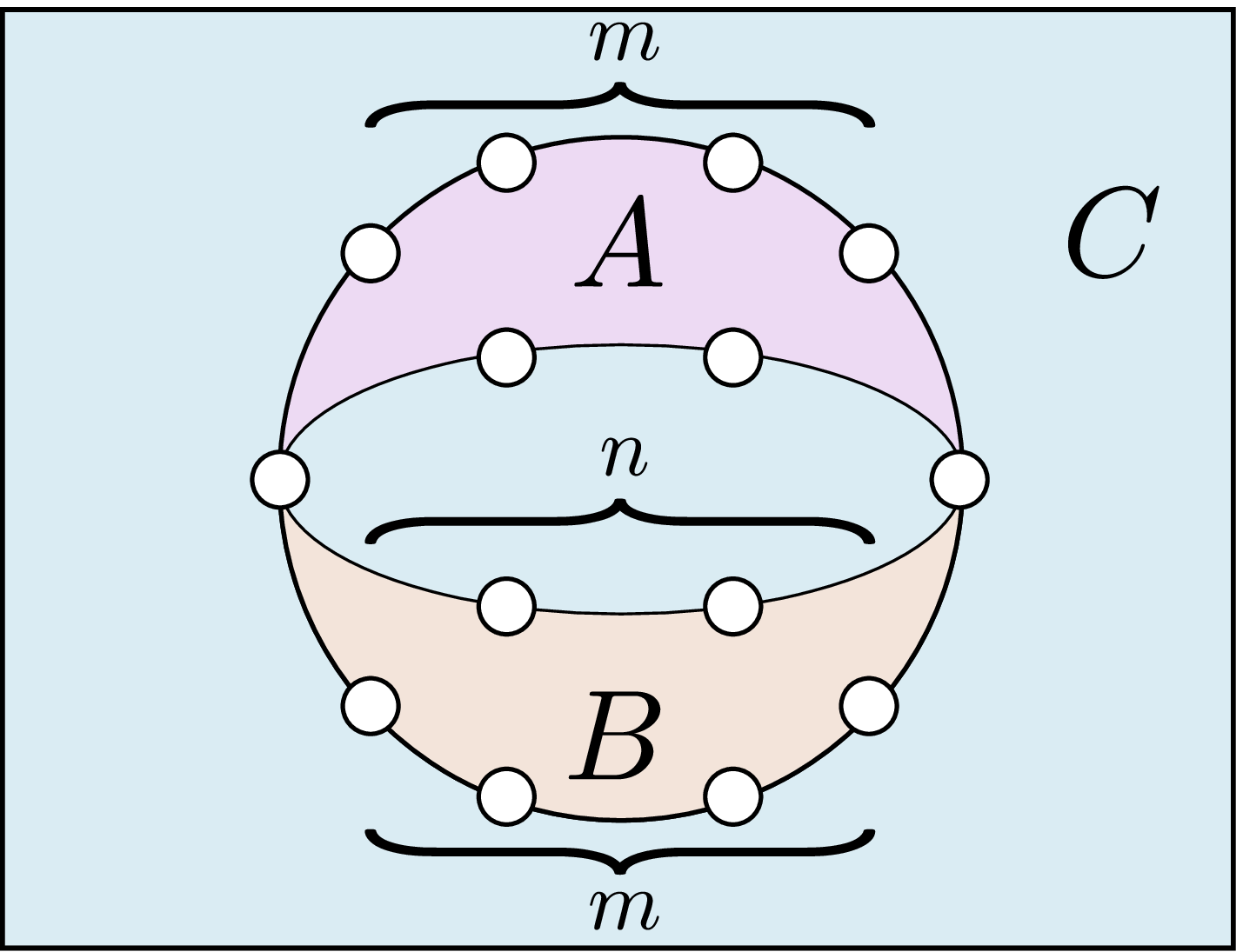}
}
\subfloat[]{%
    \includegraphics[width=0.24\textwidth]{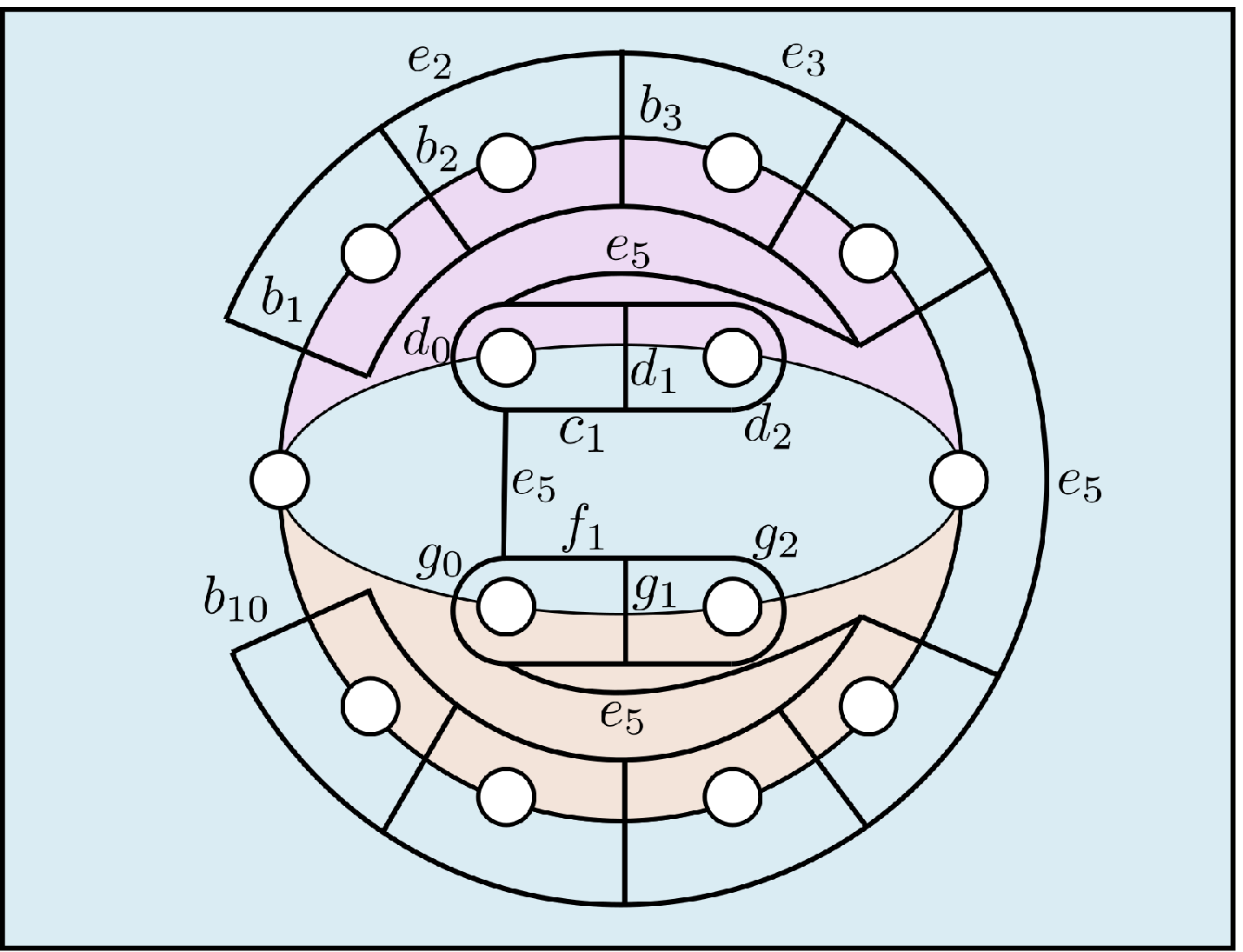}%0.36
}
    \caption{(a) Wormhole configuration for the tetrajunction calculation, viewed from the top down. We insert a total of $N = 2m + 2n + 2$ wormholes, with $m+n$ wormholes on the boundary of $A$ and $C$ (likewise for $B$ with $1$ exchanged with $2$) and a wormhole at each junction. (b) Top-down view of the anyon diagram with some anyon charge lines labelled, after performing the manipulations in the main text such that single anyon charge threads each connected component of each boundary.} \label{fig:tetrajunction}
\end{figure}

In the preceding section, we confirmed the vanishing of the Markov gap and identified a new contribution to the negativity for a particular tripartition of a topological order in which the three regions $A$, $B$, $C$ meet at a trijunction. In this section, we consider a tripartition in which $A$, $B$, and $C$ meet at tetrajunctions, as illustrated in Fig. \ref{fig:tetrajunction}(a). Here, we do not have expectations \textit{a priori} for the Markov gap nor the negativity; in particular, we would like to see whether the new contribution to the negativity found in the preceding section persists in this distinct tripartitioning.

\begin{figure*}
  \centering
  \subfloat[]{%
  \includegraphics[width=\textwidth]{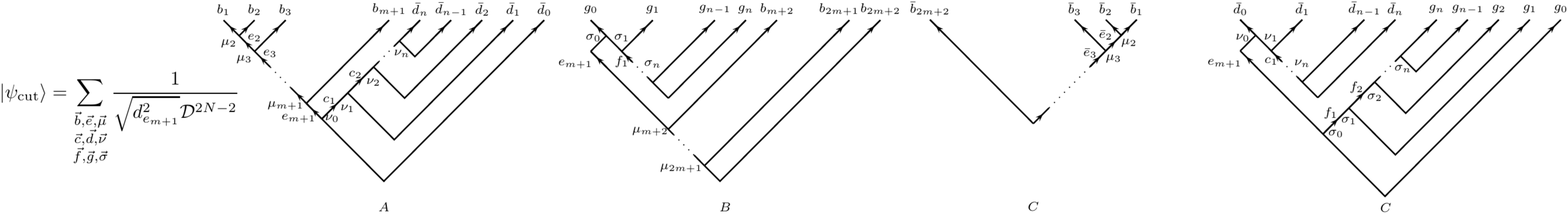}
  }
  
  \subfloat[]{%
  \includegraphics[width=0.85\textwidth]{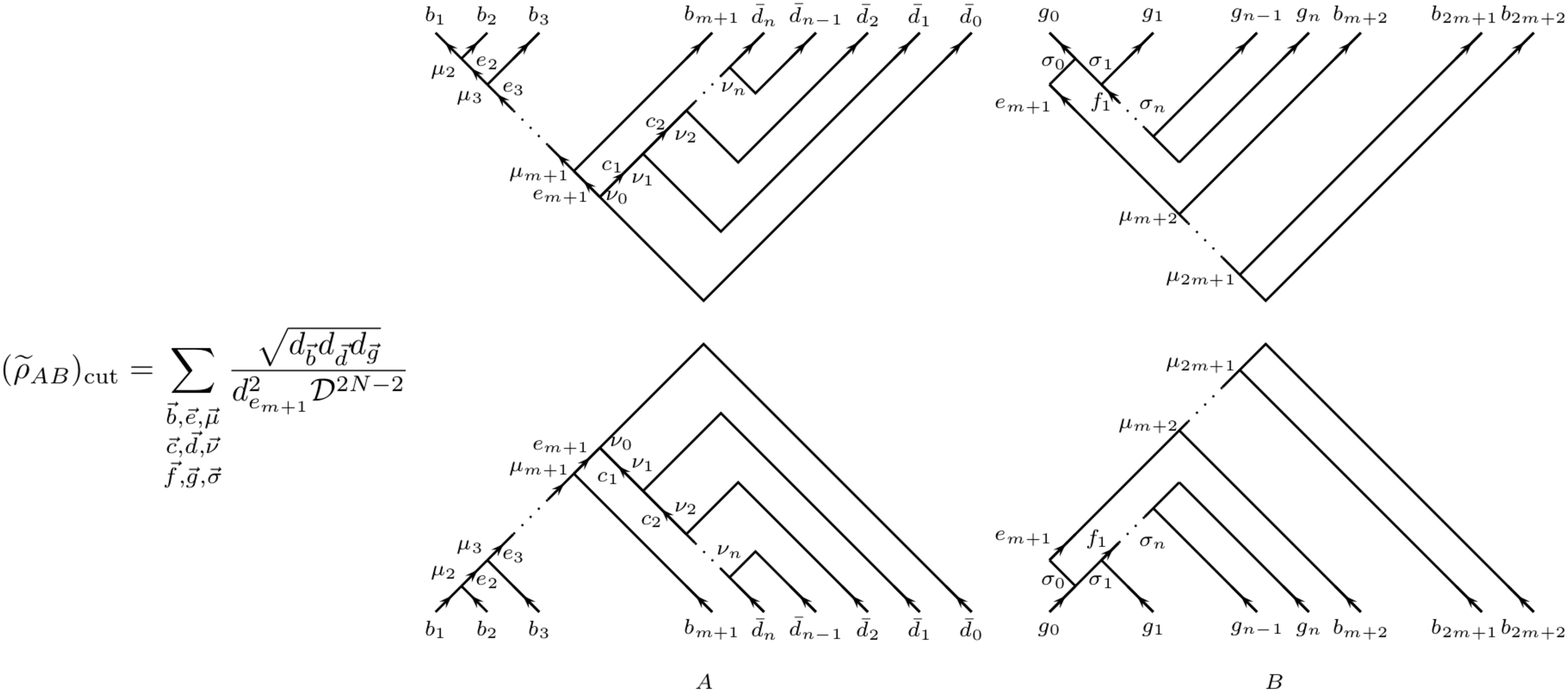}
  }
    \caption{ (a) State for the tripartition of Fig. \ref{fig:tetrajunction} after cutting along the entanglement cuts.  (b) Cut reduced density matrix  for region $AB$. } \label{fig:tetraunction-rhoAB}
\end{figure*}

As usual, we double the system and insert wormholes along the entanglement cut, including wormholes at each tetrajunction, as illustrated in Fig. \ref{fig:tetrajunction}(a). In total, we insert $N = 2m + 2n + 2$ wormholes, with $m+n$ wormholes on the boundary of $A$ and $C$ (likewise for $B$) and a wormhole at each tetrajunction. 
After performing modular $\mathcal{S}$ transformations at each wormhole, the ground state anyon diagram takes the form
%\begin{widetext}
\begin{align}
& \ket{\psi} = \mathcal{D}^{N-1}
	\begin{gathered}
\includegraphics[height=11em]{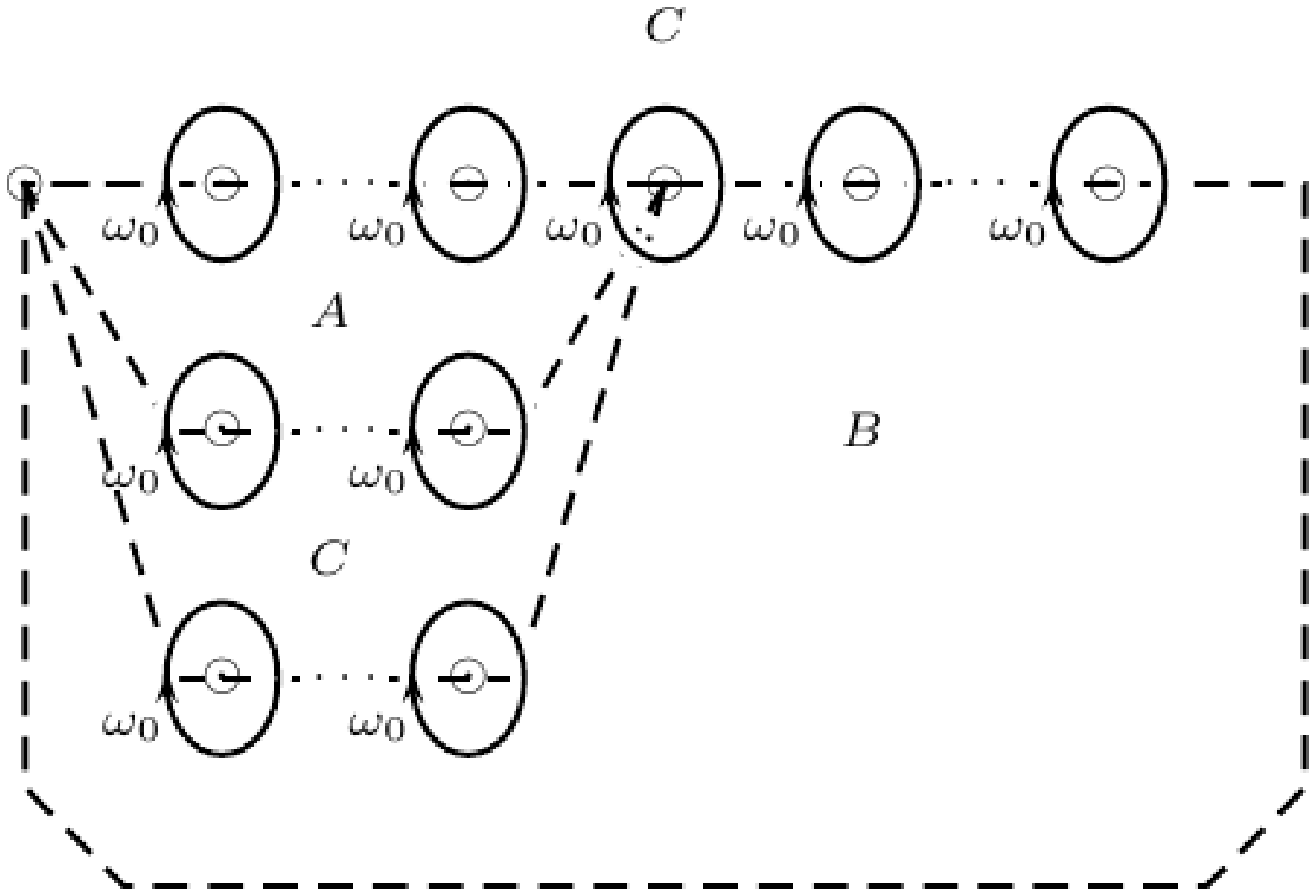}. \end{gathered} 
\end{align}
Here, we have deformed the entanglement cuts such that the wormholes lie along three lines. As in the tripartite computation, we also used the handle-slide property to eliminate one of the $\omega_0$ loops, in this case, the loop centered at one of the tetrajunctions. Making use of the definition of the $\omega_0$ loops,
\begin{align}
 \ket{\psi} &= \frac{1}{\mathcal{D}} \sum_{\vec{e} , \vec{c}, \vec{f}} \frac{d_{\vec{e}} d_{\vec{c}} d_{\vec{f}} }{\mathcal{D}^N}
\begin{gathered}
\includegraphics[height=8em]{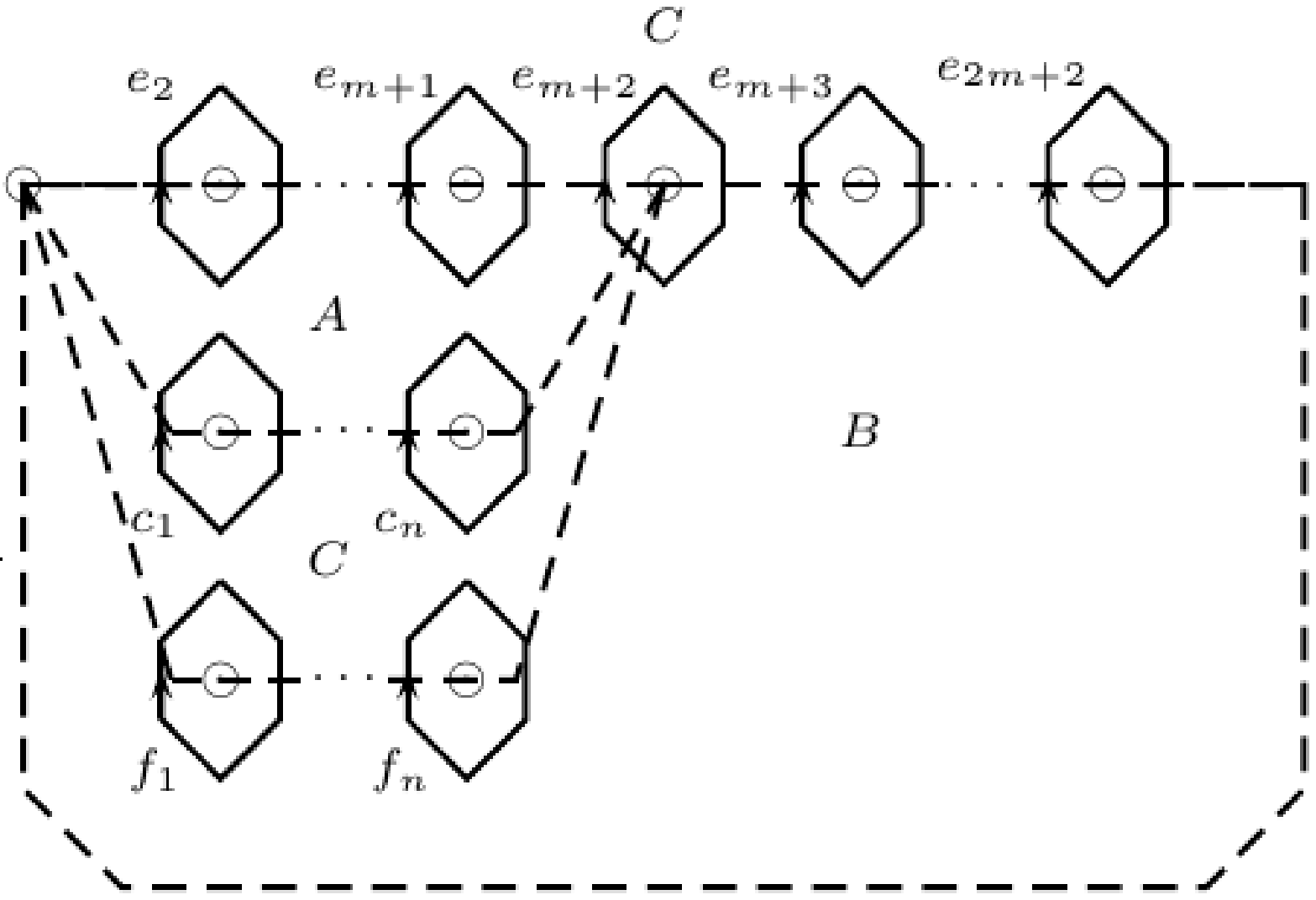}. \end{gathered} 
\end{align}
Here $\vec{e}$, $\vec{c}$, and $\vec{f}$ label the anyons circulating around the wormholes on the exterior boundary of $A\cup B$ with $C$, the interior boundary of $A$ with $C$, and the interior boundary of $B$ with $C$, respectively. From here, we can again apply the same set of manipulations as in the bipartition computation to the row of $\vec{e}$ loops to put it in a tree-like form. Then, as in the trijunction computation, we move the $e_{m+1}$ anyon line past the $\vec{c}$ and $\vec{f}$ loops using the handle-slide property and then apply resolutions of identity to join everything together into a single diagram. The resulting diagram is given by,
\begin{align}
	\ket{\psi} = &\sum_{ \substack{ \vec b , \vec e , \vec c , \\ \vec d ,  \vec f , \vec g } } \frac{\sqrt{d_{\vec b} d_{\vec d} d_{\vec g}}}{d_{e_{m+1}}^2 \mathcal{D}^{2N-2}} \\
	\notag &\times\begin{gathered}
\includegraphics[height=26em]{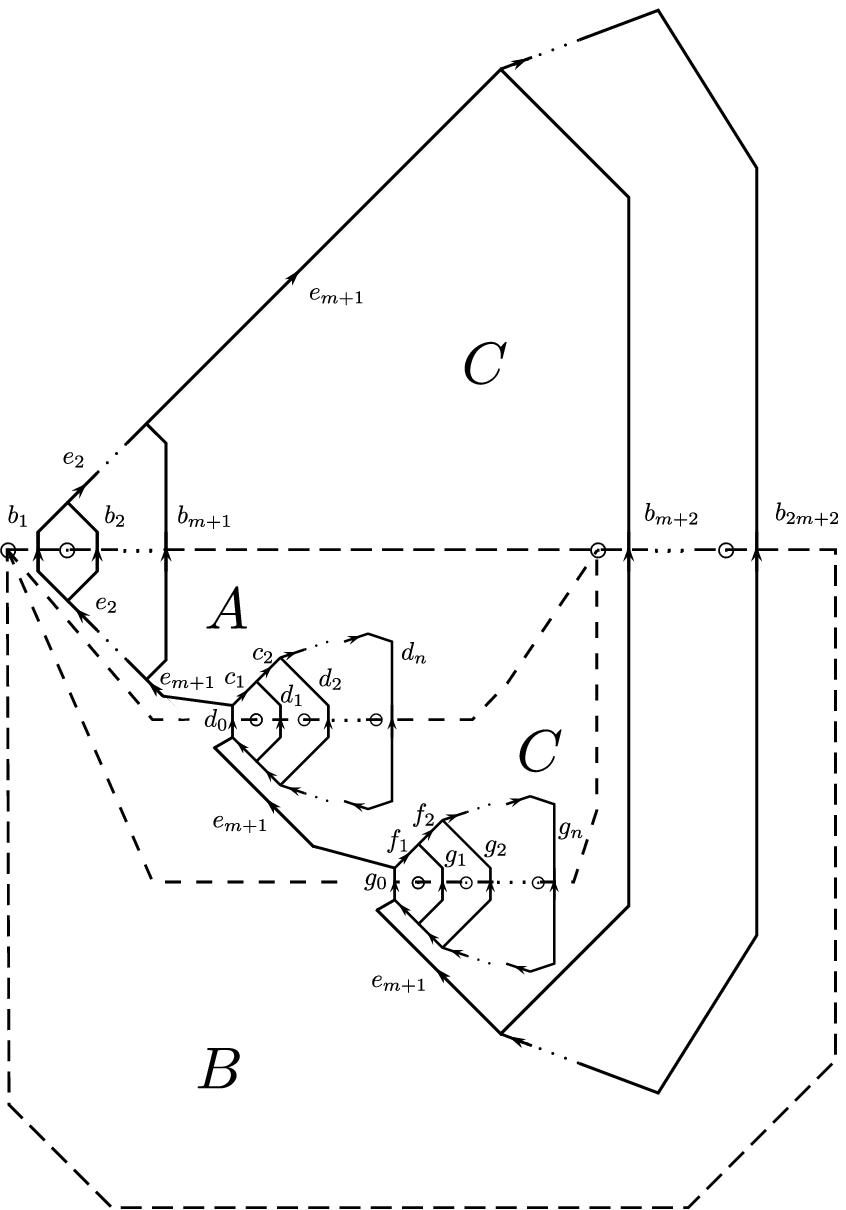}
\end{gathered}
\end{align}
In Fig. \ref{fig:tetrajunction}, we overlay this diagram on top of the partitioned subregions with their original topology for clarity. Cutting along the entanglement cuts yields the diagram in Fig. \ref{fig:tetraunction-rhoAB}(a). On tracing out $C$, we obtain the $\trho_{AB}$ given in Fig. \ref{fig:tetraunction-rhoAB}(b).

Let us briefly unpack this expression, where we have restored the fusion vertex labels. The $b$ anyons are those threading the ``outer" boundary of $A\cup B$ with $C$ while the $d$ and $g$ anyons are those threading the inner boundaries between $A$ and $B$, respectively, with $C$. As in the trijunction computation, the $\vec{e}$ charges connect the $\vec{b}$ lines and the $\vec{\mu}$ label the fusion vertices for these lines. For instance, $e_2$ splits into $b_1$ and $b_2$ at $\mu_2$, $e_3$ splits into $e_2$ and $b_3$ at $\mu_3$ and so on up to $\mu_{2m}$, at which point $\bar b_{2m+2}$ splits into $b_{2m+1}$ and $e_{2m}$. Once again, we have $b_j$ for $j \in \{1, \dots , 2m+2 \}$, $e_j$ for $j \in \{2, \dots , 2m+1 \}$, and $\mu_j$ for $j \in \{2, \dots , 2m +1\}$. We note further that the charges and vertices $\vec{b}_B = (b_{m+2} , \dots , b_{2m+2})$, $\vec{e}_B = (e_{m+2} , \dots , e_{2m+2})$, $\vec{\mu}_B = (\mu_{m+2} , \dots , \mu_{2m+1})$ appear only in the $B$ subregion, while the charges and vertices $\vec{b}_A = (b_{1} , \dots , b_{m+1})$, $\vec{e}_A = (e_{2} , \dots , e_{m})$, $\vec{\mu}_A = (\mu_{2} , \dots , \mu_{m+1})$ appear only in the $A$ subregion. The $e_{m+1}$ charge, however, once again appears in both $A$ and $B$ subregions.

Next, the $\vec{d}$ charges label the anyon lines threading the components of the ``inner" boundary between $A$ and $C$. The $\vec{c}$ charges connect the $\vec{d}$ lines and the $\vec{\nu}$ label the fusion vertices for these lines. Explicitly, $c_1$ splits at $\nu_1$ into $d_1$ and $c_2$, $c_2$ splits at $\nu_2$ into $d_2$ and $c_3$, and so on up to $\nu_{n-1}$ where $c_{n-1}$ splits into $d_{n-1}$ and $c_n$. We have $c_j$ for $j \in \{1, \dots , n-1 \}$, $d_j$ for $j \in \{0, \dots , n \}$, and $\nu_j$ for $j \in \{0, \dots , n-1 \}$. Likewise, the anyon charges and fusion vertices $\vec{f}$, $\vec{g}$, and $\vec{\sigma}$ play the same roles as $\vec{d}$, $\vec{c}$, and $\vec{\nu}$ but on the ``inner" boundary between $B$ and $C$.

As in the trijunction computations, in order to simplify the following diagrams, we use the following shorthand for the above density matrix:
\begin{align}
	(\trho_{AB})_{\mathrm{cut}} = \sum_{ \substack{ \vec b , \vec e , \vec \mu \\ \vec c , \vec d , \vec \nu \\ \vec f , \vec g , \vec \sigma } } \frac{\sqrt{d_{\vec b} d_{\vec d} d_{\vec g}}}{d_{e_{m+1}}^2 \mathcal{D}^{2N-2}} \, \begin{gathered} \includegraphics[height=11em]{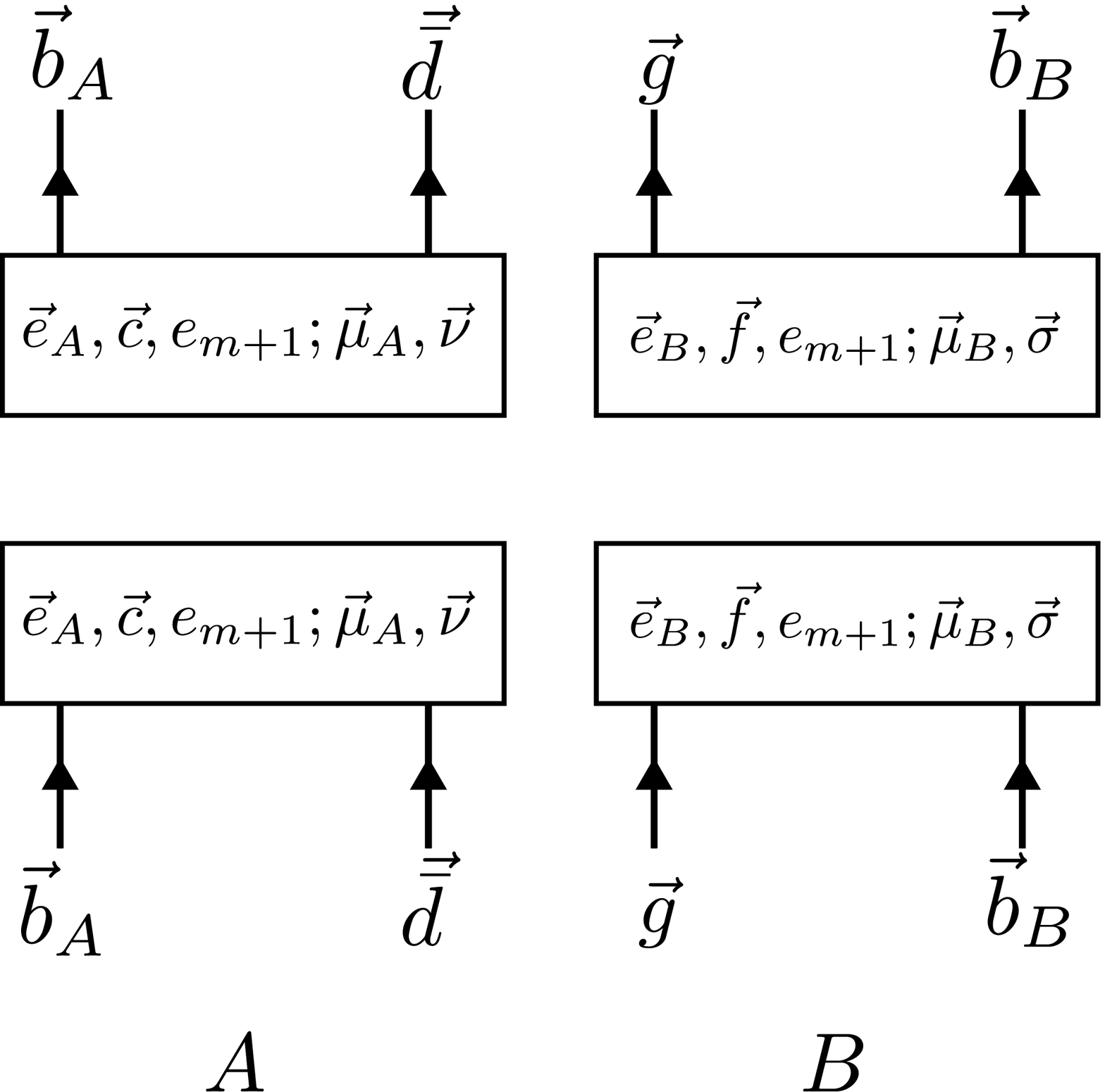} \end{gathered}
\end{align}
Again, the boxes are labeled by the internal anyon charges and fusion labels and the boxes with arrows exiting from the top and entering from the bottom correspond, respectively, to the ket and bra of the $A$ and $B$ parts of the anyon diagram.
Before proceeding to computing the desired entanglement quantities, we first note that, in contrast to the reduced density matrix for the trijunction setup, this matrix is diagonal. This is a consequence of the fact that every anyon line that exits region $A$ or $B$ necessarily passes into region $C$. Hence, tracing out $C$ sets the indices labelling the bras in $\trho_{AB}$ equal to the indices labelling the kets. 

\subsection{Mutual Information}

First, we consider the mutual information.
Using the bipartite entanglement computation, we may immediately conclude,
\begin{align}
	S_A &= -(m + n +2) \sum_b \frac{d_b^2}{\cD^2}\ln \frac{d_b}{\cD^2} - \ln \cD^2 \, ,\\
	S_B &= -(m + n +2) \sum_b \frac{d_b^2}{\cD^2}\ln \frac{d_b}{\cD^2} - \ln \cD^2 \, ,
\end{align}
since, for instance, the boundary of $A$ has $m + n + 2$ connected components. On the other hand, we cannot directly read off the entanglement entropy of $AB$. It is straightforward to find using Eq. \eqref{eq:inner-product},
\begin{align}
	\mathrm{Tr}\left[ (\trho_{AB})_{\mathrm{cut}}^\alpha \right] &= \sum_{ \substack{\vec b , \vec c , \vec f \\ \vec e , \vec d , \vec g \\ \vec \mu , \vec \nu , \vec \sigma }  } \left( \frac{d_{\vec b} d_{\vec d} d_{\vec g}}{d_{e_{m+1}}^2 \mathcal{D}^{2N-2} } \right)^\alpha,
\end{align}
and so taking the von Neumann limit,
\begin{align}
	S_{AB} = - \sum_{ \substack{ \vec b , \vec e , \vec \mu \\ \vec c , \vec d , \vec \nu \\ \vec f , \vec g , \vec \sigma } } \left( \frac{d_{\vec b} d_{\vec d} d_{\vec g}}{d_{e_{m+1}}^2 \mathcal{D}^{2N-2} } \right) \ln \left( \frac{d_{\vec b} d_{\vec d} d_{\vec g}}{d_{e_{m+1}}^2 \mathcal{D}^{2N-2} } \right).
\end{align}
Now, summing over fusion multiplicities, we find that
\begin{align}
	\sum_{\vec{c} , \vec{d} , \vec{\nu}} d_{\vec{d}} &= \sum_{\vec{c} , \vec{d}} N_{e_{m + 1}c_1}^{d_0} N_{c_1 d_1}^{c_2} \dots N_{c_{n-1} d_{n-1}}^{d_n} d_{\vec{d}} = \cD^{2n}d_{e_{m+1}},
\end{align}
\begin{align}
	\sum_{\vec{f} , \vec{g} , \vec{\sigma}} d_{\vec{g}} &= \sum_{\vec{f} , \vec{d}} N_{e_{m + 1}f_1}^{g_0} N_{f_1 g_1}^{f_2} \dots N_{f_{n-1} g_{n-1}}^{g_n} d_{\vec{g}} = \cD^{2n}d_{e_{m+1}},
\end{align}
and
\begin{align}
	\notag \sum_{\vec{b} , \vec{e} , \vec{\mu}} d_{\vec{b}} &= \sum_{\vec{b} , \vec{e}} N_{b_1 b_2}^{e_2} N_{e_2 b_3}^{e_3} \dots N_{e_{m} b_{m+1}}^{e_{m+1}}  \\
	\notag &\qquad\qquad \times N_{e_{m+1} b_{m+2}}^{e_{m+2}} \dots N_{e_{2m} b_{2m+1}}^{b_{2m+2}} d_{\vec{b}} \\
	&= \cD^{4m}d_{e_{m+1}}^2.
\end{align}
Each of these summations correspond, respectively, to the area law contributions to the entanglement entropy from the inner boundary between $A$ and $C$, the inner boundary between $B$ and $C$, and the outer boundary between $A\cup B$ and $C$. Altogether, we find,
\begin{align}
	%S_{AB} = -N \sum_{e_{m+1}} \frac{d_{e_{m+1}}^2}{\cD^2} \ln \frac{d_{e_{m+1}}}{\cD^2} - \ln \cD^2.
	S_{AB} = -N \sum_{b} \frac{d_{b}^2}{\cD^2} \ln \frac{d_{b}}{\cD^2} - \ln \cD^2.
\end{align}
We thus find that the mutual information,
\begin{align}
	I(A:B) &= -\sum_b \frac{d_b^2}{\mathcal{D}^2}\ln \frac{d_b^2}{\mathcal{D}^2} \, , %-2 \sum_b \frac{d_b^2}{\cD^2}\ln \frac{d_b}{\cD^2} - \ln \cD^2
\end{align}
does not vanish. %In particular, we note that the remaining summation is over the anyon charge $e_{m+1}$, which is again the anyon located at one of the tetrajunctions. This strongly that we should associate this contribution to the mutual information as arising from the tetrajunction.
We note, as usual, that the above expression is for the doubled phase $\mathcal{C} \times \overline{\mathcal{C}}$. Dividing by two, we obtain
\begin{align}
	I(A:B) = - \sum_b \frac{d_b^2}{\mathcal{D}^2} \ln \frac{d_b}{\mathcal{D}}
\end{align}
as the mutual information for this tripartition in the phase $\mathcal{C}$.

\subsection{Reflected Entropy}

Next, we proceed to compute the reflected entropy and the Markov gap.
Proceeding with the first part of the replica trick, it is straightforward to find for the canonical purification,
\begin{align}
	\kket{(\trho_{AB})_{\mathrm{cut}}^{\alpha/2}} = \sum_{ \substack{ \vec b , \vec e , \vec \mu \\ \vec c , \vec d , \vec \nu \\ \vec f , \vec g , \vec \sigma } } & \left( \frac{\sqrt{d_{\vec b} d_{\vec d} d_{\vec g}}}{d_{e_{m+1}}^2 \mathcal{D}^{2N-2}}\right)^{\alpha / 2} \frac{1}{\sqrt{d_{\vec b} d_{\vec d} d_{\vec g}}} \\
	\notag &\times\begin{gathered} \includegraphics[height=11em]{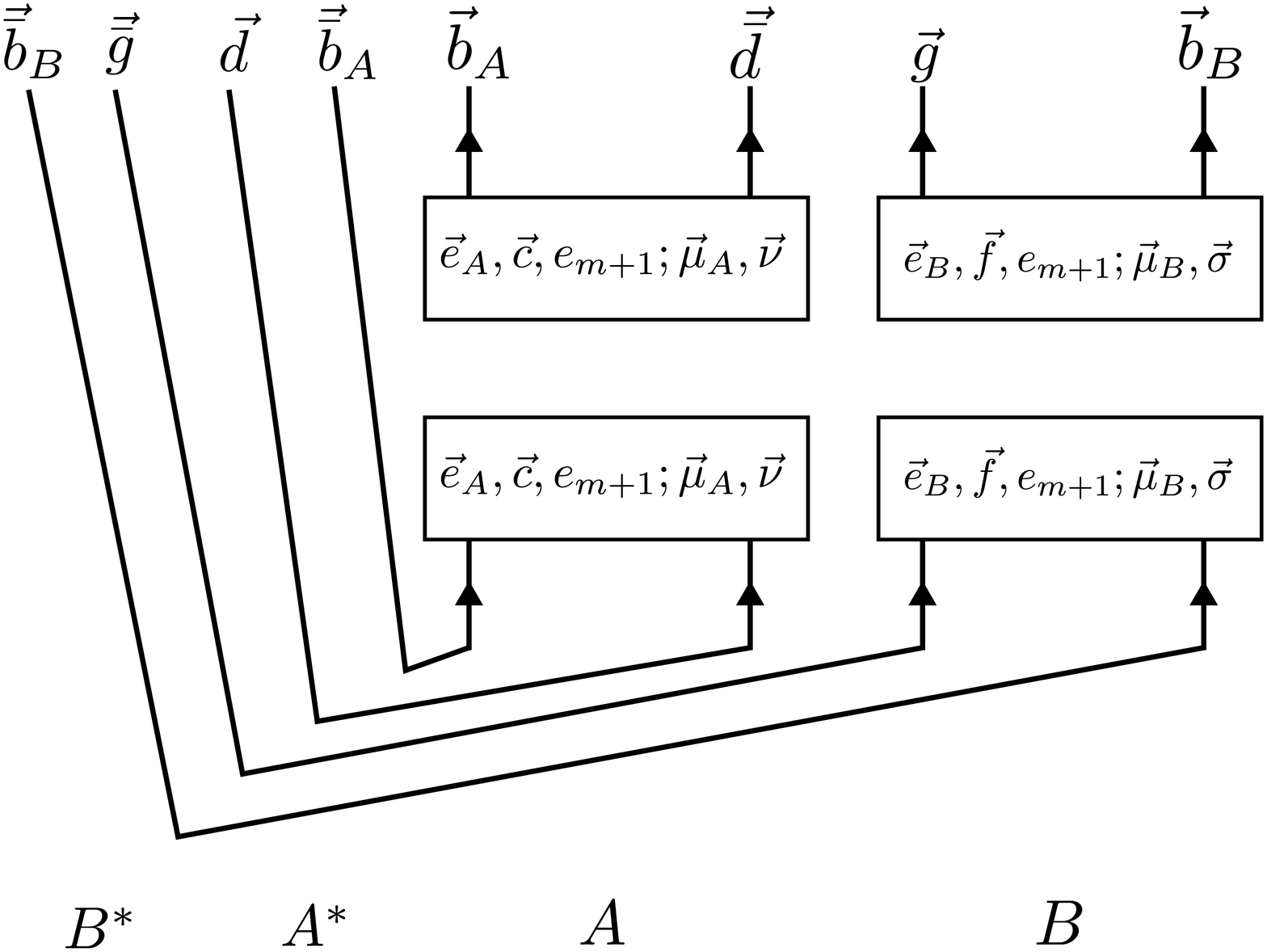} \end{gathered}
\end{align}
Following steps analogous to those in the preceding tripartition computation, we trace out $BB^*$ and then apply the second replica trick, computing,
\begin{align}
	\Tr[\rho_{AA^*}^{(\alpha)}]^\beta = \sum_{e_{m+1}} \left(\frac{B_{\alpha} A_{\alpha}}{(d_{e_{m}+1}^2 \cD^{2N-2})^\alpha} \right)^\beta,
\end{align}
where we have defined 
\begin{align}
	A_{\alpha} = \sum_{ \substack{ \vec{b}_A \vec{d} , \vec{\mu}_A \\ \vec{e}_A \vec{c} , \vec{\nu} } } (d_{\vec{b}_B} d_{\vec{d}})^\alpha 
\end{align}
and
\begin{align}
	B_{\alpha} = \sum_{ \substack{ \vec{b}_B , \vec{e}_B , \vec{\mu}_B \\ \vec{g} , \vec{f} , \vec{\sigma} } } (d_{\vec{b}_B} d_{\vec{g}})^\alpha.
\end{align}
In the replica limit $\alpha \to 1$, we have that
\begin{align}
	A_{1} = B_{1} &= d_{e_{m}+1}^2 \cD^{2m + 2n} \, .
\end{align}
Recalling that the $\vec{b_A}$ and $\vec{d}$ ($\vec{g}$ and $\vec{b}$) anyons are those piercing the boundary between regions $A$ ($B$) and $C$, we see that $A_1$ and $B_1$ contribute to the area law terms for regions $A$ and $B$ in the entanglement entropy, respectively.
Taking the von Neumann limit $\beta \to 1$, we finally obtain
\begin{align}
	S_R = -\sum_b \frac{d_b^2}{\mathcal{D}^2}\ln \frac{d_b^2}{\mathcal{D}^2} = I(A:B),
\end{align}
and so the Markov gap again vanishes. In particular, the reflected entropy receives the same universal contribution from the tetrajunctions as does the mutual information.

\subsection{Negativity}

Finally, we consider the negativity. 
Taking the partial transpose, we may compute
\begin{align}
	&(\trho_{AB}^{T_A})_{\mathrm{cut}} (\trho_{AB}^{T_A} )_{\mathrm{cut}}^\dagger = \\
	\notag &\sum_{ \substack{ \vec b , \vec e , \vec \mu \\ \vec c , \vec d , \vec \nu \\ \vec f , \vec g , \vec \sigma } } \frac{\sqrt{d_{\vec b} d_{\vec d} d_{\vec g}}}{d_{e_{m+1}}^2 \mathcal{D}^{2N-2}}  \begin{gathered} \includegraphics[height=10em]{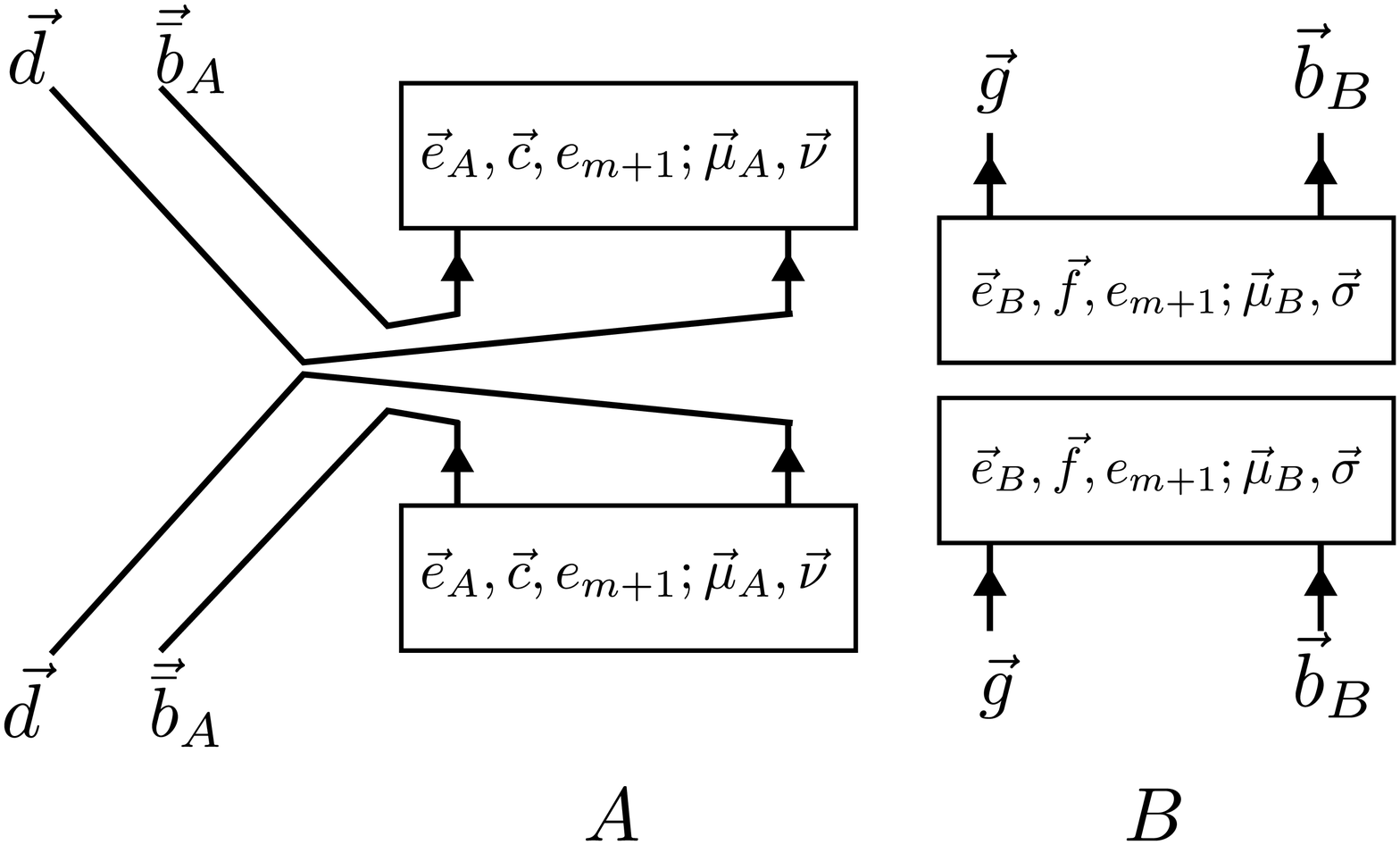} \end{gathered}.
\end{align}
Noting that the anyon diagram is diagonal in anyon and vertex labels, one may compute
\begin{align}
	\lim_{n_e \to 1} \Tr[(\trho_{AB}^{T_A})_{\mathrm{cut}} & (\trho_{AB}^{T_A})^\dagger_{\mathrm{cut}}]^{\frac{n_e}{2}} = \Tr[\trho_{AB}] = 1
\end{align}
and hence the negativity vanishes.
Ultimately, one can trace this back to the fact noted above that any anyon line connecting $A$ with $B$ necessarily passes through $C$. %Each line necessarily passes through $C$. 
Heuristically speaking, one may understand this aspect of the anyon diagram as implying that $A$ is not entangled with $B$.

%Once more, we can gain some intuition for the absence of the new contribution to the negativity found in the preceding tripartition case in the present situation by again making an analogy with the computation performed on the torus in Ref. \cite{Wen2016}. Indeed, in Ref. \cite{Wen2016} it was found that the negativity vanishes in precisely this configuration on the torus. Again, while this analogy does not provide a physical explanation for the value of the negativity, it does provide some heuristic understanding for why the additional contribution to the negativity in Eq. \eqref{eq:tripartite-negativity} in the previous tripartition setup is absent here.

\section{Trijunction with Anyon Insertions \label{sec:tripartite-insertions} }

\begin{figure}
  \centering
\subfloat[]{%
    \includegraphics[width=0.35\linewidth]{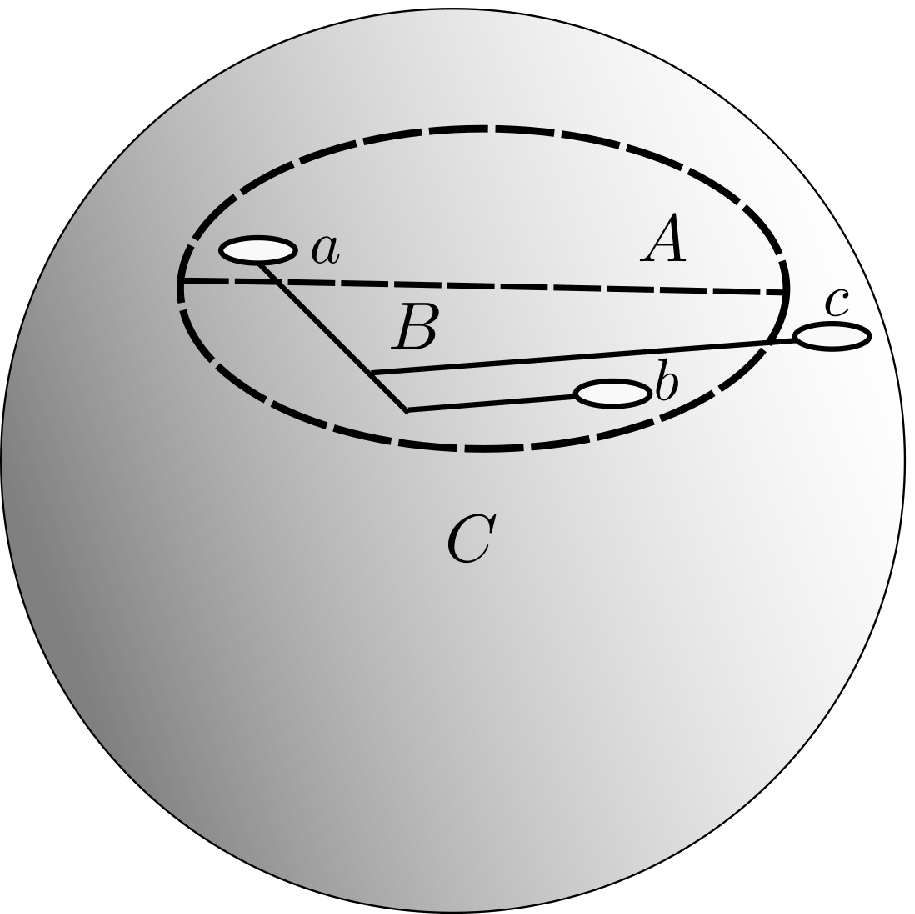}
}
\subfloat[]{%
    \includegraphics[width=0.55\linewidth]{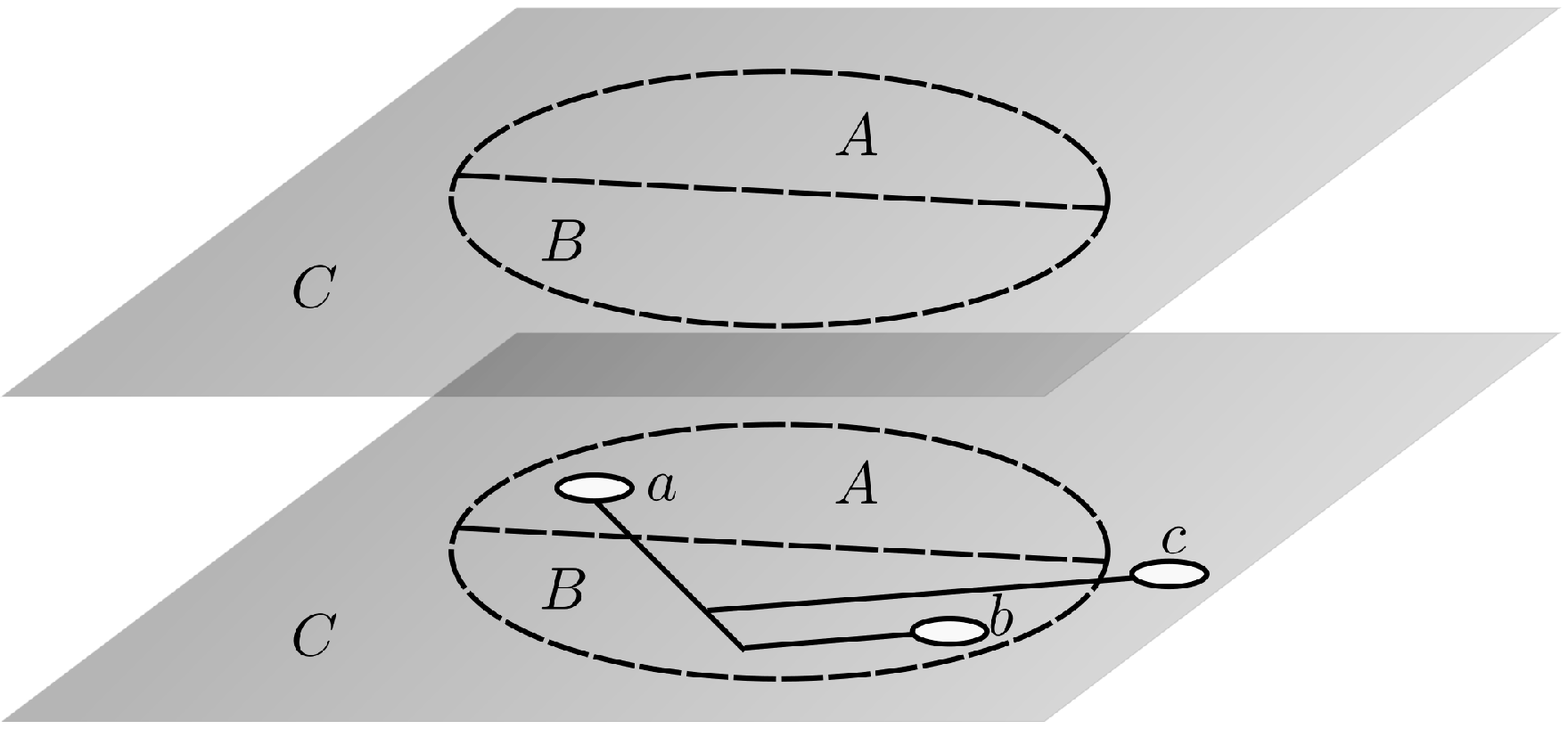}
}

\subfloat[]{%
    \includegraphics[width=0.3\textwidth]{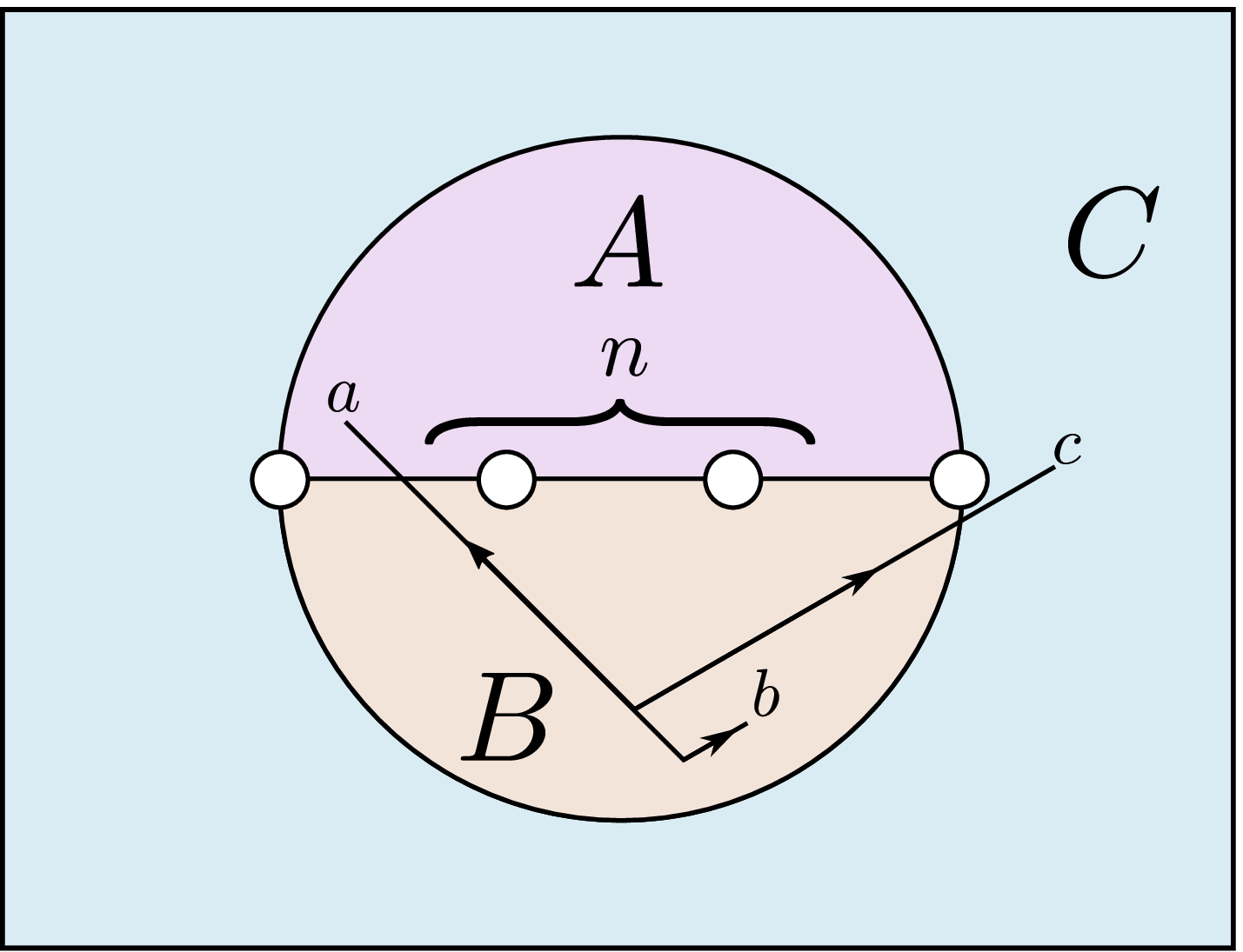}
}
    \caption{(a) Tripartition of the sphere into $A,B$ and $C$, with a trimer of anyons, $a,b$ and $c$ fusing to the identity, each localized in one of the three regions. (b) Zoom in of the sphere, after introducing a copy of the time-reversed conjugate topological phase on another sphere. (c) View from the top of the tripartition, after the wormhole insertion. We insert one wormhole at each trijunction and $n$ wormholes along the $A/B$ boundary.} \label{fig:tri-anyon-insertion}
\end{figure}

While we found the Markov gap in a topologically ordered ground state to vanish in the two tripartitions considered, it is
still possible that is non-zero in \emph{excited} states. 
To that end, we proceed to compute the Markov gap for the tripartition of Section \ref{sec:trijunction}, in the presence of an anyon trimer. %In the same way that the bipartite entanglement entropy and negativity are sensitive to the quantum dimensions of anyon pairs,
One might expect that the tripartite correlations inherent in an anyon trimer may lead to new contributions to these entanglement quantities. Indeed, while we are unable to derive a simple expressions, %closed form expression, % for general topological orders, 
we find an interesting dependence of the negativity and reflected entropy on the $F$-symbols. In particular, when each anyon in the trimer is non-Abelian, we find the Markov gap to be \emph{non-vanishing}. We compare this result with that for a single trimer, ignoring the topological liquid background, for which the reflected entropy takes a simple closed form expression.

We consider a tripartition of a topological phase on a sphere into regions $A$, $B$, and $C$, and insert three anyons, $a$, $b$, and $c$ into each region, respectively, such that they fuse to the identity, as illustrated in Fig. \ref{fig:tri-anyon-insertion}(a).  While we will ultimately restrict to multiplicity-free theories to obtain simplified expressions, we need to keep track of the fusion multiplicities at each vertex in order to carry our computation through. 
As usual, to construct the ground state anyon diagram, we introduce a second sphere with the conjugate topological phase, as indicated in Fig. \ref{fig:tri-anyon-insertion}(b). Note that we do not include any anyon insertions on this additional sphere. This means, in particular, that the contribution of the anyon insertions to the entanglement quantities we compute will \emph{not} be doubled. We then adiabatically insert wormholes between the two spheres along the entanglement cuts. 

%In order to simplify our computations, we take a slightly different approach relative to that in Section \ref{sec:trijunction}. 
Now, in our tripartition computation in the absence of anyons, we found that the negativity and R\'enyi reflected entropy are insensitive to the length of the boundary between $AB$ and $C$ -- in particular, the computation is unaffected if we do not insert any wormholes (besides those at the trijunctions) along this boundary. %In the present case, we expect the anyon insertions to yield an $O(1)$ contribution to the entanglement quantities of interest. 
Thus, in order to separate the area law and the subleading topological contributions, it is sufficient check the scaling of the negativity and reflected entropy with the number of wormholes along only the boundary between $A$ and $B$. 
With this in mind,
we insert $N=n+2$ wormholes, with one wormhole at each trijunction and $n$ wormholes along the boundary between $A$ and $B$, as depicted in Fig. \ref{fig:tri-anyon-insertion}(c).

%In the diagrammatic notation, 
Diagrammatically, we write the normalized state as
\begin{equation}
\ket{\psi} = (d_a d_b d_c)^{-1/4} \mathcal{D}^{N-1}
%%%
\begin{gathered}
\includegraphics[height=8em]{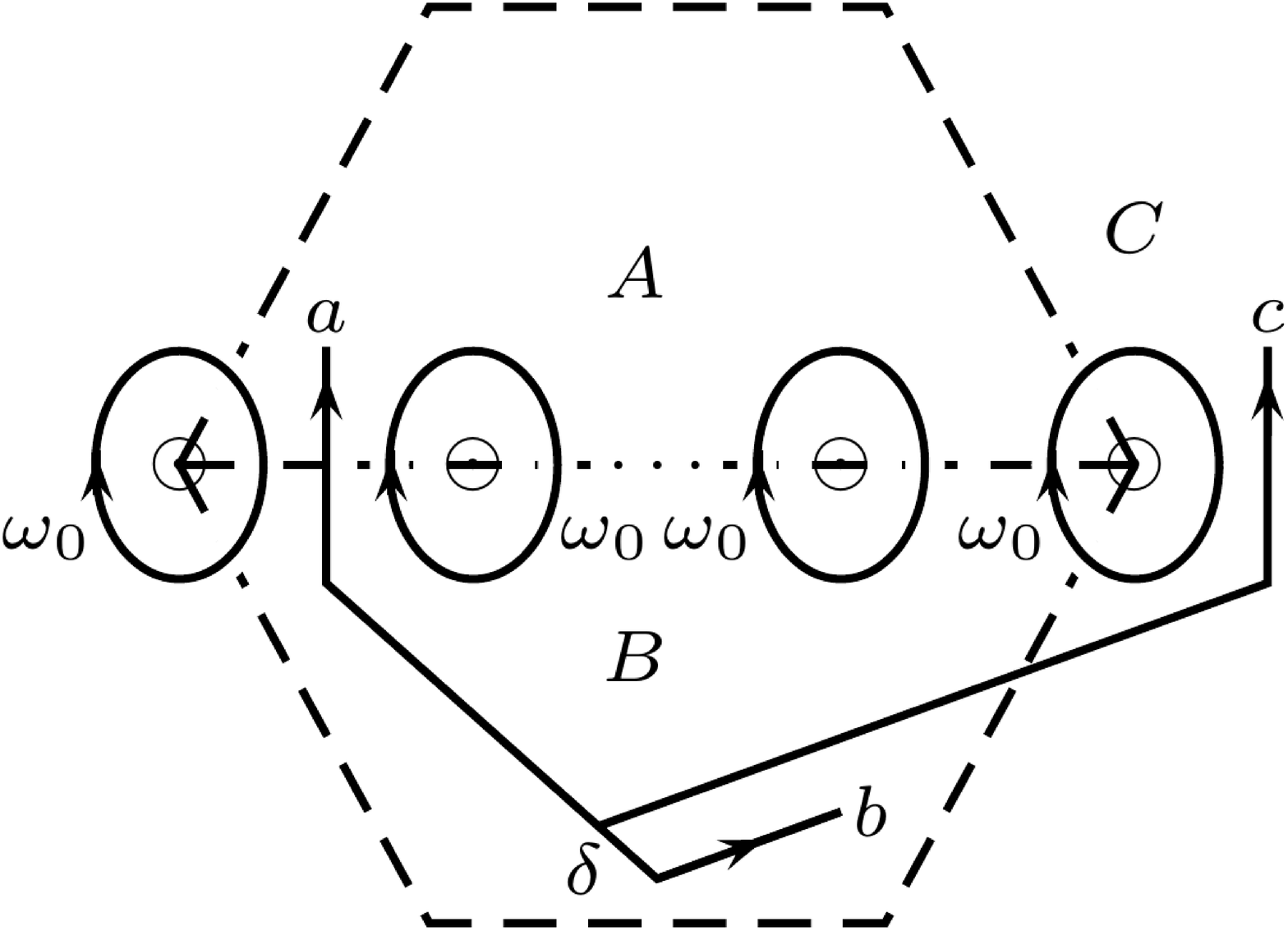} \end{gathered} 
\end{equation}
This corresponds to the bird's eye view of Fig. \ref{fig:tri-anyon-insertion}(c). The fusion vertex of the anyons $a,b,$ and $c$ is labeled by $\delta$. Since we have not inserted wormholes on the $AB/C$ boundary, the diagram -- aside from the trimer insertion -- reduces to that of the bipartite case. In particular, it is important to note that the $\omega_0$ loops are located \emph{above} the inserted anyon trimer, which we have taken to reside on the inner sphere, as depicted in Fig. \ref{fig:tri-anyon-insertion}(b). As such, we can manipulate the $\omega_0$ loops exactly in the same fashion as in the bipartite computation to obtain
\begin{align}
	\ket{\psi} = (d_a d_b d_c)^{-1/4} \sum_{\vec{b},\vec{e}} \frac{\sqrt{d_{\vec{b}}}}{\mathcal{D}^{N-1}} 	  
\begin{gathered}
\includegraphics[height=15em]{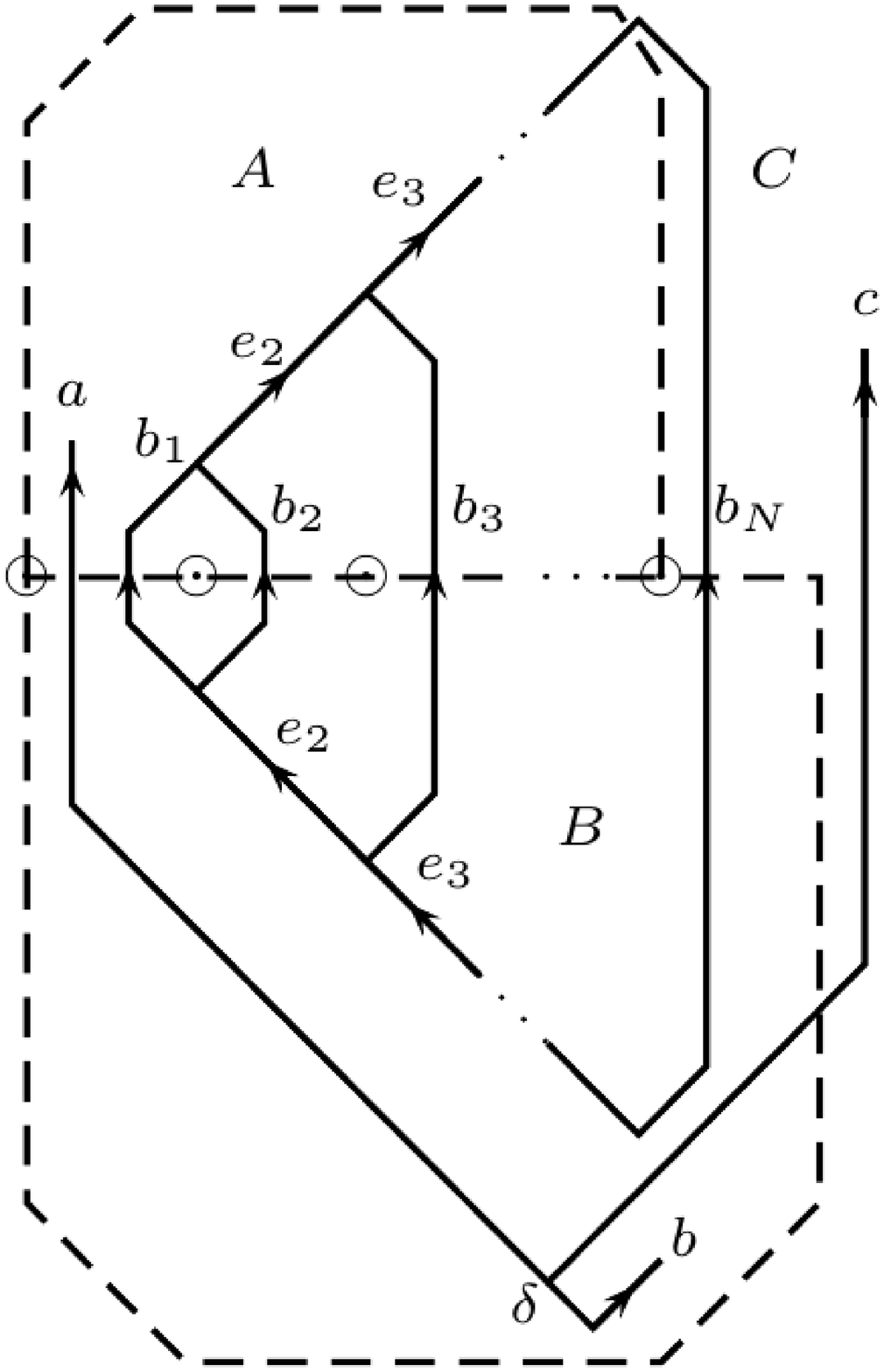} \end{gathered} 
\end{align}
where $b_j$ with $j=1,\dots , N$ and $e_j$ with $j=2 , \dots, N-2$. 
We now combine the $a$ and $c$ lines via resolutions of identity with the main part of the diagram, yielding
\begin{align}
	\ket{\psi} =  \sum_{\vec{b},\vec{e}} \frac{\sqrt{d_{\vec{b}}}}{\mathcal{D}^{N-1}} \frac{(d_a d_b d_c)^{-1/4}}{\sqrt{d_a d_c}} 	  \begin{gathered}
\includegraphics[height=15em]{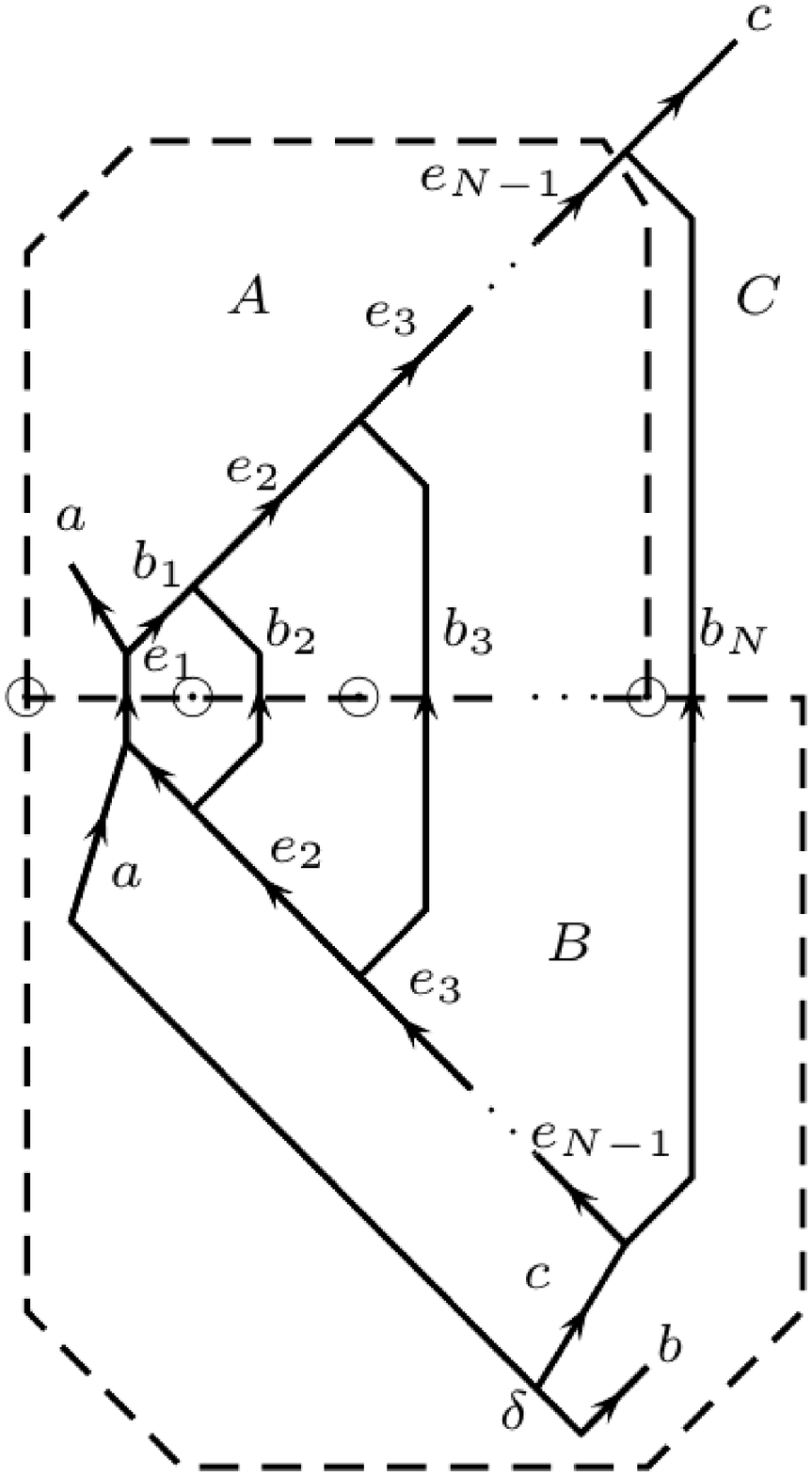} \end{gathered} 
\end{align}
In doing so, we first renamed $b_1, b_N \to e_1, e_{N-1}$ and then named  the fusion product of $a$ and $e_1$ ($c$ and $e_{N-1}$) as $b_1$ ($b_N$). So, we now have $e_j$ with $j \in \{ 1 , \dots, N-1 \}$. 

We can now cut along the boundaries between $A$, $B$, and $C$, yielding
\begin{widetext}
\begin{align}
	&\ket{\psi_{\mathrm{cut}}} = \sum_{\vec{b} , \vec{e} , \vec{\mu}} 	\frac{1}{\mathcal{D}^{N-1}} \frac{(d_a d_b d_c)^{-1/4}}{\sqrt{d_a d_c d_{e_{N-1}}}}  \begin{gathered}
\includegraphics[height=8em]{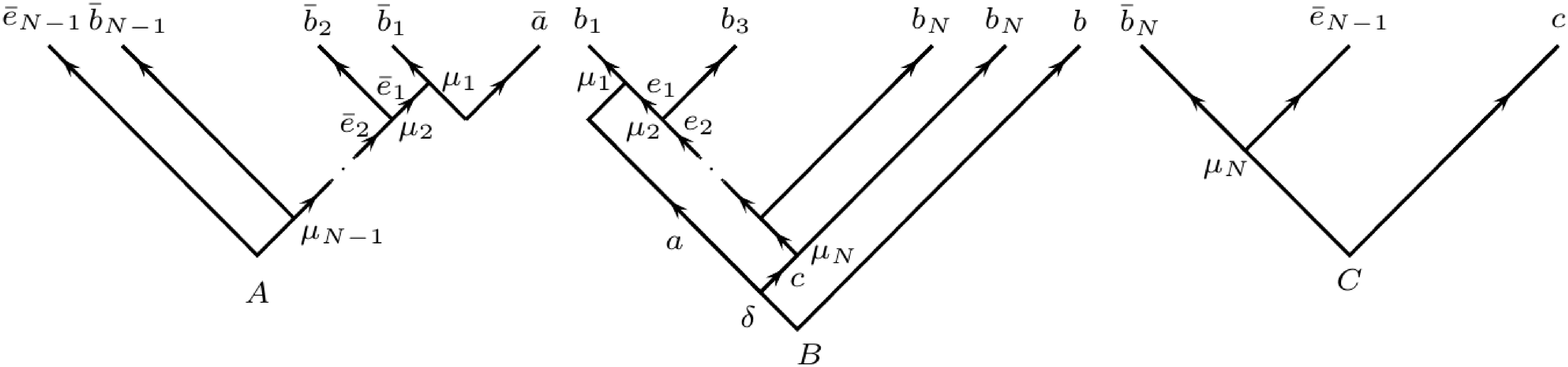}. \end{gathered} 
\end{align}
Here we have restored the vertex labels $\mu_j$, where $j$ runs from $j=1$ to $j= N$. At the vertex $\mu_1$, $b_1$ splits into $a$ and $e_1$, while at vertex $\mu_N$, $b_N$ splits into $\bar e_{N-1}$ and $c$. 
At every other vertex $\mu_j$, $b_j$ fuses with $e_{j-1}$ to yield $e_j$. On tracing out $C$ we obtain the cut reduced density matrix
\begin{align}
	\notag &(\trho_{AB})_{\mathrm{cut}} = \sum_{ \substack{\vec{b} , \vec{e} , \vec{\mu} \\ \vec{b}' , \vec{e}' , \vec{\mu}' \\ b_N e_{N-1} \mu_N } }  \frac{1}{\mathcal{D}^{2N-2}} \frac{1}{\sqrt{d_a d_b} d_a d_c} \sqrt{\frac{d_{b_N}}{d_{e_{N-1}}}}  % \\
	%&\times  
	\psscalebox{.8}{\begin{pspicture}[shift=-2](-2,-5.3)(5,.3)
        \scriptsize
        %%%
        \rput(-5.7,0){
        \psline(6.9,-.3)(7.2,-.6)\psline[ArrowInside=->](7.2,-.6)(7.8,0)\rput(7.8,0.2){$\bar{a}$}
        \psline[ArrowInside=->](6.6,-.6)(6.9,-.3)\rput(6.6,-.35){$\bar  e_1$}
        \rput(7.15,-.3){$\mu_1$}
        \psline[ArrowInside=->](6.9,-.3)(6.6,0)\rput(6.6,.2){$\bar  b_1$}
        \psline[ArrowInside=->](6.3,-.9)(6.6,-.6)\rput(6.3,-.65){$\bar  e_2$}
        \rput(6.8,-.7){$\mu_2$}
        \psline(6.6,-.6)(6.0,0)\psline[ArrowInside=->](6.3,-.3)(6.0,0)\rput(6.,.2){$\bar  b_2$}
        \psline[linestyle=dotted](6.3,-.9)(6.1,-1.1)
        \psline[ArrowInside=->](5.8,-1.4)(6.1,-1.1)
        \psline(5.8,-1.4)(4.4,0)\psline[ArrowInside=->](4.7,-.3)(4.4,0)\rput(4.45,.2){$\bar  b_{N-1}$}
        \rput(6.2,-1.55){$\mu_{N-1}$}
        \psline(5.8,-1.4)(5.5,-1.7)(3.8,0)\psline[ArrowInside=->](4.1,-.3)(3.8,0)\rput(3.75,.2){$\bar{e}_{N-1}$}}
        \rput(3,0){
        \psline[ArrowInside=->](.3,-.3)(0,0)\rput(0,.2){$b_1$}
        \psline[ArrowInside=->](.6,-.6)(.3,-.3)\rput(.6,-.35){$e_1$}
        \psline(.3,-.3)(0.,-.6)(1.4,-2)(1.7,-1.7)
        \psline[ArrowInside=->](1.4,-2)(0.,-.6)\rput(.7,-1.6){$a$}
        \psline(1.4,-2)(1.7,-2.3)(4,0)
        \psline[ArrowInside=->](3.7,-.3)(4.,0)\rput(4.,.2){$b$}
        \psline[ArrowInside=->](1.4,-2)(1.7,-1.7)\rput(1.7,-1.9){$c$}
        \rput(1.3,-2.2){$\delta$}
        \rput(2.,-1.7){$\mu_N$}
        \rput(.05,-.3){$\mu_1$}
        \psline[ArrowInside=->](.9,-.9)(.6,-.6)\rput(.9,-.65){$e_2$}
        \rput(.45,-.7){$\mu_2$}
        \psline(.6,-.6)(1.2,0)\psline[ArrowInside=->](.9,-.3)(1.2,0)\rput(1.2,.2){$b_2$}
        \psline[linestyle=dotted](.9,-.9)(1.1,-1.1)
        \psline[ArrowInside=->](1.4,-1.4)(1.1,-1.1)
        \psline(1.4,-1.4)(2.8,0)\psline[ArrowInside=->](2.5,-.3)(2.8,0)\rput(2.8,.2){$b_{N-1}$}
        \psline(1.4,-1.4)(1.7,-1.7)(3.4,0)\psline[ArrowInside=->](3.1,-.3)(3.4,0)\rput(3.4,.2){$b_N$}
        \psline[ArrowInside=->](1.7,-1.7)(1.4,-1.4)%\rput(1.55,-1.65){$e_{N-1}$}
        }
        %%% bra
        \rput(-5.7,-3.3){\rput(5.5,-2){$A$}
        \psline(6.9,-1.1)(7.2,-.8)\psline[ArrowInside=-<](7.2,-.8)(7.8,-1.4)\rput(7.8,-1.6){$\bar{a}$}
        \psline[ArrowInside=-<](6.6,-.8)(6.9,-1.1)\rput(6.6,-1.05){$\bar  e_1'$}
        \rput(7.15,-1.1){$\mu_1'$}
        \psline[ArrowInside=-<](6.9,-1.1)(6.6,-1.4)\rput(6.6,-1.6){$\bar  b_1'$}
        \psline[ArrowInside=-<](6.3,-.5)(6.6,-.8)\rput(6.3,-.75){$\bar  e_2'$}
        \rput(6.85,-.65){$\mu_2'$}
        \psline(6.6,-.8)(6.0,-1.4)\psline[ArrowInside=-<](6.3,-1.1)(6.0,-1.4)\rput(6.,-1.6){$\bar  b'_2$}
        \psline[linestyle=dotted](6.3,-.5)(6.1,-0.3)
        \psline[ArrowInside=-<](5.8,0.0)(6.1,-0.3)
        \psline(5.8,0)(4.4,-1.4)\psline[ArrowInside=-<](4.7,-1.1)(4.4,-1.4)\rput(4.45,-1.6){$\bar  b'_{N-1}$}
        \rput(6.2,.15){$\mu_{N-1}'$}
        \psline(5.8,0)(5.5,.3)(3.8,-1.4)\psline[ArrowInside=->](3.8,-1.4)(4.1,-1.1)\rput(3.75,-1.6){$\bar{e}_{N-1}$}
        }
        \rput(3,-2.4){\rput(1.7,-2.9){B}
        \psline[ArrowInside=-<](.3,-2.)(0,-2.3)\rput(0,-2.5){$b_1'$}
        \psline[ArrowInside=-<](.6,-1.7)(.3,-2.)\rput(.6,-1.95){$e_1'$}
        \psline(.3,-2.)(0.,-1.7)(1.4,-.3)(1.7,-.6)
        \psline[ArrowInside=-<](1.4,-.3)(0.,-1.7)\rput(.7,-.7){$a$}
        \psline(1.4,-.3)(1.7,0.)(4,-2.3)
        \psline[ArrowInside=-<](3.7,-2.)(4,-2.3)\rput(4.,-2.5){$b$}
        \psline[ArrowInside=-<](1.4,-.3)(1.7,-.6)\rput(1.7,-.4){$c$}
        \rput(1.3,-.1){$\delta$}
        \rput(2.,-.6){$\mu_N$}
        \rput(.05,-2.){$\mu_1'$}
        \psline[ArrowInside=-<](.9,-1.4)(.6,-1.7)\rput(.9,-1.65){$e_2'$}
        \rput(.45,-1.6){$\mu_2'$}
        \psline(.6,-1.7)(1.2,-2.3)\psline[ArrowInside=-<](.9,-2.)(1.2,-2.3)\rput(1.2,-2.5){$b_2'$}
        \psline[linestyle=dotted](.9,-1.4)(1.1,-1.2)
        \psline[ArrowInside=-<](1.4,-.9)(1.1,-1.2)
        \psline(1.4,-.9)(2.8,-2.3)\psline[ArrowInside=-<](2.5,-2.)(2.8,-2.3)\rput(2.8,-2.5){$b_{N-1}'$}
        \psline(1.4,-.9)(1.7,-.6)(3.4,-2.3)\psline[ArrowInside=-<](3.1,-2.)(3.4,-2.3)\rput(3.4,-2.5){$b_N$}
        \psline[ArrowInside=-<](1.7,-.6)(1.4,-.9)%\rput(1.55,-.65){$e_{N-1}'$}
        }
    \end{pspicture}}
\end{align}
Note that, henceforth, $\vec{b} = (b_1 , \dots , b_{N-1})$, $\vec{e} = (e_1 , \dots , e_{N-2})$, and $\vec{\mu} = (\mu_1 , \dots , \mu_{N-1})$. % only label $b_j$ with $j=1,\dots,N-1$; $e_j$ with $j=1,\dots,N-2$; and $\mu_j$ with $j=1,\dots, N-1$. 
We will sum over $b_N$, $e_{N-1}$ and $\mu_N$ separately. Note, in particular, that the above density matrix is diagonal in these indices as a consequence of the trace over $C$ .
By fusing the $a$ and $c$ lines with the main part of the diagram, we have introduced an internal loop in the $B$ part of the diagram. In order to evaluate matrix products and traces of $\trho_{AB}$, we must apply a basis change to rid ourselves of this loop. Focusing on the ``ket" part of the diagram, we first apply an $F$-move to bring the $a$ line past the $b_2$ line:
\begin{align}
	\begin{pspicture}[shift=-0.65](-.2,0)(1.1,1.7)
        \scriptsize
        \psline[ArrowInside=->](0,0)(0,1.1)\psline[ArrowInside=->](0,1.1)(0,1.5)\rput(0.,-0.15){$a$}\rput(0.,1.65){$b_1$}
        \psline[ArrowInside=->](1,0)(1,.4)\psline[ArrowInside=->](1,0.4)(1,1.5)\rput(1.,-0.15){$e_2$}\rput(1.,1.65){$b_2$}
        \psline[ArrowInside=->](1,.4)(0,1.1)\rput(.5,1.){$e_1$}
        \rput(-.2,1.1){$\mu_1$}
        \rput(1.2,.4){$\nu_2$}
    \end{pspicture} = \sum_{f\alpha\beta} [F^{b_1 b_2}_{a e_2}]_{ (e_1 \mu_1 \mu_2) (l_1 \nu_1 \tilde{\mu}_2)}  \begin{pspicture}[shift=-0.65](-0.1,-0.5)(1.2,1)
        \scriptsize
        \psline[ArrowInside=->](0.5,0.5)(0,1)\rput(0.,1.2){$b_1$}
        \psline[ArrowInside=->](0.5,0.5)(1,1)\rput(1,1.2){$b_2$}
        \psline[ArrowInside=->](0.5,0)(0.5,0.5)\rput(0.35,0.25){$l_1$}
        \rput(0.7,0.45){$\nu_1$}
        \psline[ArrowInside=-<](0.5,0.)(0,-.5)\rput(0,-0.7){$a$}
        \psline[ArrowInside=-<](0.5,0.)(1,-.5)\rput(1,-0.7){$e_2$}
        \rput(0.7,0.1){$\tilde{\mu}_2$}
    \end{pspicture}
\end{align}
By repeatedly applying $F$-moves of this form, we can bring the $a$ line down the diagram, until it yields an inner product, which can then be evaluated, yielding a factor of $\sqrt{d_a d_c / d_b}$. Explicitly, combining these $F$-moves, we find,
\begin{align}
&\psscalebox{.8}{\begin{pspicture}[shift=0](0,-1.5)(3.,.3)
	\scriptsize
        \psline[ArrowInside=->](.3,-.3)(0,0)\rput(0,.2){$b_1$}
        \psline[ArrowInside=->](.6,-.6)(.3,-.3)\rput(.6,-.35){$e_1$}
        \psline(.3,-.3)(0.,-.6)(1.4,-2)(1.7,-1.7)
        \psline[ArrowInside=->](1.4,-2)(0.,-.6)\rput(.7,-1.6){$a$}
        \psline(1.4,-2)(1.7,-2.3)(4,0)
        \psline[ArrowInside=->](3.7,-.3)(4.,0)\rput(4.,.2){$b$}
        \psline[ArrowInside=->](1.4,-2)(1.7,-1.7)\rput(1.7,-1.9){$c$}
        \rput(1.3,-2.2){$\delta$}
        \rput(2.,-1.7){$\mu_N$}
        \rput(.05,-.3){$\mu_1$}
        \psline[ArrowInside=->](.9,-.9)(.6,-.6)\rput(.9,-.65){$e_2$}
        \rput(.45,-.7){$\mu_2$}
        \psline(.6,-.6)(1.2,0)\psline[ArrowInside=->](.9,-.3)(1.2,0)\rput(1.2,.2){$b_3$}
        \psline[linestyle=dotted](.9,-.9)(1.1,-1.1)
        \psline[ArrowInside=->](1.4,-1.4)(1.1,-1.1)
        \psline(1.4,-1.4)(2.8,0)\psline[ArrowInside=->](2.5,-.3)(2.8,0)\rput(2.8,.2){$b_N$}
        \psline(1.4,-1.4)(1.7,-1.7)(3.4,0)\psline[ArrowInside=->](3.1,-.3)(3.4,0)\rput(3.4,.2){$b_N$}
        \psline[ArrowInside=->](1.7,-1.7)(1.4,-1.4)%\rput(1.55,-1.65){$e_{N-1}$}
        \end{pspicture}} = \sum_{ \substack{ l_1, \dots , l_{N-2} \\ \nu_1, \dots , \nu_{N-1} \\ \tilde{\mu}_2 , \dots , \tilde{\mu}_{N-1} } } [F^{b_1 b_2}_{a e_2}]_{ \substack{(e_1 \mu_1 \mu_2) \\ (l_1 \nu_1 \tilde{\mu}_2) } } \prod_{i=1}^{N-3} [F^{l_i b_{i+2}}_{a e_{i+2}}]_{ \substack{(e_{i+1} \tilde{\mu}_{i+1} \mu_{i+2}) \\ (l_{i+1} \nu_{i+1} \tilde{\mu}_{i+2}) } } [F^{l_{N-2} b_N}_{a c}]_{ \substack{(e_{N-1} \tilde{\mu}_{N-1} \mu_N) \\ (b \, \nu_{N-1} \delta) } } \sqrt{\frac{d_a d_c}{d_b}}
\psscalebox{.8}{\begin{pspicture}[shift=-.5](0,-1.6)(3.5,.3)
	\scriptsize
        \rput(0,0){
        \psline[ArrowInside=->](.3,-.3)(0,0)\rput(0,.2){$b_1$}
        \psline[ArrowInside=->](.6,-.6)(.3,-.3)\rput(.6,-.35){$l_2$}
        \rput(.1,-.45){$\nu_1$}
        \psline[ArrowInside=->](.3,-.3)(.6,0)\rput(.6,.2){$b_2$}
        \psline[ArrowInside=->](.9,-.9)(.6,-.6)\rput(.9,-.65){$l_3$}
        \rput(.4,-.75){$\nu_2$}
        \psline(.6,-.6)(1.2,0)\psline[ArrowInside=->](.9,-.3)(1.2,0)\rput(1.2,.2){$b_3$}
        \psline[linestyle=dotted](.9,-.9)(1.1,-1.1)
        \psline[ArrowInside=->](1.4,-1.4)(1.1,-1.1)
        \psline(1.4,-1.4)(2.8,0)\psline[ArrowInside=->](2.5,-.3)(2.8,0)\rput(2.8,.2){$b_N$}
        \rput(1.1,-1.5){$\nu_{N-1}$}
        \psline(1.4,-1.4)(1.7,-1.7)(3.4,0)\psline[ArrowInside=->](3.1,-.3)(3.4,0)\rput(3.4,.2){$b$}
        }
        \end{pspicture}}
\end{align}
As a consequence of these basis changes, we have a new set of charges $l_j$ and vertices $\nu_j$ such that at $\nu_1$, $b_1$ and $b_2$ fuse into $l_1$, at $\nu_2$, $l_1$ and $b_3$, fuse into $l_2$, and so on, until $\nu_{N-1}$, where $l_{N-1}$ and $b_N$ fuse into $\bar b$. These basis changes also required the introduction of the vertex labels $\tilde{\mu}_j$ in the intermediate steps, which appear in the $F$-symbols but not in the final anyon diagram.

Applying the same manipulations to the ``bra" part of the diagram, we obtain
\begin{align}
	\notag(\trho_{AB})_{\mathrm{cut}} & = \sum_{ \substack{ \vec{b} \vec{e} \vec{\mu} \\ \vec{b}' \vec{e}' \vec{\mu}' \\ b_N e_{N-1} } } \sum_{ \substack{ \vec{l} \vec{\nu} \\ \vec{l}' \vec{\nu}' } } \frac{1}{\mathcal{D}^{2N-2}}  \frac{1}{d_b \sqrt{d_a d_b}} \sqrt{\frac{d_{b_N}}{d_{e_{N-1}}}} L_{\vec{l},\vec{e},\vec{b}; \vec{l}',\vec{e}',\vec{b}'}^{\vec{\mu},\vec{\nu}; \vec{\mu}',\vec{\nu}'} 
	%\\
	%&\times 
	\psscalebox{.8}{  \begin{pspicture}[shift=-2](-1.4,-4.6)(5.5,.3)
        \scriptsize
        %%%
        \rput(-5.7,0){
        \psline(6.9,-.3)(7.2,-.6)\psline[ArrowInside=->](7.2,-.6)(7.8,0)\rput(7.8,0.2){$\bar{a}$}
        \psline[ArrowInside=->](6.6,-.6)(6.9,-.3)\rput(6.6,-.35){$\bar  e_1$}
        \rput(7.15,-.3){$\mu_1$}
        \psline[ArrowInside=->](6.9,-.3)(6.6,0)\rput(6.6,.2){$\bar  b_1$}
        \psline[ArrowInside=->](6.3,-.9)(6.6,-.6)\rput(6.3,-.65){$\bar  e_2$}
        \rput(6.8,-.7){$\mu_2$}
        \psline(6.6,-.6)(6.0,0)\psline[ArrowInside=->](6.3,-.3)(6.0,0)\rput(6.,.2){$\bar  b_2$}
        \psline[linestyle=dotted](6.3,-.9)(6.1,-1.1)
        \psline[ArrowInside=->](5.8,-1.4)(6.1,-1.1)
        \psline(5.8,-1.4)(4.4,0)\psline[ArrowInside=->](4.7,-.3)(4.4,0)\rput(4.45,.2){$\bar  b_{N-1}$}
        \rput(6.2,-1.55){$\mu_{N-1}$}
        \psline(5.8,-1.4)(5.5,-1.7)(3.8,0)\psline[ArrowInside=->](4.1,-.3)(3.8,0)\rput(3.75,.2){$\bar{e}_{N-1}$}}
        \rput(3,0){
        \psline[ArrowInside=->](.3,-.3)(0,0)\rput(0,.2){$b_1$}
        \psline[ArrowInside=->](.6,-.6)(.3,-.3)\rput(.6,-.35){$l_2$}
        \rput(.1,-.45){$\nu_1$}
        \psline[ArrowInside=->](.3,-.3)(.6,0)\rput(.6,.2){$b_2$}
        \psline[ArrowInside=->](.9,-.9)(.6,-.6)\rput(.9,-.65){$l_3$}
        \rput(.4,-.75){$\nu_2$}
        \psline(.6,-.6)(1.2,0)\psline[ArrowInside=->](.9,-.3)(1.2,0)\rput(1.2,.2){$b_3$}
        \psline[linestyle=dotted](.9,-.9)(1.1,-1.1)
        \psline[ArrowInside=->](1.4,-1.4)(1.1,-1.1)
        \psline(1.4,-1.4)(2.8,0)\psline[ArrowInside=->](2.5,-.3)(2.8,0)\rput(2.8,.2){$b_N$}
        \rput(1.1,-1.5){$\nu_{N-1}$}
        \psline(1.4,-1.4)(1.7,-1.7)(3.4,0)\psline[ArrowInside=->](3.1,-.3)(3.4,0)\rput(3.4,.2){$b$}
        }
        %%% bra
        \rput(-5.7,-2.4){\rput(5.5,-2){$A$}
        \psline(6.9,-1.1)(7.2,-.8)\psline[ArrowInside=-<](7.2,-.8)(7.8,-1.4)\rput(7.8,-1.6){$\bar{a}$}
        \psline[ArrowInside=-<](6.6,-.8)(6.9,-1.1)\rput(6.6,-1.05){$\bar  e_1'$}
        \rput(7.15,-1.1){$\mu_1'$}
        \psline[ArrowInside=-<](6.9,-1.1)(6.6,-1.4)\rput(6.6,-1.6){$\bar  b_1'$}
        \psline[ArrowInside=-<](6.3,-.5)(6.6,-.8)\rput(6.3,-.75){$\bar  e_2'$}
        \rput(6.85,-.65){$\mu_2'$}
        \psline(6.6,-.8)(6.0,-1.4)\psline[ArrowInside=-<](6.3,-1.1)(6.0,-1.4)\rput(6.,-1.6){$\bar  b'_2$}
        \psline[linestyle=dotted](6.3,-.5)(6.1,-0.3)
        \psline[ArrowInside=-<](5.8,0.0)(6.1,-0.3)
        \psline(5.8,0)(4.4,-1.4)\psline[ArrowInside=-<](4.7,-1.1)(4.4,-1.4)\rput(4.45,-1.6){$\bar  b'_{N-1}$}
        \rput(6.2,.15){$\mu_{N-1}'$}
        \psline(5.8,0)(5.5,.3)(3.8,-1.4)\psline[ArrowInside=->](3.8,-1.4)(4.1,-1.1)\rput(3.75,-1.6){$\bar{e}_{N-1}$}
        }
        \rput(3,-2.4){\rput(1.7,-2){$B$}
        \psline[ArrowInside=->](0,-1.4)(.3,-1.1)\rput(0,-1.6){$b_1'$}
        \psline[ArrowInside=->](.3,-1.1)(.6,-.8)\rput(.6,-1.05){$l_2'$}
        \rput(.1,-.95){$\nu_1'$}
        \psline[ArrowInside=->](.6,-1.4)(.3,-1.1)\rput(.6,-1.6){$b_2'$}
        \psline[ArrowInside=->](.6,-.8)(.9,-.5)\rput(.9,-.75){$l_3'$}
        \rput(.4,-.65){$\nu_2'$}
        \psline(.6,-.8)(1.2,-1.4)\psline[ArrowInside=->](1.2,-1.4)(.9,-1.1)\rput(1.2,-1.6){$b_3'$}
        \psline[linestyle=dotted](.9,-.5)(1.1,-0.3)
        \psline[ArrowInside=->](1.1,-0.3)(1.4,0.0)
        \psline(1.4,0)(2.8,-1.4)\psline[ArrowInside=->](2.8,-1.4)(2.5,-1.1)\rput(2.8,-1.6){$b_N$}
        \rput(1.1,.09){$\nu_{N-1}'$}
        \psline(1.4,0)(1.7,.3)(3.4,-1.4)\psline[ArrowInside=->](3.4,-1.4)(3.1,-1.1)\rput(3.4,-1.6){$b$}}
    \end{pspicture}}
\end{align}
where we have defined,
\begin{align}
	\notag L_{\vec{e},\vec{l},\vec{b}; \vec{e}',\vec{l}',\vec{b}'}^{\vec{\mu},\vec{\nu}; \vec{\mu}',\vec{\nu}'}  =  \sum_{\mu_N, \vec{\tilde{\mu}}, \vec{\tilde{\mu}}'}  & [F^{b_1 b_2}_{a e_2}]_{ \substack{(e_1 \mu_1 \mu_2) \\ (l_1 \nu_1 \tilde{\mu}_2) } } \prod_{i=1}^{N-3} [F^{l_i b_{i+2}}_{a e_{i+2}}]_{ \substack{(e_{i+1} \tilde{\mu}_{i+1} \mu_{i+2}) \\ (l_{i+1} \nu_{i+1} \tilde{\mu}_{i+2}) } } [F^{l_{N-2} b_N}_{a c}]_{ \substack{(e_{N-1} \tilde{\mu}_{N-1} \mu_N) \\ (b \, \nu_{N-1} \delta) } } \\
	\times &  [F^{b_1' b_2'}_{a e_2'}]^*_{ \substack{(e_1' \mu'_1 \mu'_2) \\ (l_1' \nu_1' \tilde{\mu}_2') } } \prod_{i=1}^{N-3} [F^{l_i' b_{i+2}'}_{a e_{i+2}'}]^*_{ \substack{(e_{i+1}' \tilde{\mu}_{i+1}' \mu_{i+2}') \\ (l_{i+1}' \nu_{i+1}' \tilde{\mu}_{i+2}') } } [F^{l_{N-2}' b_N}_{a c}]^*_{ \substack{(e_{N-1} \tilde{\mu}'_{N-1} \mu_N) \\ (b \, \nu'_{N-1} \delta) } } .
\end{align}
With the cut reduced density matrix in hand, we can proceed to compute the desired entanglement quantities. We note that, unfortunately, the above string of $F$-symbols does not readily simplify. 
\end{widetext}
%\begin{comment}
\subsection{Mutual Information}

We first start with the mutual information.
It is known that the presence of an anyon $c$ contributes $\ln d_c$ to the entanglement entropy. Indeed, we can simply use the results of Ref. \cite{Bonderson2017} to conclude that 
\begin{equation}
	I(A:B) = -2(n+1) \sum_b \frac{d_b^2}{\mathcal{D}^2} \ln \left( \frac{d_b}{\mathcal{D}^2} \right) - \ln \mathcal{D}^2 + \ln \left( \frac{d_a d_b}{d_c} \right) . \label{eq:tripartite-insertion-MI}
\end{equation}
As usual, the first term is the area law, the second the topological entanglement entropy, and the last the contribution from the anyon insertions.
As an aside, we note that the last term is related, in a perhaps trivial way, to the $F$-symbol $[F^{ab}_{ab}]_{1,(c\mu\mu)} = \sqrt{d_c / d_a d_b}$ (no sum on $\mu$). 
%For later reference, we also record here the R\'enyi mutual information:
%\begin{equation}
%	\boxed{I^{(\alpha)}\left(A:B\right) = \frac{2(n+1)}{1-\alpha} \ln \frac{\kappa_{\alpha,0}}{\mathcal{D}^{2\alpha}} - \ln \mathcal{D}^2 + \ln \left( \frac{d_a d_b}{d_c} \right) } 
%\end{equation}
%\end{comment}

%\begin{widetext}

\subsection{Negativity}

%The cut partially transposed density matrix is given by
%\begin{align}
%	(\trho_{AB})^{T_A}_{\mathrm{cut}} =  \sum_{ \substack{ \vec{b} \vec{e} \vec{\mu} \\ \vec{b}' \vec{e}' \vec{\mu}' \\ b_N e_{N-1} } } \sum_{ \substack{ \vec{l} \vec{\nu} \\ \vec{l}' \vec{\nu}' } } \frac{1}{\mathcal{D}^{2N-2}}  \frac{1}{d_b \sqrt{d_a d_b}} \sqrt{\frac{d_{b_N}}{d_{e_{N-1}}}} L_{\vec{e},\vec{l},\vec{b}; \vec{e}',\vec{l}',\vec{b}'}^{\vec{\mu},\vec{\nu}; \vec{\mu}',\vec{\nu}'} \begin{gathered}
%\includegraphics[height=14em]{figures/trijunction-puncture/rhoAB_TA.eps}
%\end{gathered}
%\end{align}
%\end{widetext}
Computation of the negativity via the replica trick is a tedious process due to the string of $F$-symbols, and so we relegate the details to Appendix \ref{sec:app-details-negativity}. In particular, we are unable to find a simple expression for the replica limit for arbitrary topological orders. However, we can take this limit for categories with no fusion multiplicity (i.e. $N_{bc}^a \in \{ 0 , 1 \}$). For such multiplicity-free theories, we find the negativity to be
\begin{align}
	%\Aboxed{
	\mathcal{E}(A:B) &= 2(n+1) \ln (\kappa_{1/2,0} / \mathcal{D}) - \ln \mathcal{D}^2 \\
	\notag + &\ln \left[ \sum_{b_N e_{N-1}} \frac{d_{b_N}}{d_b \mathcal{D}^2 } \left(\sum_{l_{N-2} } d_{l_{N-2}} |[F^{l_{N-2} b_N}_{a c}]_{ e_{N-1} b } | \right)^2 \right].
	%}
\end{align}
As usual, the first term is the area law term, the second the topological negativity, and the third term combines the contribution from the trijunction and the anyon insertions; note that it depends on $a$, $b$, and $c$, as well as $b_{N}$, $e_{N-1}$, and $l_{N-2}$, the latter three of which are the anyons residing at one of the trijunctions in the anyon diagram. Unfortunately, this last term does not appear to take a particularly symmetric form. It is interesting, however, that this contribution depends explicitly on the magnitude of the $F$-symbols, which are in fact topological invariants related to the absolute values of the braiding matrix \cite{Bonderson2021}. 
To our knowledge, such a dependence of an entanglement quantity on this piece of topological data has not been identified previously. We remark, however, that it is possible that this contribution, with its particular combination of sums over quantum dimensions and $F$-symbols may have a simpler form which could perhaps depend only on the quantum dimensions.

Once again, we must note that this expression really gives the negativity for the phase $\mathcal{C} \times \overline{\mathcal{C}} $. In order to extract the negativity for just $\mathcal{C}$ in the presence of the anyons $a$, $b$, and $c$, we must subtract off the negativity in the absence of an anyon trimer [Eq. \eqref{eq:tripartite-negativity}]. We thus obtain the main result of this section,
\begin{align}
	\begin{split}
	& \mathcal{E}(A:B) = (n+1) \ln (\kappa_{1/2,0} / \mathcal{D}) - \ln \mathcal{D} - \frac{1}{2} \ln \left[ \sum_e \frac{d_e^3}{\mathcal{D}^2} \right] \\
	+ &\ln \left[ \sum_{b_N e_{N-1}} \frac{d_{b_N}}{d_b \mathcal{D}^2 } \left(\sum_{l_{N-2} } d_{l_{N-2}} |[F^{l_{N-2} b_N}_{a c}]_{ e_{N-1} b } | \right)^2 \right] \end{split}  \label{eq:anyon-negativity}
\end{align}
This is a rather cumbersome expression. However, we can still build some intuition by checking some limiting cases. First, if either one of $a$ or $b$ is the vacuum, one finds that the negativity reduces to Eq. \eqref{eq:tripartite-negativity}. On the other hand, if $c$ is the vacuum, and so $a = \overline{b}$, then we find the negativity becomes 
\begin{equation}
	\mathcal{E}(A:B) = (n+1) \ln( \frac{\kappa_{\alpha,0}}{ \mathcal{D}}) - \ln \frac{\mathcal{D}}{d_a} + \frac{1}{2}\ln\left( \sum_{e_{m+1}} \frac{ d_{e_{m+1}}^3 }{ \mathcal{D}^2 } \right)
\end{equation}
These special cases reflect the fact that the negativity only detects entanglement between $A$ and $B$, in contrast to the mutual information Eq. \eqref{eq:tripartite-insertion-MI}, which additionally is sensitive to the entanglement between $AB$ and $C$, as manifested by the $\ln d_c$ contribution from the anyon $c$ in region $C$. 
More generally, we note that the anyonic contribution to the negativity depends only on the \emph{absolute values} of the $F$-symbols. If the underlying topological order is Abelian, then the $F$-symbols are purely phases and hence the negativity reduces to that in the absence of any anyons, Eq. \eqref{eq:tripartite-negativity}. 

\subsubsection{$SU(2)_k$ Examples}

To build some additional intuition for the above expression, let us focus on the examples of the $SU(2)_k$ topological orders, whose anyons correspond to the spin-$j/2$ representations of $SU(2)$ for $j=0,\dots, k$ and obey the fusion rules
\begin{align}
	j_1 \times j_2 = \sum_{j=|j_1 - j_2|}^{\mathrm{min}(j_1 + j_2, k - j_1 - j_2)} j.
\end{align}
Explicit expressions for the $F$-symbols may be found in, for instance, Ref. \cite{Bonderson2007}. We note that the anyons in these categories are always self-conjugate %dual 
%{\color{red} self-conjugate (?)}
and all are non-Abelian, save for $j=0$ and $j=k/2$, which are Abelian. Evaluating Eq. \eqref{eq:anyon-negativity} for $k=2,3,4$ with different combinations of $a$, $b$ and $c$, we make the following observations:
\begin{itemize}
	\item If either $a$ or $b$ is Abelian, then the contribution to the negativity from the anyon insertions vanishes.
	\item If $c$ is Abelian, then the contribution to the negativity reduces to $\ln d_a = \ln d_b$ (note that for $SU(2)_k$, the fusion of $a$ and $b$ can only yield an Abelian anyon if $a = \bar{b}$, in which case $\ln d_a = \ln d_b$).
	\item In the case where $a$, $b$, and $c$ are all non-Abelian, then the additional contribution is $\ln A$, where $A$ is less than the quantum dimensions $d_{a,b,c}$ of either $a$, $b$, or $c$. 
\end{itemize}
The above observations seem reasonable given our understanding of the negativity as a measure of bipartite entanglement between $A$ and $B$ and are in line with the special cases discussed above. In particular, when either $a$ or $b$ is Abelian, they cannot be entangled with one another, as a consequence of the lack of an internal degeneracy for Abelian anyons. Likewise, when $c$ is Abelian, $a$ and $b$ can only be entangled with one another, and so the negativity reduces to that of the usual bipartite case. Finally, when all of $a$, $b$, and $c$ are non-Abelian, then $a$ and $b$ are entangled with $c$. By monogamy of entanglement, the entanglement between $a$ and $b$ must then be less than maximal, consistent with our observations. It is interesting that these physically sensible results for the entanglement are encoded in the somewhat obscure expression Eq. \eqref{eq:tripartite-negativity}, in which the $F$-symbols appear in a non-trivial way. The fact that they do appear, however, is perhaps not surprising, given that they encode information about the fusion structure of a system of three anyons. While we cannot extract each individual component of the $F$-symbols from this expression, our result at least suggests that entanglement measures may provide a way for extracting additional data of topological orders from a single excited state beyond simply the quantum dimensions.

\subsection{Reflected Entropy}

The computation of the reflected entropy is likewise a tedious one, and so we relegate the details to the Appendix \ref{sec:app-details-reflected}. Unfortunately, even if we restrict ourselves to multiplicity-free theories, we are unable to take the von Neumann replica limit. As such, we can only quote here the R\'enyi reflected entropies:
\begin{widetext}
\begin{align}
	S_R^{(\beta)}(A:B) &= \frac{2(n+1)}{1-\beta} \ln\left( \frac{\kappa_{\beta,0}}{\mathcal{D}^{2\beta}} \right)  - \ln \mathcal{D}^2 + \frac{1}{1-\beta} \ln \left( \sum_{ l_{N-2} l'_{N-2} } \frac{(d_{l_{N-2}} d_{l'_{N-2}})^{1-\beta}}{d_c^\beta \mathcal{D}^2}  \Tr\left[ \tilde{C}^\beta_{l_{N-2} l_{N-2}' } \right]\right) \label{eq:anyon-reflected-entropy}
\end{align} 
where the matrix $\tilde{C}_{l_{N-2} l_{N-2}' }$ has matrix elements,
\begin{align}
	  \left[\tilde{C}_{l_{N-2} l_{N-2}' }\right]_{b_N,b_N''} 	  &= \sqrt{d_{b_N} d_{b_N}''} \sum_{e''_{N-1}} d_{e''_{N-1}}  [F^{a e''_{N-1} b_N}_b]_{l_{N-2}c} [F^{a e''_{N-1} b''_N}_b]_{l_{N-2}c}^*  [F^{a e''_{N-1} b_N}_b]_{l'_{N-2}c}^* [F^{a e''_{N-1} b''_N}_b]_{l'_{N-2}c} \, .
\end{align}
The first two terms correspond to the usual area law and topological contributions. As usual, in order to obtain the reflected entropy for just $\mathcal{C}$, we must divide these contributions by two. The final contribution arises from the anyon insertions. 
We were unable to find a closed form expression for the $\beta^{th}$ power of the matrix $\tilde{C}$ which prevents us from analytically continuing $\beta \to 1$. Nevertheless, we see that this new contribution depends in a non-trivial way on the $F$-symbols, similar to but distinct from the contribution we found for the negativity. In particular, we have for the Markov gap,
\begin{equation}
	S_R(A:B) - I(A:B) = \lim_{\alpha \to 1} \frac{1}{1-\alpha} \ln \left( \sum_{ l_{N-2}, l'_{N-2} } \frac{(d_{l_{N-2}} d_{l'_{N-2}})^{1-\alpha}}{d_c^\alpha \mathcal{D}^2}  \Tr\left[ \tilde{C}^\alpha_{l_{N-2} l_{N-2}'} \right]\right) - \ln\left( \frac{d_a d_b}{d_c} \right)  \,. \label{eq:anyon-renyi-markov}
\end{equation}
While we cannot evaluate this expression in general, we can build some intuition by again considering $SU(2)_k$ theories. %, as we do in the following subsection. 
\end{widetext}

\subsubsection{ $SU(2)_k$ Examples }

\begin{figure}[!tb]
\subfloat[$k=3$]{%
    \includegraphics[width=0.49\linewidth]{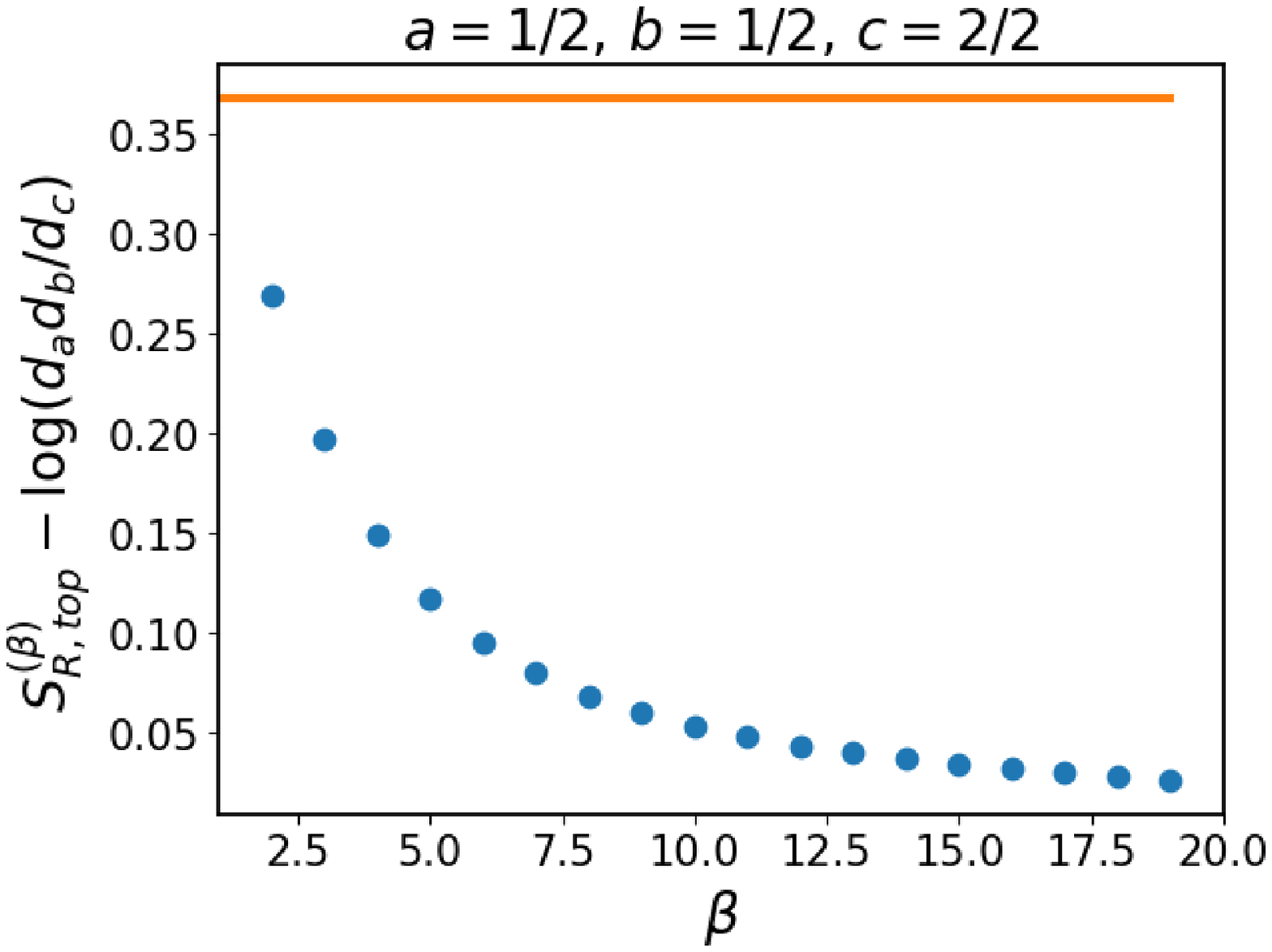}
}
\subfloat[$k=4$]{%
    \includegraphics[width=0.49\linewidth]{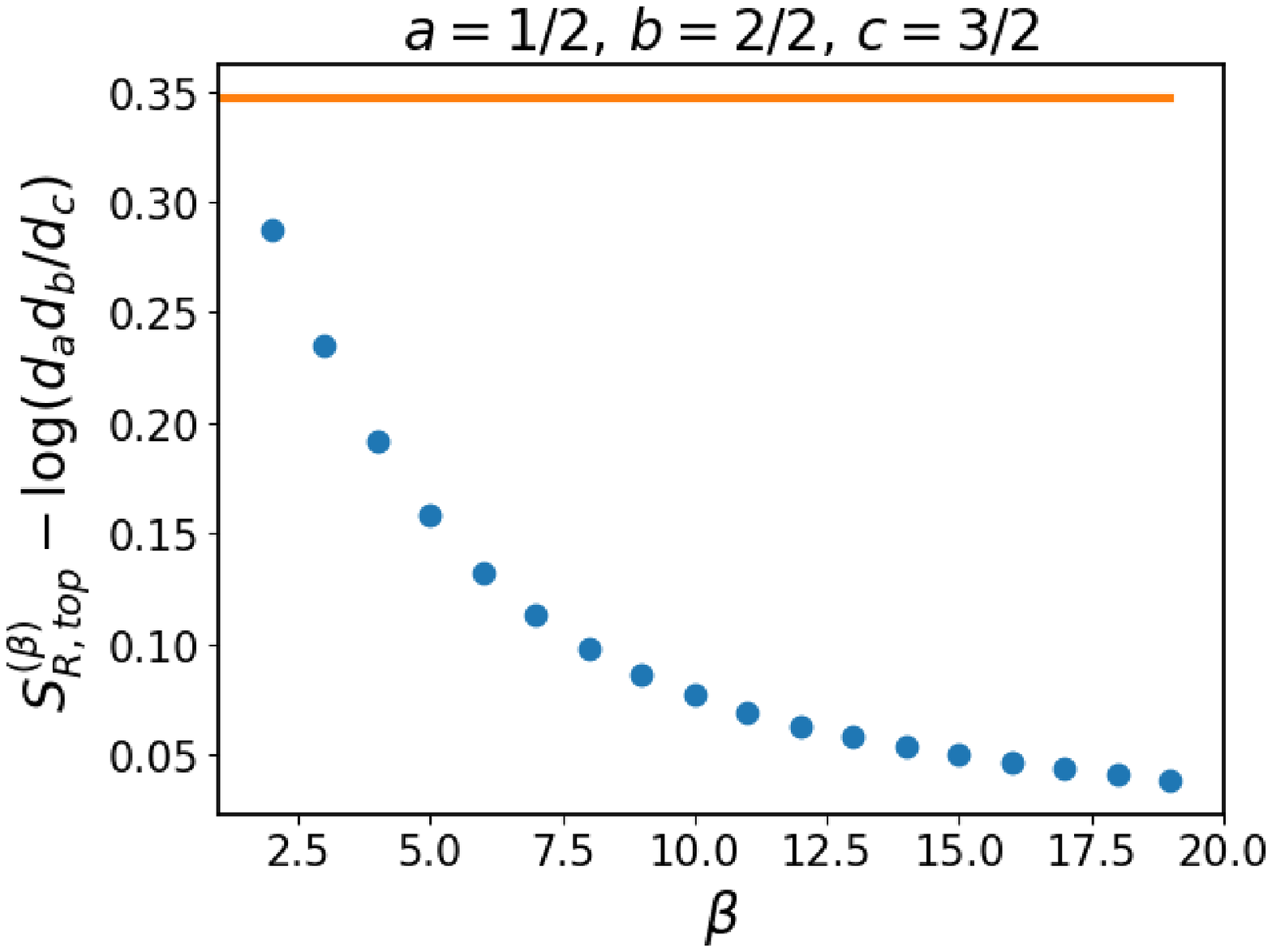}
}

\subfloat[$k=5$]{%
    \includegraphics[width=0.49\linewidth]{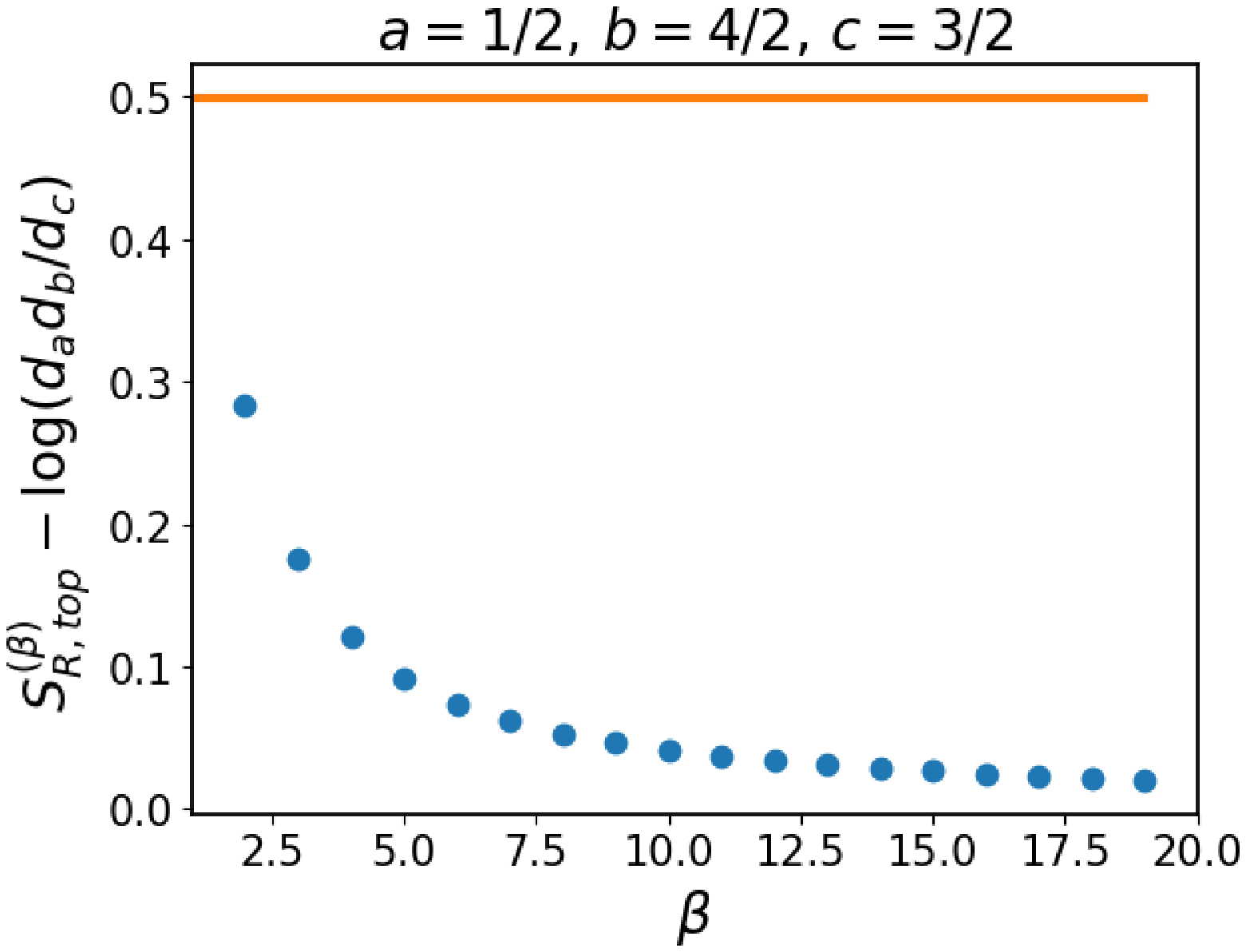}
}
  \caption{ Difference between the topological contributions to the R\'enyi reflected entropy and mutual information (indicated by the blue dots) for a tripartition in the presence of three anyons $a$, $b$, and $c$ in $SU(2)_k$ for (a) $k=3$, (b) $k=4$, and (c) $k=5$. The orange lines indicate the expected value of the Markov gap based on the reflected entropy for an isolated trimer, as given by Eq. \eqref{eq:trimer-SR}.} 
  \label{fig:anyon-insertion-markov}
\end{figure}

%Let us again consider the example of $SU(2)_k$. 
Evaluating Eq. \eqref{eq:anyon-renyi-markov} for $k=2,3,4$ for several different choices of $a$, $b$, and $c$, we find that
if at least one of $a$, $b$, and $c$ is Abelian, then Eq. \eqref{eq:anyon-renyi-markov} vanishes for all $\alpha > 1$ 
(and hence, by analytically continuing, for $\alpha=1$ 
as well) and thus the topological contributions to the reflect entropy and mutual information are the same.
On the other hand, if all of $a$, $b$, and $c$ are non-Abelian, then Eq. \eqref{eq:anyon-renyi-markov} is strictly positive for all $\alpha>1$, implying that, continuing to $\alpha=1$, 
the Markov gap is also strictly positive. We plot the R\'enyi Markov gap for several such examples in Fig. \ref{fig:anyon-insertion-markov}.
We thus appear to find that Abelian anyons do not contribute to the entanglement as measured by the reflected entropy, as is the case for the mutual information. In the case where we have a pair of non-Abelian anyons, the reflected entropy and mutual information match, as such a configuration should presumably only yield additional bipartite entanglement. The positivity of the Markov gap in the case with three non-Abelian anyons suggests that that the reflected entropy is detecting additional tripartite correlations beyond that picked up by the mutual information. 

Heuristically, one might expect the trimer insertion to contribute to tripartite entanglement, by virtue of the fact that the set of three anyons $a$, $b$, and $c$ are constrained to have a net trivial fusion channel. In contrast, the ground state anyon diagrams we have constructed are such that each component (i.e. each region between two adjacent wormhole throats) of the entanglement cut is effectively described by a superposition of anyon-antianyon Bell pairs fusing to the vacuum, as described in Section \ref{sec:bipartite}. More precisely, the ground states describable with the diagrammatic method appear to correspond to sums of triangle states (SOTS). As described in Ref. \cite{Siva2021a}, for a tripartition of a Hilbert space $\cH = \cH_A \otimes \cH_B \otimes \cH_C$, a state $\ket{\psi}$ is a SOTS if we can further decompose each sub-Hilbert space as $\cH_\alpha = \bigoplus_j \cH_{\alpha_L^j} \otimes \cH_{\alpha_R^j}$, 
such that we may write
\begin{align*}
	\ket{\psi} = \sum_j \sqrt{p_j} \ket{\psi_j}_{A_R^j B_L^j} \ket{\psi_j}_{B_R^j C_L^j} \ket{\psi_j}_{C_R^j A_L^j}
\end{align*}
where $\sum_j p_j = 1$ and $\ket{\psi_j}_{\alpha_R^j \beta_L^j}$ has support in $\cH_{\alpha_R^j} \otimes \cH_{\beta_L^j}$. It was shown in Ref. \cite{Siva2021a} that the Markov gap vanishes if and only if the state is a SOTS. The pairwise entanglement via anyon-antianyon pairs between subregions manifest in Figs. \ref{fig:general-tripartition}(b) and \ref{fig:tetrajunction}(b) matches this structure and is consistent with the vanishing of the Markov gap. We note that the SOTS of general string-net states, which provide microscopic realizations of the non-chiral states describable via the diagrammatic formalism, was already demonstrated in Ref. \cite{Siva2021b}.

%This would appear to suggest the ground states of topological orders which can be captured via this diagrammatic calculation -- namely those of the form $\mathcal{C}\times \overline{\mathcal{C}}$ -- possess primarily bipartite entanglement, providing another perspective on the vanishing of the Markov gap.

\subsection{Comparison with a Single Trimer}

Given the complicated expressions for the negativity and reflected entropy in the preceding sections, one may ask whether one can explicitly separate the contributions from the bulk and the trimer to these quantities. To that end, we consider the entanglement structure of an anyon trimer in isolation and compare with the preceding results.
We begin with a normalized state for a trimer of anyons $a$, $b$, and $c$ fusing to the identity,%\footnote{Note that this trimer differs from that appearing in the bulk computation by braiding and bending moves, which contribute unimportant overall phases.}
\begin{align}
	\ket{\psi} = \frac{1}{(d_a d_b d_c)^{1/4}} \begin{pspicture}[shift=-0.9](-0.2,0.1)(2.4,1.8)
        \scriptsize
        \psline[ArrowInside=->](1.2,.6)(2.4,1.8)
        \psline[ArrowInside=->](1.2,0.6)(.6,1.2)\rput(2.4,1.5){$c$}
        %\rput(0.48,1.15){$\mu$}
        \psline[ArrowInside=->](0.6,1.2)(0,1.8)\rput(0,1.5){$a$}
        \psline[ArrowInside=->,](0.6,1.2)(1.2,1.8)\rput(1.2,1.5){$b$}
    \end{pspicture}
\end{align}
For the sake of simplicity, we restrict to multiplicity-free theories from the outset, and so we drop vertex labels in this and the following diagrams.
Let us first compute the reflected entropy and negativity between anyons $a$ and $b$. We find for the reduced density matrix for $AB$:
\begin{align}
	\trho_{AB} &= \frac{1}{\sqrt{d_a d_b d_c}} 
		\begin{pspicture}[shift=-0.9](-0.2,0)(1.6,1.8)
        \scriptsize
        \psline[ArrowInside=->](0,0)(0.6,0.6)\rput(0,0.3){$a$}
        \psline[ArrowInside=->](1.2,0)(0.6,0.6)\rput(1.2,0.3){$b$}
        %\rput(0.77,0.65){$\mu$}
        \psline[ArrowInside=->](0.6,0.6)(0.6,1.2)\rput(0.37,0.9){$c$}
        %\rput(0.77,1.15){$\mu$}
        \psline[ArrowInside=->](0.6,1.2)(0,1.8)\rput(0,1.5){$a$}
        \psline[ArrowInside=->,](0.6,1.2)(1.2,1.8)\rput(1.2,1.5){$b$}
    \end{pspicture}
\end{align}
The canonical purification is simply given by
\begin{align}
	\kket{\sqrt{\trho}} &= \frac{1}{\sqrt{d_a d_b}} 	
	\begin{pspicture}[shift=-1.9](-2.2,-1.3)(1.3,1.9)
        \scriptsize
        \psline[ArrowInside=->](0,0)(0.6,0.6)%\rput(0,0.3){$a$}
        \psline[ArrowInside=->](1.2,0)(0.6,0.6)%\rput(1.2,0.3){$b$}
        %\rput(0.77,0.65){$\mu$}
        \psline[ArrowInside=->](0.6,0.6)(0.6,1.2)\rput(0.37,0.9){$c$}
        %\rput(0.77,1.15){$\mu$}
        \psline[ArrowInside=->](0.6,1.2)(0,1.8)\rput(0,1.5){$a$}
        \psline[ArrowInside=->,](0.6,1.2)(1.2,1.8)\rput(1.2,1.5){$b$}
        \psline(0.,0)(-.6,.6)(-.6,1.8)\rput(-.8,1.7){$\bar a$}
        \psline(1.2,0)(0.,-1.2)(-1.8,.6)(-1.8,1.8)\rput(-2.,1.7){$\bar b$}
    \end{pspicture}
\end{align}
Note that the anyon $c$ connects, and hence indicates entanglement between, $AB$ and $A^*B^*$, containing anyons $a,b$ and $\bar{a}, \bar{b}$, respectively. In order to compute the reflected entropy, we will need to change to a basis with a definite anyon charge passing instead between $AA^*$ and $BB^*$; this is accomplished via an $F$-move,
\begin{align}
	\begin{split}
	\kket{\sqrt{\trho}} &= \frac{1}{\sqrt{d_a d_b}} \sum_f [F^{ab}_{ab}]_{fc}^*
	\begin{pspicture}[shift=-1.9](-2.2,-1.3)(1.3,1.9)
        \scriptsize
        \psline[ArrowInside=-<](0.,1.2)(1.2,.6)\rput(0.6,1.1){$f$}
        \psline[ArrowInside=->](0.,0)(0.,1.2)\psline[ArrowInside=->](0.,1.2)(0,1.8)\rput(-0.2,1.7){$a$}
        \psline[ArrowInside=-<](1.2,.6)(1.2,0)\psline[ArrowInside=->,](1.2,.6)(1.2,1.8)\rput(1.,1.7){$b$}
        \psline(0.,0)(-.6,.6)(-.6,1.8)\rput(-.8,1.7){$\bar a$}
        \psline(1.2,0)(0.,-1.2)(-1.8,.6)(-1.8,1.8)\rput(-2.,1.7){$\bar b$}
    \end{pspicture}
    \end{split} \label{eq:isolated-trimer-f-move}
\end{align}
In this new basis, the anyon $f$ captures the entanglement between $AA^*$ and $BB^*$.
On tracing out $BB^*$, comprising the anyons $b$ and $\bar{b}$, and keeping careful track of bending moves, we find
\begin{align}
	\trho_{AA^*} = \sum_f |[F^{ab}_{ab}]_{fc}|^2 \frac{1}{\sqrt{d_a^2 d_f}} \begin{pspicture}[shift=-0.9](-0.2,0)(1.6,1.8)
        \scriptsize
        \psline[ArrowInside=->](0,0)(0.6,0.6)\rput(0,0.3){$\bar{a}$}
        \psline[ArrowInside=->](1.2,0)(0.6,0.6)\rput(1.2,0.3){$a$}
        \psline[ArrowInside=->](0.6,0.6)(0.6,1.2)\rput(0.37,0.9){$f$}
        \psline[ArrowInside=->](0.6,1.2)(0,1.8)\rput(0,1.5){$\bar{a}$}
        \psline[ArrowInside=->,](0.6,1.2)(1.2,1.8)\rput(1.2,1.5){$a$}
    \end{pspicture} \, .
\end{align}
We note that this particularly simple form and the cancellation of $A$ and $B$ symbols arising from bending moves arises as a consequence of our assumption of a multiplicity-free theory.
Now, using the replica trick, one readily computes
\begin{align}
	\Tr[\trho_{AA^*}^\beta] = \sum_f |[F^{ab}_{ab}]_{fc}|^{2\beta} d_f^{1-\beta}. 
\end{align}
Taking the von Neumann limit, we ultimately obtain for the reflected entropy,
\begin{align}
	S_R &= - \sum_f |[F^{ab}_{ab}]_{fc}|^2 \ln  \left( |[F^{ab}_{ab}]_{fc}|^2 / d_f \right) \label{eq:trimer-SR} \\
	\notag &= - \sum_f \left[ |[F^{ab}_{ab}]_{fc}|^2 \ln  \left( |[F^{ab}_{ab}]_{fc}|^2   \right) -  |[F^{ab}_{ab}]_{fc}|^2 \ln d_f\right].
\end{align}
By unitarity of the $F$-symbols, $|[F^{ab}_{ab}]_{fc}|^2$ is a probability distribution in $f$. The reflected entropy comprises two pieces. The first term is a Shannon entropy associated with the superposition of $f$ anyons while the second term is the topological contribution to the entropy from each $f$ anyon. This computation provides a heuristic understanding for the manner in which the reflected entropy captures tripartite correlations. As noted above, we see from Eq. \eqref{eq:isolated-trimer-f-move} that the entanglement between $a$ and $b$ is captured by the superposition of anyons $f$. This superposition is not arbitrary, however, with each contribution being weighted by $[F^{ab}_{ab}]_{fc}$, which retains knowledge of the fact that $a$ and $b$ are constrained to fuse to $c$. 
We note that the mutual information may be expressed similarly,
\begin{align}
	I = \ln \frac{d_a d_b}{d_c} = - \sum_f \delta_{1,f} \ln  \left( |[F^{ab}_{ab}]_{fc}|^2 / \delta_{1,f} \right),
\end{align}
where we used the fact that $[F^{ab}_{ab}]_{fc} = \sqrt{d_a d_b / d_c}$ and $\delta_{1,f}$ is the Kronecker delta. Interestingly, the mutual information takes the form of a relative entropy between the probability distributions $p_f = \delta_{1,f}$ and $F_f = |[F^{ab}_{ab}]_{fc}|^2$. 

We compare these results with those obtained for the Markov gap in the bulk computation in Fig. \ref{fig:anyon-insertion-markov} for the $SU(2)_k$ examples discussed above. We see that the R\'enyi reflected entropies of the bulk computation extrapolate fairly well to Eq. \eqref{eq:trimer-SR} in the von Neumann limit $\beta \to 1$. This suggests that we can safely identify the trimer contribution to the reflected entropy in the bulk computation as being given by Eq. \eqref{eq:trimer-SR}. 

While we can compute the negativity for this trimer as well, it does not match the expression from the bulk computation in Eq. \eqref{eq:anyon-negativity}. This, however, is not too surprising. In the computation of the reflected entropy for the isolated trimer, we note that the canonically purified density matrix, by construction, has a net trivial anyon charge. This implies that the quantum trace of it (and its powers) is equal to the standard trace. In turn, this means the anyonic entanglement of this state is equal to the usual entanglement entropy. This leads us to claim that the total reflected entropy of the bulk tripartition with the trimer insertion can safely be expressed as the sum of the reflected entropies of the ground state and of the trimer, separately. In contrast, the anyonic partial transpose of $\trho_{AB}$ has a net non-zero anyon charge, and so the quantum trace of it (and its powers) is not equal to the usual trace. This suggests that the anyonic negativity of the isolated trimer is not the same as the contribution of the trimer to the usual negativity. Put differently, the anyonic transpose of the anyon diagram describing the bulk topological liquid with the trimer insertion must be equivalent to the bosonic transpose, as the diagram ultimately describes a bosonic system. In contrast, the anyonic transpose of the trimer need not match the bosonic transpose, and so the trimer contribution to the negativity need not be the same in the two cases.

\section{Discussion \label{sec:discussion} }

In this work, we have employed an anyon diagrammatic approach, developed in Ref. \cite{Bonderson2017}, to compute the negativity, reflected entropy, and Markov gap for tripartitions of topologically ordered phases, both in the absence and presence of anyon insertions. We found that for the two tripartitions considered, involving either trijunctions or tetrajunctions of the three regions, the Markov gap always vanished. This is a consequence of the fact that the diagrammatic approach is only capable of computing the entanglement in topological phases of the form $\mathcal{C} \times \overline{\mathcal{C}}$, for which the minimal central charge necessarily vanishes. On the other hand, we found an interesting contribution to the negativity, which is only non-vanishing for non-Abelian orders, in the trijunction setup, while we found that the mutual information (and hence reflected entropy) have vanishing area law terms but a novel constant contribution for the tetrajunction setup. 
%We illustrated how these contributions can be understood as arising from the trijunctions/tetrajunctions by making an analogy with the corresponding entanglement computations for tripartitions of the same topological on a torus. 
Finally, we found that the Markov gap is non-vanishing upon the insertion of a non-Abelian anyon trimer and depends, as does the negativity, in a non-trivial way on the $F$-symbols. Looking forward, there are several avenues for further exploration.

An advantage of this diagrammatic approach is that it allows for simple computations of entanglement quantities for arbitrary multipartitions. However, it is not able to capture contributions like the central charge, which lies beyond the purely topological data captured by a modular tensor category, and is expected to appear in the Markov gap for chiral topological orders. It is thus an important problem to develop more general techniques that allow for computing the reflected entropy, negativity, and other quantities for generic (chiral) topological phases; this constitutes the subject of forthcoming work \cite{Liu2023}.

As we noted in the main text, we cannot rule out that the dependence of the negativity and reflected entropy on the $F$-symbols in the presence of an anyon trimer is a trivial one, in that some simplifications may reduce these expressions to simple functions of the quantum dimensions. Nevertheless, it has been shown recently that modular tensor categories are \emph{not} uniquely defined by their modular data \cite{Mignard2017,Delaney2018,Bonderson2019}, and so this raises the question whether it is possible to find an entanglement measure, multipartition, and anyon configuration which allows for accessing beyond modular data. 

Finally, it would interesting to better understand the physical implications of a non-vanishing Markov gap in anyonic systems. In the situations considered here, where we investigated a topological phase whose constituent, local degrees of freedom are bosonic, the Markov gap distinguishes between W-state and GHZ-state like correlations \cite{Siva2021a}. %However, one could compute the Markov gap in few-body systems of anyons, in the same way the negativity was computed in Ref. \cite{Shapourian2020}. 
For pure states of three qubits, it is known that W-states and GHZ-states constitute the two unique patterns of three-party entanglement \cite{Dur2000}. It would be interesting to understand if the same holds for three anyon systems, or if there are more classes of three-party entanglement, and if the Markov gap can distinguish them.

\textit{Note added:} Upon completion of this work, we became aware of the related work Ref. \cite{Yin2023}, which investigates the negativity and a complementary mixed-state entanglement quantity, the realignment negativity, using a surgery approach. Subsequently, the related Ref. \cite{Liu2023a} appeared, with a scheme to distinguish Abelian and non-Abelian orders.

\begin{acknowledgments}

We thank Jonah Kudler-Flam, Yuya Kusuki, and Yuhan Liu for collaboration on related projects.
SR thanks Hassan Shapourian for discussions.
This work is supported 
by the National Science Foundation under 
Award No.~DMR-2001181, 
a Simons Investigator Grant from
the Simons Foundation (Award No.~566116),
and
the Gordon and Betty Moore Foundation through Grant
GBMF8685 toward the Princeton theory program.
This work was performed in part at Aspen Center for Physics, which is supported by National Science Foundation grant PHY-1607611. 

\end{acknowledgments}

\appendix

\section{Review of Anyon Models \label{sec:app-anyon-models} }

In this Appendix, we provide a brief review of modular tensor category and the diagrammatic formalism for describing systems of anyons on a sphere, following Refs. \cite{Kitaev2006a,Bonderson2007,Bonderson2017}. %In the following subsection, we discuss the new elements required to describe anyons on a manifold of non-zero genus. This formalism was first discussed in Refs. \cite{Pfeifer2012,Pfeifer2014}; here we follow the conventions of Ref. \cite{Bonderson2017}. 
The basic objects of an anyon model $\mathcal{C}$ are anyons, labeled as $\{ a, b, c , \dots \}$, which are defined to obey an associate, commutative fusion algebra:
\begin{align}
	a \times b = \sum_c N_{ab}^c c.
\end{align}
Here $N_{ab}^c$ is the fusion multiplicity, indicating the number of ways that $a$ and $b$ can fuse to yield $c$. For a given $a$, if $\sum_c N_{ab}^c > 1$ for some $b$, we say that $a$ is non-Abelian; otherwise, $a$ is Abelian. We require that there exists a unique trivial or vacuum charge, denoted by $0$, which satisfies $N_{a0}^c = \delta_{ac}$ for all $a$ and $c$. Each anyon $a$ must also have a conjugate or anti-anyon $\bar{a}$ such that $N_{a \bar{a}}^0 \neq 0$.

The \emph{quantum dimension} $d_a$ of anyon $a$ is defined to be the largest eigenvalue of the multiplicity matrix $N_a$, with matrix elements $[N_a]_{bc} = N_{ab}^c$. Non-Abelian (Abelian) anyons have $d_a > 1$ ($d_a = 1$); heuristically speaking, the quantum dimension thus provides a measure of the non-local degeneracy provided by each anyon. We define the total quantum dimension of $\mathcal{C}$ as
\begin{align}
	\cD = \sqrt{\sum_a d_a^2}.
\end{align}
We have, in particular, the relation
\begin{align}
	d_a d_b = \sum_c N_{ab}^c \label{eq:quantum-dimension-relation}
\end{align}
which relates the quantum dimensions of two anyons $a$ and $b$ with those of all their fusion channels.

\subsection{Basis and Fusion}
The above fusion structure implies that the Hilbert space of a system of anyons is constrained and hence does not factorize. The states in this space are most easily represented in a diagrammatic representation as follows. We define the splitting space $V_c^{ab}$ as the space of two anyons $a$ and $b$ with total anyon charge $c$. States in this space are represented as
\begin{align}
	\ket{a,b;c,\mu} = \left( \frac{d_c}{d_a d_b} \right)^{1/4}     \begin{pspicture}[shift=-0.39](-0.1,0)(1.2,1)
        \scriptsize
        \psline[ArrowInside=->](0.5,0.5)(0,1)\rput(0.1,0.7){$a$}
        \psline[ArrowInside=->](0.5,0.5)(1,1)\rput(0.95,0.7){$b$}
        \psline[ArrowInside=->](0.5,0)(0.5,0.5)\rput(0.65,0.1){$c$}
        \rput(0.65,0.45){$\mu$}
    \end{pspicture}
\end{align}
where the vertex label $\mu = 1 , \dots , N_{ab}^c$ enumerates the ways $c$ can split into $a$ and $b$. The fusion space $V_{ab}^c$ is defined as the dual space of $V_c^{ab}$ and is spanned by states 
\begin{align}
	\bra{a,b;c,\mu} = \left( \frac{d_c}{d_a d_b} \right)^{1/4}    \begin{pspicture}[shift=-0.38](0,0)(1.3,1)
        \scriptsize
        \psline[ArrowInside=->](0,0)(0.5,0.5)\rput(0.1,0.32){$a$}
        \psline[ArrowInside=->](1,0)(0.5,0.5)\rput(0.95,0.32){$b$}
        \psline[ArrowInside=->](0.5,0.5)(0.5,1)\rput(0.65,0.9){$c$}
        \rput(0.7,0.55){$\mu$}
    \end{pspicture} \, .
\end{align}
We can then construct the Hilbert space of three anyons $a$, $b$, and $c$ with total charge $d$ as $V^{abc}_d \cong \bigoplus_e V^{ab}_e \otimes V_d^{ec}$, which is spanned by the basis of states
\begin{align}
	\ket{a,b;e,\mu}\ket{e,c;d,\nu} = \left(\frac{d_d}{d_a d_b d_c}\right)^{1/4}
	\begin{pspicture}[shift=-.8](0,-1.7)(2.,0.2)
        \scriptsize
        \psline[ArrowInside=-<](0,0)(0.5,-0.5)\rput(0.,0.2){$a$}
        \psline[ArrowInside=->](.5,-.5)(1.,0)\rput(1.,0.2){$b$}
        \psline[ArrowInside=-<](0.5,-0.5)(1,-1)\rput(2.,0.2){$c$}
        \psline[ArrowInside=-<](1,-1)(1.5,-1.5)\rput(1.5,-1.7){$d$}
        \psline[ArrowInside=->](1,-1)(2,0)
        \rput(0.5,-.7){$\mu$}
        \rput(1.,-1.2){$\nu$}
        \rput(.9,-.6){$e$}
    \end{pspicture}
\end{align}
We can equivalently construct this space as $V^{abc}_d \cong \bigoplus_f V^{bc}_f \otimes V_d^{af}$ with basis
\begin{align}
	\ket{b,c;f,\alpha}\ket{a,f;d,\beta} = \left(\frac{d_d}{d_a d_b d_c}\right)^{1/4}
		\begin{pspicture}[shift=-.8](0,-1.7)(2.,0.2)
        \scriptsize
        \psline[ArrowInside=-<](0,0)(1.,-1.)\rput(0.,0.2){$a$}
        \psline[ArrowInside=->](1.5,-.5)(1.,0)\rput(1.,0.2){$b$}
        \psline[ArrowInside=->](1.5,-0.5)(2,0)\rput(2.,0.2){$c$}
        \psline[ArrowInside=-<](1,-1)(1.5,-1.5)\rput(1.5,-1.7){$d$}
        \psline[ArrowInside=->](1,-1)(1.5,-0.5)
        \rput(1.5,-.7){$\alpha$}
        \rput(.95,-1.25){$\beta$}
        \rput(1.1,-.6){$f$}
    \end{pspicture}
\end{align}
These constructions are isomorphic, and the associated bases of states are related by an $F$-move,
\begin{align}
		\begin{pspicture}[shift=-.8](0,-1.7)(2.,0.2)
        \scriptsize
        \psline[ArrowInside=-<](0,0)(0.5,-0.5)\rput(0.,0.2){$a$}
        \psline[ArrowInside=->](.5,-.5)(1.,0)\rput(1.,0.2){$b$}
        \psline[ArrowInside=-<](0.5,-0.5)(1,-1)\rput(2.,0.2){$c$}
        \psline[ArrowInside=-<](1,-1)(1.5,-1.5)\rput(1.5,-1.7){$d$}
        \psline[ArrowInside=->](1,-1)(2,0)
        \rput(0.5,-.7){$\mu$}
        \rput(1.,-1.2){$\nu$}
        \rput(.9,-.6){$e$}
    \end{pspicture} = 
    \sum_{f , \alpha , \beta} [F^{abc}_d]_{(e,\mu,\nu) (f,\alpha, \beta)} \begin{pspicture}[shift=-.8](0,-1.7)(2.,0.2)
        \scriptsize
        \psline[ArrowInside=-<](0,0)(1.,-1.)\rput(0.,0.2){$a$}
        \psline[ArrowInside=->](1.5,-.5)(1.,0)\rput(1.,0.2){$b$}
        \psline[ArrowInside=->](1.5,-0.5)(2,0)\rput(2.,0.2){$c$}
        \psline[ArrowInside=-<](1,-1)(1.5,-1.5)\rput(1.5,-1.7){$d$}
        \psline[ArrowInside=->](1,-1)(1.5,-0.5)
        \rput(1.5,-.7){$\alpha$}
        \rput(.95,-1.25){$\beta$}
        \rput(1.1,-.6){$f$}
    \end{pspicture}
\end{align}
where (for a unitary modular tensor category) the $F^{abc}_d$ are unitary matrices known as the $F$-symbols and must satisfy a consistency equation known as the pentagon equation. The space of states with $n$ anyons is then constructed iteratively as $V^{\vec{a}}_c \equiv V^{a_1 \dots a_n}_d \cong \bigoplus_{\vec{b}} V^{a_1 a_2}_{b_2} \otimes V_{b_3}^{b_2 a_3} \otimes \dots \otimes V_c^{b_{n-1} a_n}$ and is spanned by the states
\begin{align}
	\ket{\vec{a},\vec{b};\vec{\mu},c} = \left( \frac{d_c}{d_{\vec{a}}} \right)^{1/4}
	    \begin{pspicture}[shift=-.75](-0.2,0)(2.2,2.0)
        \scriptsize
        \psline[ArrowInside=->](0.5,1)(0,1.5)\rput(0,1.7){$a_1$}
        \psline[ArrowInside=->](0.5,1)(1,1.5)\rput(1,1.7){$a_2$}
        \psline[ArrowInside=->](1,0.5)(2,1.5)\rput(2,1.7){$a_n$}
        \psline[ArrowInside=->](1,0)(1,0.5)\rput(1.15,0.25){$c$}
        \psline[ArrowInside=->](0.7,0.8)(0.5,1)\rput(0.5,0.75){$b_2$}
        \psline[ArrowInside=->](1,0.5)(0.8,0.7)\rput(0.6,0.5){$b_{n-1}$}
        \rput(0.2,1){$\mu_2$}
        \rput(1.3,0.5){$\mu_n$}
        \rput(0.75,0.75){.}
    \end{pspicture}
\end{align}
where the intermediate anyons $\vec{b}$ and vertices $\vec{\mu}$ are permitted by fusion.
Here we have used the shorthand notation $\vec{b} = (b_2 , \dots , b_{n-1})$ and further defined
\begin{align}
	d_{\vec{a}} = d_{a_1} \dots d_{a_n} = \sum_c N^c_{a_1 \dots a_n} d_c \, .
\end{align}

There exists an isomorphism between the splitting and fusion spaces. This allows us to ``bend" the anyon worldlines via $A$-moves and $B$-moves,
\begin{align}
	\begin{pspicture}[shift=-0.39](-0.6,0)(1.2,1)
        \scriptsize
        \psline[ArrowInside=->](-0.5,0.5)(0,1)\rput(-.6,0.7){$\bar{a}$}
        \psline[ArrowInside=->](0.5,0.5)(0,1)\rput(0.1,0.7){$a$}
        \psline[ArrowInside=->](0.5,0.5)(1,1)\rput(0.95,0.7){$b$}
        \psline[ArrowInside=->](0.5,0)(0.5,0.5)\rput(0.65,0.1){$c$}
        \rput(0.65,0.45){$\mu$}
    \end{pspicture} &= \sum_\mu [A^{ab}_c]_{\mu\nu} 
    \begin{pspicture}[shift=-0.38](-.1,0)(1.3,1)
        \scriptsize
        \psline[ArrowInside=->](0,0)(0.5,0.5)\rput(0.1,0.32){$\bar{a}$}
        \psline[ArrowInside=->](1,0)(0.5,0.5)\rput(0.95,0.32){$b$}
        \psline[ArrowInside=->](0.5,0.5)(0.5,1)\rput(0.65,0.9){$c$}
        \rput(0.7,0.55){$\nu$}
    \end{pspicture} \\
  \begin{pspicture}[shift=-0.39](0.,0)(1.7,1)
        \scriptsize
        \psline[ArrowInside=->](0.5,0.5)(0,1)\rput(0.1,0.7){$a$}
        \psline[ArrowInside=->](0.5,0.5)(1,1)\rput(0.95,0.7){$b$}
        \psline[ArrowInside=->](1.5,0.5)(1,1)\rput(1.6,0.7){$\bar{b}$}
        \psline[ArrowInside=->](0.5,0)(0.5,0.5)\rput(0.65,0.1){$c$}
        \rput(0.65,0.45){$\mu$}
    \end{pspicture} &= \sum_\mu [B^{ab}_c]_{\mu\nu} 
    \begin{pspicture}[shift=-0.38](-.1,0)(1.3,1)
        \scriptsize
        \psline[ArrowInside=->](0,0)(0.5,0.5)\rput(0.1,0.32){$a$}
        \psline[ArrowInside=->](1,0)(0.5,0.5)\rput(0.95,0.32){$\bar{b}$}
        \psline[ArrowInside=->](0.5,0.5)(0.5,1)\rput(0.65,0.9){$c$}
        \rput(0.7,0.55){$\nu$}
    \end{pspicture}
\end{align}
where the $A$ and $B$ symbols are unitary matrices defined in terms of the $F$-symbols as
\begin{align}
	[A^{ab}_c]_{\mu\nu} &= \sqrt{\frac{d_a d_b}{d_c}} \varkappa_a^* [F^{\bar{a}a b}_b]^*_{1,(c,\mu,\nu)} \\
	[B^{ab}_c]_{\mu\nu} &= \sqrt{\frac{d_a d_b}{d_c}} [F^{a a \bar{b}}_b]^*_{(c,\mu,\nu),1},
\end{align}
where $\varkappa_a$ is the Frobenius-Schur indicator of $a$. Making use of these bends, one can define an $F$-move for a diagram with two legs pointing up and down:
\begin{align}
	\begin{pspicture}[shift=-0.65](-.2,0)(1.2,1.7)
        \scriptsize
        \psline[ArrowInside=->](0,0)(0,1.1)\psline[ArrowInside=->](0,1.1)(0,1.5)\rput(0.,-0.15){$c$}\rput(0.,1.65){$a$}
        \psline[ArrowInside=->](1,0)(1,.4)\psline[ArrowInside=->](1,0.4)(1,1.5)\rput(1.,-0.15){$d$}\rput(1.,1.65){$b$}
        \psline[ArrowInside=->](1,.4)(0,1.1)\rput(.5,1.){$e$}
        \rput(-.1,1.1){$\mu$}
        \rput(1.1,.4){$\nu$}
    \end{pspicture} = \sum_{f\alpha\beta} [F^{ab}_{cd}]_{ (e \mu \nu) (f \alpha \beta)}  \begin{pspicture}[shift=-0.65](-0.1,-0.5)(1.2,1)
        \scriptsize
        \psline[ArrowInside=->](0.5,0.5)(0,1)\rput(0.,1.2){$a$}
        \psline[ArrowInside=->](0.5,0.5)(1,1)\rput(1,1.2){$b$}
        \psline[ArrowInside=->](0.5,0)(0.5,0.5)\rput(0.35,0.25){$f$}
        \rput(0.65,0.45){$\alpha$}
        \psline[ArrowInside=-<](0.5,0.)(0,-.5)\rput(0,-0.7){$c$}
        \psline[ArrowInside=-<](0.5,0.)(1,-.5)\rput(1,-0.7){$d$}
        \rput(0.65,0.1){$\beta$}
    \end{pspicture}
\end{align}
where
\begin{align}
	[F^{ab}_{cd}]_{(e \mu \nu) (f \alpha \beta)} = \sqrt{\frac{d_e d_f}{d_a d_d}} [F^{ceb}_f]^*_{(e\mu\alpha)(f\nu\beta)}
\end{align}
is also a unitary matrix.

\subsection{Inner Products and Traces}
Inner products of states in fusion and splitting spaces are evaluated diagrammatically by simply stacking the corresponding diagrams. Explicitly, the orthonormality condition
\begin{align}
	\braket{a', b';c',\mu'|a,b;c,\mu} = \delta_{aa'} \delta_{bb'} \delta_{cc'} \delta_{\mu\mu'} \mathbf{1}_c
\end{align}
becomes
\begin{align}
	\left(\frac{d_c d_{c'}}{d_a d_b d_{a'} d_{b'}}\right)^{1/4}
	\begin{pspicture}[shift=-1.](0,-1)(1.3,1)
        \scriptsize
        \psline[ArrowInside=->](0,0)(0.5,0.5)\rput(0.1,0.32){$a'$}
        \psline[ArrowInside=->](1,0)(0.5,0.5)\rput(0.95,0.32){$b'$}
        \psline[ArrowInside=->](0.5,0.5)(0.5,1)\rput(0.65,0.9){$c'$}
        \rput(0.7,0.55){$\mu'$}
    \rput(0,-1.){
        \psline[ArrowInside=->](0.5,0.5)(0,1)\rput(0.1,0.7){$a$}
        \psline[ArrowInside=->](0.5,0.5)(1,1)\rput(0.95,0.7){$b$}
        \psline[ArrowInside=->](0.5,0)(0.5,0.5)\rput(0.65,0.1){$c$}
        \rput(0.65,0.45){$\mu$}
    }
    \end{pspicture}
    = \delta_{aa'} \delta_{bb'} \delta_{cc'} \delta_{\mu\mu'}
    \begin{pspicture}[shift=-1.05](-.1,-1.05)(.1,1.05)
        \scriptsize
        \psline[ArrowInside=->](0,-1)(0,1)\rput(0.2,-1){$c$}
    \end{pspicture} \label{eq:inner-product}
\end{align}
In particular, for a space of multiple anyons fusing to the identity, we have that
\begin{align}
\begin{gathered}
\includegraphics[height=11em]{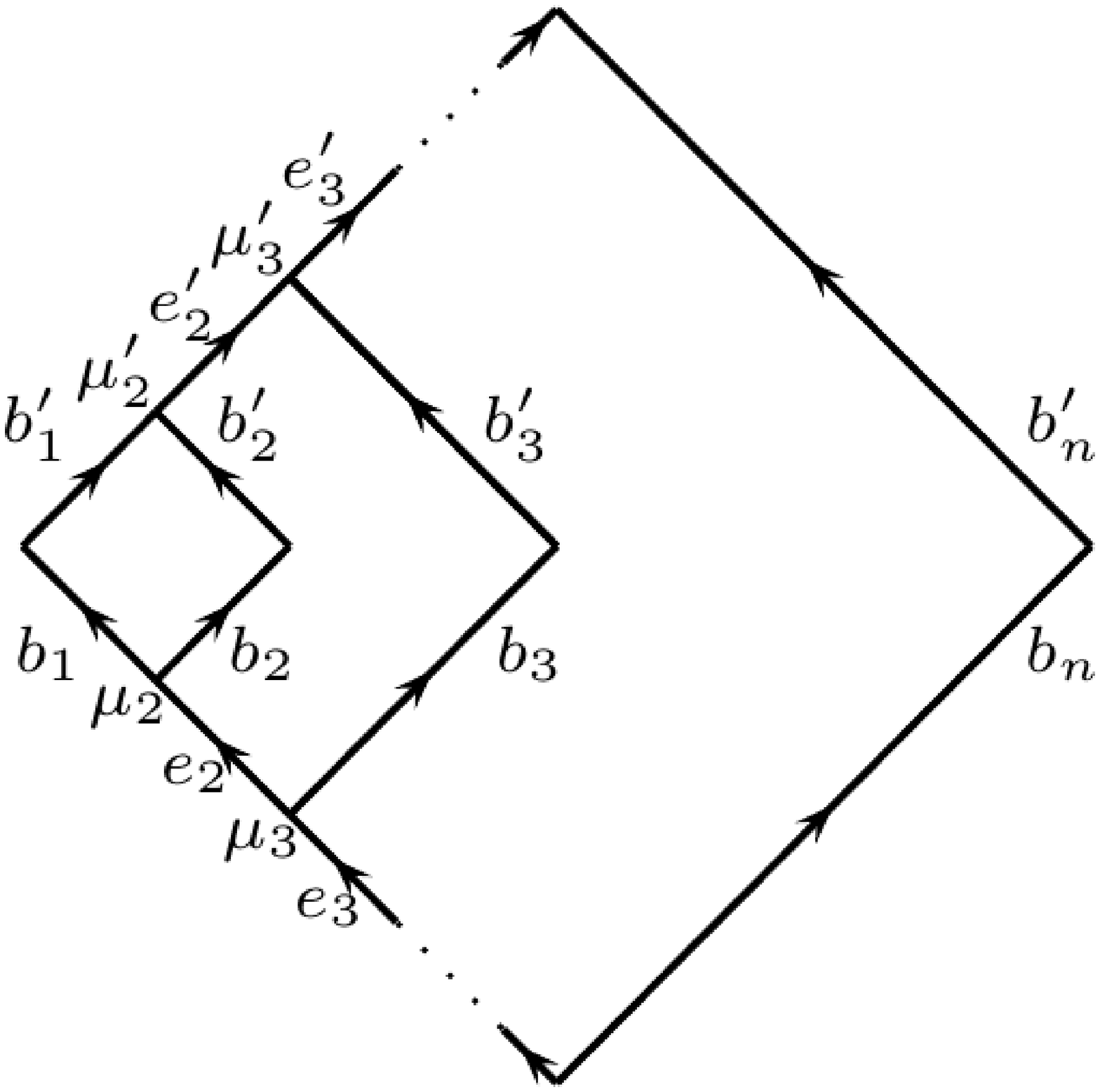} \end{gathered} 
    = \sqrt{d_{\vec{b}}} \delta_{\vec{b} ,\vec{b}'}  \delta_{\vec{e} ,\vec{e}'} \delta_{\vec{\mu}, \vec{\mu}'} \label{eq:overlap}
\end{align}
which is obtained through repeated application of Eq. \eqref{eq:inner-product}. We make repeated use of this relation in our computations. Note that, although we have considered anyon states in the canonical basis (i.e. all lines branch off a single ``stem"), the same relation holds for inner products of multiple anyon states in any other basis in a tree-like form. This follows by applying $F$-moves to the bra and ket in the above relation and making use of unitarity of the $F$-symbols.

In general, diagrams with only closed loops correspond to complex numbers while diagrams with open lines correspond either to states or operators.
With that said, the space of operators $V_{a_1 \dots a_n}^{a_1' \dots a_n'} = \bigoplus V^c_{a_1 \dots a_n} \otimes V_c^{a_1' \dots a_n'} $ acting on anyons $a_1, \dots , a_n$ is spanned by
\begin{align}
	    \ket{\vec{a}',\vec{b}';\vec{\mu}',c}&\bra{\vec{a},\vec{b};\vec{\mu},c} =\left(\frac{d_c^2}{d_{\vec{a}}d_{\vec{a}'}}\right)^{1/4}
    \begin{pspicture}[shift=-1.5](-0.2,-1.5)(2.2,1.5)
        \scriptsize
        \psline[ArrowInside=->](0.5,1)(0,1.5)\rput(0,1.7){$a'_1$}
        \psline[ArrowInside=->](0.5,1)(1,1.5)\rput(1,1.7){$a'_2$}
        \psline[ArrowInside=->](1,0.5)(2,1.5)\rput(2,1.7){$a'_n$}
        \psline[ArrowInside=->](1,0)(1,0.5)\rput(1.15,0.25){$c$}
        \psline[ArrowInside=->](0.7,0.8)(0.5,1)\rput(0.5,0.75){$b'_2$}
        \psline[ArrowInside=->](1,0.5)(0.8,0.7)\rput(0.6,0.5){$b'_{n-1}$}
        \rput(0.2,1){$\mu'_2$}
        \rput(1.3,0.5){$\mu'_n$}
        \rput(0.75,0.75){.}
        \rput(0,0.5){
        \psline[ArrowInside=->](0,-1.5)(0.5,-1)\rput(0.05,-1.7){$a_1$}
        \psline[ArrowInside=->](1,-1.5)(0.5,-1)\rput(1,-1.7){$a_2$}
        \psline[ArrowInside=->](2,-1.5)(1,-0.5)\rput(2,-1.7){$a_n$}
        \psline[ArrowInside=->](0.5,-1)(0.7,-0.8)\rput(0.5,-0.75){$b_2$}
        \psline[ArrowInside=->](0.8,-0.7)(1,-0.5)\rput(0.6,-0.45){$b_{n-1}$}
        \rput(0.25,-1){$\mu_2$}
        \rput(1.3,-0.5){$\mu_n$}
        \rput(0.75,-0.75){.}}
    \end{pspicture}.
\end{align}
In particular, a resolution of the identity acting on a pair of anyons,
\begin{align}
	\mathbf{1}_{ab} = \sum_{c,\mu} \ket{a,b;c,\mu}\bra{a,b;c,\mu}
\end{align}
is provided by
\begin{align}
		\begin{pspicture}[shift=-0.65](-.2,0)(1.1,1.5)
        \scriptsize
        \psline[ArrowInside=->](0,0)(0,1.5)\rput(0.,-0.15){$a$}
        \psline[ArrowInside=->](1,0)(1,1.5)\rput(1.,-0.15){$b$}
    \end{pspicture}
    =
    \sum_{c , \mu} \sqrt{\frac{d_c}{d_a d_b}} \begin{pspicture}[shift=-0.65](-0.1,-0.5)(1.2,1)
        \scriptsize
        \psline[ArrowInside=->](0.5,0.5)(0,1)\rput(0.,1.2){$a$}
        \psline[ArrowInside=->](0.5,0.5)(1,1)\rput(1,1.2){$b$}
        \psline[ArrowInside=->](0.5,0)(0.5,0.5)\rput(0.35,0.25){$c$}
        \rput(0.65,0.45){$\mu$}
        \psline[ArrowInside=-<](0.5,0.)(0,-.5)\rput(0,-0.7){$a$}
        \psline[ArrowInside=-<](0.5,0.)(1,-.5)\rput(1,-0.7){$b$}
        \rput(0.65,0){$\mu$}
    \end{pspicture} \label{eq:resolution-identity}
\end{align}

Next, we come to the trace. As usual, we have that
\begin{align}
	\Tr[\ket{\vec{a}',\vec{b}';\vec{\mu}',c}&\bra{\vec{a},\vec{b};\vec{\mu},c}] = \delta_{aa'} \delta_{bb'} \delta_{cc'} \delta_{\mu\mu'}.
\end{align}
In the diagrammatic formalism, we define the \emph{quantum trace}, $\qTr$, as the operation in which we contract all the output legs of an operator with the corresponding input legs. Denoting a generic operator $X\in V^{a_1 \dots a_n}_{a'_1 \dots a'_n}$ as,
\begin{align}
\label{eq:operator}
X =
\psscalebox{.7}{
 \pspicture[shift=-0.8](-1,-1.)(1,0.9)
  \small
%%%%% Box:
  \psframe[linewidth=0.9pt,linecolor=black,border=0](-0.8,-0.5)(0.8,0.5)
  \rput[bl]{0}(-0.15,-0.1){$X$}
  \rput[bl]{0}(-0.22,0.7){$\mathbf{\ldots}$}
  \rput[bl]{0}(-0.22,-0.75){$\mathbf{\ldots}$}
%%%%% Line connections:
  \psset{linewidth=0.9pt,linecolor=black,arrowscale=1.5,arrowinset=0.15}
  \psline(0.6,0.5)(0.6,1)   \rput(0.6,1.2){$a_n'$}
  \psline(-0.6,0.5)(-0.6,1) \rput(-0.6,1.2){$a_1'$}
  \psline(0.6,-0.5)(0.6,-1) \rput(0.6,-1.2){$a_n$}
  \psline(-0.6,-0.5)(-0.6,-1) \rput(-0.6,-1.2){$a_1$}
%%%%% Arrows:
  \psline{->}(0.6,0.5)(0.6,0.9)
  \psline{->}(-0.6,0.5)(-0.6,0.9)
  \psline{-<}(0.6,-0.5)(0.6,-0.9)
  \psline{-<}(-0.6,-0.5)(-0.6,-0.9)
\endpspicture
},
\end{align}
we define the trace operation as 
\begin{align}
\qTr X 
= \sum_{a_1 , \ldots , a_n}
\psscalebox{.7}{
\pspicture[shift=-1.1](-1.0,-1.2)(2.3,1.1)
  \small
%%%%% Labels:
  \psframe[linewidth=0.9pt,linecolor=black,border=0](-0.8,-0.5)(0.8,0.5)
  \rput[bl]{0}(-0.15,-0.1){$X$}
  \rput[bl]{0}(-0.4,0.7){$\mathbf{\ldots}$}
  \rput[bl]{0}(-0.22,-0.75){$\mathbf{\ldots}$}
  \rput[bl]{0}(1.52,0){$\mathbf{\ldots}$}
%%%%% Line connections:
  \psset{linewidth=0.9pt,linecolor=black,arrowscale=1.5,arrowinset=0.15}
%%%%% Arcs
  \psarc(1.0,0.5){0.4}{0}{180}
  \psarc(1.0,-0.5){0.4}{180}{360}
  \psarc(0,0.5){0.6}{90}{180}
  \psarc(0,-0.5){0.6}{180}{270}
  \psarc(1.5,0.5){0.6}{0}{90}
  \psarc(1.5,-0.5){0.6}{270}{360}
%%%%% Line connections:
  \psline(1.4,-0.5)(1.4,0.5)
  \psline(0,1.1)(1.5,1.1)
  \psline(0,-1.1)(1.5,-1.1)
  \psline(2.1,-0.5)(2.1,0.5)
%%%%% Arrows
  \psline{->}(1.4,0.2)(1.4,-0.1)
  \psline{->}(2.1,0.2)(2.1,-0.1)
%%%%% Leg Labels:
  \rput[bl](-1.07,0.6){$a_1$}
  \rput[bl](0.1,0.6){$a_n$}
 \endpspicture
}
.
\end{align}
Diagrams in which two legs with distinct anyons are contracted evaluate to zero.
The spherical property ensures that contracting all the legs to the right is equivalent to contracting all the legs to the left. 
Similarly, for an operator acting on multiple anyons (with net zero charge),
\begin{align}
\begin{gathered}
\includegraphics[height=11em]{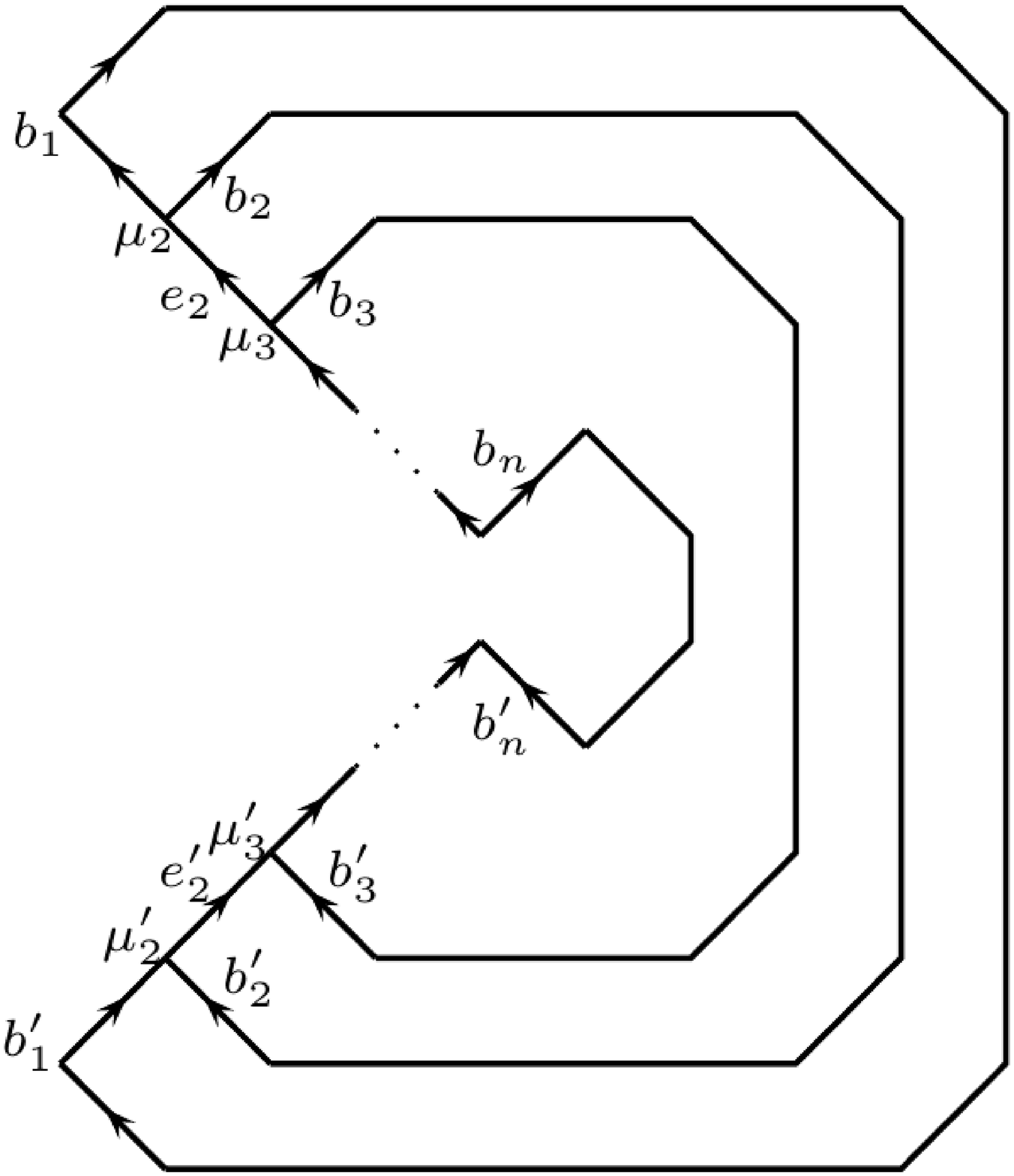} \end{gathered}
    = \sqrt{d_{\vec{b}}} \delta_{\vec{b} ,\vec{b}'}  \delta_{\vec{e} ,\vec{e}'} \delta_{\vec{\mu}, \vec{\mu}'} \label{eq:trace}
\end{align}
Again, this relation is employed frequently in our calculations and holds when the operator on the left hand side is expressed in other bases, aside from the canonical basis.
The quantum trace and regular trace are in general not equivalent but are instead related as
\begin{align}
	\Tr X &= \sum_c \frac{1}{d_c} \qTr X_c \\
	\qTr X &= \sum_c d_c \Tr X_c
\end{align}
where $X_c$ is the projection of operator $X$ to the $c$ fusion channel.

The partial trace is defined similarly and is obtained by contracting a subset of lines located at either the left or right edge of the diagram. Explicitly, suppose we have an operator $X\in V^{a_1 \dots a_n; b_1 \dots b_m}_{a'_1 \dots a'_n; b_1' \dots b_m'}$ and we wish to trace out region $B$, corresponding to the anyons $b_i$. This operation is defined diagrammatically as
\begin{align}
\qTr_B X 
= \sum_{b_1 , \ldots , b_n}
\psscalebox{.8}{
\pspicture[shift=-1.1](-2.6,-1.2)(2.5,1.1)
  \small
%%%%% Labels:
  \psframe[linewidth=0.9pt,linecolor=black,border=0](-2.25,-0.5)(0.8,0.5)
  \rput[bl]{0}(-0.725,-0.1){$X$}
  \rput[bl]{0}(-0.4,0.7){$\mathbf{\ldots}$}
  \rput[bl]{0}(-0.22,-0.75){$\mathbf{\ldots}$}
  \rput[bl]{0}(1.52,0){$\mathbf{\ldots}$}
%%%%% Line connections:
  \psset{linewidth=0.9pt,linecolor=black,arrowscale=1.5,arrowinset=0.15}
%%%%% Arcs
  \psarc(1.0,0.5){0.4}{0}{180}
  \psarc(1.0,-0.5){0.4}{180}{360}
  \psarc(0,0.5){0.6}{90}{180}
  \psarc(0,-0.5){0.6}{180}{270}
  \psarc(1.5,0.5){0.6}{0}{90}
  \psarc(1.5,-0.5){0.6}{270}{360}
%%%%% Line connections:
  \psline(1.4,-0.5)(1.4,0.5)
  \psline(0,1.1)(1.5,1.1)
  \psline(0,-1.1)(1.5,-1.1)
  \psline(2.1,-0.5)(2.1,0.5)
%%%%% Arrows
  \psline{->}(1.4,0.2)(1.4,-0.1)
  \psline{->}(2.1,0.2)(2.1,-0.1)
%%%%% Leg Labels:
  \rput[bl](2.25,0.){$b_1$}
  \rput[bl](.95,0.){$b_n$}
%%%%% Line connections:
  \psset{linewidth=0.9pt,linecolor=black,arrowscale=1.5,arrowinset=0.15}
  \psline(-0.9,0.5)(-0.9,1)   \rput(-0.9,1.2){$a_n'$}
  \psline(-2.1,0.5)(-2.1,1) \rput(-2.1,1.2){$a_1'$}
  \psline(-0.9,-0.5)(-0.9,-1) \rput(-0.9,-1.2){$a_n$}
  \psline(-2.1,-0.5)(-2.1,-1) \rput(-2.1,-1.2){$a_1$}
%%%%% Arrows:
  \psline{->}(-0.9,0.5)(-0.9,0.9)
  \psline{->}(-2.1,0.5)(-2.1,0.9)
  \psline{-<}(-0.9,-0.5)(-0.9,-0.9)
  \psline{-<}(-2.1,-0.5)(-2.1,-0.9)
 \endpspicture
}
.
\end{align}
The partial trace vanishes unless $b_i = b_i'$. If the anyons to be traced out are in th middle of the diagram, they must first be moved to the left or right edges of the diagram via braiding operations.
Again, the quantum and regular partial traces are not equivalent, but rather are related as
\begin{align}
	\Tr_B X &= \sum_{c,f} \frac{d_f}{d_c} \left[\qTr X_c\right]_f \\
	\qTr_B X &= \sum_c \frac{d_c}{d_f} \left[\Tr X_c\right]_f .
\end{align}
In the case where the initial net fusion channel $c$ and the net fusion channel after taking the partial trace are both trivial, the two traces end up being equivalent. This will always be the case in the situations considered in the main text.

\subsection{Braiding}
Thus far we have only specified the fusion structure of our anyon model. We can also braid anyons, a process implemented by the braid operator,
The braiding operator is represented diagrammatically as
\begin{align}
	R_{ab} = 
	\begin{pspicture}[shift=-0.6](-1.2,0)(0.,1.5)
        \scriptsize
        % braid
        \psline(0,0)(-1,1.5)
        \psline[border=2pt](-1,0)(0,1.5)
        \psline[ArrowInside=->](0,0)(-0.33,0.5)\rput(-1,.3){$b$}
        \psline[ArrowInside=->](-1,0)(-0.66,0.5)\rput(0,.3){$a$}
    \end{pspicture}
    =
    \sum_{c\mu\nu} \sqrt{\frac{d_c}{d_a d_b}} [R^{ab}_c]_{\mu\nu} 
    \begin{pspicture}[shift=-0.65](-0.1,-0.5)(1.2,1)
        \scriptsize
        \psline[ArrowInside=->](0.5,0.5)(0,1)\rput(0.,1.2){$a$}
        \psline[ArrowInside=->](0.5,0.5)(1,1)\rput(1,1.2){$b$}
        \psline[ArrowInside=->](0.5,0)(0.5,0.5)\rput(0.35,0.25){$c$}
        \rput(0.65,0.45){$\mu$}
        \psline[ArrowInside=-<](0.5,0.)(0,-.5)\rput(0,-0.7){$b$}
        \psline[ArrowInside=-<](0.5,0.)(1,-.5)\rput(1,-0.7){$a$}
        \rput(0.65,0){$\nu$}
    \end{pspicture}
\end{align}
where the braiding matrices $R^{ab}_c$ are unitary matrices which satisfy a set of consistency relations known as the hexagon equations. From the braiding matrix, we may define the modular $\mathcal{S}$-matrix as 
\begin{align}
	S_{ab} = \qTr[R_{ab} R_{ba}]
\end{align}
The $\mathcal{S}$ matrix is unitary and contains the quantum dimensions of the anyons:
\begin{align}
	\mathcal{S}_{0a} = \frac{d_a}{\cD}.
\end{align}
Using the $\mathcal{S}$-matrix we may define the $\omega_a$ loop, which appears frequently in the calculations of the main text and is defined as
\begin{equation}
\omega_a
\begin{pspicture}[shift=-.3](-.6,-.4)(.5,.3)
  \scriptsize
  \psellipse[linecolor=black,border=0](0,0)(.5,.3)
  \psline[ArrowInside=-<](-.1,-.279)(.075,-.2825)
  \end{pspicture}
  =\sum_x \mathcal{S}_{0a}\mathcal{S}_{ax}^*
\begin{pspicture}[shift=-.3](-.6,-.4)(.6,.4)
\scriptsize
  \psellipse[linecolor=black,border=0](0,0)(.5,.3)  \rput(0,-.5){$x$}
  \psline[ArrowInside=-<](-.1,-.279)(.075,-.2825)
  \end{pspicture}.
\end{equation}
It acts a projector on all charges threading the loop,
\begin{equation}
\begin{pspicture}[shift=-.4](-.9,-.7)(.6,.5)
\scriptsize
  \psellipse[linecolor=black,border=0](0,-.1)(.5,.2)
  \psline[ArrowInside=-<](-.05,-.284)(.15,-.277)
  \psline(0,-.5)(0,-.35)
  \psline[border=1.5pt](0,-.22)(0,.5)
  \psline[ArrowInside=->](0,-.1)(0,.1)
  \rput(0,-.6){$b$}
  \rput(-.75,0){$\omega_a$}
  \end{pspicture}
  = \delta_{ab}
  \begin{pspicture}[shift=-.4](-.2,-.5)(.4,.5)
  \scriptsize
  \psline[ArrowInside=->](0,-.5)(0,.5)
  \rput(.,-.6){$b$}
  \end{pspicture}. \label{eq:omegaa-proj}
\end{equation}
Finally, we make note of the following important relation satisfied by $\omega_0$ loops encircling non-contractible cycles, known as the \textit{handle-slide} property:
\begin{equation}\label{eq:handle-slide}
    \begin{pspicture}[shift=-1](-1,-1)(1,1)
        \scriptsize
        \psellipse(0,0)(.5,.4)\psline[ArrowInside=->](0,-.39)(-.15,-.368)
        %\psline(-0.5,0)(0,0.5)(0.5,0)(0,-0.5)(-0.5,0)(0,0.5)
        \rput(0.45,-0.4){$\omega_0$}\rput(0,0){$\otimes$}
		\psline(0,-1)(0,-0.8)(-0.8,0)(0,0.8)(0,1)
		\psline[ArrowInside=->](0,-0.8)(-0.8,0)\rput(-0.6,-0.6){$a$}
    \end{pspicture}
    =
    \begin{pspicture}[shift=-1](-1,-1)(1,1)
        \scriptsize
        \psellipse(0,0)(.5,.4)\psline[ArrowInside=->](0,-.39)(-.15,-.368)
        %\psline(-0.5,0)(0,0.5)(0.5,0)(0,-0.5)(-0.5,0)(0,0.5)
        \rput(-0.5,-0.4){$\omega_0$}\rput(0,0){$\otimes$}
		\psline(0,-1)(0,-0.8)(0.8,0)(0,0.8)(0,1)
		\psline[ArrowInside=->](0,-0.8)(0.8,0)\rput(0.6,-0.6){$a$}
    \end{pspicture},
\end{equation}
which can be checked by using resolutions of identity to move the $a$ charge line from one side to the other. This says that we can pass a charge line through a contractible cycle, if that cycle is enclosed by an $\omega_0$ loop. 

\section{Computational Details \label{sec:app-details} }

In this appendix, we collect the intermediate steps involved in the computation of the negativity and reflected entropy in the case of the tripartition with anyon insertions, discussed in Section \ref{sec:tripartite-insertions}. As in our other computations, we use a shorthand for the anyon diagrams for the density matrix:
\begin{align}
	\notag(\trho_{AB})_{\mathrm{cut}} = \sum_{ \substack{ \vec{b} \vec{e} \vec{\mu} \\ \vec{b}' \vec{e}' \vec{\mu}' \\ b_N e_{N-1} } } & \sum_{ \substack{ \vec{l} \vec{\nu} \\ \vec{l}' \vec{\nu}' } } \frac{1}{\mathcal{D}^{2N-2}}  \frac{1}{d_b \sqrt{d_a d_b}} \sqrt{\frac{d_{b_N}}{d_{e_{N-1}}}} L_{\vec{l},\vec{e},\vec{b}; \vec{l}',\vec{e}',\vec{b}'}^{\vec{\mu},\vec{\nu}; \vec{\mu}',\vec{\nu}'} \\
	&\times   	\begin{gathered}
\includegraphics[height=10em]{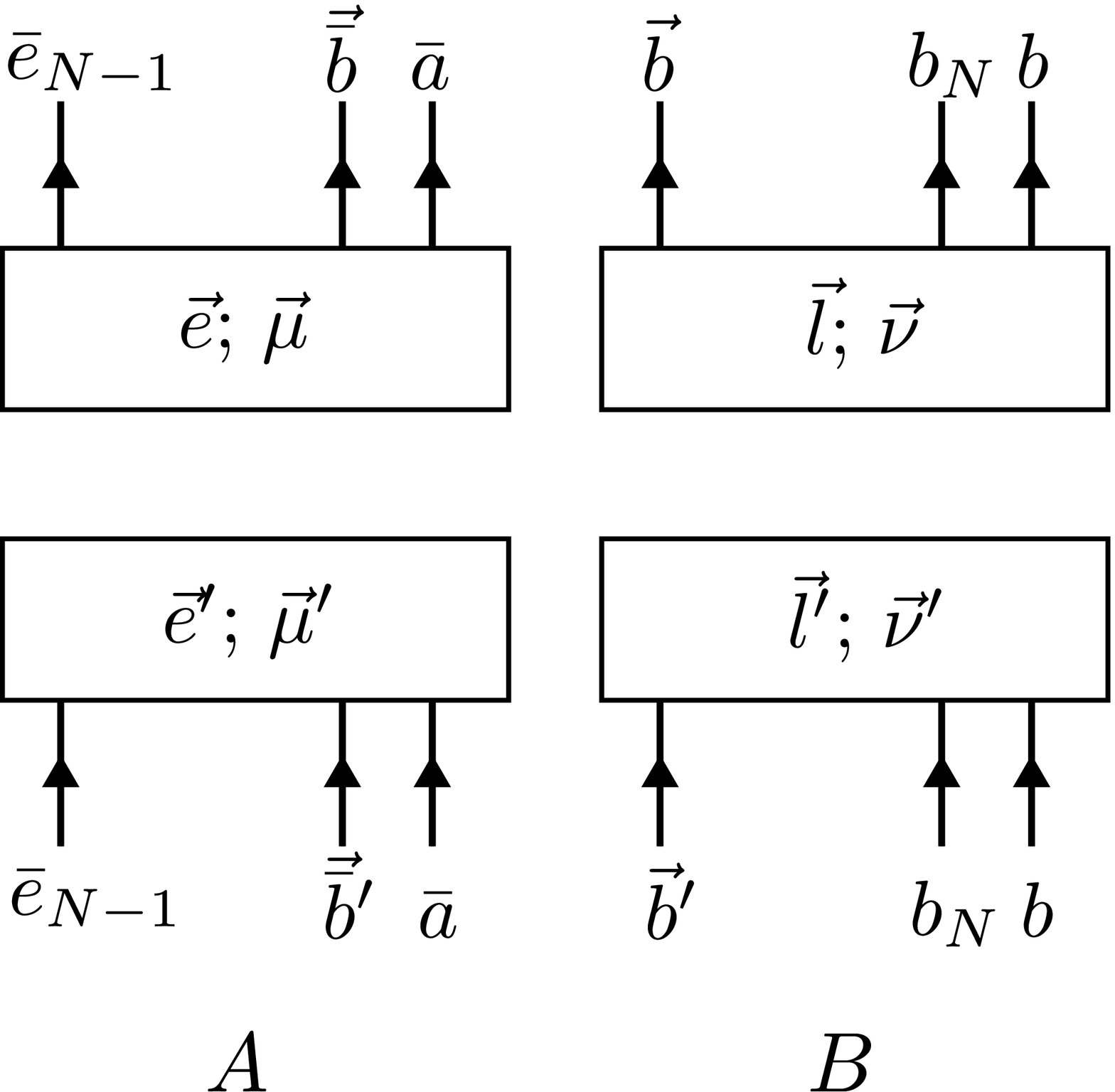}
\end{gathered}
\end{align}
We recall $\vec{b} = (b_1 , \dots , b_{N-1})$, $\vec{l} = (l_1 , \dots , l_{N-2}$, $\vec{\mu} = (\mu_2 , \dots , \mu_{N-1})$, $\vec{\nu} = (\nu_1 , \dots , \nu_{N-1})$, and $\vec{\tilde{\mu}} = (\tilde{\mu}_2 , \dots , \tilde{\mu}_{N-1})$. 

\subsection{Negativity Calculation Details \label{sec:app-details-negativity} }
The partially transposed density matrix is given by
\begin{align}
	\notag(\trho_{AB})_{\mathrm{cut}}^{T_A} = \sum_{ \substack{ \vec{b} \vec{e} \vec{\mu} \\ \vec{b}' \vec{e}' \vec{\mu}' \\ b_N e_{N-1} } } & \sum_{ \substack{ \vec{l} \vec{\nu} \\ \vec{l}' \vec{\nu}' } } \frac{1}{\mathcal{D}^{2N-2}}  \frac{1}{d_b \sqrt{d_a d_b}} \sqrt{\frac{d_{b_N}}{d_{e_{N-1}}}} L_{\vec{l},\vec{e},\vec{b}; \vec{l}',\vec{e}',\vec{b}'}^{\vec{\mu},\vec{\nu}; \vec{\mu}',\vec{\nu}'} \\
	&\times   	\begin{gathered}
\includegraphics[height=10em]{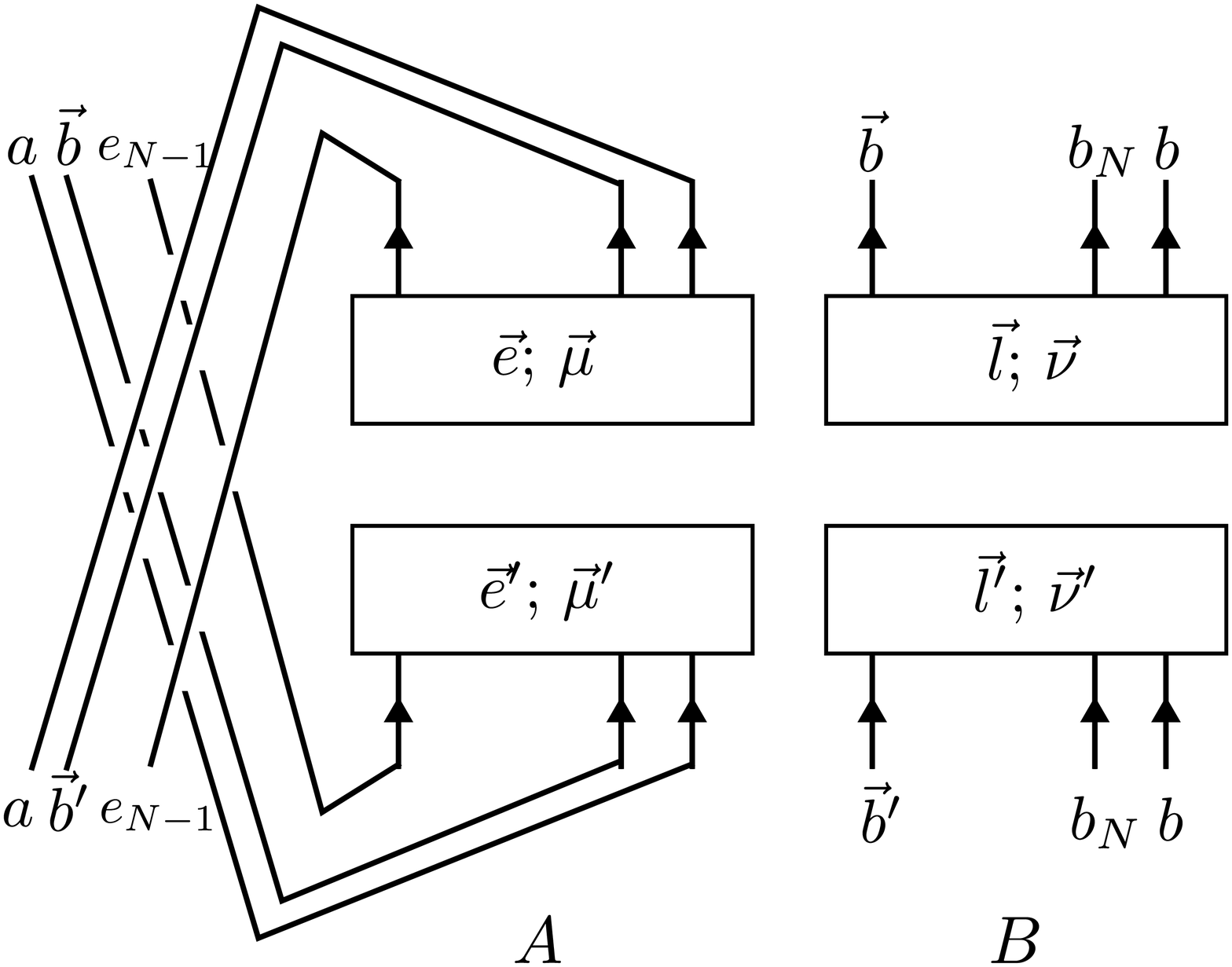}
\end{gathered}
\end{align}
In order to perform the replica trick, we first compute
\begin{align}
	\notag &(\trho_{AB})_{\mathrm{cut}}^{T_A}(\trho_{AB}^{T_A})^\dagger_{\mathrm{cut}} = \\
	\notag & \sum_{ \substack{ \vec{b} \vec{e}' \vec{\mu}' \\ \vec{b}' \vec{e}'' \vec{\mu}'' \\ b_N e_{N-1} } }  \sum_{ \substack{ \vec{l} \vec{\nu}  \\ \vec{l}''' \vec{\nu}'''  } } \frac{d_{\vec{b}} d_{\vec{b}'} d_{b_N}^2 d_{e_{N-1}}}{d_b^2 \mathcal{D}^{4N-4}} \frac{1}{\sqrt{d_a d_b d_{\vec{b}} d_{\vec{b}'} d_{b_N} d_{e_{N-1}}}} \\
	& \times \sum_{\substack{\vec{e} \vec{\mu} \\ \vec{l}' \vec{\nu}' }} L_{\vec{e},\vec{l},\vec{b}; \vec{e}',\vec{l}',\vec{b}'}^{\vec{\mu},\vec{\nu}; \vec{\mu}',\vec{\nu}'} L_{\vec{e}'',\vec{l}',\vec{b}'; \vec{e},\vec{l}''',\vec{b}}^{\vec{\mu}'',\vec{\nu}'; \vec{\mu},\vec{\nu}'''} 	\\
	\notag & \times \begin{gathered}
\includegraphics[height=10em]{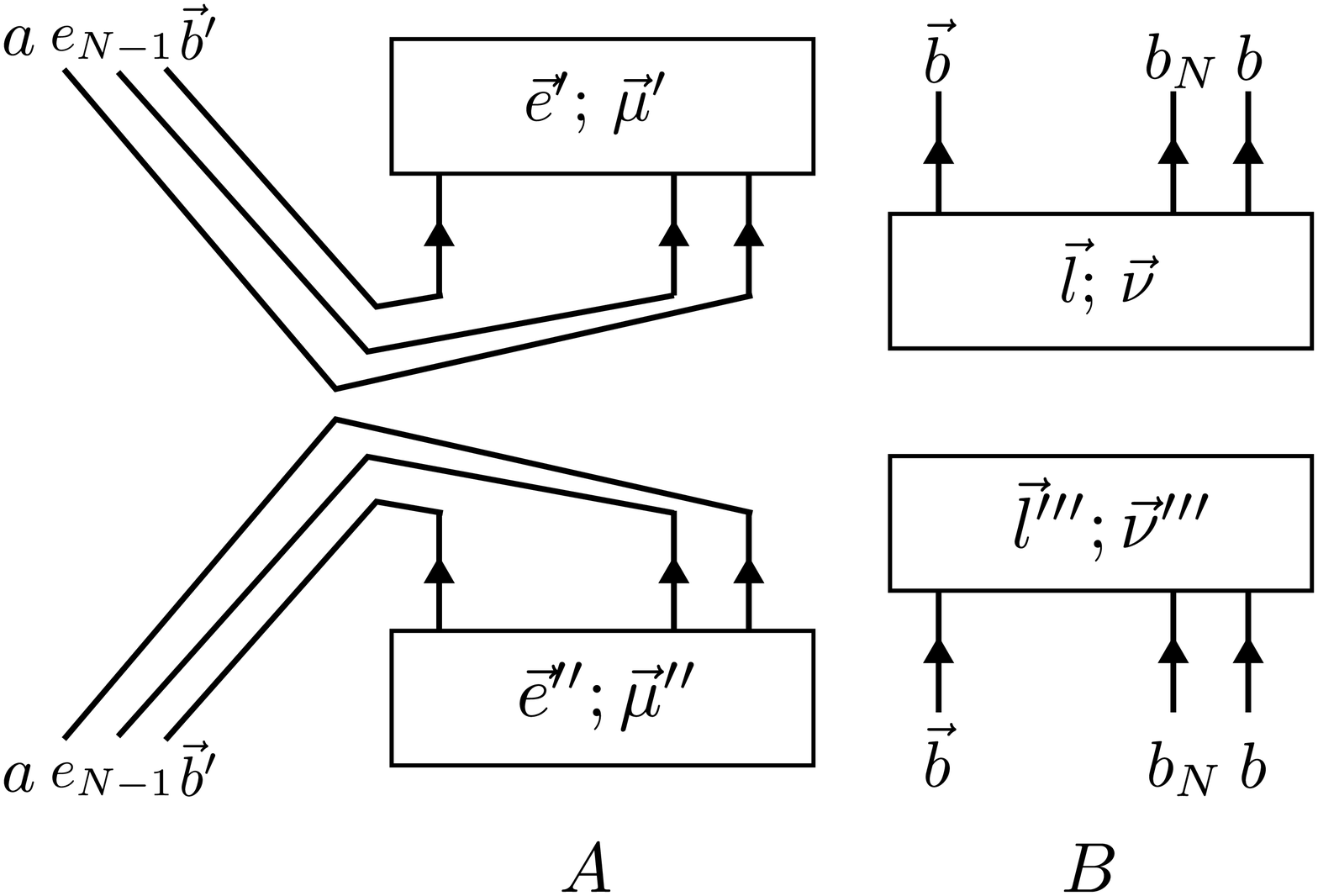}
\end{gathered}
\end{align}
where we have made use of the fact $(L_{\vec{l},\vec{e},\vec{b}; \vec{l}',\vec{e}',\vec{b}'}^{\vec{\mu},\vec{\nu}; \vec{\mu}',\vec{\nu}'})^* = L_{\vec{l}',\vec{e}',\vec{b}'; \vec{l},\vec{e},\vec{b}}^{\vec{\mu}',\vec{\nu}' ; \vec{\mu},\vec{\nu}}$. Note that while the density matrix is diagonal in the external anyon charge labels $\vec{b}$, $b_N$, and $e_{N-1}$, it is not diagonal in the internal charge lines $\vec{e}$ and $\vec{l}$. For convenience, we define,
\begin{align}
	H^{\vec{\mu}',\vec{\nu};\vec{\mu}''\vec{\nu}'''}_{\vec{e}',\vec{l};\vec{e}'',\vec{l}'''} \equiv \sum_{\substack{\vec{e}, \vec{\mu} \\ \vec{l}', \vec{\nu}' }} L_{\vec{e},\vec{l},\vec{b}; \vec{e}',\vec{l}',\vec{b}'}^{\vec{\mu},\vec{\nu}; \vec{\mu}',\vec{\nu}'} L_{\vec{e}'',\vec{l}',\vec{b}'; \vec{e},\vec{l}'',\vec{b}}^{\vec{\mu}'',\vec{\nu}'; \vec{\mu},\vec{\nu}'''}
\end{align}
To reduce clutter, we have left the dependence of $H$ on $\vec{b},\vec{b}',e_{N-1},\mu_{N}$ and $b_N$ implicit. Now, using unitarity of the $F$-symbols to evaluate sums such as,
\begin{align}
	\sum_{e_1,\mu_1,\mu_2} [F^{b_1 b_2}_{a e_2}]_{ \substack{(e_1 \mu_1 \mu_2) \\ (l_1 \nu_1 \tilde{\mu}_2) } } [F^{b_1 b_2}_{a e_2}]^*_{ \substack{(e_1 \mu_1 \mu_2) \\ (l'''_1 \nu'''_1 \tilde{\mu}'''_2) } } = \delta_{l_1,l'''_1} \delta_{\nu_1 , \nu'''_1} \delta_{\tilde{\mu}_2,\tilde{\mu}'''_2}
\end{align}
we find,
\begin{align}
		H^{\vec{\mu}',\vec{\nu};\vec{\mu}''\vec{\nu}'''}_{\vec{e}',\vec{l};\vec{e}'',\vec{l}'''} = \sum_{\mu_N, \mu'_N}  [O_{\mu_N,\mu_N'}]_{\vec{e}'\vec{\mu}';\vec{e}''\vec{\mu}''} [J_{\mu_N,\mu_N'}]_{\vec{l}\vec{\nu};\vec{l}'''\vec{\nu}'''}
\end{align}
where we have defined
\begin{widetext}
\begin{align}
	[O_{\mu_N,\mu_N'}]_{\vec{e}'\vec{\mu}';\vec{e}''\vec{\mu}''} = \sum_{\vec{l}' \vec{\nu}'} 
	& \sum_{\vec{\tilde{\mu}}'} [F^{b_1' b_2'}_{a e_2'}]^*_{ \substack{(e_1' \mu'_1 \mu'_2) \\ (l_1' \nu_1' \tilde{\mu}_2') } } \prod_{i=1}^{N-3} [F^{l_i' b_{i+2}'}_{a e_{i+2}'}]^*_{ \substack{(e_{i+1}' \tilde{\mu}_{i+1}' \mu_{i+2}') \\ (l_{i+1}' \nu_{i+1}' \tilde{\mu}_{i+2}') } } [F^{l_{N-2}' b_N}_{a c}]^*_{ \substack{(e_{N-1} \tilde{\mu}'_{N-1} \mu_N) \\ (b \, \nu'_{N-1} \delta) } } \\
	\notag \times & \sum_{\vec{\tilde{\mu}}''} [F^{b'_1 b'_2}_{a e''_2}]_{ \substack{(e''_1 \mu''_1 \mu''_2) \\ (l'_1 \nu'_1 \tilde{\mu}''_2) } } \prod_{i=1}^{N-3} [F^{l_i' b'_{i+2}}_{a e''_{i+2}}]_{ \substack{(e''_{i+1} \tilde{\mu}''_{i+1} \mu''_{i+2}) \\ (l'_{i+1} \nu'_{i+1} \tilde{\mu}''_{i+2}) } } [F^{l'_{N-2} b_N}_{a c}]_{ \substack{(e_{N-1} \tilde{\mu}''_{N-1} \mu'_N) \\ (b \, \nu'_{N-1} \delta) } } \, ,
\end{align}
and
\begin{align}
	[J_{\mu_N,\mu_N'}]_{\vec{l}\vec{\nu};\vec{l}'''\vec{\nu}'''} = \sum_{\tilde{\mu}_{N-1}} [F^{l_{N-2} b_N}_{a c}]_{ \substack{(e_{N-1} \tilde{\mu}_{N-1} \mu_N) \\ (b \, \nu_{N-1} \delta) } }  [F^{l_{N-2} b_N}_{a c}]^*_{ \substack{(e_{N-1} \tilde{\mu}_{N-1} \mu'_N) \\ (b \, \nu'''_{N-1} \delta) } } \prod_{i=1}^{N-2} \delta_{l_i,l_i'''}\delta_{\nu_i,\nu_i'''} \, .
\end{align}
We have left the dependence of $O$ and $J$ on $\vec{b},\vec{b}',e_{N-1}$ and $b_N$ implicit. Note that the matrix $J_{\mu_N,\mu_N'}$ depends only on anyon charges and fusion vertices at the trijunction. Proceeding with the replica trick, one finds for $n_e \in 2\mathbb{Z}$,
\begin{align}
	\Tr[(\trho_{AB}^{T_A}(\trho_{AB}^{T_A})^\dagger)^{n_e/2}] &= \sum_{ \substack{ \vec{b} \vec{b}'  \\ b_N e_{N-1} } }  \left( \frac{d_{b_N}^2  d_{\vec{b}} d_{\vec{b}'}}{d_b^2 \mathcal{D}^{4N-4}} \right)^{n_e/2}  \sum_{ \substack{   \vec{e}' , \vec{\mu}' \\ \vec{l} , \vec{\nu} } } [H^{n_e/2}]^{\vec{\mu}',\vec{\nu};\vec{\mu}'\vec{\nu}}_{\vec{e}',\vec{l};\vec{e}',\vec{l}}
\end{align}
Defining the matrix $E^{l_{N-2} l'_{N-2}}$ with elements
\begin{align}
	E^{l_{N-2} l'_{N-2}}_{\nu_{N-1} \tilde{\mu}'_{N-1} ; \nu'_{N-1} \tilde{\mu}''_{N-1}} = \sum_{\substack{ \mu_N^{(1)} \mu_N^{(2)} \\ \nu_{N-2}' \tilde{\mu}_{N-1}  }} & [F^{l_{N-2} b_N}_{a c}]_{ \substack{(e_{N-1} \tilde{\mu}_{N-1} \mu^{(1)}_N) \\ (b \, \nu_{N-1} \delta) } }  [F^{l'_{N-2} b_N}_{a c}]^*_{ \substack{(e_{N-1} \tilde{\mu}'_{N-1} \mu^{(1)}_N) \\ (b \, \nu'_{N-1} \delta) } } \\ 
	\notag & \times [F^{l'_{N-2} b_N}_{a c}]_{ \substack{(e_{N-1} \tilde{\mu}''_{N-1} \mu^{(2)}_N) \\ (b \, \nu'_{N-1} \delta) } }  [F^{l_{N-2} b_N}_{a c}]^*_{ \substack{(e_{N-1} \tilde{\mu}_{N-1} \mu^{(2)}_N) \\ (b \, \nu_{N-1} \delta) } },
\end{align}
%\end{widetext}
%we may write 
one finds, after a series of algebraic manipulations employing the unitarity of the $F$-symbols, that
\begin{align}
	&\sum_{ \substack{   \vec{e}' , \vec{\mu}' \\ \vec{l} , \vec{\nu} } } [H^{n_e/2}]^{\vec{\mu}',\vec{\nu};\vec{\mu}'\vec{\nu}}_{\vec{e}',\vec{l};\vec{e}',\vec{l}}  = \sum_{ \substack{l_1 , \dots , l_{N-2} \\ \nu_1, \dots , \nu_{N-2} } } \sum_{ \substack{l'_1 , \dots , l'_{N-2} \\ \nu'_1, \dots , \nu'_{N-2} } } \Tr[(E_{l_{N-2},l'_{N-2}})^{n_e/2}].
\end{align}
Altogether, 
\begin{align}
	\Tr[(\trho_{AB}^{T_A}(\trho_{AB}^{T_A})^\dagger)^{{n_e/2}}] &= \sum_{ \substack{ b_N e_{N-1} } }  \left( \frac{d_{b_N}^2 }{d_b^2 \mathcal{D}^{4N-4}} \right)^{n_e/2}  \sum_{ \substack{ \vec{b} \\ l_1 , \dots , l_{N-2} \\ \nu_1, \dots , \nu_{N-2} } } \sum_{ \substack{ \vec{b}' \\ l'_1 , \dots , l'_{N-2} \\ \nu'_1, \dots , \nu'_{N-2} } } (d_{\vec{b}} d_{\vec{b}}')^{n_e/2} \Tr[(E^{l_{N-2},l'_{N-2}})^{n_e/2}] \, .
\end{align}
\end{widetext}
Now, in order to extract the area law term, we use the definition Eq. \eqref{eq:K-matrix} to write
\begin{align}
	 \sum_{ \substack{ \vec{b} \\ l_1 , \dots , l_{N-2} \\ \nu_1, \dots , \nu_{N-2} } } d_{\vec{b}}^{n_e/2} %&= \sum_{ \substack{ \vec{b} \\ l_1 , \dots , l_{N-2} }} N_{b_1 b_2}^{l_1} N_{l_1 b_3}^{l_2} \dots N^{l_{N-2}}_{l_{N-3} b_{N-1}} d_{b_1}^{n_e/2} \dots d_{b_{N-1}}^{n_e/2} \\
	 %&= \sum_{ \substack{ b_0 \vec{b} \\ l_1 , \dots , l_{N-2} }} N_{0,b_1}^{b_0} N_{b_0 b_2}^{l_1} N_{l_1 b_3}^{l_2} \dots N^{l_{N-2}}_{l_{N-3} b_{N-1}} d_{b_1}^{n_e/2} \dots d_{b_{N-1}}^{n_e/2} \\
	 &= [K_{n_e/2}^{N-1}]_{0 l_{N-2}},
\end{align}
where we have used $N_{0 b_1}^{b_0} = \delta_{b_0,b_1}$. Again, as in the computation of the negativity in the absence of anyons, we may use the diagonal form of $K_{n_e/2}$, to write
\begin{align}
	[K_{n_e/2}^{N-1}]_{0 l_{N-1}} %&= \sum_\mu \kappa_{{n_e/2},\mu}^{N-1} [v_{{n_e/2},\mu}]_{0}[v_{{n_e/2},\mu}]_{l_{N-2}}^* \\
	%&= \kappa_{{n_e/2},0}^{N-1} (d_{l_{N-1}}/\mathcal{D}^2) +  \sum_{\mu \neq 0} \kappa_{{n_e/2},\mu}^{N-1} [v_{{n_e/2},\mu}]_{0}[v_{{n_e/2},\mu}]_{l_{N-2}}^* \\
	&=  \kappa_{{n_e/2},0}^{N-1} \frac{d_{l_{N-2}}}{\mathcal{D}^2}  e^{F(N-1,l_{N-2},{n_e/2})} 
\end{align}
In the thermodynamic limit $N = n + 2 \to \infty$, we have that $F(N-1,l_{N-2},{n_e/2}) \to 0$, and so
\begin{align}
	\notag \Tr[(\trho_{AB}^{T_A}(\trho_{AB}^{T_A})^\dagger)^{{n_e/2}}] &\to \sum_{ \substack{ b_N , e_{N-1} \\ l_{N-2} , l'_{N-2} } } \left( \frac{d_{b_N}^2 }{d_b^2 \mathcal{D}^{4N-4}} \right)^{n_e/2}  \kappa_{{n_e/2},0}^{2N-2} \\
	&\times\frac{d_{l_{N-2}} d_{l'_{N-2}}}{\mathcal{D}^4} \Tr[(E^{l_{N-2},l'_{N-2}})^{n_e/2}] \, .
\end{align}
It is not clear how to take the replica limit $n_e \to 1$, as it is not immediately apparent what the square root of the matrix $E$ is.
However, if we assume a multiplicity free theory, $E$ reduces to a scalar:
\begin{align}
	E^{l_{N-2} l'_{N-2}} = |[F^{l_{N-2} b_N}_{a c}]_{ e_{N-1} b } |^2 |[F^{l'_{N-2} b_N}_{a c}]_{ e_{N-1} b }|^2.
\end{align}
Taking the replica limit $n_e \to 1$, we find for the negativity in the thermodynamic limit,
\begin{align}
	&\mathcal{E}(A:B) = 2(n+1) \ln (\kappa_{1/2,0} / \mathcal{D}) - \ln \mathcal{D}^2 \\
	\notag &+ \ln \left[ \sum_{b_N e_{N-1}} \frac{d_{b_N}}{d_b \mathcal{D}^2 } \left(\sum_{l_{N-2} } d_{l_{N-2}} |[F^{l_{N-2} b_N}_{a c}]_{ e_{N-1} b } | \right)^2 \right],
\end{align}
recalling that the total number of wormholes $N = n+2$. We thus arrive at Eq. \eqref{eq:anyon-negativity} of the main text.

\subsection{Reflected Entropy Calculation Details \label{sec:app-details-reflected} }

We perform the double replica trick to compute the reflected entropy. 
After performing some tedious algebra using the unitarity of the $F$-symbols, one finds,
\begin{align}
	\begin{split}
	\sum_{\vec{b}' \vec{e}' \vec{\mu}' \vec{l}' \vec{\nu}'} & L_{\vec{l},\vec{e},\vec{b}; \vec{l}',\vec{e}',\vec{b}'}^{\vec{\mu},\vec{\nu}; \vec{\mu}',\vec{\nu}'} d_{\vec{b}'} L_{\vec{l}',\vec{e}',\vec{b}'; \vec{l}'',\vec{e}'',\vec{b}''}^{\vec{\mu}',\vec{\nu}'; \vec{\mu}'',\vec{\nu}''} \\
	&= \mathcal{D}^{2N-4} \frac{d_b d_{e_{N-1}}}{d_{c}} L_{\vec{l},\vec{e},\vec{b}; \vec{l}'',\vec{e}'',\vec{b}'}^{\vec{\mu},\vec{\nu}; \vec{\mu}'',\vec{\nu}''} \, .
	\end{split}
\end{align}
Making use of this, we find for $\alpha \in 2\mathbb{Z}$,
\begin{widetext}
\begin{align}
	(\rho_{AB})_{\mathrm{cut}}^{\frac{\alpha}{2}} &= \sum_{ \substack{ \vec{b} \vec{e} \vec{\mu} \\ \vec{b}' \vec{e}' \vec{\mu}' \\ b_N e_{N-1} } } \sum_{ \substack{ \vec{l} \vec{\nu} \\ \vec{l}' \vec{\nu}' } } \frac{1}{d_b d_c^{\frac{\alpha}{2}-1} \mathcal{D}^{2N-2} \mathcal{D}^{\alpha-2} }  \frac{(d_{b_N} d_{e_{N-1}})^{\frac{\alpha}{2}-1}}{ \sqrt{d_a d_b}} \sqrt{\frac{d_{b_N}}{d_{e_{N-1}}}}   (L^\frac{\alpha}{2}_{b_N e_{N-1}})_{\vec{l},\vec{e},\vec{b}; \vec{l}',\vec{e}',\vec{b}'}^{\vec{\mu},\vec{\nu}; \vec{\mu}',\vec{\nu}'} \begin{gathered}
\includegraphics[height=11em]{figures/tri-ins-boxes.eps}
\end{gathered}
\end{align}
Taking the canonical purification and tracing out $BB^*$, we obtain 
\begin{align}
	\rho_{AA^*}^{(\alpha)} &= \sum_{ \substack{\vec{b} \vec{e} \vec{\mu} \\ \vec{b}' \vec{e}' \vec{\mu}' \\ \vec{e}'' \vec{\mu}'' \\ \vec{e}''' \vec{\mu}''' \\ e_{N-1} e_{N-1}'' } } \left(\frac{1}{d_b \mathcal{D}^{2N-2} \mathcal{D}^{\alpha-2} d_c^{\frac{\alpha}{2}-1} }\right)^2 \frac{\sqrt{d_{\vec{b}} d_{\vec{b}'}}}{d_a \sqrt{d_{e_{N-1}} d_{e_{N-1}''}} }
	 [T_{\vec{b} \vec{b}'}]_{ \vec{e} \vec{e}' e_{N-1} ;  \vec{e}'' \vec{e}''' e_{N-1}''} \begin{gathered}
\includegraphics[height=14em]{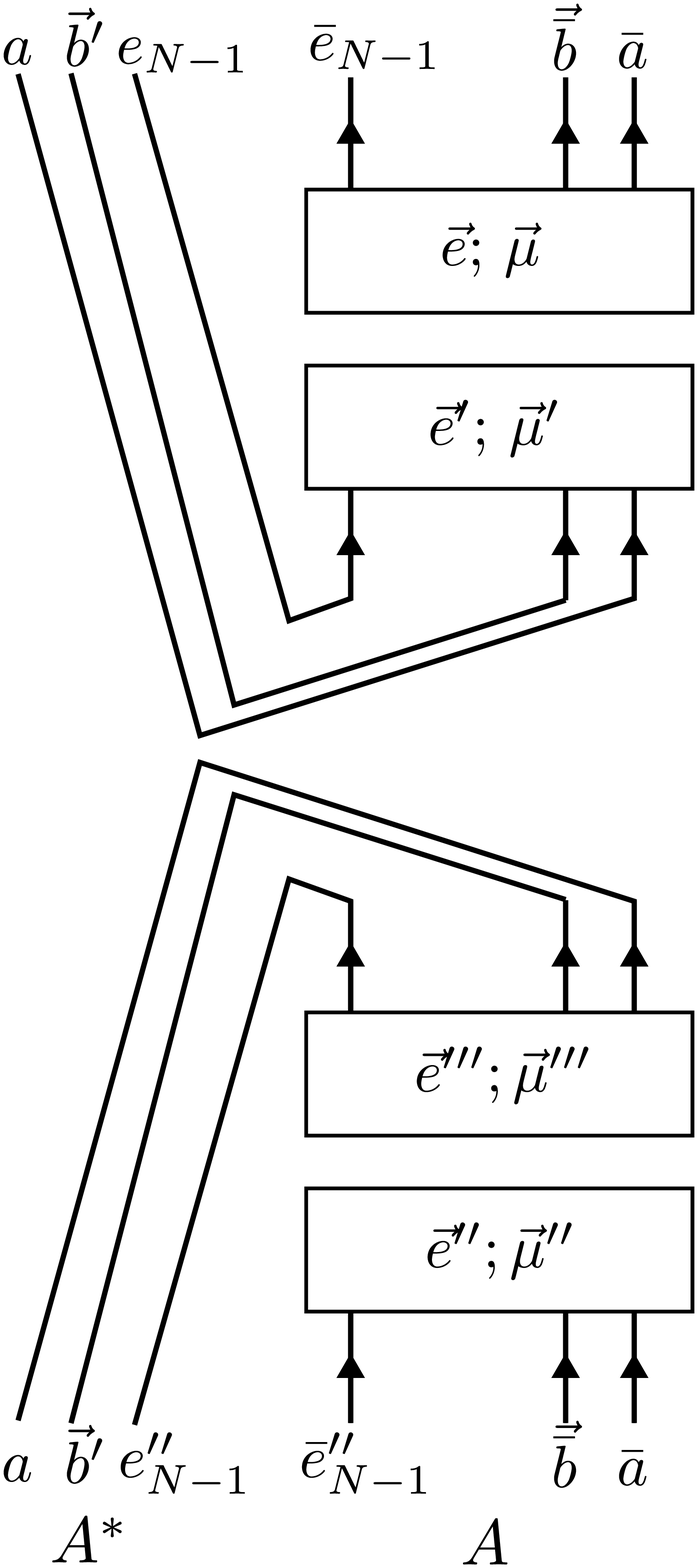}
\end{gathered}
\end{align}
where $T_{\vec{b} \vec{b}' ; \alpha}$ is a matrix with elements
\begin{align}
	[T_{\vec{b} \vec{b}' ; \alpha}]_{ \vec{e} \vec{e}' e_{N-1} ;  \vec{e}'' \vec{e}''' e_{N-1}''} = \sum_{ \substack{\vec{l} \vec{\nu} \\ \vec{l}' \vec{\nu}' \\ b_N} }  [L_{b_N e_{N-1}}]_{\vec{l},\vec{e},\vec{b}; \vec{l}',\vec{e}',\vec{b}'}^{\vec{\mu},\vec{\nu}; \vec{\mu}',\vec{\nu}'} ([L_{b_N e_{N-1}'' } ]_{\vec{l},\vec{e}'',\vec{b}; \vec{l}',\vec{e}''',\vec{b}'}^{\vec{\mu}'',\vec{\nu}; \vec{\mu}''',\vec{\nu}'})^* d_{b_N}^{\alpha} (d_{e_{N-1}} d_{e_{N-1}''})^{\frac{\alpha}{2}-1} \, .
\end{align}
We then find
\begin{align}
	\Tr[\rho_{AA^*}^{(\alpha)}]^\beta &= \sum_{ \vec{b} \vec{b}' } \left( \frac{d_{\vec{b}} d_{\vec{b}'} }{(d_b \mathcal{D}^{2N-2} \mathcal{D}^{\alpha-2})^2 d_c^{\frac{\alpha}{2}-1}} \right)^{\beta}  \Tr[T_{\vec{b} \vec{b}';\alpha}^\beta] \, .
\end{align}
Defining 
\begin{align}
\begin{split}
	C_{b_N b_N'' l_{N-2} l_{N-2}'}^{\nu_{N-1} \nu_{N-1}' \nu_{N-1}'' \nu_{N-1}''' } = \sum_{e''_{N-1}} d_{e''_{N-1}}^{\alpha-2} \sum_{\mu_N  \mu''_N  } \sum_{ \tilde{\mu}_{N-1} \tilde{\mu}_{N-1}'' } & [F^{l_{N-2} b_N}_{a c}]_{ \substack{(e''_{N-1} \tilde{\mu}_{N-1} \mu_N) \\ (b \, \nu_{N-1} \delta) } }^*  [F^{l_{N-2} b''_N}_{a c}]_{ \substack{(e''_{N-1} \tilde{\mu}_{N-1} \mu''_N) \\ (b \, \nu''_{N-1} \delta) } } \\
	\times 	 &  [F^{l'_{N-2} b_N}_{a c}]_{ \substack{(e''_{N-1} \tilde{\mu}''_{N-1} \mu_N) \\ (b \, \nu'_{N-1} \delta) } }  [F^{l_{N-2}' b''_N}_{a c}]^*_{ \substack{(e''_{N-1} \tilde{\mu}''_{N-1} \mu''_N) \\ (b \, \nu'''_{N-1} \delta) } },
\end{split}
\end{align}
one finds
\begin{align}
\begin{split}
	[T^2_{\vec{b} \vec{b}';\alpha}]_{ \vec{e} \vec{e}' e_{N-1} ;  \vec{e}'''' \vec{e}''''' e_{N-1}''''} =  \sum_{ \substack{\vec{l} \vec{\nu}  \vec{l}' \vec{\nu}' \\ b_N b_N'' \\ \nu_{N-1} \nu_{N-1}' \\  \nu_{N-1}'' \nu_{N-1}''' } } & [L_{b_N e_{N-1}}]_{\vec{l},\vec{e},\vec{b}; \vec{l}',\vec{e}',\vec{b}'}^{\vec{\mu},\vec{\nu} \nu_{N-1}; \vec{\mu}',\vec{\nu}' \nu_{N-1}'}  C_{b_N b_N'' l_{N-2} l_{N-2}'}^{\nu_{N-1} \nu_{N-1}' \nu_{N-1}'' \nu_{N-1}''' }  \\
		\times &  ([L_{b''_N e_{N-1}'''' } ]_{\vec{l},\vec{e}'''',\vec{b}; \vec{l}',\vec{e}''''',\vec{b}'}^{\vec{\mu}'''',\vec{\nu} \nu_{N-1}''; \vec{\mu}''''',\vec{\nu}' \nu_{N-1}'''})^* d_{b_N}^{\alpha} d_{b_N''}^{2} (d_{e_{N-1}} d_{e_{N-1}''''})^{\frac{\alpha}{2}-1}
\end{split}
\end{align}
where we have split the sum over $\vec{\nu} = (\nu_1 \dots \nu_{N-1})$ into a sum over $\vec{\nu} = (\nu_1 \dots \nu_{N-2})$ and $\nu_{N-1}$. Thus, defining the matrix $\tilde{C}_{l_{N-2} l_{N-2}';\alpha}$ with elements
\begin{align}
	\left[\tilde{C}_{l_{N-2} l_{N-2}';\alpha}\right]_{b_N ; b_N''}^{\nu_{N-1} \nu_{N-1}' ; \nu_{N-1}'' \nu_{N-1}''' }  \equiv d_{b_N}^{\frac{\alpha}{2}} C_{b_N b_N'' l_{N-2} l_{N-2}'}^{\nu_{N-1} \nu_{N-1}' \nu_{N-1}'' \nu_{N-1}''' }  d_{b''_N}^{\frac{\alpha}{2}}
\end{align}
we have
\begin{align}
	\Tr[T_{\vec{b} \vec{b}';\alpha}^\beta] &= \sum_{ \substack{\vec{l} \vec{\nu} \\ \vec{l}' \vec{\nu}'}} \Tr\left[ \tilde{C}_{l_{N-2} l_{N-2}';\alpha}^\beta \right] \, .
\end{align}
Hence,
\begin{align}
	\Tr [\rho_{AA^*}^{(\alpha)}]^\beta =  \sum_{ \vec{b} \vec{b}' } \left( \frac{d_{\vec{b}} d_{\vec{b}'} }{(d_b \mathcal{D}^{2N-2} \mathcal{D}^{\alpha-2} d_c^{\frac{\alpha}{2}-1})^2} \right)^{\beta}  \sum_{ \substack{\vec{l} \vec{\nu} \\ \vec{l}' \vec{\nu}'}} \Tr\left[ \tilde{C}^\beta_{l_{N-2} l_{N-2}' ; \alpha} \right]\, .
\end{align}
We first take the limit $\alpha \to 1$:
\begin{align}
	\Tr [\rho_{AA^*}^{(1)}]^\beta =  \sum_{ \vec{b} \vec{b}' } \left( \frac{d_c d_{\vec{b}} d_{\vec{b}'} }{(d_b \mathcal{D}^{2N-3} )^2} \right)^{\beta}  \sum_{ \substack{\vec{l} \vec{\nu} \\ \vec{l}' \vec{\nu}'}} \Tr\left[ \tilde{C}^\beta_{l_{N-2} l_{N-2}' ; 1} \right]\, .
\end{align}
Using the definition of Eq. \eqref{eq:K-matrix}, we may write
\begin{align}
	 \sum_{ \substack{ \vec{b} \\ l_1 , \dots , l_{N-2} \\ \nu_1, \dots , \nu_{N-2} } } d_{\vec{b}}^\beta  &= [K_\beta^{N-1}]_{0 l_{N-2}} = \kappa_{\beta,0}^{N-1} \frac{d_{l_{N-2}}}{\mathcal{D}^2}  e^{F(N-1,l_{N-2},\beta)} 
\end{align}
So, in the limit $N = n+2 \to \infty$, we find
\begin{align}
	S_R^{(\beta)}(A:B) &= \frac{2N-2}{1-\beta} \ln\left( \frac{\kappa_{\beta,0}}{\mathcal{D}^{2\beta}} \right)  - \ln \mathcal{D}^2 + \frac{1}{1-\beta} \ln \left( \sum_{ l_{N-2} l'_{N-2} } \frac{d_c^\beta d_{l_{N-2}} d_{l'_{N-2}}}{d_b^{2\beta} \mathcal{D}^2}  \Tr\left[ \tilde{C}^\beta_{l_{N-2} l_{N-2}' ; 1/2} \right]\right)
\end{align}
The first term is the area law term, the second the usual TEE, and the last a new contribution from the anyon insertions.
For a multiplicity free theory, we have that
\begin{align}
	C_{b_N b_N'' l_{N-2} l_{N-2}';\alpha} = \sum_{e''_{N-1}} d_{e''_{N-1}}^{\alpha-2}  & [F^{l_{N-2} b_N}_{a c}]_{ \substack{e''_{N-1}  b  } }^* [F^{l_{N-2} b''_N}_{a c}]_{ e''_{N-1}  b  }  [F^{l'_{N-2} b_N}_{a c}]_{  e''_{N-1}  b }  [F^{l_{N-2}' b''_N}_{a c}]^*_{ e''_{N-1}  b } \, ,
\end{align}
and so
\begin{align}
	  \left[\tilde{C}_{l_{N-2} l_{N-2}' ; 1}\right]_{b_N,b_N''} %&= d_{b_N}^{1/2} C_{b_N b_N'' l_{N-2} l_{N-2}'; 1/2} d_{b''_N}^{1/2} \\
	  %&= d_{b_N}^{1/2} d_{b''_N}^{1/2} \sum_{e''_{N-1}} d_{e''_{N-1}}^{-1}  [F^{l_{N-2} b_N}_{a c}]_{ \substack{e''_{N-1}  b  } }^* [F^{l_{N-2} b''_N}_{a c}]_{ e''_{N-1}  b  }  [F^{l'_{N-2} b_N}_{a c}]_{  e''_{N-1}  b }  [F^{l_{N-2}' b''_N}_{a c}]^*_{ e''_{N-1}  b } \\
	  &= d_{b_N}^{1/2} d_{b''_N}^{1/2} \sum_{e''_{N-1}} \frac{d_{e''_{N-1}} d_b^2}{d_{l_{N-2}} d_{l'_{N-2}} d_c^2}  [F^{a e''_{N-1} b_N}_b]_{l_{N-2}c} [F^{a e''_{N-1} b''_N}_b]_{l_{N-2}c}^*  [F^{a e''_{N-1} b_N}_b]_{l'_{N-2}c}^* [F^{a e''_{N-1} b''_N}_b]_{l'_{N-2}c}
\end{align}
Note that in the expression in the main text [Eq. \eqref{eq:anyon-renyi-markov}], we have pulled out the factor of $d_b^2/d_c^2$ from the matrix $\tilde{C}$, defining $\tilde{C}_{l_{N-2} l_{N-2}'} = (d_c/d_b)^2 \tilde{C}_{l_{N-2} l_{N-2}' ; 1}$.

\end{widetext}

\bibliography{references}

\end{document}